\documentclass[longauth]{aaEC}

\usepackage{graphicx}
\usepackage{natbib}
\usepackage{scalerel}

\usepackage[table]{xcolor}

\usepackage[varg]{txfonts}
\usepackage[pdfencoding=auto,psdextra]{hyperref}
\hypersetup{
    colorlinks=true,
    linkcolor=blue,
    filecolor=magenta,      
    urlcolor=blue,
    citecolor=blue
}
\makeatletter
\renewcommand*\aa@pageof{, page \thepage{} of \pageref*{LastPage}}
\makeatother

\usepackage[utf8]{inputenc}

\usepackage[switch, modulo]{lineno}
\usepackage{orcidlink} %this has to come after hyperref
\newcommand{\orcid}[1]{\orcidlink{#1}}

\usepackage{euclid}

\usepackage{color}

\usepackage{amsmath}	% Advanced maths commands
\usepackage{amssymb}	% Extra maths symbols
\usepackage{mathtools}   % FJC: extra math
\usepackage{upgreek}
\renewcommand{\pi}{\uppi}
\definecolor{darkgreen}{rgb}{0.0,0.5,0.0}
\definecolor{darkred}{rgb}{0.5,0.0,0.0}
\definecolor{brown}{rgb}{0.65,.16,0.16}
\definecolor{grey}{rgb}{0.4,0.5,0.6}

\newcommand{\msun}{\,M_\odot}
\newcommand{\sersic}{\mathrm{S\Acute{e}rsic}}
\newcommand{\beq}{\begin{equation}}
\newcommand{\eeq}{\end{equation}}

\begin{document}
%
% Put the title of your paper here:
%
   \title{\Euclid\/. V. The Flagship galaxy mock catalogue: \\ a comprehensive simulation for the \Euclid\/ mission}

%\title{Euclid}
%\subtitle{V. The Flagship galaxy mock catalogue: a comprehensive simulation for the Euclid mission}    

%% please do not edit the author list -- contact ECEB Bureau for changes

%%%% please do not edit the author list -- contact ECEB Bureau for changes
%\newcommand{\orcid}[1]{} %% if already defined in aa.cls: comment, or use renewcommand			   
\author{Euclid Collaboration: F.~J.~Castander\orcid{0000-0001-7316-4573}\thanks{\email{fjc@ice.csic.es}}\inst{\ref{aff1},\ref{aff2}}
\and P.~Fosalba\orcid{0000-0002-1510-5214}\inst{\ref{aff1},\ref{aff2}}
\and J.~Stadel\orcid{0000-0001-7565-8622}\inst{\ref{aff3}}
\and D.~Potter\orcid{0000-0002-0757-5195}\inst{\ref{aff3}}
\and J.~Carretero\orcid{0000-0002-3130-0204}\inst{\ref{aff4},\ref{aff5}}
\and P.~Tallada-Cresp\'{i}\orcid{0000-0002-1336-8328}\inst{\ref{aff4},\ref{aff5}}
\and L.~Pozzetti\orcid{0000-0001-7085-0412}\inst{\ref{aff6}}
\and M.~Bolzonella\orcid{0000-0003-3278-4607}\inst{\ref{aff6}}
\and G.~A.~Mamon\orcid{0000-0001-8956-5953}\inst{\ref{aff7},\ref{aff8}}
\and L.~Blot\orcid{0000-0002-9622-7167}\inst{\ref{aff9},\ref{aff10}}
\and K.~Hoffmann\inst{\ref{aff1}}
\and M.~Huertas-Company\orcid{0000-0002-1416-8483}\inst{\ref{aff11},\ref{aff12},\ref{aff13},\ref{aff14}}
\and P.~Monaco\orcid{0000-0003-2083-7564}\inst{\ref{aff15},\ref{aff16},\ref{aff17},\ref{aff18}}
\and E.~J.~Gonzalez\orcid{0000-0002-0226-9893}\inst{\ref{aff19},\ref{aff5},\ref{aff20}}
\and G.~De~Lucia\orcid{0000-0002-6220-9104}\inst{\ref{aff16}}
\and C.~Scarlata\orcid{0000-0002-9136-8876}\inst{\ref{aff21}}
\and M.-A.~Breton\inst{\ref{aff1},\ref{aff22},\ref{aff10}}
\and L.~Linke\orcid{0000-0002-2622-8113}\inst{\ref{aff23}}
\and C.~Viglione\inst{\ref{aff2},\ref{aff1}}
\and S.-S.~Li\orcid{0000-0001-9952-7408}\inst{\ref{aff24}}
\and Z.~Zhai\inst{\ref{aff25},\ref{aff26},\ref{aff27},\ref{aff28}}
\and Z.~Baghkhani\orcid{0000-0002-6632-2614}\inst{\ref{aff1}}
\and K.~Pardede\orcid{0000-0002-7728-8220}\inst{\ref{aff29},\ref{aff30},\ref{aff31},\ref{aff18}}
\and C.~Neissner\orcid{0000-0001-8524-4968}\inst{\ref{aff19},\ref{aff5}}
\and R.~Teyssier\orcid{0000-0001-7689-0933}\inst{\ref{aff32}}
\and M.~Crocce\orcid{0000-0002-9745-6228}\inst{\ref{aff1},\ref{aff2}}
\and I.~Tutusaus\orcid{0000-0002-3199-0399}\inst{\ref{aff33}}
\and L.~Miller\orcid{0000-0002-3376-6200}\inst{\ref{aff34}}
\and G.~Congedo\orcid{0000-0003-2508-0046}\inst{\ref{aff35}}
\and A.~Biviano\orcid{0000-0002-0857-0732}\inst{\ref{aff16},\ref{aff18}}
\and M.~Hirschmann\orcid{0000-0002-3301-3321}\inst{\ref{aff36},\ref{aff16}}
\and A.~Pezzotta\orcid{0000-0003-0726-2268}\inst{\ref{aff37}}
\and H.~Aussel\orcid{0000-0002-1371-5705}\inst{\ref{aff38}}
\and H.~Hoekstra\orcid{0000-0002-0641-3231}\inst{\ref{aff24}}
\and T.~Kitching\orcid{0000-0002-4061-4598}\inst{\ref{aff39}}
\and W.~J.~Percival\orcid{0000-0002-0644-5727}\inst{\ref{aff27},\ref{aff28},\ref{aff40}}
\and L.~Guzzo\orcid{0000-0001-8264-5192}\inst{\ref{aff41},\ref{aff42}}
\and Y.~Mellier\inst{\ref{aff8},\ref{aff7}}
\and P.~A.~Oesch\orcid{0000-0001-5851-6649}\inst{\ref{aff43},\ref{aff44},\ref{aff45}}
\and R.~A.~A.~Bowler\orcid{0000-0003-3917-1678}\inst{\ref{aff46}}
\and S.~Bruton\orcid{0000-0002-6503-5218}\inst{\ref{aff21}}
\and V.~Allevato\orcid{0000-0001-7232-5152}\inst{\ref{aff47}}
\and V.~Gonzalez-Perez\orcid{0000-0001-9938-2755}\inst{\ref{aff48}}
\and M.~Manera\orcid{0000-0003-4962-8934}\inst{\ref{aff49},\ref{aff19}}
\and S.~Avila\orcid{0000-0001-5043-3662}\inst{\ref{aff19}}
\and A.~Kov\'acs\orcid{0000-0002-5825-579X}\inst{\ref{aff50},\ref{aff51}}
\and N.~Aghanim\orcid{0000-0002-6688-8992}\inst{\ref{aff52}}
\and B.~Altieri\orcid{0000-0003-3936-0284}\inst{\ref{aff53}}
\and A.~Amara\inst{\ref{aff54}}
\and L.~Amendola\orcid{0000-0002-0835-233X}\inst{\ref{aff55}}
\and S.~Andreon\orcid{0000-0002-2041-8784}\inst{\ref{aff42}}
\and N.~Auricchio\orcid{0000-0003-4444-8651}\inst{\ref{aff6}}
\and M.~Baldi\orcid{0000-0003-4145-1943}\inst{\ref{aff56},\ref{aff6},\ref{aff57}}
\and A.~Balestra\orcid{0000-0002-6967-261X}\inst{\ref{aff58}}
\and S.~Bardelli\orcid{0000-0002-8900-0298}\inst{\ref{aff6}}
\and R.~Bender\orcid{0000-0001-7179-0626}\inst{\ref{aff37},\ref{aff59}}
\and C.~Bodendorf\inst{\ref{aff37}}
\and D.~Bonino\orcid{0000-0002-3336-9977}\inst{\ref{aff60}}
\and E.~Branchini\orcid{0000-0002-0808-6908}\inst{\ref{aff61},\ref{aff62},\ref{aff42}}
\and M.~Brescia\orcid{0000-0001-9506-5680}\inst{\ref{aff63},\ref{aff47},\ref{aff64}}
\and J.~Brinchmann\orcid{0000-0003-4359-8797}\inst{\ref{aff65},\ref{aff66}}
\and S.~Camera\orcid{0000-0003-3399-3574}\inst{\ref{aff67},\ref{aff68},\ref{aff60}}
\and V.~Capobianco\orcid{0000-0002-3309-7692}\inst{\ref{aff60}}
\and C.~Carbone\orcid{0000-0003-0125-3563}\inst{\ref{aff69}}
\and S.~Casas\orcid{0000-0002-4751-5138}\inst{\ref{aff70}}
\and M.~Castellano\orcid{0000-0001-9875-8263}\inst{\ref{aff71}}
\and S.~Cavuoti\orcid{0000-0002-3787-4196}\inst{\ref{aff47},\ref{aff64}}
\and A.~Cimatti\inst{\ref{aff72}}
\and C.~J.~Conselice\orcid{0000-0003-1949-7638}\inst{\ref{aff46}}
\and L.~Conversi\orcid{0000-0002-6710-8476}\inst{\ref{aff73},\ref{aff53}}
\and Y.~Copin\orcid{0000-0002-5317-7518}\inst{\ref{aff74}}
\and L.~Corcione\orcid{0000-0002-6497-5881}\inst{\ref{aff60}}
\and F.~Courbin\orcid{0000-0003-0758-6510}\inst{\ref{aff75}}
\and H.~M.~Courtois\orcid{0000-0003-0509-1776}\inst{\ref{aff76}}
\and A.~Da~Silva\orcid{0000-0002-6385-1609}\inst{\ref{aff77},\ref{aff78}}
\and H.~Degaudenzi\orcid{0000-0002-5887-6799}\inst{\ref{aff43}}
\and A.~M.~Di~Giorgio\orcid{0000-0002-4767-2360}\inst{\ref{aff79}}
\and J.~Dinis\orcid{0000-0001-5075-1601}\inst{\ref{aff77},\ref{aff78}}
\and M.~Douspis\orcid{0000-0003-4203-3954}\inst{\ref{aff52}}
\and F.~Dubath\orcid{0000-0002-6533-2810}\inst{\ref{aff43}}
\and C.~A.~J.~Duncan\inst{\ref{aff46}}
\and X.~Dupac\inst{\ref{aff53}}
\and S.~Dusini\orcid{0000-0002-1128-0664}\inst{\ref{aff80}}
\and A.~Ealet\orcid{0000-0003-3070-014X}\inst{\ref{aff74}}
\and M.~Farina\orcid{0000-0002-3089-7846}\inst{\ref{aff79}}
\and S.~Farrens\orcid{0000-0002-9594-9387}\inst{\ref{aff38}}
\and S.~Ferriol\inst{\ref{aff74}}
\and S.~Fotopoulou\orcid{0000-0002-9686-254X}\inst{\ref{aff81}}
\and N.~Fourmanoit\inst{\ref{aff82}}
\and M.~Frailis\orcid{0000-0002-7400-2135}\inst{\ref{aff16}}
\and E.~Franceschi\orcid{0000-0002-0585-6591}\inst{\ref{aff6}}
\and P.~Franzetti\inst{\ref{aff69}}
\and S.~Galeotta\orcid{0000-0002-3748-5115}\inst{\ref{aff16}}
\and W.~Gillard\orcid{0000-0003-4744-9748}\inst{\ref{aff82}}
\and B.~Gillis\orcid{0000-0002-4478-1270}\inst{\ref{aff35}}
\and C.~Giocoli\orcid{0000-0002-9590-7961}\inst{\ref{aff6},\ref{aff83}}
\and P.~G\'omez-Alvarez\orcid{0000-0002-8594-5358}\inst{\ref{aff84},\ref{aff53}}
\and B.~R.~Granett\orcid{0000-0003-2694-9284}\inst{\ref{aff42}}
\and A.~Grazian\orcid{0000-0002-5688-0663}\inst{\ref{aff58}}
\and F.~Grupp\inst{\ref{aff37},\ref{aff59}}
\and S.~V.~H.~Haugan\orcid{0000-0001-9648-7260}\inst{\ref{aff85}}
\and M.~S.~Holliman\inst{\ref{aff35}}
\and W.~Holmes\inst{\ref{aff86}}
\and I.~Hook\orcid{0000-0002-2960-978X}\inst{\ref{aff87}}
\and F.~Hormuth\inst{\ref{aff88}}
\and A.~Hornstrup\orcid{0000-0002-3363-0936}\inst{\ref{aff89},\ref{aff90}}
\and P.~Hudelot\inst{\ref{aff7}}
\and K.~Jahnke\orcid{0000-0003-3804-2137}\inst{\ref{aff91}}
\and M.~Jhabvala\inst{\ref{aff92}}
\and B.~Joachimi\orcid{0000-0001-7494-1303}\inst{\ref{aff93}}
\and E.~Keih\"anen\orcid{0000-0003-1804-7715}\inst{\ref{aff94}}
\and S.~Kermiche\orcid{0000-0002-0302-5735}\inst{\ref{aff82}}
\and A.~Kiessling\orcid{0000-0002-2590-1273}\inst{\ref{aff86}}
\and M.~Kilbinger\orcid{0000-0001-9513-7138}\inst{\ref{aff38}}
\and R.~Kohley\inst{\ref{aff53}}
\and B.~Kubik\orcid{0009-0006-5823-4880}\inst{\ref{aff74}}
\and M.~K\"ummel\orcid{0000-0003-2791-2117}\inst{\ref{aff59}}
\and M.~Kunz\orcid{0000-0002-3052-7394}\inst{\ref{aff95}}
\and H.~Kurki-Suonio\orcid{0000-0002-4618-3063}\inst{\ref{aff96},\ref{aff97}}
\and O.~Lahav\orcid{0000-0002-1134-9035}\inst{\ref{aff93}}
\and R.~Laureijs\inst{\ref{aff98}}
\and D.~Le~Mignant\orcid{0000-0002-5339-5515}\inst{\ref{aff99}}
\and S.~Ligori\orcid{0000-0003-4172-4606}\inst{\ref{aff60}}
\and P.~B.~Lilje\orcid{0000-0003-4324-7794}\inst{\ref{aff85}}
\and V.~Lindholm\orcid{0000-0003-2317-5471}\inst{\ref{aff96},\ref{aff97}}
\and I.~Lloro\inst{\ref{aff100}}
\and D.~Maino\inst{\ref{aff41},\ref{aff69},\ref{aff101}}
\and E.~Maiorano\orcid{0000-0003-2593-4355}\inst{\ref{aff6}}
\and O.~Mansutti\orcid{0000-0001-5758-4658}\inst{\ref{aff16}}
\and O.~Marggraf\orcid{0000-0001-7242-3852}\inst{\ref{aff102}}
\and K.~Markovic\orcid{0000-0001-6764-073X}\inst{\ref{aff86}}
\and N.~Martinet\orcid{0000-0003-2786-7790}\inst{\ref{aff99}}
\and F.~Marulli\orcid{0000-0002-8850-0303}\inst{\ref{aff103},\ref{aff6},\ref{aff57}}
\and R.~Massey\orcid{0000-0002-6085-3780}\inst{\ref{aff104}}
\and D.~C.~Masters\orcid{0000-0001-5382-6138}\inst{\ref{aff105}}
\and S.~Maurogordato\inst{\ref{aff106}}
\and H.~J.~McCracken\orcid{0000-0002-9489-7765}\inst{\ref{aff7}}
\and E.~Medinaceli\orcid{0000-0002-4040-7783}\inst{\ref{aff6}}
\and S.~Mei\orcid{0000-0002-2849-559X}\inst{\ref{aff107}}
\and M.~Melchior\inst{\ref{aff108}}
\and M.~Meneghetti\orcid{0000-0003-1225-7084}\inst{\ref{aff6},\ref{aff57}}
\and E.~Merlin\orcid{0000-0001-6870-8900}\inst{\ref{aff71}}
\and G.~Meylan\inst{\ref{aff75}}
\and J.~J.~Mohr\orcid{0000-0002-6875-2087}\inst{\ref{aff59},\ref{aff37}}
\and M.~Moresco\orcid{0000-0002-7616-7136}\inst{\ref{aff103},\ref{aff6}}
\and L.~Moscardini\orcid{0000-0002-3473-6716}\inst{\ref{aff103},\ref{aff6},\ref{aff57}}
\and E.~Munari\orcid{0000-0002-1751-5946}\inst{\ref{aff16},\ref{aff18}}
\and R.~Nakajima\inst{\ref{aff102}}
\and R.~C.~Nichol\orcid{0000-0003-0939-6518}\inst{\ref{aff54}}
\and S.-M.~Niemi\inst{\ref{aff98}}
\and C.~Padilla\orcid{0000-0001-7951-0166}\inst{\ref{aff19}}
\and K.~Paech\orcid{0000-0003-0625-2367}\inst{\ref{aff37}}
\and S.~Paltani\orcid{0000-0002-8108-9179}\inst{\ref{aff43}}
\and F.~Pasian\orcid{0000-0002-4869-3227}\inst{\ref{aff16}}
\and J.~A.~Peacock\orcid{0000-0002-1168-8299}\inst{\ref{aff35}}
\and K.~Pedersen\inst{\ref{aff109}}
\and V.~Pettorino\inst{\ref{aff98}}
\and S.~Pires\orcid{0000-0002-0249-2104}\inst{\ref{aff38}}
\and G.~Polenta\orcid{0000-0003-4067-9196}\inst{\ref{aff110}}
\and M.~Poncet\inst{\ref{aff111}}
\and L.~A.~Popa\inst{\ref{aff112}}
\and F.~Raison\orcid{0000-0002-7819-6918}\inst{\ref{aff37}}
\and R.~Rebolo\inst{\ref{aff11},\ref{aff113}}
\and A.~Renzi\orcid{0000-0001-9856-1970}\inst{\ref{aff114},\ref{aff80}}
\and J.~Rhodes\orcid{0000-0002-4485-8549}\inst{\ref{aff86}}
\and G.~Riccio\inst{\ref{aff47}}
\and E.~Romelli\orcid{0000-0003-3069-9222}\inst{\ref{aff16}}
\and M.~Roncarelli\orcid{0000-0001-9587-7822}\inst{\ref{aff6}}
\and C.~Rosset\orcid{0000-0003-0286-2192}\inst{\ref{aff107}}
\and E.~Rossetti\orcid{0000-0003-0238-4047}\inst{\ref{aff56}}
\and R.~Saglia\orcid{0000-0003-0378-7032}\inst{\ref{aff59},\ref{aff37}}
\and D.~Sapone\orcid{0000-0001-7089-4503}\inst{\ref{aff115}}
\and M.~Schirmer\orcid{0000-0003-2568-9994}\inst{\ref{aff91}}
\and P.~Schneider\orcid{0000-0001-8561-2679}\inst{\ref{aff102}}
\and T.~Schrabback\orcid{0000-0002-6987-7834}\inst{\ref{aff23}}
\and M.~Scodeggio\inst{\ref{aff69}}
\and A.~Secroun\orcid{0000-0003-0505-3710}\inst{\ref{aff82}}
\and G.~Seidel\orcid{0000-0003-2907-353X}\inst{\ref{aff91}}
\and S.~Serrano\orcid{0000-0002-0211-2861}\inst{\ref{aff2},\ref{aff116},\ref{aff1}}
\and C.~Sirignano\orcid{0000-0002-0995-7146}\inst{\ref{aff114},\ref{aff80}}
\and G.~Sirri\orcid{0000-0003-2626-2853}\inst{\ref{aff57}}
\and L.~Stanco\orcid{0000-0002-9706-5104}\inst{\ref{aff80}}
\and J.-L.~Starck\orcid{0000-0003-2177-7794}\inst{\ref{aff38}}
\and A.~N.~Taylor\inst{\ref{aff35}}
\and H.~I.~Teplitz\orcid{0000-0002-7064-5424}\inst{\ref{aff105}}
\and I.~Tereno\inst{\ref{aff77},\ref{aff117}}
\and R.~Toledo-Moreo\orcid{0000-0002-2997-4859}\inst{\ref{aff118}}
\and F.~Torradeflot\orcid{0000-0003-1160-1517}\inst{\ref{aff5},\ref{aff4}}
\and A.~Tsyganov\inst{\ref{aff119}}
\and L.~Valenziano\orcid{0000-0002-1170-0104}\inst{\ref{aff6},\ref{aff120}}
\and T.~Vassallo\orcid{0000-0001-6512-6358}\inst{\ref{aff59},\ref{aff16}}
\and A.~Veropalumbo\orcid{0000-0003-2387-1194}\inst{\ref{aff42},\ref{aff62},\ref{aff121}}
\and Y.~Wang\orcid{0000-0002-4749-2984}\inst{\ref{aff105}}
\and J.~Weller\orcid{0000-0002-8282-2010}\inst{\ref{aff59},\ref{aff37}}
\and A.~Zacchei\orcid{0000-0003-0396-1192}\inst{\ref{aff16},\ref{aff18}}
\and G.~Zamorani\orcid{0000-0002-2318-301X}\inst{\ref{aff6}}
\and F.~M.~Zerbi\inst{\ref{aff42}}
\and J.~Zoubian\inst{\ref{aff82}}
\and E.~Zucca\orcid{0000-0002-5845-8132}\inst{\ref{aff6}}
\and C.~Baccigalupi\orcid{0000-0002-8211-1630}\inst{\ref{aff18},\ref{aff16},\ref{aff17},\ref{aff30}}
\and F.~Bernardeau\inst{\ref{aff122},\ref{aff7}}
\and A.~Boucaud\orcid{0000-0001-7387-2633}\inst{\ref{aff107}}
\and E.~Bozzo\orcid{0000-0002-8201-1525}\inst{\ref{aff43}}
\and C.~Burigana\orcid{0000-0002-3005-5796}\inst{\ref{aff123},\ref{aff120}}
\and M.~Calabrese\orcid{0000-0002-2637-2422}\inst{\ref{aff124},\ref{aff69}}
\and P.~Casenove\inst{\ref{aff111}}
\and G.~Castignani\orcid{0000-0001-6831-0687}\inst{\ref{aff6}}
\and C.~Colodro-Conde\inst{\ref{aff11}}
\and D.~Di~Ferdinando\inst{\ref{aff57}}
\and J.~A.~Escartin~Vigo\inst{\ref{aff37}}
\and G.~Fabbian\orcid{0000-0002-3255-4695}\inst{\ref{aff125},\ref{aff126},\ref{aff127}}
\and F.~Finelli\orcid{0000-0002-6694-3269}\inst{\ref{aff6},\ref{aff120}}
\and J.~Gracia-Carpio\inst{\ref{aff37}}
\and S.~Ili\'c\orcid{0000-0003-4285-9086}\inst{\ref{aff128},\ref{aff33}}
\and P.~Liebing\inst{\ref{aff39}}
\and S.~Marcin\inst{\ref{aff108}}
\and M.~Martinelli\orcid{0000-0002-6943-7732}\inst{\ref{aff71},\ref{aff129}}
\and S.~Matthew\orcid{0000-0001-8448-1697}\inst{\ref{aff35}}
\and N.~Mauri\orcid{0000-0001-8196-1548}\inst{\ref{aff72},\ref{aff57}}
\and M.~P\"ontinen\orcid{0000-0001-5442-2530}\inst{\ref{aff96}}
\and C.~Porciani\orcid{0000-0002-7797-2508}\inst{\ref{aff102}}
\and Z.~Sakr\orcid{0000-0002-4823-3757}\inst{\ref{aff55},\ref{aff33},\ref{aff130}}
\and V.~Scottez\inst{\ref{aff8},\ref{aff131}}
\and E.~Sefusatti\orcid{0000-0003-0473-1567}\inst{\ref{aff16},\ref{aff18},\ref{aff17}}
\and J.~Steinwagner\inst{\ref{aff37}}
\and M.~Tenti\orcid{0000-0002-4254-5901}\inst{\ref{aff57}}
\and M.~Viel\orcid{0000-0002-2642-5707}\inst{\ref{aff18},\ref{aff16},\ref{aff30},\ref{aff17},\ref{aff132}}
\and M.~Wiesmann\orcid{0009-0000-8199-5860}\inst{\ref{aff85}}
\and Y.~Akrami\orcid{0000-0002-2407-7956}\inst{\ref{aff133},\ref{aff134}}
\and S.~Anselmi\orcid{0000-0002-3579-9583}\inst{\ref{aff80},\ref{aff114},\ref{aff10}}
\and M.~Archidiacono\orcid{0000-0003-4952-9012}\inst{\ref{aff41},\ref{aff101}}
\and F.~Atrio-Barandela\orcid{0000-0002-2130-2513}\inst{\ref{aff135}}
\and E.~Aubourg\orcid{0000-0002-5592-023X}\inst{\ref{aff107},\ref{aff136}}
\and A.~Balaguera-Antolinez\orcid{0000-0001-5028-3035}\inst{\ref{aff11},\ref{aff113}}
\and M.~Ballardini\orcid{0000-0003-4481-3559}\inst{\ref{aff137},\ref{aff6},\ref{aff138}}
\and D.~Bertacca\orcid{0000-0002-2490-7139}\inst{\ref{aff114},\ref{aff58},\ref{aff80}}
\and M.~Bethermin\orcid{0000-0002-3915-2015}\inst{\ref{aff139},\ref{aff99}}
\and A.~Blanchard\orcid{0000-0001-8555-9003}\inst{\ref{aff33}}
\and H.~B\"ohringer\orcid{0000-0001-8241-4204}\inst{\ref{aff37},\ref{aff140},\ref{aff141}}
\and S.~Borgani\orcid{0000-0001-6151-6439}\inst{\ref{aff15},\ref{aff18},\ref{aff16},\ref{aff17}}
\and T.~Bouvard\orcid{0009-0002-7959-312X}\inst{\ref{aff142}}
\and R.~Cabanac\orcid{0000-0001-6679-2600}\inst{\ref{aff33}}
\and A.~Calabro\orcid{0000-0003-2536-1614}\inst{\ref{aff71}}
\and B.~Camacho~Quevedo\orcid{0000-0002-8789-4232}\inst{\ref{aff2},\ref{aff1}}
\and G.~Canas-Herrera\orcid{0000-0003-2796-2149}\inst{\ref{aff98},\ref{aff143}}
\and A.~Cappi\inst{\ref{aff6},\ref{aff106}}
\and F.~Caro\inst{\ref{aff71}}
\and C.~S.~Carvalho\inst{\ref{aff117}}
\and T.~Castro\orcid{0000-0002-6292-3228}\inst{\ref{aff16},\ref{aff17},\ref{aff18},\ref{aff132}}
\and K.~C.~Chambers\orcid{0000-0001-6965-7789}\inst{\ref{aff144}}
\and S.~Contarini\orcid{0000-0002-9843-723X}\inst{\ref{aff37}}
\and T.~Contini\orcid{0000-0003-0275-938X}\inst{\ref{aff33}}
\and A.~R.~Cooray\orcid{0000-0002-3892-0190}\inst{\ref{aff145}}
\and M.~Costanzi\orcid{0000-0001-8158-1449}\inst{\ref{aff15},\ref{aff16},\ref{aff18}}
\and O.~Cucciati\orcid{0000-0002-9336-7551}\inst{\ref{aff6}}
\and S.~Davini\orcid{0000-0003-3269-1718}\inst{\ref{aff62}}
\and B.~De~Caro\inst{\ref{aff69}}
\and S.~de~la~Torre\inst{\ref{aff99}}
\and G.~Desprez\inst{\ref{aff146}}
\and A.~D\'iaz-S\'anchez\orcid{0000-0003-0748-4768}\inst{\ref{aff147}}
\and J.~J.~Diaz\inst{\ref{aff12}}
\and S.~Di~Domizio\orcid{0000-0003-2863-5895}\inst{\ref{aff61},\ref{aff62}}
\and H.~Dole\orcid{0000-0002-9767-3839}\inst{\ref{aff52}}
\and S.~Escoffier\orcid{0000-0002-2847-7498}\inst{\ref{aff82}}
\and M.~Ezziati\orcid{0009-0003-6065-1585}\inst{\ref{aff99}}
\and A.~G.~Ferrari\orcid{0009-0005-5266-4110}\inst{\ref{aff72},\ref{aff57}}
\and P.~G.~Ferreira\orcid{0000-0002-3021-2851}\inst{\ref{aff34}}
\and I.~Ferrero\orcid{0000-0002-1295-1132}\inst{\ref{aff85}}
\and A.~Finoguenov\orcid{0000-0002-4606-5403}\inst{\ref{aff96}}
\and A.~Fontana\orcid{0000-0003-3820-2823}\inst{\ref{aff71}}
\and F.~Fornari\orcid{0000-0003-2979-6738}\inst{\ref{aff120}}
\and L.~Gabarra\orcid{0000-0002-8486-8856}\inst{\ref{aff34}}
\and K.~Ganga\orcid{0000-0001-8159-8208}\inst{\ref{aff107}}
\and J.~Garc\'ia-Bellido\orcid{0000-0002-9370-8360}\inst{\ref{aff133}}
\and T.~Gasparetto\orcid{0000-0002-7913-4866}\inst{\ref{aff16}}
\and E.~Gaztanaga\orcid{0000-0001-9632-0815}\inst{\ref{aff1},\ref{aff2},\ref{aff148}}
\and F.~Giacomini\orcid{0000-0002-3129-2814}\inst{\ref{aff57}}
\and F.~Gianotti\orcid{0000-0003-4666-119X}\inst{\ref{aff6}}
\and A.~H.~Gonzalez\orcid{0000-0002-0933-8601}\inst{\ref{aff149}}
\and G.~Gozaliasl\orcid{0000-0002-0236-919X}\inst{\ref{aff150},\ref{aff96}}
\and A.~Hall\orcid{0000-0002-3139-8651}\inst{\ref{aff35}}
\and W.~G.~Hartley\inst{\ref{aff43}}
\and H.~Hildebrandt\orcid{0000-0002-9814-3338}\inst{\ref{aff151}}
\and J.~Hjorth\orcid{0000-0002-4571-2306}\inst{\ref{aff152}}
\and A.~D.~Holland\inst{\ref{aff153},\ref{aff154}}
\and O.~Ilbert\orcid{0000-0002-7303-4397}\inst{\ref{aff99}}
\and S.~Joudaki\orcid{0000-0001-8820-673X}\inst{\ref{aff148}}
\and E.~Jullo\orcid{0000-0002-9253-053X}\inst{\ref{aff99}}
\and J.~J.~E.~Kajava\orcid{0000-0002-3010-8333}\inst{\ref{aff155},\ref{aff156}}
\and V.~Kansal\orcid{0000-0002-4008-6078}\inst{\ref{aff157},\ref{aff158}}
\and D.~Karagiannis\orcid{0000-0002-4927-0816}\inst{\ref{aff159},\ref{aff160}}
\and C.~C.~Kirkpatrick\inst{\ref{aff94}}
\and J.~Le~Graet\orcid{0000-0001-6523-7971}\inst{\ref{aff82}}
\and L.~Legrand\orcid{0000-0003-0610-5252}\inst{\ref{aff161}}
\and J.~Lesgourgues\orcid{0000-0001-7627-353X}\inst{\ref{aff70}}
\and T.~I.~Liaudat\orcid{0000-0002-9104-314X}\inst{\ref{aff136}}
\and A.~Loureiro\orcid{0000-0002-4371-0876}\inst{\ref{aff162},\ref{aff163}}
\and J.~Macias-Perez\orcid{0000-0002-5385-2763}\inst{\ref{aff164}}
\and M.~Magliocchetti\orcid{0000-0001-9158-4838}\inst{\ref{aff79}}
\and C.~Mancini\orcid{0000-0002-4297-0561}\inst{\ref{aff69}}
\and F.~Mannucci\orcid{0000-0002-4803-2381}\inst{\ref{aff165}}
\and R.~Maoli\orcid{0000-0002-6065-3025}\inst{\ref{aff166},\ref{aff71}}
\and C.~J.~A.~P.~Martins\orcid{0000-0002-4886-9261}\inst{\ref{aff167},\ref{aff65}}
\and L.~Maurin\orcid{0000-0002-8406-0857}\inst{\ref{aff52}}
\and R.~B.~Metcalf\orcid{0000-0003-3167-2574}\inst{\ref{aff103},\ref{aff6}}
\and M.~Migliaccio\inst{\ref{aff168},\ref{aff169}}
\and M.~Miluzio\inst{\ref{aff53}}
\and A.~Mora\orcid{0000-0002-1922-8529}\inst{\ref{aff170}}
\and C.~Moretti\orcid{0000-0003-3314-8936}\inst{\ref{aff30},\ref{aff132},\ref{aff16},\ref{aff18},\ref{aff17}}
\and G.~Morgante\inst{\ref{aff6}}
\and S.~Nadathur\orcid{0000-0001-9070-3102}\inst{\ref{aff148}}
\and L.~Nicastro\orcid{0000-0001-8534-6788}\inst{\ref{aff6}}
\and Nicholas~A.~Walton\orcid{0000-0003-3983-8778}\inst{\ref{aff125}}
\and M.~Oguri\orcid{0000-0003-3484-399X}\inst{\ref{aff171},\ref{aff172}}
\and L.~Patrizii\inst{\ref{aff57}}
\and V.~Popa\orcid{0000-0002-9118-8330}\inst{\ref{aff112}}
\and A.~Pourtsidou\orcid{0000-0001-9110-5550}\inst{\ref{aff35},\ref{aff173}}
\and P.~Reimberg\orcid{0000-0003-3410-0280}\inst{\ref{aff8}}
\and I.~Risso\orcid{0000-0003-2525-7761}\inst{\ref{aff121}}
\and P.-F.~Rocci\inst{\ref{aff52}}
\and R.~P.~Rollins\orcid{0000-0003-1291-1023}\inst{\ref{aff35}}
\and B.~Rusholme\orcid{0000-0001-7648-4142}\inst{\ref{aff174}}
\and M.~Sahl\'en\orcid{0000-0003-0973-4804}\inst{\ref{aff175}}
\and A.~G.~S\'anchez\orcid{0000-0003-1198-831X}\inst{\ref{aff37}}
\and J.~Schaye\orcid{0000-0002-0668-5560}\inst{\ref{aff24}}
\and J.~A.~Schewtschenko\inst{\ref{aff35}}
\and A.~Schneider\orcid{0000-0001-7055-8104}\inst{\ref{aff3}}
\and M.~Schultheis\inst{\ref{aff106}}
\and M.~Sereno\orcid{0000-0003-0302-0325}\inst{\ref{aff6},\ref{aff57}}
\and F.~Shankar\orcid{0000-0001-8973-5051}\inst{\ref{aff176}}
\and A.~Shulevski\orcid{0000-0002-1827-0469}\inst{\ref{aff177},\ref{aff178},\ref{aff179}}
\and A.~Silvestri\orcid{0000-0001-6904-5061}\inst{\ref{aff143}}
\and P.~Simon\inst{\ref{aff102}}
\and A.~Spurio~Mancini\orcid{0000-0001-5698-0990}\inst{\ref{aff180},\ref{aff39}}
\and S.~A.~Stanford\orcid{0000-0003-0122-0841}\inst{\ref{aff181}}
\and K.~Tanidis\inst{\ref{aff34}}
\and C.~Tao\orcid{0000-0001-7961-8177}\inst{\ref{aff82}}
\and N.~Tessore\orcid{0000-0002-9696-7931}\inst{\ref{aff93}}
\and G.~Testera\inst{\ref{aff62}}
\and M.~Tewes\orcid{0000-0002-1155-8689}\inst{\ref{aff102}}
\and S.~Toft\orcid{0000-0003-3631-7176}\inst{\ref{aff90},\ref{aff45},\ref{aff44}}
\and S.~Tosi\orcid{0000-0002-7275-9193}\inst{\ref{aff61},\ref{aff62}}
\and A.~Troja\orcid{0000-0003-0239-4595}\inst{\ref{aff114},\ref{aff80}}
\and M.~Tucci\inst{\ref{aff43}}
\and C.~Valieri\inst{\ref{aff57}}
\and J.~Valiviita\orcid{0000-0001-6225-3693}\inst{\ref{aff96},\ref{aff97}}
\and D.~Vergani\orcid{0000-0003-0898-2216}\inst{\ref{aff6}}
\and F.~Vernizzi\orcid{0000-0003-3426-2802}\inst{\ref{aff122}}
\and G.~Verza\orcid{0000-0002-1886-8348}\inst{\ref{aff182},\ref{aff127}}
\and P.~Vielzeuf\orcid{0000-0003-2035-9339}\inst{\ref{aff82}}
\and J.~R.~Weaver\orcid{0000-0003-1614-196X}\inst{\ref{aff183}}
\and L.~Zalesky\orcid{0000-0001-5680-2326}\inst{\ref{aff144}}
\and P.~Dimauro\orcid{0000-0001-7399-2854}\inst{\ref{aff71},\ref{aff184}}
\and P.-A.~Duc\orcid{0000-0003-3343-6284}\inst{\ref{aff139}}
\and Y.~Fang\inst{\ref{aff59}}
\and A.~M.~N.~Ferguson\inst{\ref{aff35}}
\and C.~M.~Gutierrez\orcid{0000-0001-7854-783X}\inst{\ref{aff185}}
\and I.~Kova{\v{c}}i{\'{c}}\orcid{0000-0001-6751-3263}\inst{\ref{aff186}}
\and S.~Kruk\orcid{0000-0001-8010-8879}\inst{\ref{aff53}}
\and A.~M.~C.~Le~Brun\orcid{0000-0002-0936-4594}\inst{\ref{aff10}}
\and A.~Montoro\orcid{0000-0003-4730-8590}\inst{\ref{aff1},\ref{aff2}}
\and C.~Murray\inst{\ref{aff107}}
\and L.~Pagano\orcid{0000-0003-1820-5998}\inst{\ref{aff137},\ref{aff138}}
\and D.~Paoletti\orcid{0000-0003-4761-6147}\inst{\ref{aff6},\ref{aff120}}
\and E.~Sarpa\orcid{0000-0002-1256-655X}\inst{\ref{aff30},\ref{aff132},\ref{aff17}}
\and A.~Viitanen\orcid{0000-0001-9383-786X}\inst{\ref{aff94},\ref{aff71}}
\and J.~Mart\'{i}n-Fleitas\orcid{0000-0002-8594-569X}\inst{\ref{aff170}}
\and L.~Y.~A.~Yung\orcid{0000-0003-3466-035X}\inst{\ref{aff92},\ref{aff187}}}
										   
%%%% please do not edit the affiliation list -- contact ECEB Bureau for changes
\institute{Institute of Space Sciences (ICE, CSIC), Campus UAB, Carrer de Can Magrans, s/n, 08193 Barcelona, Spain\label{aff1}
\and
Institut d'Estudis Espacials de Catalunya (IEEC),  Edifici RDIT, Campus UPC, 08860 Castelldefels, Barcelona, Spain\label{aff2}
\and
Department of Astrophysics, University of Zurich, Winterthurerstrasse 190, 8057 Zurich, Switzerland\label{aff3}
\and
Centro de Investigaciones Energ\'eticas, Medioambientales y Tecnol\'ogicas (CIEMAT), Avenida Complutense 40, 28040 Madrid, Spain\label{aff4}
\and
Port d'Informaci\'{o} Cient\'{i}fica, Campus UAB, C. Albareda s/n, 08193 Bellaterra (Barcelona), Spain\label{aff5}
\and
INAF-Osservatorio di Astrofisica e Scienza dello Spazio di Bologna, Via Piero Gobetti 93/3, 40129 Bologna, Italy\label{aff6}
\and
Institut d'Astrophysique de Paris, UMR 7095, CNRS, and Sorbonne Universit\'e, 98 bis boulevard Arago, 75014 Paris, France\label{aff7}
\and
Institut d'Astrophysique de Paris, 98bis Boulevard Arago, 75014, Paris, France\label{aff8}
\and
Kavli Institute for the Physics and Mathematics of the Universe (WPI), University of Tokyo, Kashiwa, Chiba 277-8583, Japan\label{aff9}
\and
Laboratoire Univers et Th\'eorie, Observatoire de Paris, Universit\'e PSL, Universit\'e Paris Cit\'e, CNRS, 92190 Meudon, France\label{aff10}
\and
Instituto de Astrof\'isica de Canarias, Calle V\'ia L\'actea s/n, 38204, San Crist\'obal de La Laguna, Tenerife, Spain\label{aff11}
\and
Instituto de Astrof\'isica de Canarias (IAC); Departamento de Astrof\'isica, Universidad de La Laguna (ULL), 38200, La Laguna, Tenerife, Spain\label{aff12}
\and
Universit\'e PSL, Observatoire de Paris, Sorbonne Universit\'e, CNRS, LERMA, 75014, Paris, France\label{aff13}
\and
Universit\'e Paris-Cit\'e, 5 Rue Thomas Mann, 75013, Paris, France\label{aff14}
\and
Dipartimento di Fisica - Sezione di Astronomia, Universit\`a di Trieste, Via Tiepolo 11, 34131 Trieste, Italy\label{aff15}
\and
INAF-Osservatorio Astronomico di Trieste, Via G. B. Tiepolo 11, 34143 Trieste, Italy\label{aff16}
\and
INFN, Sezione di Trieste, Via Valerio 2, 34127 Trieste TS, Italy\label{aff17}
\and
IFPU, Institute for Fundamental Physics of the Universe, via Beirut 2, 34151 Trieste, Italy\label{aff18}
\and
Institut de F\'{i}sica d'Altes Energies (IFAE), The Barcelona Institute of Science and Technology, Campus UAB, 08193 Bellaterra (Barcelona), Spain\label{aff19}
\and
Instituto de Astronomia Teorica y Experimental (IATE-CONICET), Laprida 854, X5000BGR, C\'ordoba, Argentina\label{aff20}
\and
Minnesota Institute for Astrophysics, University of Minnesota, 116 Church St SE, Minneapolis, MN 55455, USA\label{aff21}
\and
Institut de Ciencies de l'Espai (IEEC-CSIC), Campus UAB, Carrer de Can Magrans, s/n Cerdanyola del Vall\'es, 08193 Barcelona, Spain\label{aff22}
\and
Universit\"at Innsbruck, Institut f\"ur Astro- und Teilchenphysik, Technikerstr. 25/8, 6020 Innsbruck, Austria\label{aff23}
\and
Leiden Observatory, Leiden University, Einsteinweg 55, 2333 CC Leiden, The Netherlands\label{aff24}
\and
Department of Astronomy, School of Physics and Astronomy, Shanghai Jiao Tong University, Shanghai 200240, China\label{aff25}
\and
Shanghai Key Laboratory for Particle Physics and Cosmology, Shanghai 200240, China\label{aff26}
\and
Waterloo Centre for Astrophysics, University of Waterloo, Waterloo, Ontario N2L 3G1, Canada\label{aff27}
\and
Department of Physics and Astronomy, University of Waterloo, Waterloo, Ontario N2L 3G1, Canada\label{aff28}
\and
INFN Gruppo Collegato di Parma, Viale delle Scienze 7/A 43124 Parma, Italy\label{aff29}
\and
SISSA, International School for Advanced Studies, Via Bonomea 265, 34136 Trieste TS, Italy\label{aff30}
\and
International Centre for Theoretical Physics (ICTP), Strada Costiera 11, 34151 Trieste, Italy\label{aff31}
\and
Department of Astrophysical Sciences, Peyton Hall, Princeton University, Princeton, NJ 08544, USA\label{aff32}
\and
Institut de Recherche en Astrophysique et Plan\'etologie (IRAP), Universit\'e de Toulouse, CNRS, UPS, CNES, 14 Av. Edouard Belin, 31400 Toulouse, France\label{aff33}
\and
Department of Physics, Oxford University, Keble Road, Oxford OX1 3RH, UK\label{aff34}
\and
Institute for Astronomy, University of Edinburgh, Royal Observatory, Blackford Hill, Edinburgh EH9 3HJ, UK\label{aff35}
\and
Institute of Physics, Laboratory for Galaxy Evolution, Ecole Polytechnique F\'ed\'erale de Lausanne, Observatoire de Sauverny, CH-1290 Versoix, Switzerland\label{aff36}
\and
Max Planck Institute for Extraterrestrial Physics, Giessenbachstr. 1, 85748 Garching, Germany\label{aff37}
\and
Universit\'e Paris-Saclay, Universit\'e Paris Cit\'e, CEA, CNRS, AIM, 91191, Gif-sur-Yvette, France\label{aff38}
\and
Mullard Space Science Laboratory, University College London, Holmbury St Mary, Dorking, Surrey RH5 6NT, UK\label{aff39}
\and
Perimeter Institute for Theoretical Physics, Waterloo, Ontario N2L 2Y5, Canada\label{aff40}
\and
Dipartimento di Fisica "Aldo Pontremoli", Universit\`a degli Studi di Milano, Via Celoria 16, 20133 Milano, Italy\label{aff41}
\and
INAF-Osservatorio Astronomico di Brera, Via Brera 28, 20122 Milano, Italy\label{aff42}
\and
Department of Astronomy, University of Geneva, ch. d'Ecogia 16, 1290 Versoix, Switzerland\label{aff43}
\and
Niels Bohr Institute, University of Copenhagen, Jagtvej 128, 2200 Copenhagen, Denmark\label{aff44}
\and
Cosmic Dawn Center (DAWN)\label{aff45}
\and
Jodrell Bank Centre for Astrophysics, Department of Physics and Astronomy, University of Manchester, Oxford Road, Manchester M13 9PL, UK\label{aff46}
\and
INAF-Osservatorio Astronomico di Capodimonte, Via Moiariello 16, 80131 Napoli, Italy\label{aff47}
\and
Departamento de F\'isica Te\'orica, Facultad de Ciencias, Universidad Aut\'onoma de Madrid, 28049 Cantoblanco, Madrid, Spain\label{aff48}
\and
Serra H\'unter Fellow, Departament de F\'{\i}sica, Universitat Aut\`onoma de Barcelona, E-08193 Bellaterra, Spain\label{aff49}
\and
Konkoly Observatory, HUN-REN CSFK, MTA Centre of Excellence, Budapest, Konkoly Thege Mikl\'os {\'u}t 15-17. H-1121, Hungary\label{aff50}
\and
MTA-CSFK Lend\"ulet Large-Scale Structure Research Group, Konkoly-Thege Mikl\'os \'ut 15-17, H-1121 Budapest, Hungary\label{aff51}
\and
Universit\'e Paris-Saclay, CNRS, Institut d'astrophysique spatiale, 91405, Orsay, France\label{aff52}
\and
ESAC/ESA, Camino Bajo del Castillo, s/n., Urb. Villafranca del Castillo, 28692 Villanueva de la Ca\~nada, Madrid, Spain\label{aff53}
\and
School of Mathematics and Physics, University of Surrey, Guildford, Surrey, GU2 7XH, UK\label{aff54}
\and
Institut f\"ur Theoretische Physik, University of Heidelberg, Philosophenweg 16, 69120 Heidelberg, Germany\label{aff55}
\and
Dipartimento di Fisica e Astronomia, Universit\`a di Bologna, Via Gobetti 93/2, 40129 Bologna, Italy\label{aff56}
\and
INFN-Sezione di Bologna, Viale Berti Pichat 6/2, 40127 Bologna, Italy\label{aff57}
\and
INAF-Osservatorio Astronomico di Padova, Via dell'Osservatorio 5, 35122 Padova, Italy\label{aff58}
\and
Universit\"ats-Sternwarte M\"unchen, Fakult\"at f\"ur Physik, Ludwig-Maximilians-Universit\"at M\"unchen, Scheinerstrasse 1, 81679 M\"unchen, Germany\label{aff59}
\and
INAF-Osservatorio Astrofisico di Torino, Via Osservatorio 20, 10025 Pino Torinese (TO), Italy\label{aff60}
\and
Dipartimento di Fisica, Universit\`a di Genova, Via Dodecaneso 33, 16146, Genova, Italy\label{aff61}
\and
INFN-Sezione di Genova, Via Dodecaneso 33, 16146, Genova, Italy\label{aff62}
\and
Department of Physics "E. Pancini", University Federico II, Via Cinthia 6, 80126, Napoli, Italy\label{aff63}
\and
INFN section of Naples, Via Cinthia 6, 80126, Napoli, Italy\label{aff64}
\and
Instituto de Astrof\'isica e Ci\^encias do Espa\c{c}o, Universidade do Porto, CAUP, Rua das Estrelas, PT4150-762 Porto, Portugal\label{aff65}
\and
Faculdade de Ci\^encias da Universidade do Porto, Rua do Campo de Alegre, 4150-007 Porto, Portugal\label{aff66}
\and
Dipartimento di Fisica, Universit\`a degli Studi di Torino, Via P. Giuria 1, 10125 Torino, Italy\label{aff67}
\and
INFN-Sezione di Torino, Via P. Giuria 1, 10125 Torino, Italy\label{aff68}
\and
INAF-IASF Milano, Via Alfonso Corti 12, 20133 Milano, Italy\label{aff69}
\and
Institute for Theoretical Particle Physics and Cosmology (TTK), RWTH Aachen University, 52056 Aachen, Germany\label{aff70}
\and
INAF-Osservatorio Astronomico di Roma, Via Frascati 33, 00078 Monteporzio Catone, Italy\label{aff71}
\and
Dipartimento di Fisica e Astronomia "Augusto Righi" - Alma Mater Studiorum Universit\`a di Bologna, Viale Berti Pichat 6/2, 40127 Bologna, Italy\label{aff72}
\and
European Space Agency/ESRIN, Largo Galileo Galilei 1, 00044 Frascati, Roma, Italy\label{aff73}
\and
Universit\'e Claude Bernard Lyon 1, CNRS/IN2P3, IP2I Lyon, UMR 5822, Villeurbanne, F-69100, France\label{aff74}
\and
Institute of Physics, Laboratory of Astrophysics, Ecole Polytechnique F\'ed\'erale de Lausanne (EPFL), Observatoire de Sauverny, 1290 Versoix, Switzerland\label{aff75}
\and
UCB Lyon 1, CNRS/IN2P3, IUF, IP2I Lyon, 4 rue Enrico Fermi, 69622 Villeurbanne, France\label{aff76}
\and
Departamento de F\'isica, Faculdade de Ci\^encias, Universidade de Lisboa, Edif\'icio C8, Campo Grande, PT1749-016 Lisboa, Portugal\label{aff77}
\and
Instituto de Astrof\'isica e Ci\^encias do Espa\c{c}o, Faculdade de Ci\^encias, Universidade de Lisboa, Campo Grande, 1749-016 Lisboa, Portugal\label{aff78}
\and
INAF-Istituto di Astrofisica e Planetologia Spaziali, via del Fosso del Cavaliere, 100, 00100 Roma, Italy\label{aff79}
\and
INFN-Padova, Via Marzolo 8, 35131 Padova, Italy\label{aff80}
\and
School of Physics, HH Wills Physics Laboratory, University of Bristol, Tyndall Avenue, Bristol, BS8 1TL, UK\label{aff81}
\and
Aix-Marseille Universit\'e, CNRS/IN2P3, CPPM, Marseille, France\label{aff82}
\and
Istituto Nazionale di Fisica Nucleare, Sezione di Bologna, Via Irnerio 46, 40126 Bologna, Italy\label{aff83}
\and
FRACTAL S.L.N.E., calle Tulip\'an 2, Portal 13 1A, 28231, Las Rozas de Madrid, Spain\label{aff84}
\and
Institute of Theoretical Astrophysics, University of Oslo, P.O. Box 1029 Blindern, 0315 Oslo, Norway\label{aff85}
\and
Jet Propulsion Laboratory, California Institute of Technology, 4800 Oak Grove Drive, Pasadena, CA, 91109, USA\label{aff86}
\and
Department of Physics, Lancaster University, Lancaster, LA1 4YB, UK\label{aff87}
\and
Felix Hormuth Engineering, Goethestr. 17, 69181 Leimen, Germany\label{aff88}
\and
Technical University of Denmark, Elektrovej 327, 2800 Kgs. Lyngby, Denmark\label{aff89}
\and
Cosmic Dawn Center (DAWN), Denmark\label{aff90}
\and
Max-Planck-Institut f\"ur Astronomie, K\"onigstuhl 17, 69117 Heidelberg, Germany\label{aff91}
\and
NASA Goddard Space Flight Center, Greenbelt, MD 20771, USA\label{aff92}
\and
Department of Physics and Astronomy, University College London, Gower Street, London WC1E 6BT, UK\label{aff93}
\and
Department of Physics and Helsinki Institute of Physics, Gustaf H\"allstr\"omin katu 2, 00014 University of Helsinki, Finland\label{aff94}
\and
Universit\'e de Gen\`eve, D\'epartement de Physique Th\'eorique and Centre for Astroparticle Physics, 24 quai Ernest-Ansermet, CH-1211 Gen\`eve 4, Switzerland\label{aff95}
\and
Department of Physics, P.O. Box 64, 00014 University of Helsinki, Finland\label{aff96}
\and
Helsinki Institute of Physics, Gustaf H{\"a}llstr{\"o}min katu 2, University of Helsinki, Helsinki, Finland\label{aff97}
\and
European Space Agency/ESTEC, Keplerlaan 1, 2201 AZ Noordwijk, The Netherlands\label{aff98}
\and
Aix-Marseille Universit\'e, CNRS, CNES, LAM, Marseille, France\label{aff99}
\and
NOVA optical infrared instrumentation group at ASTRON, Oude Hoogeveensedijk 4, 7991PD, Dwingeloo, The Netherlands\label{aff100}
\and
INFN-Sezione di Milano, Via Celoria 16, 20133 Milano, Italy\label{aff101}
\and
Universit\"at Bonn, Argelander-Institut f\"ur Astronomie, Auf dem H\"ugel 71, 53121 Bonn, Germany\label{aff102}
\and
Dipartimento di Fisica e Astronomia "Augusto Righi" - Alma Mater Studiorum Universit\`a di Bologna, via Piero Gobetti 93/2, 40129 Bologna, Italy\label{aff103}
\and
Department of Physics, Institute for Computational Cosmology, Durham University, South Road, DH1 3LE, UK\label{aff104}
\and
Infrared Processing and Analysis Center, California Institute of Technology, Pasadena, CA 91125, USA\label{aff105}
\and
Universit\'e C\^{o}te d'Azur, Observatoire de la C\^{o}te d'Azur, CNRS, Laboratoire Lagrange, Bd de l'Observatoire, CS 34229, 06304 Nice cedex 4, France\label{aff106}
\and
Universit\'e Paris Cit\'e, CNRS, Astroparticule et Cosmologie, 75013 Paris, France\label{aff107}
\and
University of Applied Sciences and Arts of Northwestern Switzerland, School of Engineering, 5210 Windisch, Switzerland\label{aff108}
\and
Department of Physics and Astronomy, University of Aarhus, Ny Munkegade 120, DK-8000 Aarhus C, Denmark\label{aff109}
\and
Space Science Data Center, Italian Space Agency, via del Politecnico snc, 00133 Roma, Italy\label{aff110}
\and
Centre National d'Etudes Spatiales -- Centre spatial de Toulouse, 18 avenue Edouard Belin, 31401 Toulouse Cedex 9, France\label{aff111}
\and
Institute of Space Science, Str. Atomistilor, nr. 409 M\u{a}gurele, Ilfov, 077125, Romania\label{aff112}
\and
Departamento de Astrof\'isica, Universidad de La Laguna, 38206, La Laguna, Tenerife, Spain\label{aff113}
\and
Dipartimento di Fisica e Astronomia "G. Galilei", Universit\`a di Padova, Via Marzolo 8, 35131 Padova, Italy\label{aff114}
\and
Departamento de F\'isica, FCFM, Universidad de Chile, Blanco Encalada 2008, Santiago, Chile\label{aff115}
\and
Satlantis, University Science Park, Sede Bld 48940, Leioa-Bilbao, Spain\label{aff116}
\and
Instituto de Astrof\'isica e Ci\^encias do Espa\c{c}o, Faculdade de Ci\^encias, Universidade de Lisboa, Tapada da Ajuda, 1349-018 Lisboa, Portugal\label{aff117}
\and
Universidad Polit\'ecnica de Cartagena, Departamento de Electr\'onica y Tecnolog\'ia de Computadoras,  Plaza del Hospital 1, 30202 Cartagena, Spain\label{aff118}
\and
Centre for Information Technology, University of Groningen, P.O. Box 11044, 9700 CA Groningen, The Netherlands\label{aff119}
\and
INFN-Bologna, Via Irnerio 46, 40126 Bologna, Italy\label{aff120}
\and
Dipartimento di Fisica, Universit\`a degli studi di Genova, and INFN-Sezione di Genova, via Dodecaneso 33, 16146, Genova, Italy\label{aff121}
\and
Institut de Physique Th\'eorique, CEA, CNRS, Universit\'e Paris-Saclay 91191 Gif-sur-Yvette Cedex, France\label{aff122}
\and
INAF, Istituto di Radioastronomia, Via Piero Gobetti 101, 40129 Bologna, Italy\label{aff123}
\and
Astronomical Observatory of the Autonomous Region of the Aosta Valley (OAVdA), Loc. Lignan 39, I-11020, Nus (Aosta Valley), Italy\label{aff124}
\and
Institute of Astronomy, University of Cambridge, Madingley Road, Cambridge CB3 0HA, UK\label{aff125}
\and
School of Physics and Astronomy, Cardiff University, The Parade, Cardiff, CF24 3AA, UK\label{aff126}
\and
Center for Computational Astrophysics, Flatiron Institute, 162 5th Avenue, 10010, New York, NY, USA\label{aff127}
\and
Universit\'e Paris-Saclay, CNRS/IN2P3, IJCLab, 91405 Orsay, France\label{aff128}
\and
INFN-Sezione di Roma, Piazzale Aldo Moro, 2 - c/o Dipartimento di Fisica, Edificio G. Marconi, 00185 Roma, Italy\label{aff129}
\and
Universit\'e St Joseph; Faculty of Sciences, Beirut, Lebanon\label{aff130}
\and
Junia, EPA department, 41 Bd Vauban, 59800 Lille, France\label{aff131}
\and
ICSC - Centro Nazionale di Ricerca in High Performance Computing, Big Data e Quantum Computing, Via Magnanelli 2, Bologna, Italy\label{aff132}
\and
Instituto de F\'isica Te\'orica UAM-CSIC, Campus de Cantoblanco, 28049 Madrid, Spain\label{aff133}
\and
CERCA/ISO, Department of Physics, Case Western Reserve University, 10900 Euclid Avenue, Cleveland, OH 44106, USA\label{aff134}
\and
Departamento de F{\'\i}sica Fundamental. Universidad de Salamanca. Plaza de la Merced s/n. 37008 Salamanca, Spain\label{aff135}
\and
IRFU, CEA, Universit\'e Paris-Saclay 91191 Gif-sur-Yvette Cedex, France\label{aff136}
\and
Dipartimento di Fisica e Scienze della Terra, Universit\`a degli Studi di Ferrara, Via Giuseppe Saragat 1, 44122 Ferrara, Italy\label{aff137}
\and
Istituto Nazionale di Fisica Nucleare, Sezione di Ferrara, Via Giuseppe Saragat 1, 44122 Ferrara, Italy\label{aff138}
\and
Universit\'e de Strasbourg, CNRS, Observatoire astronomique de Strasbourg, UMR 7550, 67000 Strasbourg, France\label{aff139}
\and
Ludwig-Maximilians-University, Schellingstrasse 4, 80799 Munich, Germany\label{aff140}
\and
Max-Planck-Institut f\"ur Physik, Boltzmannstr. 8, 85748 Garching, Germany\label{aff141}
\and
Thales~Services~S.A.S., 290 All\'ee du Lac, 31670 Lab\`ege, France\label{aff142}
\and
Institute Lorentz, Leiden University, Niels Bohrweg 2, 2333 CA Leiden, The Netherlands\label{aff143}
\and
Institute for Astronomy, University of Hawaii, 2680 Woodlawn Drive, Honolulu, HI 96822, USA\label{aff144}
\and
Department of Physics \& Astronomy, University of California Irvine, Irvine CA 92697, USA\label{aff145}
\and
Department of Astronomy \& Physics and Institute for Computational Astrophysics, Saint Mary's University, 923 Robie Street, Halifax, Nova Scotia, B3H 3C3, Canada\label{aff146}
\and
Departamento F\'isica Aplicada, Universidad Polit\'ecnica de Cartagena, Campus Muralla del Mar, 30202 Cartagena, Murcia, Spain\label{aff147}
\and
Institute of Cosmology and Gravitation, University of Portsmouth, Portsmouth PO1 3FX, UK\label{aff148}
\and
Department of Astronomy, University of Florida, Bryant Space Science Center, Gainesville, FL 32611, USA\label{aff149}
\and
Department of Computer Science, Aalto University, PO Box 15400, Espoo, FI-00 076, Finland\label{aff150}
\and
Ruhr University Bochum, Faculty of Physics and Astronomy, Astronomical Institute (AIRUB), German Centre for Cosmological Lensing (GCCL), 44780 Bochum, Germany\label{aff151}
\and
DARK, Niels Bohr Institute, University of Copenhagen, Jagtvej 155, 2200 Copenhagen, Denmark\label{aff152}
\and
Centre for Electronic Imaging, Open University, Walton Hall, Milton Keynes, MK7~6AA, UK\label{aff153}
\and
XCAM Limited, 2 Stone Circle Road, Northampton, NN3 8RF, UK\label{aff154}
\and
Department of Physics and Astronomy, Vesilinnantie 5, 20014 University of Turku, Finland\label{aff155}
\and
Serco for European Space Agency (ESA), Camino bajo del Castillo, s/n, Urbanizacion Villafranca del Castillo, Villanueva de la Ca\~nada, 28692 Madrid, Spain\label{aff156}
\and
ARC Centre of Excellence for Dark Matter Particle Physics, Melbourne, Australia\label{aff157}
\and
Centre for Astrophysics \& Supercomputing, Swinburne University of Technology, Victoria 3122, Australia\label{aff158}
\and
School of Physics and Astronomy, Queen Mary University of London, Mile End Road, London E1 4NS, UK\label{aff159}
\and
Department of Physics and Astronomy, University of the Western Cape, Bellville, Cape Town, 7535, South Africa\label{aff160}
\and
ICTP South American Institute for Fundamental Research, Instituto de F\'{\i}sica Te\'orica, Universidade Estadual Paulista, S\~ao Paulo, Brazil\label{aff161}
\and
Oskar Klein Centre for Cosmoparticle Physics, Department of Physics, Stockholm University, Stockholm, SE-106 91, Sweden\label{aff162}
\and
Astrophysics Group, Blackett Laboratory, Imperial College London, London SW7 2AZ, UK\label{aff163}
\and
Univ. Grenoble Alpes, CNRS, Grenoble INP, LPSC-IN2P3, 53, Avenue des Martyrs, 38000, Grenoble, France\label{aff164}
\and
INAF-Osservatorio Astrofisico di Arcetri, Largo E. Fermi 5, 50125, Firenze, Italy\label{aff165}
\and
Dipartimento di Fisica, Sapienza Universit\`a di Roma, Piazzale Aldo Moro 2, 00185 Roma, Italy\label{aff166}
\and
Centro de Astrof\'{\i}sica da Universidade do Porto, Rua das Estrelas, 4150-762 Porto, Portugal\label{aff167}
\and
Dipartimento di Fisica, Universit\`a di Roma Tor Vergata, Via della Ricerca Scientifica 1, Roma, Italy\label{aff168}
\and
INFN, Sezione di Roma 2, Via della Ricerca Scientifica 1, Roma, Italy\label{aff169}
\and
Aurora Technology for European Space Agency (ESA), Camino bajo del Castillo, s/n, Urbanizacion Villafranca del Castillo, Villanueva de la Ca\~nada, 28692 Madrid, Spain\label{aff170}
\and
Center for Frontier Science, Chiba University, 1-33 Yayoi-cho, Inage-ku, Chiba 263-8522, Japan\label{aff171}
\and
Department of Physics, Graduate School of Science, Chiba University, 1-33 Yayoi-Cho, Inage-Ku, Chiba 263-8522, Japan\label{aff172}
\and
Higgs Centre for Theoretical Physics, School of Physics and Astronomy, The University of Edinburgh, Edinburgh EH9 3FD, UK\label{aff173}
\and
Caltech/IPAC, 1200 E. California Blvd., Pasadena, CA 91125, USA\label{aff174}
\and
Theoretical astrophysics, Department of Physics and Astronomy, Uppsala University, Box 515, 751 20 Uppsala, Sweden\label{aff175}
\and
School of Physics \& Astronomy, University of Southampton, Highfield Campus, Southampton SO17 1BJ, UK\label{aff176}
\and
ASTRON, the Netherlands Institute for Radio Astronomy, Postbus 2, 7990 AA, Dwingeloo, The Netherlands\label{aff177}
\and
Kapteyn Astronomical Institute, University of Groningen, PO Box 800, 9700 AV Groningen, The Netherlands\label{aff178}
\and
Anton Pannekoek Institute for Astronomy, University of Amsterdam, Postbus 94249, 1090 GE Amsterdam, The Netherlands\label{aff179}
\and
Department of Physics, Royal Holloway, University of London, TW20 0EX, UK\label{aff180}
\and
Department of Physics and Astronomy, University of California, Davis, CA 95616, USA\label{aff181}
\and
Center for Cosmology and Particle Physics, Department of Physics, New York University, New York, NY 10003, USA\label{aff182}
\and
Department of Astronomy, University of Massachusetts, Amherst, MA 01003, USA\label{aff183}
\and
Observatorio Nacional, Rua General Jose Cristino, 77-Bairro Imperial de Sao Cristovao, Rio de Janeiro, 20921-400, Brazil\label{aff184}
\and
Instituto de Astrof\'\i sica de Canarias, c/ Via Lactea s/n, La Laguna E-38200, Spain. Departamento de Astrof\'\i sica de la Universidad de La Laguna, Avda. Francisco Sanchez, La Laguna, E-38200, Spain\label{aff185}
\and
Sterrenkundig Observatorium, Universiteit Gent, Krijgslaan 281 S9, 9000 Gent, Belgium\label{aff186}
\and
Space Telescope Science Institute, 3700 San Martin Dr, Baltimore, MD 21218, USA\label{aff187}}

\date{\today}

% 
% Put your abstract here
%

\abstract{We present the Flagship galaxy mock, a simulated catalogue of billions of galaxies designed to support the scientific exploitation of the \Euclid mission. \Euclid is a medium-class mission of the European Space Agency optimised to determine the properties of dark matter and dark energy on the largest scales of the Universe. It probes structure formation over more than 10 billion years primarily from the combination of weak gravitational lensing and galaxy clustering data. The breath of \Euclid's data will also foster a wide variety of scientific analyses.
The Flagship simulation was developed to provide a realistic approximation to the galaxies that will be observed by \Euclid and used in its scientific analyses. We ran a state-of-the-art $N$-body simulation with four trillion particles, producing a lightcone on the fly. From the dark matter particles, we produced a catalogue of 16 billion haloes in one octant of the sky in the lightcone up to redshift $z=3$. We then populated these haloes with mock galaxies using a halo occupation distribution and abundance matching approach, calibrating the free parameters of the galaxy mock against observed correlations and other basic galaxy properties. Modelled galaxy properties include luminosity and flux in several bands, redshifts, positions and velocities, spectral energy distributions, shapes and sizes, stellar masses, star formation rates, metallicities, emission line fluxes, and lensing properties. We selected a final sample of 3.4 billion galaxies with a magnitude cut of $\HE <26$, where we are complete. We have performed a comprehensive set of validation tests to check the similarity to observational data and theoretical models. In particular, our catalogue is able to closely reproduce the main characteristics of the weak lensing and galaxy clustering samples to be used in the mission's main cosmological analysis. Moreover, given its depth and completeness, this new galaxy mock also provides the community with a powerful tool for developing a wide range of scientific analyses beyond the \Euclid mission.
}

%
% Provide up to five key words:
%

%
% Add short versions of title and author list for page headings
%
   \titlerunning{\Euclid galaxy mock catalogue}
   \authorrunning{Euclid Collaboration: F.J.~Castander, et al.}

   \maketitle
%
%-------------------------------------------------------------------
%
%
%   Start the main text of your paper here
%

\section{\label{sec:Intro}Introduction}

%{\bf General intro: surveys and Euclid}

The discovery of the accelerated expansion of the Universe has driven a large observational effort to study its cause and nature \citep{Albrecht2006,Amendola18}. This phenomenon, usually referred to as dark energy, can be the result of a hypothesised fluid with negative pressure or the inadequacy of our current understanding of gravitation. 
Large observational surveys are needed to sample enough volume and a high number density of sources to properly characterize the Universe's evolution from its expansion rate and the growth rate of its structures. Current large surveys such as the Dark Energy Survey (DES; \citealp{Abbott2023}), the Hyper Suprime Cam Subaru Strategic 
 Program (HSC-SSP; \citealp{Aihara:18}), the Kilo-Degree Survey (KiDS; \citealp{Heymans2021}) and the Dark Energy Spectroscopic Instrument (DESI; \citealp{Dey2019}) are providing data that in combination with cosmic microwave background (CMB) data place strong constraints on cosmological parameters \citep[e.g.,][]{Planck:2018vyg}. Forthcoming surveys from the ground, such as the Vera C. Rubin Observatory (LSST; \citealp{Ivezic2019}), or from space, like \Euclid \citep{Laureijs:11,EuclidSkyOverview} and the \emph{Nancy Grace Roman} Space Telescope \citep{Akeson2019}, will collect more and higher-quality data that will allow us to determine the cosmological parameters, and in particular the equation of state parameter of dark energy, to unprecedented precision. 
With the gain in statistical precision in the measurements from these surveys, the control of systematic errors from the combination of different cosmological probes has become key to achieving the expected accuracy. 

Within this framework, the European Space Agency approved the \Euclid mission to carry out a comprehensive survey of most of the extragalactic sky from space. The \Euclid mission is thoroughly described in~\cite{EuclidSkyOverview}. \Euclid will use gravitational lensing and galaxy clustering as the main probes to study cosmology. 
It will carry out a wide and a deep survey of approximately 14\,000 and 50\,deg$^2$, respectively. The wide survey (EWS) will reach a magnitude limit of $\IE\sim 24.5$, and the deep survey (EDS) will push approximately two magnitudes fainter~\citep{Scaramella:22}. In its wide survey (EWS), \Euclid will take images of billions of galaxies to determine their shapes and also obtain slitless spectra of tens of millions of galaxies to determine their redshifts, using its two main science instruments: the
visible imaging instrument~\citep[VIS,][]{EuclidSkyVIS} and the Near In-
frared Spectrometer and Photometer~\citep[NISP,][]{EuclidSkyNISP}.
The combination of weak gravitational lensing and galaxy clustering will provide stringent cosmological constraints~\citep{Blanchard:20}.

%{\bf Cosmo needs for mocks:}

Current and future cosmological surveys need simulations. In their definition stages, simulations are needed to define the survey requirements, to optimize its design and to plan the survey. Once a survey is running, simulations are needed to analyse the data and interpret the results. In the case of \Euclid, there is a strong simulation effort to prepare the science exploitation of the data. From a programmatic point of view, the effort has focused on the simulations needed to support the mission reviews and explore the optimisation of the mission's scientific reach. For that purpose, \Euclid has undertaken science performance verification exercises in which comprehensive analyses of the mission are performed to check the compliance with the scientific requirements. 
With the \Euclid launch on 1 July 2023, the simulation focus has now turned to enabling the science exploitation of the first data releases.
    
%{\bf methods for mocks}

The optimal science exploitation of the new generation of galaxy surveys, such as \Euclid, demands the development of large-volume and high-mass resolution numerical simulations that reproduce the large-scale galaxy distribution that these new surveys will observe with high fidelity. Not only do these help to assess the performance with a realism that cannot be achieved otherwise, but such simulations are also an essential tool for the development of the data processing and science analysis pipelines. Given the computational cost, so far, most synthetic galaxy catalogues have been developed out of $N$-body simulations where only gravity is used to follow the evolution of structure~\citep[e.g.,][]{Bertschinger:91,Couchman:95,Stadel:01,Harnois-Deraps:13,Menon:15,Habib:16,Potter:17,Ishiyama:21,Garrison:21,Springel:21}. For a recent review on cosmological $N$-body simulations, see \citet{angulo:22}. Galaxies are introduced in these simulations populating the dark matter haloes using different methods including semi-analytical models~\citep[SAM; e.g.,][]{White:78,White:91,Kauffmann:93,Kauffmann:99,Somerville:99,Benson:00,Cole:00,Hatton:03,Springel:05,Hirschmann2016,Lagos2018,DeLucia2024}, other empirical and phenomenological models like halo occupation distribution models~\citep[HOD; e.g.,][]{Cooray:02,Jing:98,Benson:00, Seljak:00,Peacock:00,Scoccimarro:01,Berlind:02,Bullock:02}, abundance matching~\citep[AM; e.g.,][]{Klypin:99,Kravtsov:04,Tasitsiomi:04}, and sub-halo abundance matching \cite[SHAM; e.g.,][] {CSSTmocks}. With the improvement of computing capabilities, hydrodynamical simulations (for a recent review, see \citealt{Vogelsberger:20}) are now starting to be feasible for simulating cosmologically relevant volumes (\citealt{Dolag:16, Pillepich:18, Schaye:23}; see also the CAMELS project, the largest compilation of hydrodynamic simulations to date: \citealt{Villaescusa:23}), and are sometimes used to train and inform phenomenological methods to populate $N$-body simulations.

The production of simulated galaxy catalogues is a prolific line of development. Several cosmological surveys have produced simulations tailored to their observed samples but there are also more general-purpose simulated mocks. These galaxy catalogues include those based on the Uchuu simulation~\citep[e.g.,][]{Ereza:23,Dong-Paez:24} used for the SDSS data analysis, catalogues produced for the DESI survey~\citep[e.g.,][]{Smith:22,BAM,AbacusDESI}, catalogues developed within the Rubin-LSST DESC collaboration~\citep[e.g.,][]{DESCDC2a,DESCDC2b}, catalogues produced for the Chinese Space-station Survey Telescope~\citep[CSST,][]{CSSTmocks}, and general purpose galaxy catalogues~\citep[e.g.,][]{Cardinal,UniverseMachine}.   

%{\bf Flagship approach}

Within \Euclid, we have developed the Flagship simulation, a comprehensive simulation, in terms of including a vast number of consistent galaxy properties, to help optimise the mission and prepare its scientific analysis and exploitation.
The scientific goals of the mission from the main cosmological probes, weak gravitational lensing and galaxy clustering, set the requirements of the simulation in term of mass resolution, volume, and redshift coverage. Given that modelling \Euclid with full hydrodynamic simulations over the volume surveyed by the mission is computationally not possible at the moment, the approach we followed to create mock surveys was to develop a state-of-the-art $N$-body simulation and populate the gravitationally bound dark matter structures (haloes) with galaxies in a way that best matches observational data, placing special care into simulating consistently the weak lensing and clustering properties to enable combined probes analyses. 

%{\bf Flagship 1 and Flagship 2}

The first production was the \Euclid Flagship~1 simulation~\citep[FS1 hereafter,][]{Potter:17}. The $N$-body simulation was run on the Piz Daint supercomputer at the Swiss National Supercomputing Centre in 2016. A lightcone of dark matter (DM) particles was generated on the fly, replicating the simulated box (with periodic boundary conditions) as a way to fill the full lightcone volume of the \Euclid survey. In this scheme, we place the observer in one corner of the central box within the lightcone volume. The \texttt{ROCKSTAR} halo finder~\citep{Behroozi:13} was run on the DM particle distribution to generate a halo catalogue, which is now publicly available at the CosmoHub data distribution platform\footnote{\url{https://cosmohub.pic.es/catalogs/157}} \citep{Carretero:17,Tallada:20}. From the halo catalogue, we produced a galaxy catalogue that was used by the Euclid Collaboration to perform some early science analyses and performance assessments of the mission as a whole.
To improve the scope of the Flagship simulation, a second version, called Flagship~2 (Flagship or abbreviated as FS2 hereafter) was run in 2020. There were several improvements with respect to the first version. The mass resolution and the maximum redshift covered by the lightcone output were increased in order to improve the completeness of the resulting catalogue to encompass all the galaxies expected to be detected by \Euclid. 
In particular, the mass resolution increased by a factor of 2 in FS2, to reach a particle mass $m_{\rm p} = 10^9$ $h^{-1}\, \si{\solarmass}$, which in turn allows for modelling galaxies about one magnitude fainter than with FS1 at all redshifts (see Sect.~\ref{sec:galcat}), and the lightcone was extended from $z=2.3$ in FS1 to $z=3$ in FS2. We also changed the way in which the spectral energy distributions were assigned in FS2 to make the resulting photometric properties closer to those observed.
Similarly to the first version, a lightcone was produced on the fly (i.e., as the simulation run) and a halo catalogue was generated in post-processing with \texttt{ROCKSTAR}. Using HOD and AM techniques combined with relations between observational properties, we generated a galaxy catalogue containing around five billion objects covering one octant of the sky (see Sect.~\ref{sec:galcat} for further details). Positions, velocities, physical properties, lensing quantities, and photometry in multiple bands were computed for all galaxies, totalling 399 parameters per galaxy generating a catalogue of 5.9 terabytes that can be accessed through the  CosmoHub platform,\footnote{\url{https://cosmohub.pic.es}} which is hosted by the  \Euclid mission Spanish Science Data Center. This catalogue has been the  baseline input simulation for the \Euclid mission pipelines and a key ingredient for its scientific preparation before the satellite launch. In this regard, additional galaxy mocks, which will not be discussed in this paper, have been constructed within the Euclid Consortium, to address more probe-specific scientific questions and account for the variance due to modelling uncertainties.

%{\bf Paper outline}

This paper describes in detail the production of the second version of the \Euclid Flagship galaxy catalogue (FS2). Upon publication of this paper, we expect to make a public release of the latest version of the catalogue (version 2.1.10), which will be available at the Cosmohub web portal. This publication will serve as a reference for its usage. Although designed for the \Euclid mission, the catalogue can be very useful for many other studies and future galaxy surveys given its breadth in terms of number of galaxies simulated (e.g., 3.4 billion galaxies for a magnitude limited sample with $\HE<26$), volume covered (one octant of the sky up to $z=3$), and the wide range of galaxy properties computed that, in particular, allow us to model the galaxy clustering of both photometric and spectroscopic galaxy samples along with their weak lensing observables consistently down to sub-arcminute scales. The paper is structured as follows. In Sect.~\ref{sec:darkmatter}, we describe the production and main characteristics of the $N$-body FS2 dark matter simulation. In Sect.~\ref{sec:lens}, we present the computation of the lensing properties. In Sect.~\ref{sec:halocat}, we explain the production of the halo catalogue. In Sect.~\ref{sec:galcat}, we describe in detail the computation of the galaxy properties. The validation of the properties of the galaxy catalogue against observational constraints and theoretical models is presented in Sect.~\ref{sec:validation}. Finally, we summarise our findings and present our conclusions in Sect.~\ref{sec:summary}. Unless otherwise stated, all magnitudes reported in this paper are in the AB system.   

%--------------------------------------------------------------------
\section{Dark matter simulation \label{sec:darkmatter}}

\subsection{The Flagship run}

The \Euclid Flagship $N$-body dark matter simulation 
features a simulation box of 3600 \si{\hMpc} on a side with $16\,000^3$ particles, leading to a particle mass of $m_{\rm p} = 10^9$ $h^{-1}\, \si{\solarmass}$. This four trillion particle simulation is the largest $N$-body simulation performed to date and matches the basic science requirements of the mission as it allows us to accurately resolve $10^{11} h^{-1}\msun$ haloes which host the faintest galaxies \Euclid will observe (i.e., model a complete sample down to the \Euclid flux limit) and samples a cosmological volume comparable to the one that the satellite will survey. The simulation was performed using the \texttt{PKDGRAV3} code~\citep{Potter:16} on the Piz Daint supercomputer at the Swiss National Supercomputing Centre (CSCS). The simulation was run with a softening length of 4.5 $h^{-1}\,{\rm kpc}$.
It uses the \Euclid reference cosmology, with the following values for the matter density $\Omega_\mathrm{m} = 0.319$, baryon density $\Omega_\mathrm{b} = 0.049$, dark energy density (in the form of a cosmological constant) $\Omega_{\Lambda} = 0.681 - \Omega_\mathrm{r} - \Omega_{\nu}$, with a radiation density $\Omega_\mathrm{r} = 0.00005509$, and a contribution from massive neutrinos $\Omega_{\nu} = 0.00140343$ which is derived from the minimum neutrino mass possible ($0.0587$ eV) given the measured mixing angles and assuming a normal hierarchy 
\citep{DeSalas:18}. 
Besides, the values of the other cosmological parameters are: the equation of state parameter of dark energy $w_\mathrm{de} = -1.0$, the reduced Hubble parameter at redshift $z=0$ (i.e., Hubble constant), $h = 0.67$, the scalar spectral index of the initial fluctuations $n_\mathrm{s} = 0.96$, and the primordial power spectrum amplitude $A_\mathrm{s} = 2.1 \times 10^{-9}$ (i.e., $\sigma_8 \simeq 0.813$ derived) at $k_{\rm pivot} = 0.05 \,\rm Mpc^{-1}$. 

Using the \Euclid reference cosmology allows comparison to many other smaller simulations from $N$-body codes as well from approximate techniques that also use these reference values within the collaboration. The initial conditions were realised at $z = 99$ with first-order Lagrangian perturbation theory (1LPT) displacements from a uniform particle grid. The transfer functions for the density field and the velocity field were generated at this initial redshift by \texttt{CLASS}~\citep{Lesgourgues:11} and \texttt{CONCEPT}~\citep{Dakin:22}. As back-scaling was not used, all linear contributions from radiation, massive neutrinos, and metric perturbations (in the $N$-body gauge) were included via a lookup table and applied as a small corrective particle-mesh (PM) force at each time step \citep{Fidler:19}. This ensures a match to the linear evolution of the matter density field at all redshifts when including these additional linear terms (see Fig.~\ref{fig:flagship-pk}). 

The main $N$-body data product was produced on-the-fly during the simulation and is a continuous full-sky particle lightcone (to $z=3$), where each particle was output exactly when the shrinking light surface sweeps by it. The full-sky coverage was achieved by replicating the simulation box. The resulting catalogue contains 31 trillion particle positions and peculiar velocities (700 TB of data), and it was used to compute the roughly 125 billion \texttt{ROCKSTAR} main haloes and full-sky lensing maps with a \texttt{HEALPix} tessellation resolution $N_{\rm side} = 8192$~\citep{Gorski:05}, corresponding to $0\farcm{43}$ per pixel. Note that the mock galaxy catalogue described below was computed only in one octant of the sky. The chosen mock area was primarily based on computational efficiency, and it was deemed adequate to model the FS2 WIDE survey footprint for the first year data release (DR1), that will cover about 2500 deg$^2$, for which the mock was mainly designed.

\subsection{\texttt{PKDGRAV3} power spectrum validation}

Prior to performing the Flagship simulation, the \texttt{PKDGRAV3} $N$-body code was validated against the well-established \texttt{GADGET3}~\citep{springel:08}, \texttt{GADGET4}~\citep{Springel:21}, \texttt{ABACUS}~\citep{Garrison:21}, and \texttt{RAMSES}~\citep{Teyssier:02} codes, which each use very different methods to solve the Poisson equation as well as different methods to integrate the equations of motion. The results of these comparisons are given in \cite{Schneider:16} and more recently in \citeauthor{Springel:21} (\citeyear{Springel:21}, Fig.~50) and \citeauthor{Garrison:19} (\citeyear{Garrison:19}, Fig.~5). All codes agree at the 1\% level up to $k = 10\;h\,$Mpc$^{-1}$. Convergence of the power spectrum as a function of the particle mass and simulation box size was also investigated. Conservatively, a particle mass of $m_{\rm p} = 10^{9} h^{-1} M_{\odot}$ is required to ensure 1 \% convergence of the power spectrum up to $k = 10\;h\,$Mpc$^{-1}$. Simulation boxes larger than 1~$h^{-1}\,$Gpc are sufficient to ensure convergence in the power spectrum~\citep[e.g.,][]{Klypin:18}. However, the requirements of a light cone to $z=3$, with as little replication of the volume as possible, lead to a box of 3600 $h^{-1}$Mpc on a side. This simulated box contained 4 trillion dark-matter particles, which yields the mass resolution desired and also corresponds to the limiting number allowed by the Piz Daint supercomputer given \texttt{PKDGRAV3}'s memory requirements (which are about 64 bytes/particle, including the tree structure and all buffers for analysis and management of the Input/Output). 

\begin{figure}
\centering
\includegraphics[width=\columnwidth]{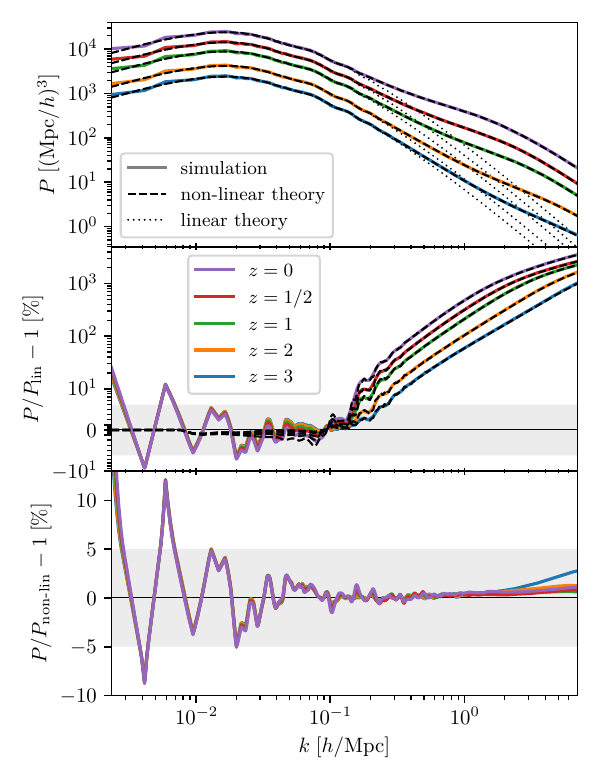}
\caption{\label{fig:flagship-pk} Nonlinear power spectrum at various redshifts compared to linear theory from \texttt{CLASS} \citep{Lesgourgues:11}. The use of the forward-scaling method in the $N$-body gauge \citep{Fidler:19} allows the $N$-body simulation to agree with linear theory at all redshifts including the effects from weak field GR, radiation, and massive neutrinos. The narrow $\pm 5\%$ band is plotted using a linear scale, making those fluctuations more visible and allowing also for negative values. Fluctuations at low $k$ are statistical and reflect the sample variance of the particular realisation. At high $k$ the enhanced power is due to nonlinear growth of structure not captured by the linear theory. Note that this nonlinear contribution also includes the slight $-2\%$ dip at $k \approx 10^{-1} \;h\,{\rm Mpc}^{-1}$, corresponding precisely to the first BAO peak in the power spectrum. The forward-scaling method used for the Flagship simulation guarantees a match to the linear theory of \texttt{CLASS} at all redshifts, which was not possible to achieve with the traditional back-scaling technique (which only guarantees this at $z = 0$ in the presence of massive neutrinos and radiation). The bottom panel compares the power spectra to the \Euclid Emulator nonlinear power spectrum \citep{EE2:21}. }
\end{figure}

In Fig.~\ref{fig:flagship-pk}, we compare the power spectrum measured from the \Euclid Flagship $N$-body dark matter simulation to linear theory computed by \texttt{CLASS} \citep{Lesgourgues:11} and to the \Euclid Emulator nonlinear power spectrum \citep{EE2:21}. 
Comparison to the \texttt{BACCO} emulator \citep{Angulo:21} shows a very similar level of agreement. The `spikes' at lower $k$ are due to the cosmic variance present in the realisation of the \Euclid Flagship $N$-body simulation. When comparing to other models of the nonlinear power spectrum, such as \texttt{Halofit} \citep{Takahashi:12} and \texttt{HMCode2020} \citep{hmcode:21}, the comparison is not quite as good, with deviations over redshift at $k > 0.1\;h\,{\rm Mpc}^{-1}$ extending to $\pm 5\%$, and notably, these models do not accurately capture the nonlinear form of the baryonic wiggles at $k$ approximately $0.1$--$0.4\;h\,{\rm Mpc}^{-1}$.

\subsection{Dark matter clustering}

\begin{figure*}
    \centering
    \vspace{-10 pt}
    % \subcaptionbox{}{
    \includegraphics[width=0.45\textwidth]{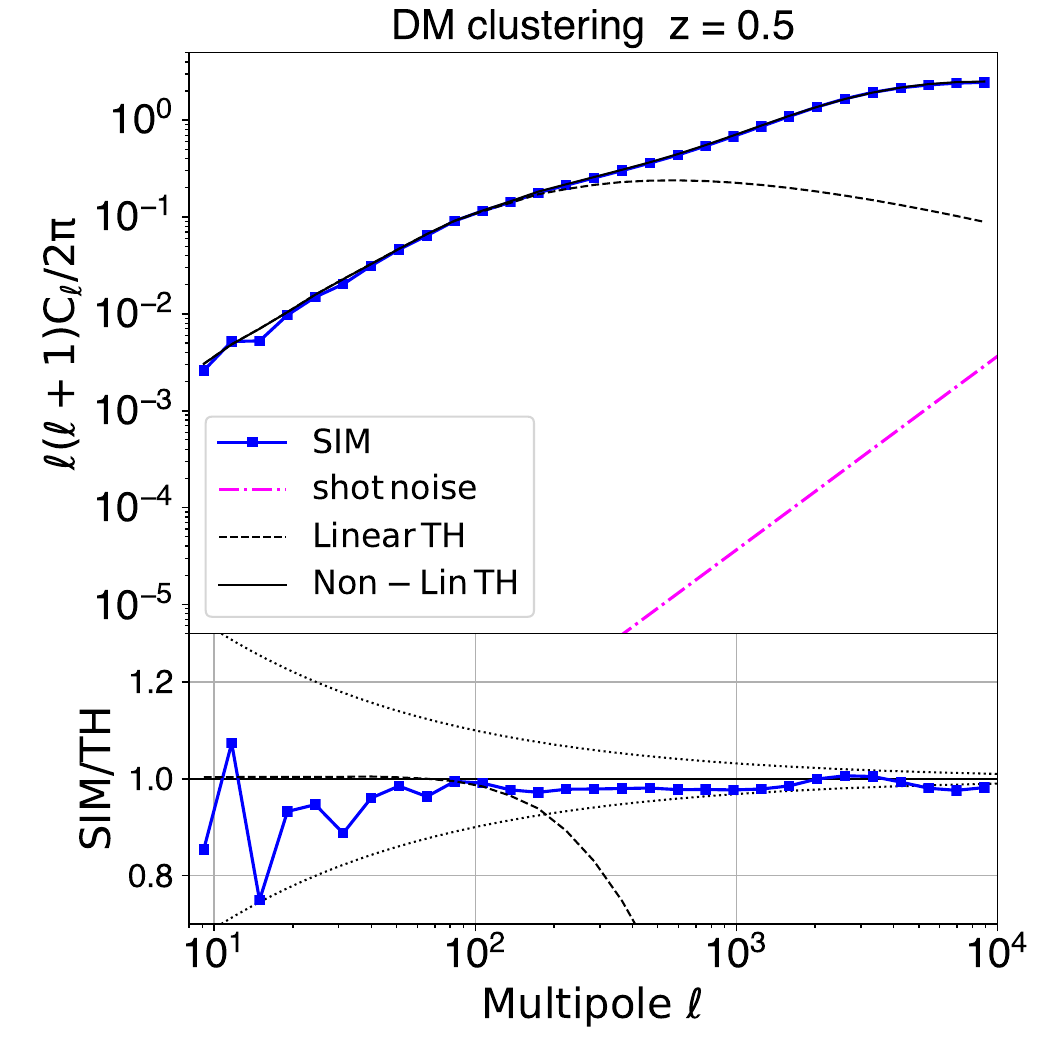}%
     \qquad
      % \subcaptionbox{}{
      \includegraphics[width=0.45\textwidth]{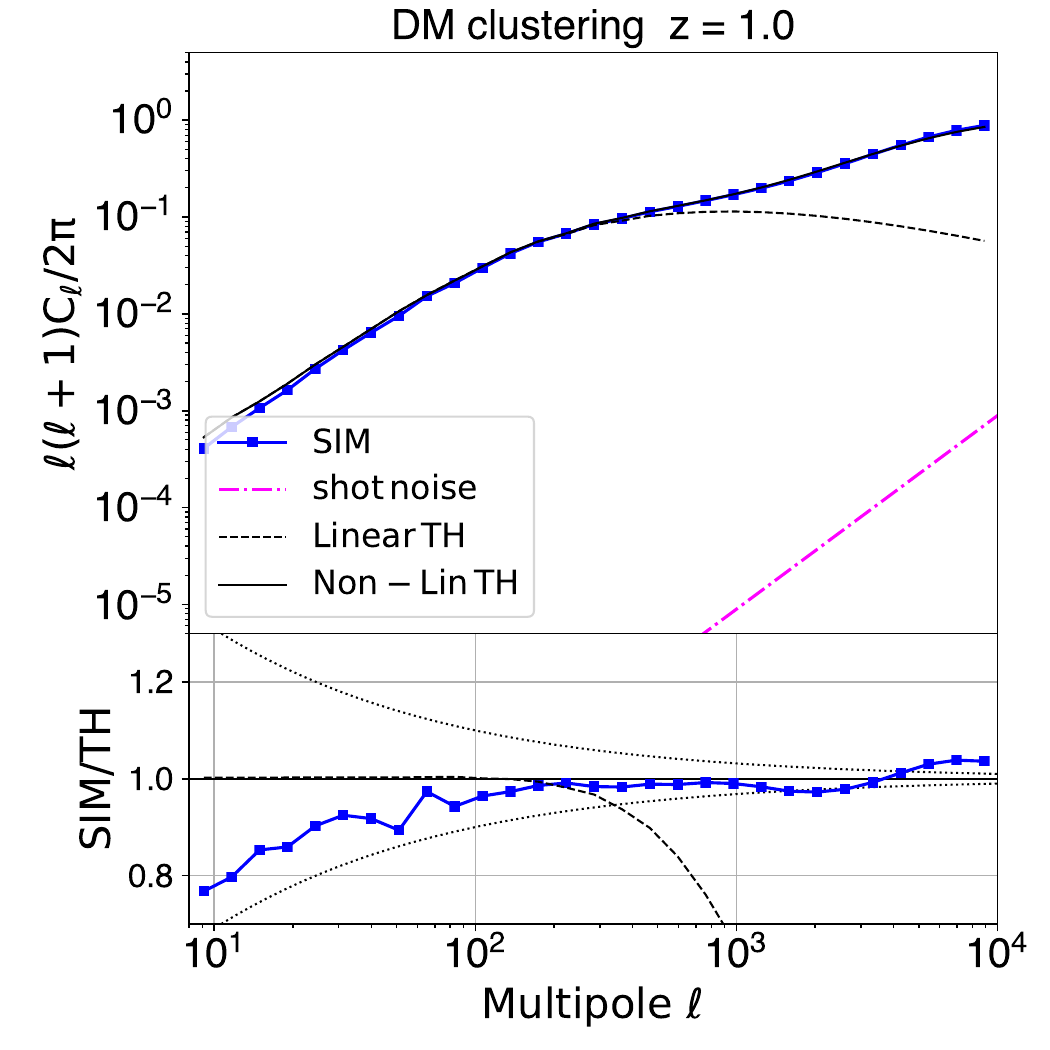}%
     \qquad
    % \subcaptionbox{}{
    \includegraphics[width=0.45\textwidth]{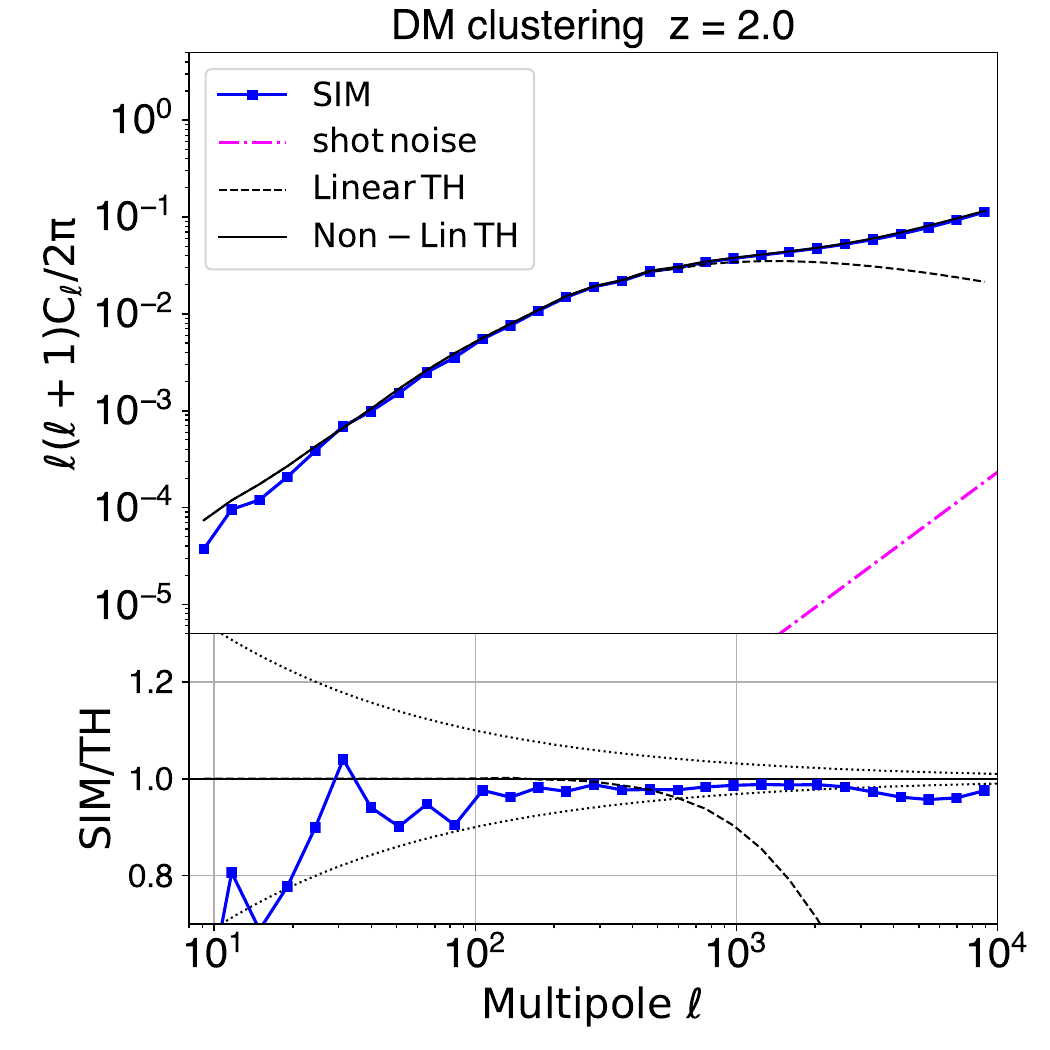}
    \qquad
    % \subcaptionbox{}
    \includegraphics[width=0.45\textwidth]{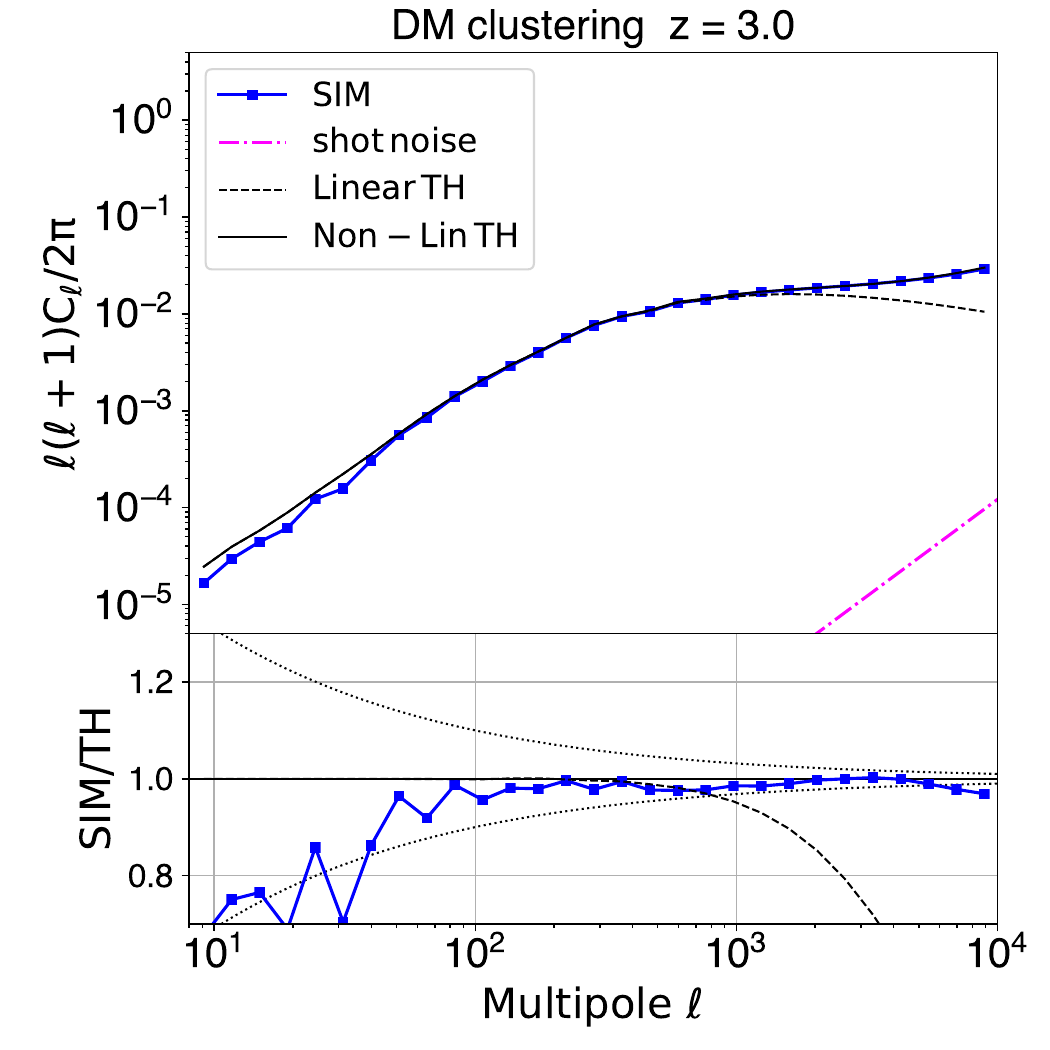}%

      %\vspace{-10 pt}
    \caption{Angular power spectrum of the dark matter field in the lightcone across different redshifts. Plots show the clustering in the following z-bins: $0.49 < z <  0.51$ (top left), $0.99 < z < 1.00$ (top right), $2.02 < z< 2.08$ (bottom left) and $2.94 < z < 3.06$ (bottom right). The clustering in the simulation (symbols) is compared against linear (dashed) and nonlinear (solid) theoretical predictions \citep{Takahashi:12}. Particle shot-noise is also shown for reference (dot-dashed line).  Residuals with respect to nonlinear theory are displayed in the lower panels, along with sample variance error envelopes (dotted). 
    }
    \label{fig:clsmatter}
\end{figure*}

Galaxy clustering is one of the main probes of the \Euclid mission. In order to validate this probe in the galaxy mock, we first
study whether the clustering of the dark matter distribution in the lightcone behaves according to theoretical expectations. Figure~\ref{fig:clsmatter} shows the angular power spectrum of dark matter in thin redshift shells (in the \texttt{HEALPix} tessellation, see Sect.~\ref{sec:lens} for details), across the full depth of the Flagship $N$-body lightcone, i.e., $0.5 < z < 3$. Measurements in the simulation make use of the {\tt PolSpice} code (see \citealt{Szapudi01, Chon04, Fosalba04})\footnote{ \url{https://www2.iap.fr/users/hivon/software/PolSpice}} which corrects for the effect of angular masks in our finite-sky analysis.

Results show that measurements in the simulation agree with linear theory expectations on large scales (low multipoles) and nonlinear theory \citep{Takahashi:12} down to very small scales (high multipoles, $\ell \sim 10^4$), within sample variance errors (see dotted envelopes in lower panels for each redshift bin). We note that particle shot-noise is negligible ($< 1\%$ for all multipoles) given the high particle density, around 90\,particles/$(h^{-1}\,{\rm Mpc})^3$, in the lightcone. The agreement between Flagship measurements and the Euclid Emulator2 (EE2) predictions is expected to be at a very similar level of agreement, as discussed in \citet[][see in particular their Figure 13]{EE2:21} where they show that the EE2 and halofit agree within 3$\%$ up to very small (nonlinear) scales, $k < 4\;h\,{\rm Mpc}^{-1}$, for $z<2$.

%--------------------------------------------------------------------
\section{\texttt{HEALPix} lensing mass maps\label{sec:lens}}

Following the approach presented in~\cite{Fosalba:08} and~\cite{Fosalba:15b}, we construct a lightcone simulation by replicating the simulation box (and translating it) around the observer.  
Given the large box-size used for the Flagship simulation, $L_{\rm box}=3600\, \si{\hMpc}$, this approach allows us to build
all-sky lensing outputs without repetition up to $z\sim2.0$ and with one replication up to our maximum redshift, $z_{\rm max}=3$. 
Then, we decompose the dark matter lightcone into a set of all-sky concentric spherical shells of given width, $\Delta_r$, around the observer, what we call the `onion universe'. Each dark matter `onion shell' is then projected onto a 2D pixelised map using the \texttt{HEALPix} tessellation~\citep{Gorski:05}.  
For the lensing maps presented in this paper we have chosen a shell width of $\Delta_r \approx 95.22$
megayears in lookback time, up to $z=10$ (and finer at higher redshifts), 
and an angular resolution of $\Delta_{\theta} \approx \sqrt{3/\pi} \, 3600/
N_{\rm side} \approx \ang{;0.43;}$, for the \texttt{HEALPix} map resolution
$N_{\rm side} =8192$ that we use. 

By combining the dark matter `onion shells'  that make up the
lightcone, we can easily derive lensing observables, as explained in \cite{Fosalba:08}.  This approach, based on approximating the
observables by a discrete sum of  2D dark-matter density maps
multiplied by the appropriate lensing weights, agrees with the much more complex and CPU time consuming ray-tracing technique in the Born approximation limit, i.e., in the limit where lensing deflections are calculated using unperturbed light paths (see e.g, \citealt{Hilbert:2020}).

\begin{figure*}
\includegraphics[width=\textwidth]{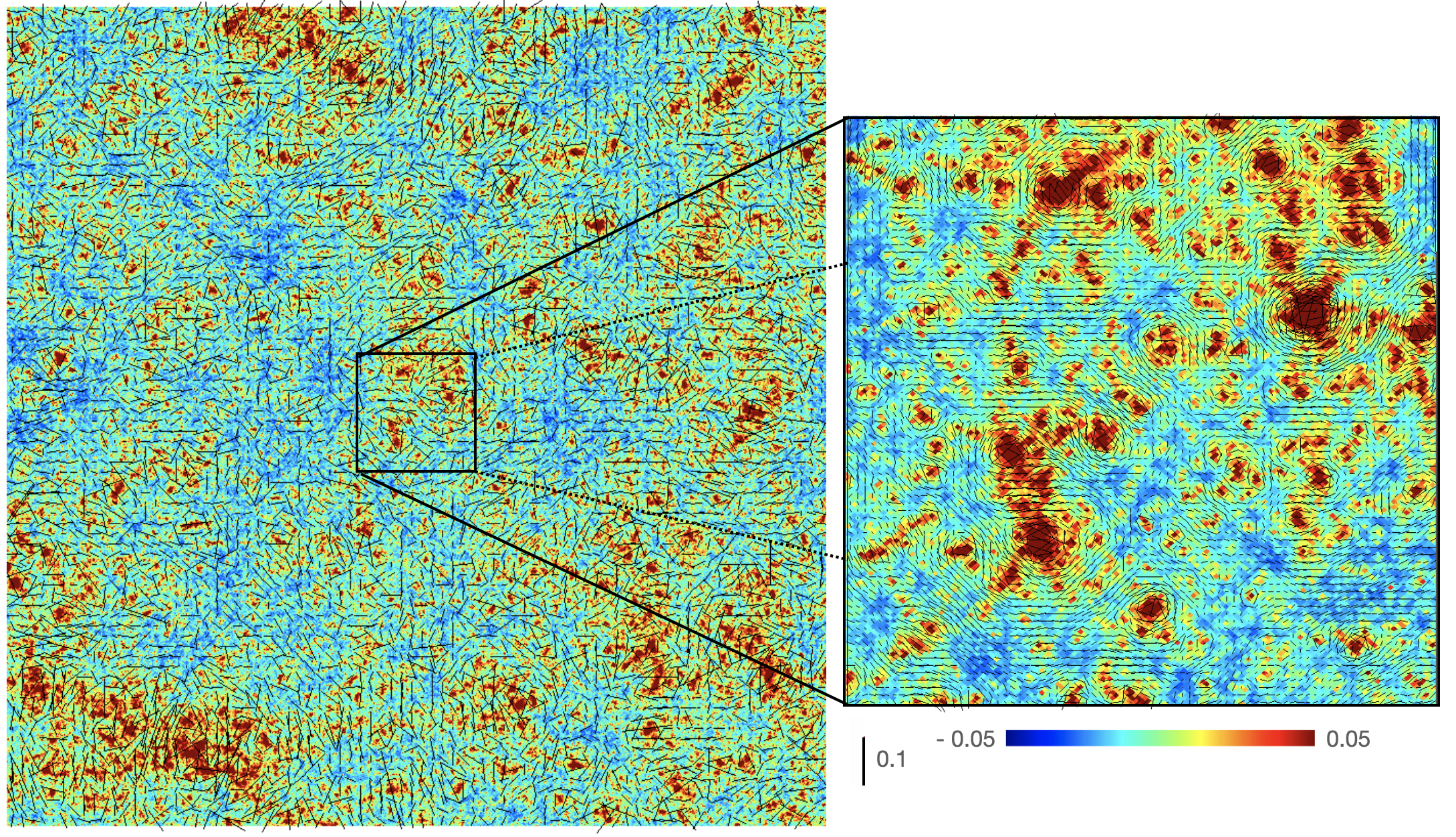}
    \caption{Lensing convergence (colour coded) and shear field (sticks) for sources at $z_{\rm s}=1$ over a patch of $50\,{\rm deg}^2$ (left) and in a zoom-in of the central $1\,{\rm deg}^2$ area (right). The convergence field colour bar displays values within the range $\pm 3\,\sigma$, where $\sigma$ is the rms value of the full-sky map. The stick at the bottom of the zoom-in image shows a reference amplitude for the shear sticks overlaid on that area of the mass map.}
    \label{fig:lensingmap}
\end{figure*}

Following this technique we are able to produce all-sky maps of the convergence field (which is simply related to the lensing potential in harmonic space), as well
as maps for other lensing fields obtained from covariant derivatives of the lensing potential, such as the deflection angle, shear, flexion, etc. 
Figure~\ref{fig:lensingmap} shows the all-sky map of the convergence field, $\kappa$, for the Flagship simulation, for sources at $z_s=1$, with a pixel resolution of \ang{;0.43;}. The colour scale shown spans over the range $-\sigma < \kappa< 3\, \sigma$, where $\sigma$ is the root mean square (rms) fluctuation of the full-sky convergence map.

The angular power spectrum of the convergence field in the Born approximation reads (for a flat LCDM cosmology),
\begin{equation}
C_{\ell}^{\kappa} =
{\frac{9H_0^4\Omega_{\rm m}^2}{4c^4}} \, \int \, \mathrm{d}r \, P(\ell/r,r) 
\frac{(r_{\rm s}-r)^2}{r_{\rm s}^2~a^2} \;,
\label{eq:clkappa}
\end{equation}
where $\ell$ is the multipole order, $H_0 = 100\, h\, \kmsMpc$ is the Hubble constant, $c$ is the speed of light,  and $r_{\rm s}$ is the comoving distance to the lensing sources (we assume all sources are located at the same redshift in this approximation) where $P(\ell/r,r)$ is the 3D density power spectrum in the simulation at a given comoving distance $r$ from the observer.

\begin{figure}
    \centering
    \vspace{-10 pt}
    \includegraphics[width=0.95\hsize]{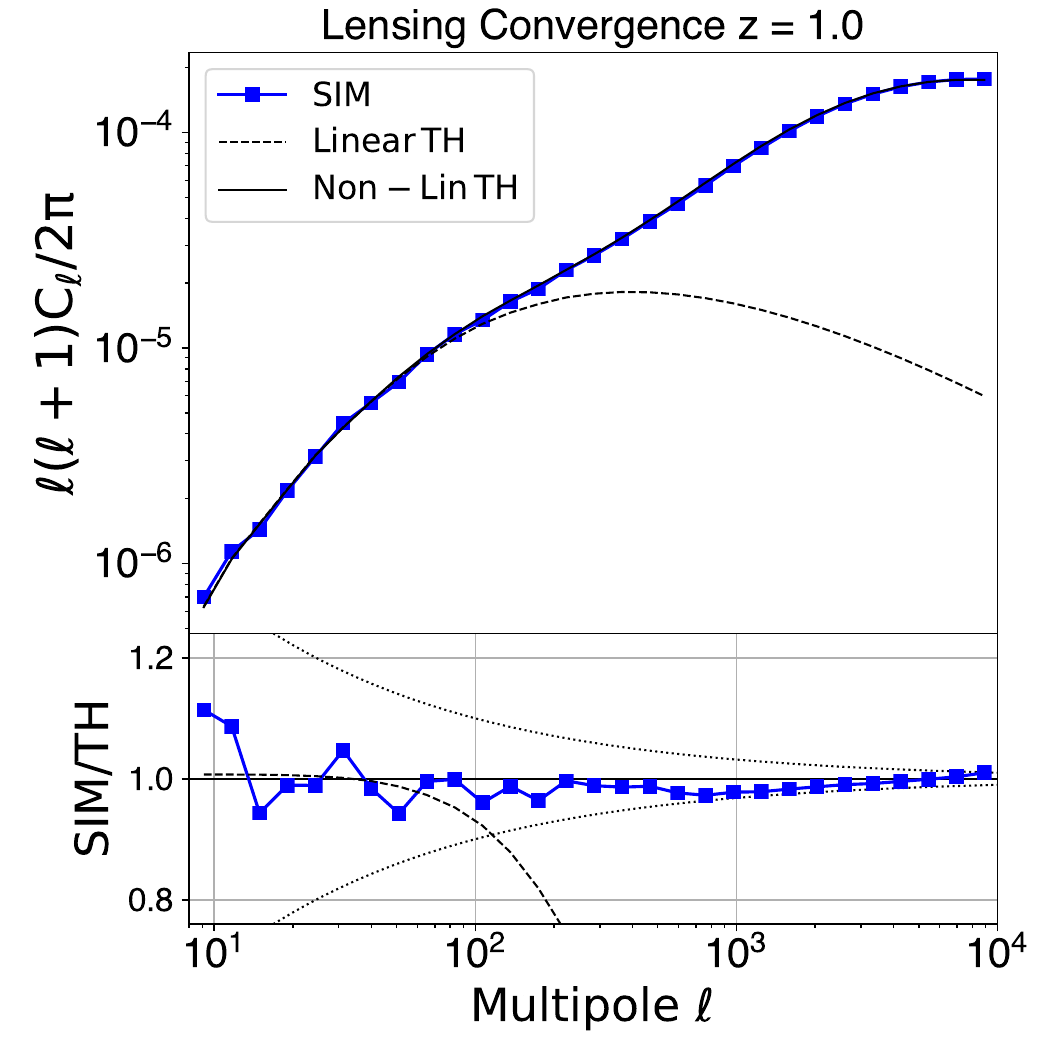}%
      
      \includegraphics[width=0.95\hsize]{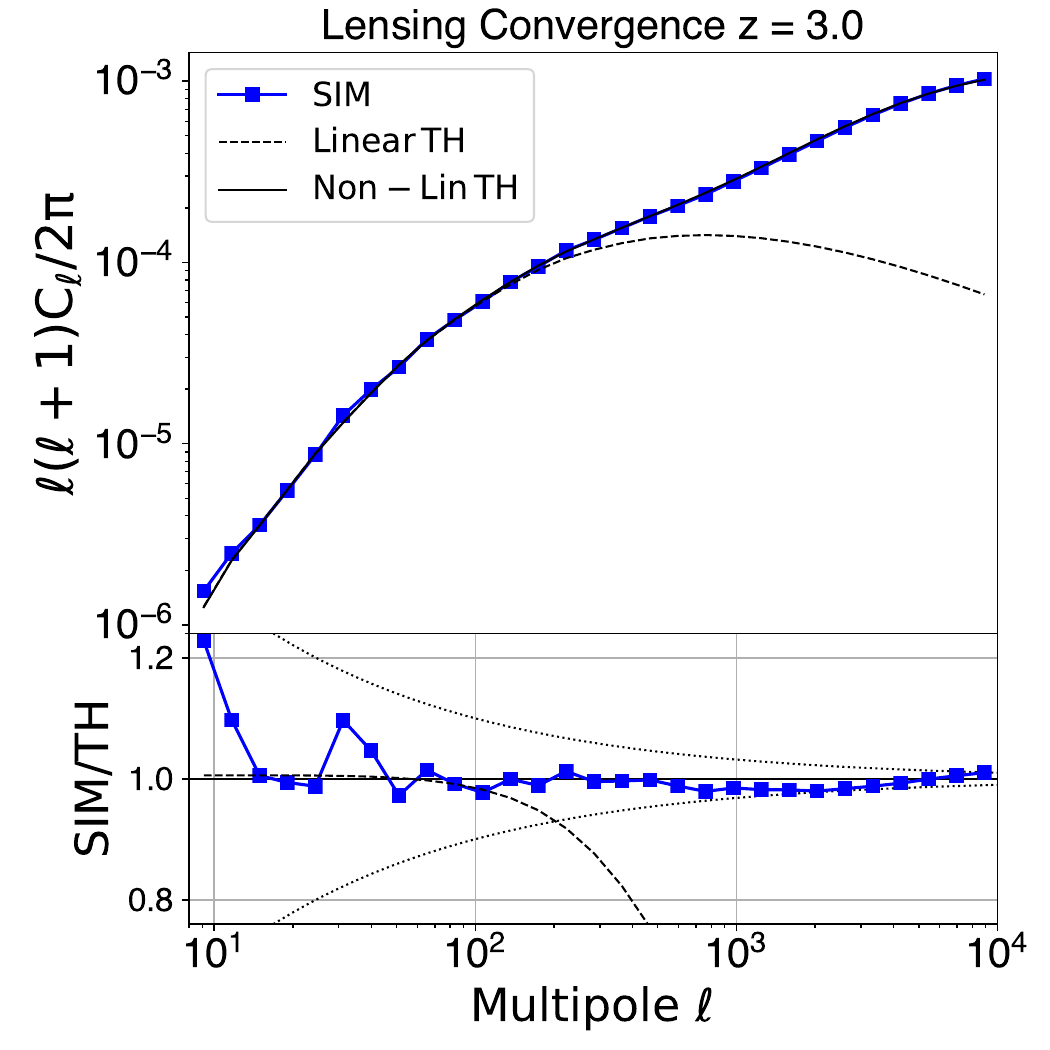}%

    \caption{Angular power spectrum of the convergence field at $z_{\rm s}=1$ (top) and $z_{\rm s}=3$ (bottom). Plots show measurements in the simulation (symbols) compared against linear theory (dashed line) and nonlinear (solid line) fits to simulations \citep{Takahashi:12}. The lower panels show the ratio between the simulation and nonlinear theory. Sample variance error envelopes are displayed as dotted lines.
    }%
    \label{fig:clskappa}%
\end{figure}

In this approach, we can take the spherical transform of the
lensing potential all-sky map to obtain the corresponding maps for the other weak-lensing observables through simple relations in harmonic space (see~\citealt{Hu:00}). 
In particular, the convergence field, $\kappa$, is related to the lensing potential, $\phi$, through the 2D equivalent to the usual (3D) Poisson equation, which in spherical harmonic decomposition reads
\begin{equation}
\kappa_{\ell m} = -\frac{1}{2}\ell(\ell+1) \phi_{\ell m}\;.
\end{equation}
One can thus use this expression to derive the lensing potential
at each source plane (or 2D lightcone map), and obtain other
lensing observables, such as deflection and shear, through their relation to the lensing potential in harmonic space \citep[see][for details]{Fosalba:15b}.
As a basic validation of the mass maps, Fig.~\ref{fig:clskappa} shows the measurement of the convergence angular power spectrum in the simulation compared with theory predictions, for two different source redshifts across the lightcone. Overall there is good agreement between the mass map clustering compared to theory in the full range of scales (multipoles) shown, given the sample variance errors (see figure caption for details).

%--------------------------------------------------------------------

\section{Halo catalogue
\label{sec:halocat}}

The dark matter haloes were identified directly on the lightcone particle data using the \texttt{ROCKSTAR} halo finder~\citep{Behroozi:13}. 
\texttt{ROCKSTAR} is a phase space-linking friends-of-friends method that is able to find the hierarchy of substructure from parent dark matter haloes to the smallest subhaloes. \texttt{ROCKSTAR} is also a high-performance parallel halo finder; however, it is not capable of handling such a massive (10 trillion particle) simulation in its standard (public) version. In order to use it for finding haloes in the FS2 particle lightcone data, we had to split the data into computational `bricks' of $375\,h^{-1}\,{\rm Mpc}\times375\,h^{-1}\,{\rm Mpc}\times 625\, h^{-1}\,{\rm Mpc}$, each with an extended `ghost' region of 5 $h^{-1}\,{\rm Mpc}$ on each side to avoid discontinuities. The full particle lightcone comprises 3448 such computational bricks, each of which could be computed independently on a cloud of (56 core) servers at the University of Zurich. One complication in the processing is that the data in the lightcone is over a variable expansion factor, $a$, as a function of depth in the lightcone, changing from 1.0 at the centre to 0.25 ($z=3.0$) at the edge. This fact must be accounted for when converting the particle momenta to physical peculiar velocities.\footnote{In \texttt{PKDGRAV3}'s internal units, particle momenta are computed as $p = a^2 \dot{x}$ and physical peculiar velocities as $v = a \dot{x}$.} \texttt{ROCKSTAR} was written to compute halo catalogues from a set of simulation snapshots, each at a fixed expansion factor, and not a lightcone and, therefore, needed to be modified to handle particles with radially dependent expansion factor. \texttt{ROCKSTAR} uses these peculiar velocities for both linking (where the linking length in velocity space is adapted from halo to halo) as well as for `unbinding', the process of removing particles from a halo that are deemed not to be gravitationally bound to it (in isolation). Once all haloes within a brick have been found, the parent haloes with centres in the ghost region and subhaloes of such parent haloes are removed from the catalogue so that the individual bricks fit seamlessly together.  

The science reach of \Euclid depends on how much volume it can sample and how many tracers it can use for cosmological analysis. Its design was optimised to obtain the most stringent constraints on cosmological parameters. For its weak lensing analysis, it reaches magnitude limits of $\IE \simeq 24.5$. 
The default minimum number of particles for \texttt{ROCKSTAR} to define a halo is set to 20. However, we set the minimum number of particles to define a halo to 10 to be complete at the \Euclid weak lensing magnitude limit (see Sect.~\ref{sect:incompleteness}). As we use such small and poorly resolved haloes,  
we correct the halo masses of haloes with few particles to avoid incompleteness and discreteness effects in the halo mass function (Sect.~\ref{sec:mass_corrections}). The final all-sky halo catalogue contains 126 billion main haloes. The galaxy catalogue is generated from the halo catalogue in just one octant of the lightcone that contains 15.8 billion main haloes.

\subsection{Halo mass function
\label{sec:halo_mass_function}}

The \texttt{ROCKSTAR} halo finder produces different estimates of the halo mass. These estimates include: the mass, $M_{\rm fof}$, of the particles linked together with a friends-of-friends algorithm of linking length $b=0.2$; the mass, $M_{\rm vir}$, contained within the virial radius; the sum of the mass, $M_{\rm bound}$, of the bound particles within the virial radius; the mass, $M_{\rm 200b}$, of the particles within an overdensity of 200 relative to the background density; and the mass, $M_{\rm 200c}$, of the particles within an overdensity of 200 relative to the critical density. Appendix~\ref{app:hmf_defs} provides a comparison of the halo mass function (HMF) for the different halo mass definitions.

Based on the similar behaviour of the HMF with the different mass estimates (except for the friends-of-friends mass definition), we decided to choose the $M_{\rm bound}$ definition as a default choice to build the galaxy catalogue. As our method of assigning galaxy luminosities is based on AM, the particular choice of mass estimate is not important.

\begin{figure}
\center
\includegraphics[width=\columnwidth]{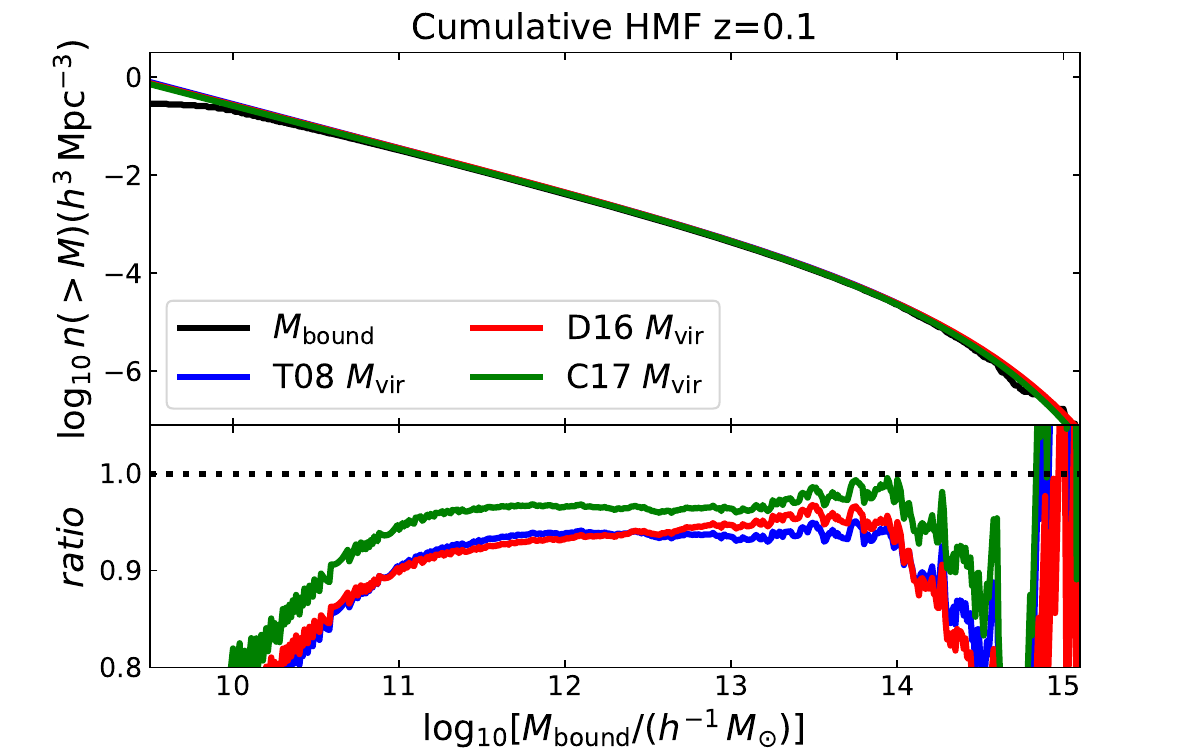}
\caption{\label{fig:cumHMF_bound_others_z01}
Cumulative $M_{\rm bound}$ HMF of the Flagship haloes compared to the T08, D16 and C17 halo mass functions at redshift $z=0.1$. 
In the top panel we show the cumulative HMFs (colours indicated in the legend). The lower panel shows the ratio of the cumulative $M_{\rm bound}$ HMF to the other cumulative HMFs. The dotted line serves as a reference point when cumulative HMFs are the same. 
}
\end{figure}

We compare the $M_{\rm bound}$ HMF to other HMFs in the literature. We use as main reference the~\cite{Tinker:08} HMF, hereafter T08, as it has been widely used in the literature for HMF comparison. We also compare it to the HMFs of~\cite{Despali:16}, D16, and of~\cite{Comparat:17}, C17. We use the {\tt hmf}\footnote{\url{https://hmf.readthedocs.io/en/latest/}}~\citep{Murray:2013} and {\tt COLOSSUS}\footnote{\url{http://www.benediktdiemer.com/code/colossus/}}~\citep{Diemer:18} packages to compute the HMFs using the reference Flagship cosmological parameters. Figure~\ref{fig:cumHMF_bound_others_z01} shows the cumulative HMFs at low redshift, $z=0.1$ in the top panel, and the ratio of the $M_{\rm bound}$ HMF to the other HMFs in the lower panel. We compute the T08, D16 and C17 HMFs for the $M_{\rm vir}$ mass definition. Our $M_{\rm bound}$ is almost the same as $M_{\rm vir}$ for the most massive haloes but differs for the lowest mass haloes due to the unbinding of particles. There is an overall offset of around 3--7\% lower abundance in the $M_{\rm bound}$ HMF compared to the other predictions for the same cosmology. For halo masses
below $M_{\rm bound} = 10^{11.5}\, h^{-1}\,\si{\solarmass}$, equivalent to $\sim$300 particles, the $M_{\rm bound}$ HMF starts to be incomplete. 
The differential HMF shows the same trends as the cumulative HMF. The lightcone has little volume at $z=0.1$ and therefore the HMF is very noisy above  $M_{\rm bound} = 10^{13.5}\, h^{-1}\,\si{\solarmass}$.

\begin{figure}
\center
\includegraphics[width=\columnwidth]{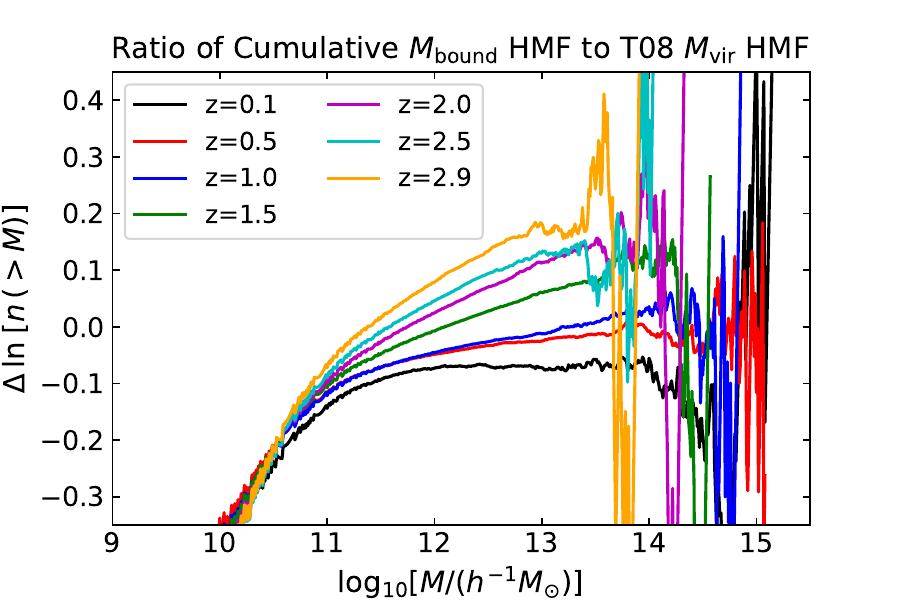}
\caption{\label{fig:ratio_bound_to_Tinker_allz}
Natural logarithm of the ratio of the cumulative $M_{\rm bound}$ HMF to the T08 $M_{\rm vir}$ HMF at several redshifts indicated in the figure legend. 
The normalisation between the two varies mostly within a 10\% difference range. The slope of the Flagship cumulative $M_{\rm bound}$ HMF is shallower than the one of the T08 $M_{\rm vir}$ HMF at high redshift.}
\end{figure}

Figure~\ref{fig:ratio_bound_to_Tinker_allz} shows the ratio of the cumulative $M_{\rm bound}$ HMF to the T08 $M_{\rm vir}$ HMF at several redshifts spanning the redshift range of the simulation lightcone. While the slopes of the HMFs in the mass range $10^{11.5}\, h^{-1}\,\si{\solarmass} \lesssim M_{\rm bound} \lesssim 10^{13.5}\, h^{-1}\,\si{\solarmass}$ are similar at low redshift, $z\lesssim 1$, at higher redshift, the slope of the $M_{\rm bound}$ HMF progressively gets shallower than the T08 $M_{\rm vir}$ HMF. The ratio of the abundance at a given $M_{\rm bound}$ halo mass compared to T08 abundance at the same halo mass also increases with redshift. Part of the difference may be due to the different power spectrum transfer function used in the Flagship run compared to the input we have given to the \texttt{hmf} code to compute the T08 HMF, generated with \texttt{CAMB}~\citep{Lewis:00} for the same cosmology.
We have performed the same comparison to the D16 and C17 HMFs (not shown in the figure) finding qualitatively the same result.
Nevertheless, the photometric quantities we compute for the mock galaxies depend on an abundance matching technique and therefore are not affected by small changes in the HMF.
The observed luminosity function is recovered for the galaxies by construction in the AM technique despite any mismatch or incompleteness observed in the mass function of the dark matter haloes.

\subsection{Mass corrections\label{sec:mass_corrections}}

As we will see later, we assign galaxy luminosities to central galaxies using a halo mass-luminosity relation derived from abundance matching between the cumulative HMF and the cumulative galaxy luminosity function. We push the detection of dark matter haloes to the very low limit of ten particles, making the effects of discreteness very noticeable at the low mass end of the HMF. Furthermore, as mentioned above, below halo masses of $\logten [{ M}/(h^{-1}\,\si{\solarmass})] \lesssim 11.5$ our HMF starts to be incomplete. In order to produce galaxy luminosities that are not discrete and incomplete, we need to correct the HMF for these two effects. 

\subsubsection{Incompleteness correction}\label{sect:incompleteness}
In order to reach the faint absolute magnitudes that \Euclid will observe, we need to detect haloes down to the corresponding low masses. For a \Euclid magnitude limit of $\IE = 24.5$, we need to reach absolute magnitudes around $M_{\IE} -5\logten h \simeq -13.0$ to be complete at redshifts $z\gtrsim0.1$. 
To reach this absolute magnitude limit, we need to reach a mass limit of ${M}\simeq 2\times10^{10} h^{-1} M_{\odot}$. Given the resolution of the simulation, this mass corresponds to 20 particles. As we re-scale the halo masses to account for the HMF incompleteness, we need to push down to haloes identified with at least 10 particles.  With the rejection of unbound particles, the $M_{\rm bound}$ halo definition can have even fewer particles contributing to the halo mass.
At this particle mass threshold, the halo catalogue is not complete. Nevertheless, as the two-point correlation of haloes is approximately independent of mass at low masses, the two-point correlation properties of all haloes detected will not differ from the one it would have had if the catalogue had been complete. That way, we can reassign the halo masses with abundance matching and assume that we are complete to the halo abundance given by the lowest number of particles and that the two-point clustering, which we use to calibrate the galaxy mock, will not change.

\begin{figure}
\center
\includegraphics[width=\columnwidth]{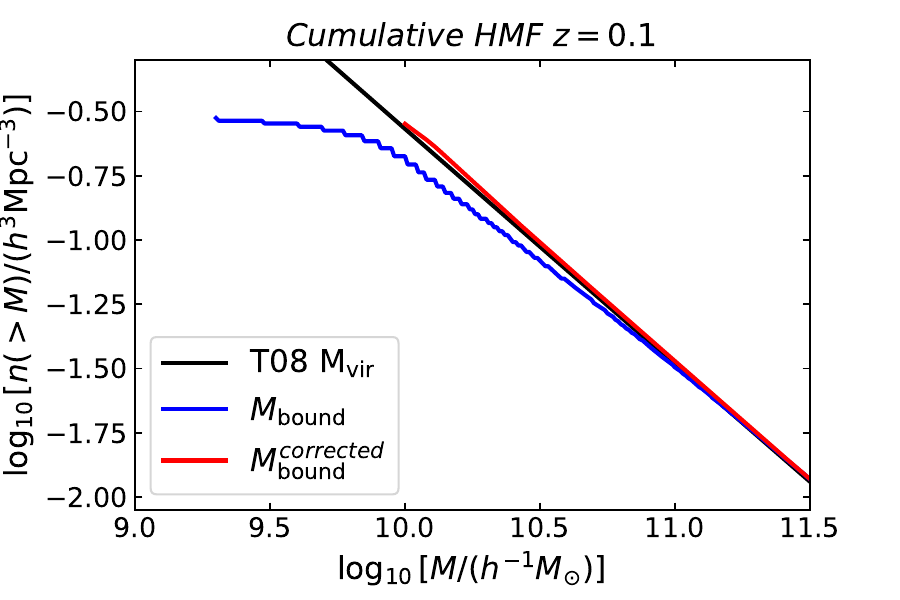}
\caption{\label{fig:cumHMF_z01_AM_incompleteness} Cumulative HMF at redshift $z\sim0.1$ for the $M_{\rm bound}$ definition (blue), the T08 $M_{\rm vir}$ HMF (black) and the one resulting from the mass reassignment procedure (red) that corrects for incompleteness and discreteness effects.}
\end{figure}

We correct for incompleteness by reassigning the halo masses in the following way. 
We assume that the slope of the cumulative HMF at low masses is the same as the T08 cumulative HMF. Given that the Flagship halo abundance for the $M_{\rm bound}$ definition is somewhat lower than the T08, we adjust the abundance at a mass $\logten [{M}/(h^{-1}\,\si{\solarmass})] = 11.5$, which corresponds to approximately 300 particles per halo. Above this mass threshold, there seems to be no incompleteness due to the low number of particles (see Fig.~\ref{fig:cumHMF_z01_AM_incompleteness}).
We reassign the halo masses below this threshold to have the same abundance that a fiducial HMF constructed with the same faint-end slope of the T08 cumulative HMF and normalised to the Flagship cumulative HMF at $\logten [{M}/(h^{-1}\,\si{\solarmass})] = 11.5$.
The process is captured in Fig.~\ref{fig:cumHMF_z01_AM_incompleteness} where the original cumulative HMF at redshift $z\sim0.1$ is shown in blue, the T08 HMF is shown in orange, and the resulting cumulative halo mass after the abundance matching procedure is shown in red. The new cumulative HMF has the same faint-end slope as the T08 HMF and the normalisation of the original HMF by construction.

While this procedure is conceptually simple, implementing it directly into the mock generation is too slow, as one needs to compute the observed cumulative HMF and to invert the T08 cumulative HMF for each galaxy. We therefore developed a faster way of implementing this correction. First, we compute the relation between the original $M_{\rm bound}$ halo mass and the abundance-matched halo mass. 
In order to be able to compute this correction efficiently, we fit this relation with five parameters at all redshifts in intervals of 0.1 in redshift. We then fit each parameter as a function of redshift. We also fit the offsets of the cumulative HMF to the T08 values at $\logten \,[{M}/(h^{-1}\,\si{\solarmass})] = 11.5$.

\subsubsection{Discreteness correction}

We correct for discreteness by assuming that the cumulative HMF is well defined at the mass values corresponding to a given integer number of particles. We proceed as follows. In a given volume $V$ for each number of particles $p_i$, we have $N_i$ haloes with those $p_i$ particles corresponding to a halo mass $M_i = p_i\,m_\mathrm{p}$. The cumulative abundance for this halo mass is $n_i (\geqslant M_i) = \sum_{i=1}^{\infty} N_i / V$, that is, the number of haloes with mass larger or equal than $M_i$ divided by the volume. For all haloes with $p_i$ particles, we want to reassign their masses to distribute them according to a power-law distribution in abundance between masses $M_i$ and $M_{i+1}$. For low masses, it is a good approximation to assume that the cumulative HMF behaves as a power law, 
\begin{equation}
    {\rm CHMF} (\geqslant M_i) \simeq 10^{\alpha} M_i^{\beta}\;,
\end{equation} 
in a small range of masses. For each halo with $p_i$ particles and mass $M_i$, we draw a random number $u$ uniformly distributed in the range $[0,1]$ and use this number to obtain a halo mass $M$ from the normalised cumulative halo mass distribution between $M_i$ and $M_{i+1}$,
\begin{equation}
    u = \frac{{\rm CHMF}(\geqslant M) - {\rm CHMF}(\geqslant M_{i+1})}{{\rm CHMF}(\geqslant M_i) - {\rm CHMF}(\geqslant M_{i+1})}\;,
\end{equation}
assigning a halo mass
\begin{equation}
    M = \left[ u (M_i^{\beta} - M_{i+1}^{\beta}) + M_{i+1}^{\beta}\right]^{\frac{1}{\beta}}\;.
\end{equation}
So, in order to make the halo mass distribution continuous, we only need to know the slope $\beta$ of the cumulative HMF as a function of mass and also as a function of redshift (as it evolves). We produce functional fits to the value of the slope to be able to have the slope at any value of the mass and redshift faster than by computing the cumulative HMF in every step. 

\subsection{Halo clustering}

\begin{figure*}
    \centering
    \vspace{-10 pt}
    % \subcaptionbox{}{
    \includegraphics[width=0.3\textwidth]{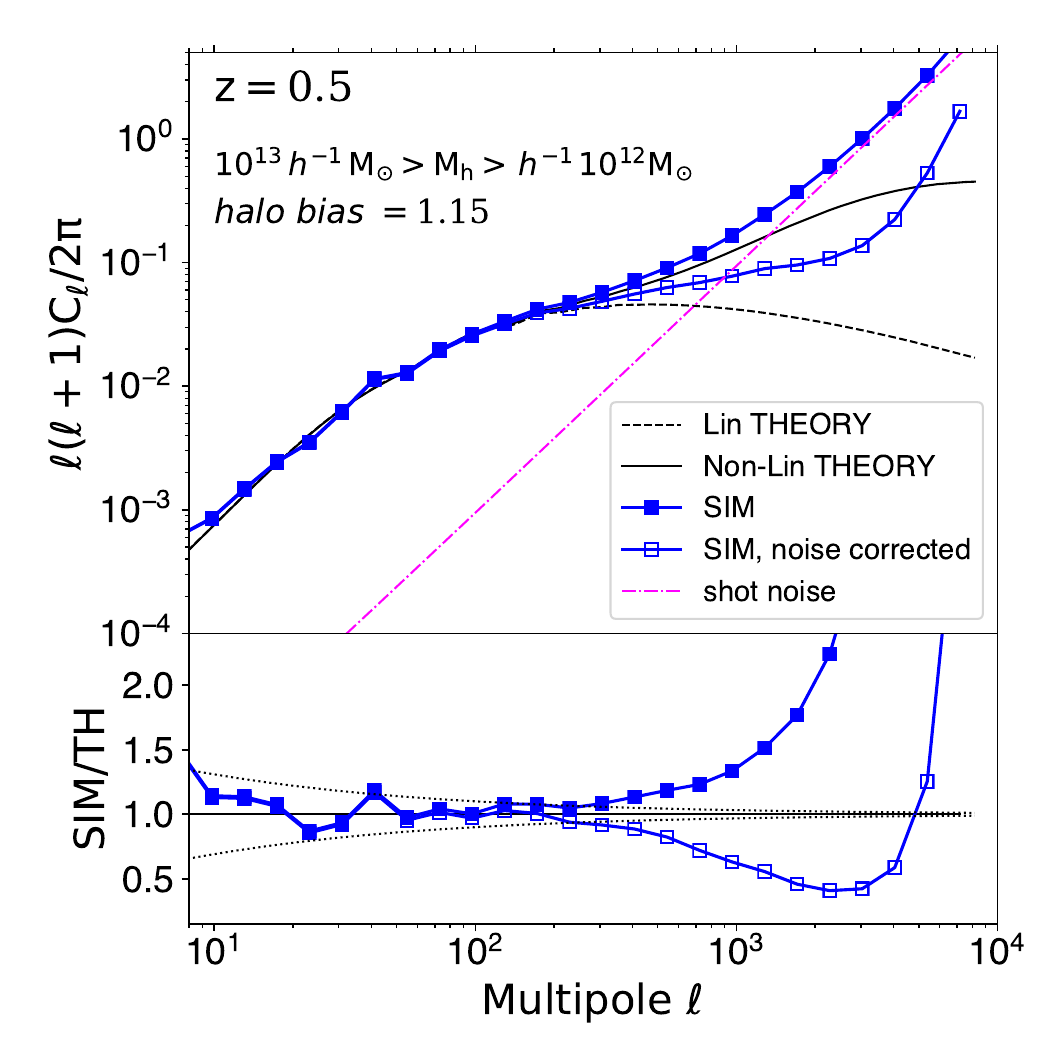}%
     \qquad
      % \subcaptionbox{}{
      \includegraphics[width=0.3\textwidth]{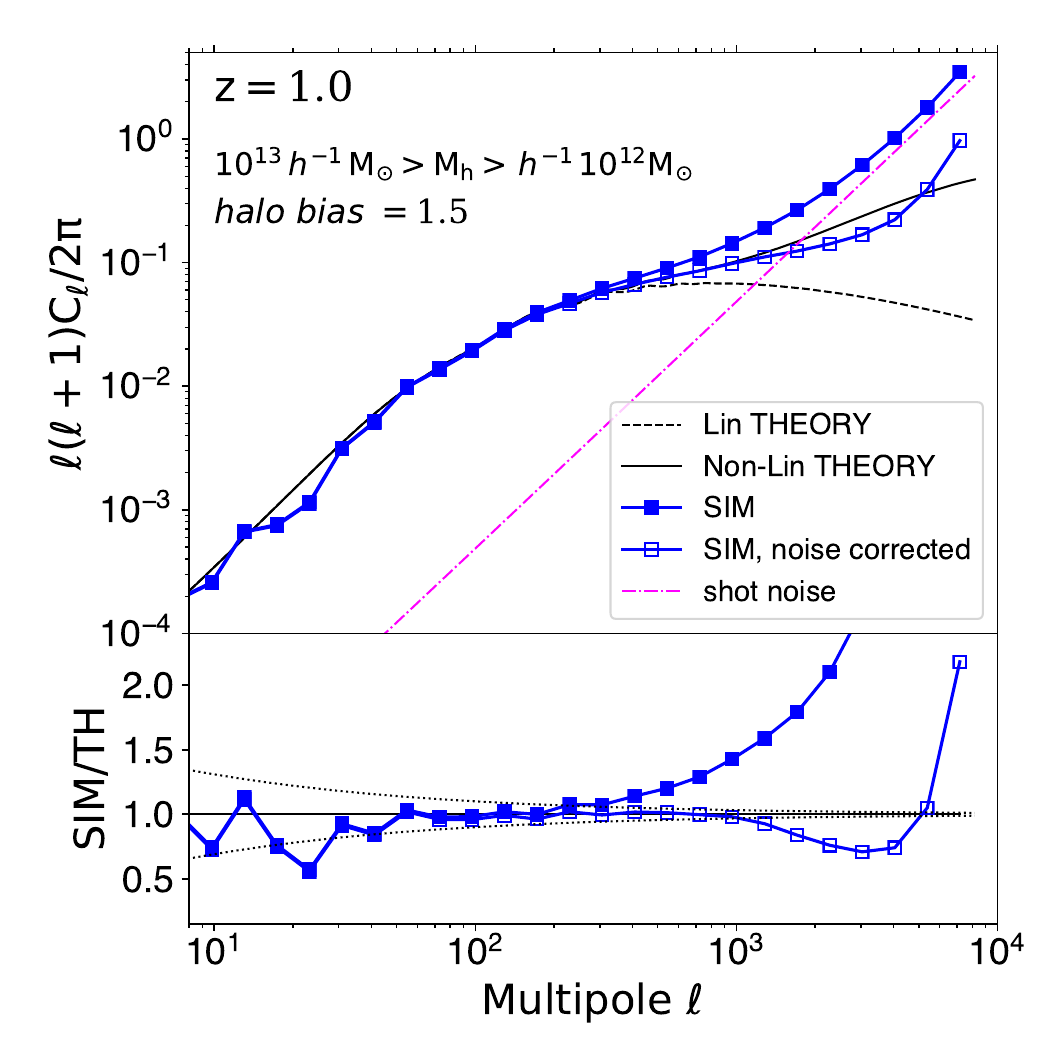}%
     \qquad
    % \subcaptionbox{}{
    \includegraphics[width=0.3\textwidth]{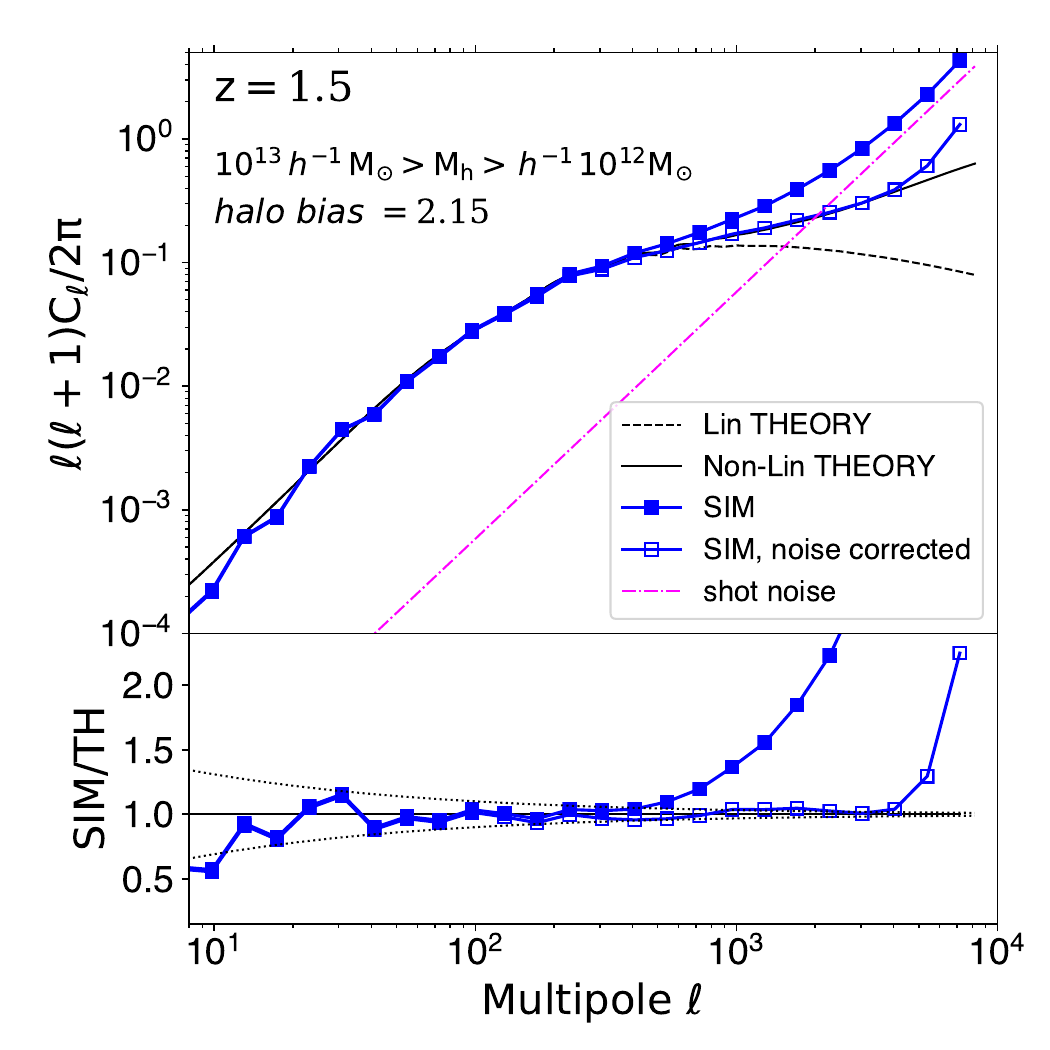}
    \qquad
    % \subcaptionbox{}{
    \includegraphics[width=0.3\textwidth]{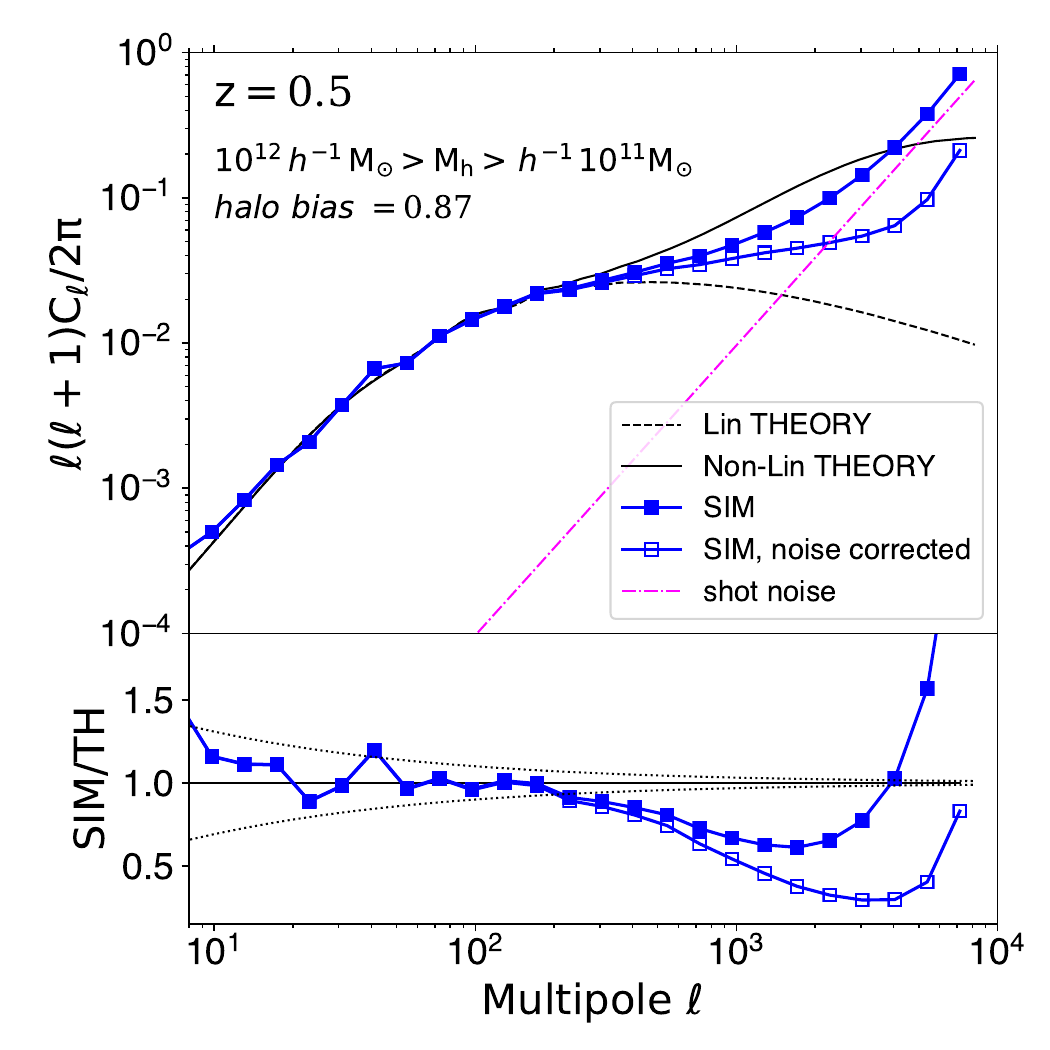}%
     \qquad
      % \subcaptionbox{}
      \includegraphics[width=0.3\textwidth]{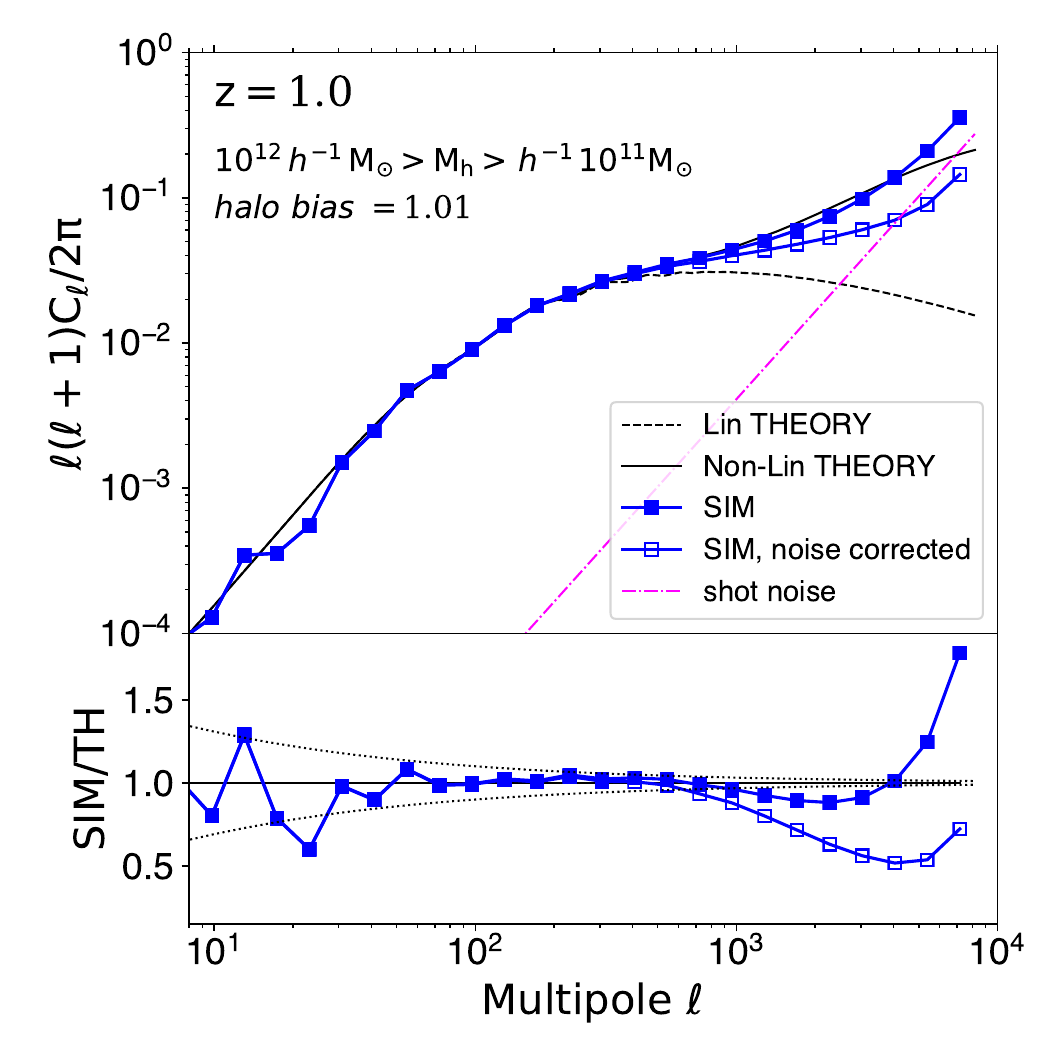}%
     \qquad
    % \subcaptionbox{}{
    \includegraphics[width=0.3\textwidth]{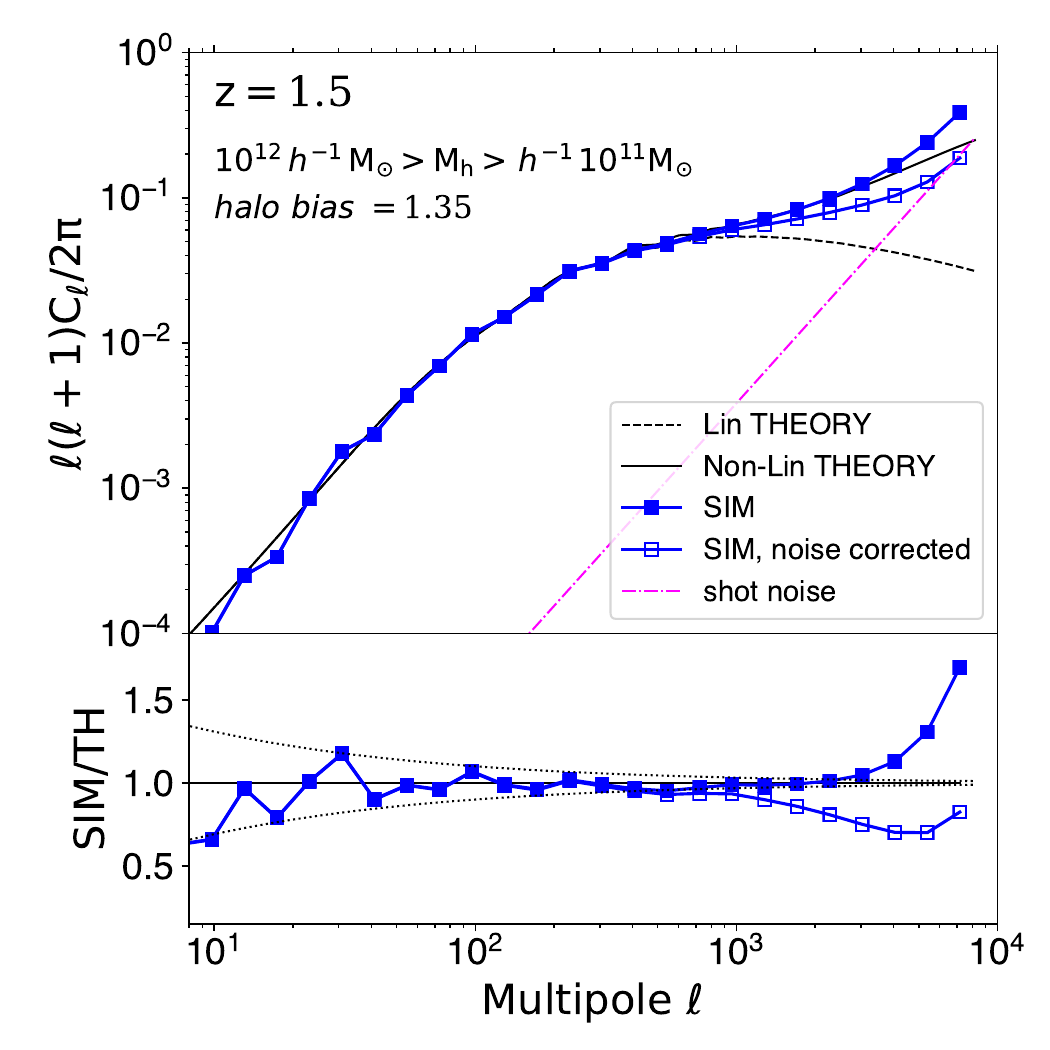}
      %\vspace{-10 pt}
    \caption{Angular power spectra of the dark matter haloes in the lightcone. Top panels show clustering for haloes of mass  $10^{13} h^{-1}M_\odot > M_{\rm h} > 10^{12} h^{-1}M_\odot$, whereas bottom panels display the corresponding results for  $10^{12} h^{-1}M_\odot > M_{\rm h} > 10^{11} h^{-1}M_\odot$. Lower panels show ratios with respect to nonlinear theory, with dotted lines displaying the  envelope for sample variance errors.  
    }%
    \label{fig:clshaloes}%
\end{figure*}

Figure~\ref{fig:clshaloes} shows the angular clustering of haloes in harmonic space at different redshifts ($z = 0.5, 1, 1.5$ from left to right) and for two mass bins ($10^{13} h^{-1}M_\odot > M_{\rm h} > 10^{12} h^{-1}M_\odot$, and  $10^{12} h^{-1}M_\odot > M_{\rm h} > 10^{11} h^{-1}M_\odot$, top and bottom rows, respectively).  The theory predictions correspond to the dark matter clustering globally re-scaled with an estimate of the linear halo bias, fitted to match the measured clustering on the lowest multipoles. Clustering corrected with a simple model for the shot-noise is also shown. The resulting (corrected) clustering is in good agreement with linear theory expectations at low multipoles, i.e, $\ell < 200$ at $z=0.5$, and $\ell < 500$ at $z=1.5$, within sample variance errors (see lower panels). These limiting multipoles are only approximate given that the scales beyond which linear matter growth and the linear halo bias model break down depend both on redshift and halo mass. We also note we have assumed that haloes are Poisson distributed which is not strictly the case \citep{Ginzburg17}. However a proper correction of the shot-noise for halo clustering is beyond the scope of this paper.

%--------------------------------------------------------------------

\section{From haloes to galaxies
\label{sec:galcat}} 

Galaxies were generated in the Flagship simulation using a prescription that includes HOD and AM techniques and observed relations between galaxy properties. The prescription follows the methodology used in populating the MICE Grand Challenge simulation with galaxies (\citealt{Carretero:15}, C15 hereafter). The starting point is the Flagship halo catalogue described above.
We use the \texttt{SciPIC} algorithm, described in Appendix~\ref{app:SciPIC}, to compute the galaxy properties. We run the pipeline at the Spanish \Euclid Science Data Center.
In the following subsections, we describe the different steps of the galaxy catalogue production.

\subsection{Galaxy luminosities \label{subsec:luminosities}}

Following the HOD philosophy, haloes are populated with central and satellite galaxies. Our prescription starts with a hybrid HOD and AM method, that computes the number of satellites in each halo and assigns the galaxy luminosities. Galaxy clustering measurements are used to determine the parameters and the relations implemented. 

The method has the following steps. First, it assumes that each halo is composed of one central galaxy and some satellites. We use a simple HOD in which the mean number of satellites (the satellite occupation) depends only on the halo mass as a power law. For all haloes with masses larger than the minimum halo mass ($M_{\rm h} \geq M_{\rm min}$), the number of central galaxies and the mean number of satellite galaxies are given by 
\begin{equation}
\begin{multlined}
N_{\rm cen} = 1 \;, \\
\ave{N_{\rm sat}} = \left( \frac{M}{M_1} \right)^{\alpha}\;. \\
\label{eq:HOD}
\end{multlined}
\end{equation}
We assign the number of satellite galaxies in each halo drawing a realisation of a Poisson distribution with mean $\ave{N_{\rm sat}}$. 
We parameterise $M_1$ as a multiplicative factor, $f$ times $M_{\rm min}$:
\begin{equation}
\label{eq:M1}
M_1 = f\, M_{\rm min}.
\end{equation}
In MICE, we calibrated the factor $f$ as a function of halo mass to match the SDSS two-dimensional projected clustering constraints of~\cite{Zehavi:11} at low redshift. In Flagship, we adopt a constant value for the multiplicative factor in Eq.~\eqref{eq:M1} and fix it to $f=15$, which is approximately the mean value we used in MICE~\citep{Carretero:15}. We further choose the multiplicative factor not to depend on redshift given the weak constraints on galaxy clustering at high redshift. We also set the exponent in Eq.~\eqref{eq:HOD} to a fixed value, $\alpha = 1$, with no redshift dependence either. We will show in Sect.~\ref{sec:validation} that our galaxy mock gives results that are in good agreement with a large set of observational data.

We then use AM to assign the galaxy luminosities. In order to obtain a relation between the halo mass and the central galaxy luminosity, we first calculate the cumulative density of galaxies (central and satellites), $n_{\rm gal}$, as a function of the halo mass.\footnote{This function gives, at a given halo mass, the number (per unit volume) of centrals and satellites that can be found at that mass threshold. For example, for our values of the $f$ and $\alpha$ HOD parameters, haloes of $M = 10^{14} h^{-1}\,\si{\solarmass}$ will have a mean of 1.67 galaxies (1 central + 0.67 satellites) for a mass threshold $M_{\rm min} = 10^{13} h^{-1}\,\si{\solarmass}$. These same haloes will have a mean of 7.67 galaxies (1 central + 6.67 satellites) for a mass threshold $M_{\rm min} = 10^{12} h^{-1}\,\si{\solarmass}$ and a mean of 67.67 galaxies (1 central + 66.67 satellites) for a mass threshold $M_{\rm min} = 10^{11} h^{-1}\,\si{\solarmass}$. That is, the same halo will have a different number of satellites depending on which mass threshold is considered.}
We compute this function by integrating the CHMF, taking into account the HOD assignment,   
\begin{equation}
\label{eq:CGF}
n_{\rm gal} (>M_{\rm min}) = \int^\infty_{M_{\rm min}} n(M')\,  \left[ 1 + \left( \frac{M'}{f\, M_{\rm min}} \right)^{\alpha} \right] \, \diff M',
\end{equation}
where we have used Eqs.~\eqref{eq:HOD} and~\eqref{eq:M1} to compute the number of galaxies per halo at the mass threshold, $M_{\rm min}$. Note that both $n_{\rm gal}$ and $n$ are densities, that is, number of galaxies per unit volume. We refer to this function defined in Eq.~\eqref{eq:CGF} as the cumulative galaxy function (CGF). 

\begin{figure}
\center
\includegraphics[width=\columnwidth]{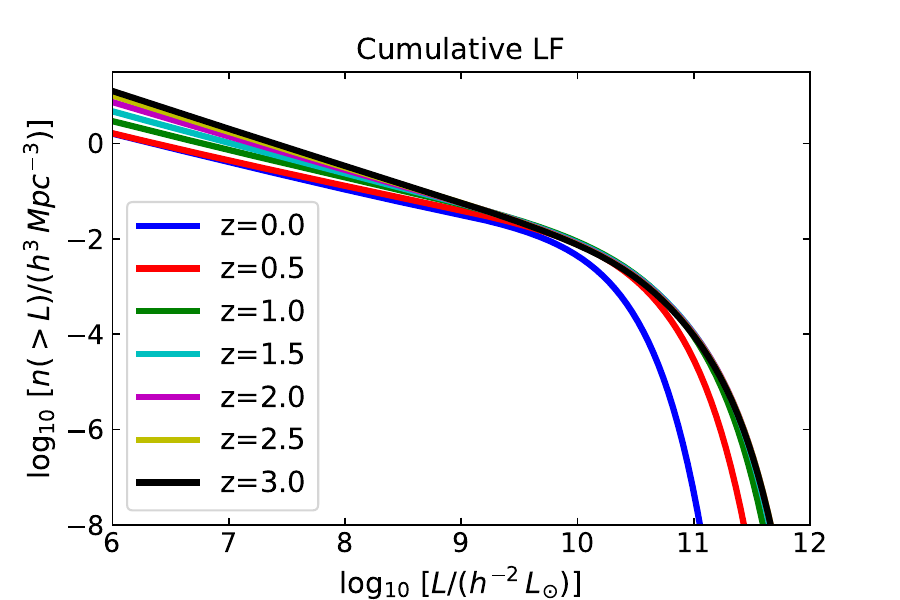}
\caption{\label{fig:cLF_allz} Cumulative luminosity function  of the galaxy population in the $r01$ band at different redshifts used in our abundance matching procedure.}
\end{figure}

The adopted galaxy luminosity function (LF) for the AM is a variation of the function given by \cite{Dahlen:05}, which is based on multi-band data taken in the Great Observatories Origins Deep Survey~\citep[GOODS;][]{Giavalisco:04}. The GOODS LF is parameterised in several optical and near-infrared bands up to redshift $z=2$. We extrapolate it to higher redshifts, $z=3$, and transform it to our reference band. The total luminosity function is the sum of the LFs of three populations, each with its own evolution. As most of our calibration is inherited from the MICE catalogue which was performed at low redshift using SDSS data, and in particular the New York University Value Added Galaxy Catalogue (NYU-VAGC; \citealt{Blanton:05c}), we choose as our reference filter the SDSS $r01$ band,\footnote{$^{0.1}r$ in the notation of ~\cite{Blanton:03}.} which is the SDSS $r$ filter redshifted to $z=0.1$~\citep{Blanton:03}. Figure~\ref{fig:cLF_allz} shows the cumulative LF function in the $r01$ band for several redshifts. 

Several studies have shown that when generating a galaxy mock catalogue with AM using the observed cumulative LF, the resulting galaxy clustering amplitude for the most luminous galaxies is higher than observed. The clustering amplitude for these luminous galaxies can be lowered if scatter is applied in the luminosity assignment~\citep[e.g.][]{Tasitsiomi:04, More:09a, Behroozi:10, Reddick:13, Carretero:15}. Given that the highest luminosity range is dominated by an exponential decay, the introduced scatter results in assigning 
higher central galaxy luminosities to haloes of lower masses, thus reducing the amplitude of the clustering at the high luminosity range. Nevertheless, at lower luminosities where the LF is mainly dominated by a power law, the inclusion of this scatter has no effect in the overall shape of this function.  Besides obtaining a more realistic clustering for the most luminous galaxies, introducing this scatter also reflects the fact that galaxy formation is a stochastic process. Taking this into account, we apply a scatter to the galaxy luminosities resulting from the AM step. We define an unscattered LF, $\Phi_{\rm unscat} (L)$, that when convolved with a Gaussian scatter in the logarithm of the luminosity (function $G$ in Eq.~\ref{eq:LF_scatter}) gives the observed LF, $\Phi_{\rm obs}(L)$,
\begin{equation}
\label{eq:LF_scatter}
\Phi_{\rm obs}(L) = \int^{\infty}_{-\infty} \Phi_{\rm unscat}(L') \;G(L - L') \, \diff L'\;, 
\end{equation}
where the Gaussian function $G$ has a mean of zero and a standard deviation of $\Delta \logten [L/(h^{-2} L_{\odot})]$. 
To obtain the unscattered LF, we need to solve for $\Phi_{\rm unscat}(L)$ in the convolution equation (\ref{eq:LF_scatter}). Instead, to gain computing efficiency, we approximate the effect of the convolution with an exponential decay factor:
\begin{equation}
\label{eq:LF_scatter_approx}
\begin{multlined}
\Phi_{\rm unscat}(L,z,\Delta \logten L)  \\= \Phi_{\rm obs}(L,z) \, \exp \left[- \left( \frac{L}{L_{\rm sct}(z,\Delta \logten L)} \right)^{\beta_{\rm sct}(z,\Delta \logten L)} \right], 
\end{multlined}
\end{equation}
where we fit the parameters $L_{\rm sct}$ and $\beta_{\rm sct}$ as a function of redshift, $z$, and the value of the scatter, $\Delta \logten [L/(h^{-2} L_{\odot})]$. 
We compute the cumulative unscattered LF as 
\begin{equation}
\label{eq:CLF_unscat}
n_{\rm gal} (>L) = \int^{\infty}_{L} \Phi_{\rm unscat}(L') \, \diff L',
\end{equation}
where we have omitted the dependence on redshift and the luminosity scatter. We can establish a relation between halo mass and luminosity by applying AM between the cumulative number of galaxies function (Eq.~\ref{eq:CGF}) and the cumulative unscattered LF (Eq.~\ref{eq:CLF_unscat}):
\begin{equation}
\label{eq:AM_ML}
n_{\rm gal} (>M_{\rm halo},z) = n_{\rm gal} (>L,z,\Delta \logten [L/(h^{-2} L_{\odot})]) \;.
\end{equation}

\begin{figure}
\center
\includegraphics[width=\columnwidth]{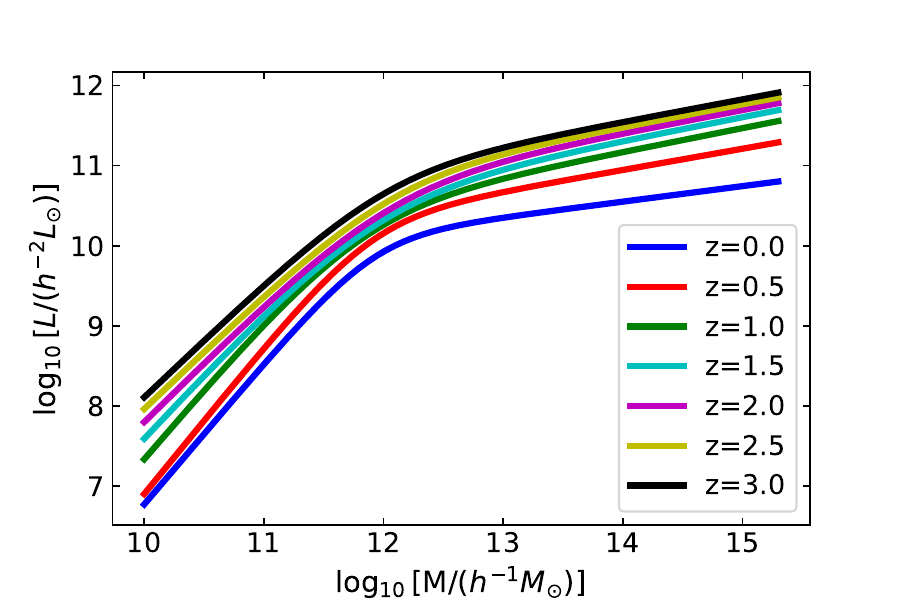}
\caption{\label{fig:AM_GFLF_allz} Relation between halo mass and luminosity at various redshifts resulting from abundance matching the cumulative galaxy function and the cumulative unscattered luminosity function following Eq.~\eqref{eq:AM_ML}.  }
\end{figure}

Figure~\ref{fig:AM_GFLF_allz} shows the resulting AM relations for a few redshift values from Eq.~\eqref{eq:AM_ML} for our assumed value of $\Delta \logten [L/(h^{-2}\,L_{\odot})] = 0.12$. The relation is fitted as a double power law with parameters fitted as a function of halo mass and redshift.  
The luminosities of the central galaxies are assigned from their halo mass and their redshift using the relation coming from Eq.~\eqref{eq:AM_ML} and shown in Fig.~\ref{fig:AM_GFLF_allz} for a few redshift values. We then add a random realisation of the scatter in the logarithm of the luminosity to its value assuming a Gaussian distribution. We take the standard deviation of this distribution to be $\Delta \logten [L/(h^{-2}\,L_{\odot})] = 0.12$, which is similar to the value applied in other studies 
and produces a consistent clustering signal in our catalogue at high luminosities and at low redshift. 

\begin{figure}
\center
\includegraphics[width=\columnwidth]{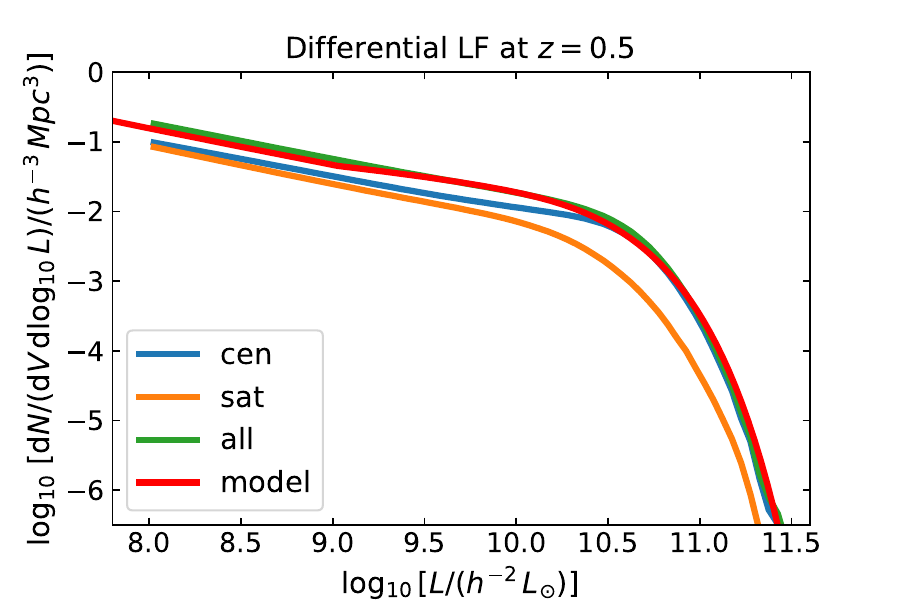}
\caption{\label{fig:dLF_cen_sat_all_z05} Differential luminosity function for central and satellite galaxies and all galaxies at redshift $z=0.5$ in the Flagship catalogue. We also show the model LF as in Fig.\ref{fig:cLF_allz} but in its differential form.
}
\end{figure}

In order to assign luminosities to the satellites, we first compute the global LF for
these galaxies by subtracting the LF for the centrals from the total LF used to compute the AM. Figure~\ref{fig:dLF_cen_sat_all_z05} provides a visual representation of this step. It shows the model LF used to compute the AM relation between mass and luminosity and the measured LF in the Flagship catalogue in a thin redshift slice centred at $z=0.5$ for all the galaxies and for centrals and satellites only. Once we have the global LF for the satellites (orange line in Fig.~\ref{fig:dLF_cen_sat_all_z05}), we assume that it is the result of the sum of the individual LFs of satellites within each halo for all haloes. We model the cumulative LF of the satellites within each halo using a modified four-parameter Schechter function of the form
\begin{equation}
\label{eq:CLF_satellites}
N_{\rm sat}^{\rm halo} (>L) = A \,\left( \frac{L}{L_*(L_{\rm cen})} \right)^{\alpha} \exp \left[ - \left( \frac{L}{L_*(L_{\rm cen})} \right)^{\beta} \right]\;,
\end{equation}
where $A$ is the normalisation factor that ensures that the halo contains the number of satellites predicted by the realisation of the HOD with luminosities higher than the minimum luminosity assigned for that halo, $L_\mathrm{min}$, 
$\alpha$ is the faint slope that we fix to $\alpha = -0.5$, and $\beta$ is a parameter that controls the steepness of the bright-end exponential cut-off. Furthermore, $L_*$ is the characteristic luminosity that depends linearly on the luminosity of the central galaxy, $L_* = a\, L_{\rm cen} + b$. Taking into account this model for each halo, we fit the three parameters, $a$, $b$, and $\beta$, to make the sum of the satellite LF of all the haloes match the global LF. We obtain the values $a=10^{-0.5}$, $b=0$ and $\beta=1.5$ from this process. Finally, we assign the luminosities for the satellites in each halo by randomly drawing the luminosities following Eq.~\eqref{eq:CLF_satellites}.

\subsection{Galaxy positions and colours \label{subsec:positions}}

Before allocating positions and velocities to the galaxies, we assign them a $(g01-r01)_{\rm HOD}$\footnote{This $(g01-r01)_{\rm HOD}$ colour is named {\tt g01r01\_hod} in the Flagship catalogue available in CosmoHub: \url{https://cosmohub.pic.es}} colour, defined as the colour computed using the $g$ and $r$ SDSS filters redshifted to $z=0.1$~\citep{Blanton:03}. This colour will be used together with the luminosity and redshift to assign the spectral energy distribution to the galaxies (see Sect.~\ref{subsec:galaxySED}). 
We define three colour populations and assign individual colours to galaxies depending on which population they belong to. The abundance of these colour populations in each halo is defined by the HOD prescription. We follow the same procedure as C15 to assign the $(g01-r01)_{\rm HOD}$ colour that depends on the galaxy type. We have extended the C15 fits to fainter luminosities as the Flagship catalogue contains haloes of lower mass and thus reaches fainter luminosities. A full description of the method is given in C15. We summarise it here, noting the extensions and improvements made. 

\begin{figure}
\center
\includegraphics[width=\columnwidth]{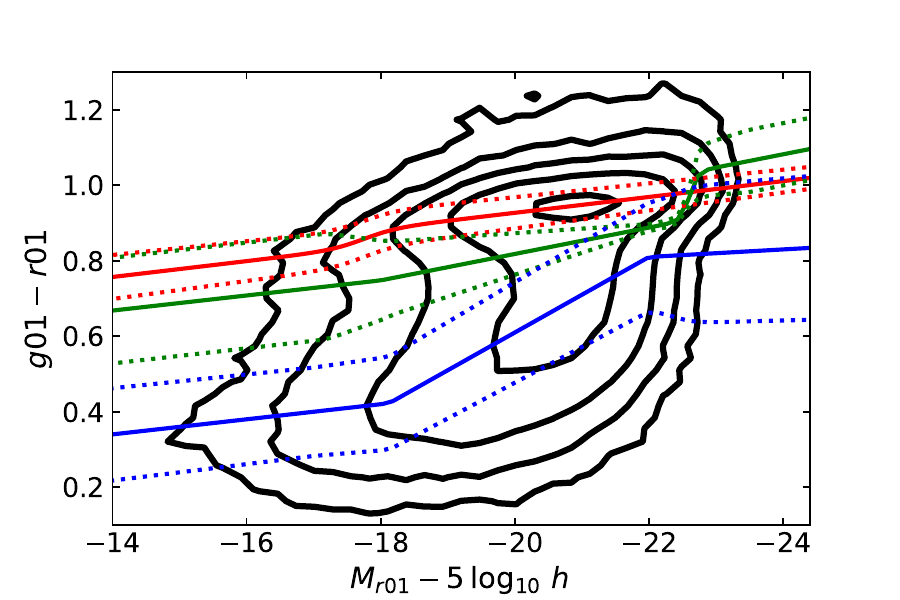}
\caption{\label{fig:CMD_NYUVAC_contours_lines} Colour-magnitude diagram of the SDSS NYU-VAGC. The black lines show contours of the same density of galaxies in this space. The red, green, and blue solid lines show the mean of the distributions of the three Gaussian fits to the ($g01-r01$) colour distributions as a function of magnitude. The dotted lines show the position of the mean $\pm$ the standard deviation of the Gaussian distributions.  }
\end{figure}

\begin{figure}
\center
\includegraphics[width=\columnwidth]{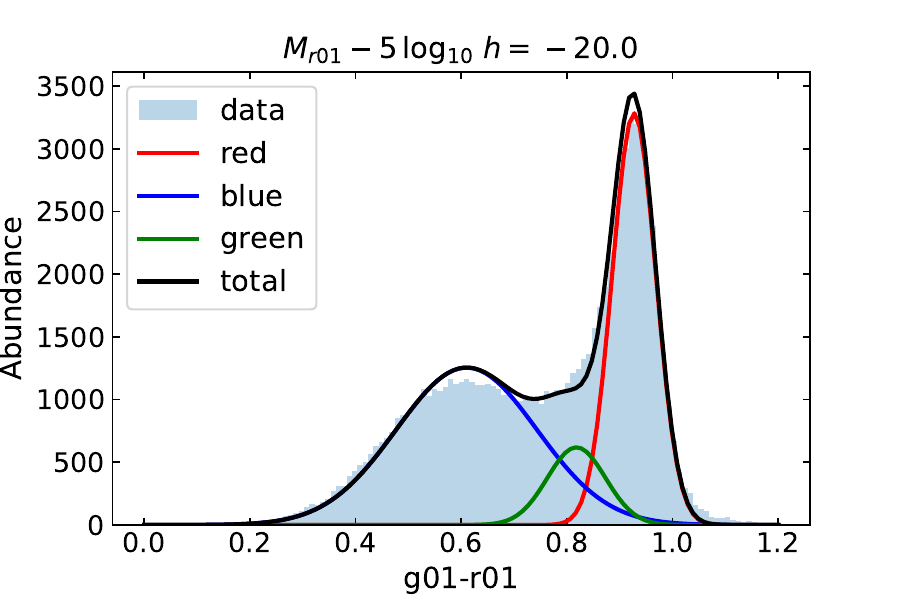}
\caption{\label{fig:CMD_cut_mr-20_rgb} Colour distribution of galaxies in the SDSS NYU-VAGC at absolute magnitude $M_{r01} -5\, \logten h = -20.0$. The light blue histograms represents the data in the catalogue. The red, green, and blue lines are the fits to the three Gaussian populations and the black line the sum of the three.}
\end{figure}

We start from the colour-magnitude diagram (CMD) using the low redshift  NYU-VAGC SDSS galaxy catalogue~\citep{Blanton:05b}, which we show in Fig.~\ref{fig:CMD_NYUVAC_contours_lines}. We fit the $g01-r01$ colour distribution as a function of luminosity in this colour-magnitude diagram to three galaxy populations: blue, green, and red. Similarly to what was done in~\cite{Baldry:04}, \cite{Skibba:09}, and \cite{Carretero:15}, we assume that the three populations are characterised by Gaussian distributions. We fit their means and standard deviations as a function of luminosity. Figure~\ref{fig:CMD_NYUVAC_contours_lines} includes these fits as solid and dotted lines, respectively, in the colour-magnitude diagram. The red, green, and blue lines correspond to the red, green, and blue populations, respectively. Note that at high luminosities, there appears to be a small extra component redder than the main Gaussian-distributed red population. For computational convenience, we absorb this extra red population within the green population, which has no galaxies at high luminosities. This is the same procedure that was used for the MICE catalogue~\citep{Carretero:15}.
As an example of these fits, Fig.~\ref{fig:CMD_cut_mr-20_rgb} presents these three colour distributions at the absolute magnitude $M_{r01} - 5\,\logten h = -20.0$. Note that the mean and standard deviations of the three populations are fit by smooth global functions that depend on the absolute magnitude and not at each value of the absolute magnitude separately.

\begin{figure}
\center
\includegraphics[width=\columnwidth]{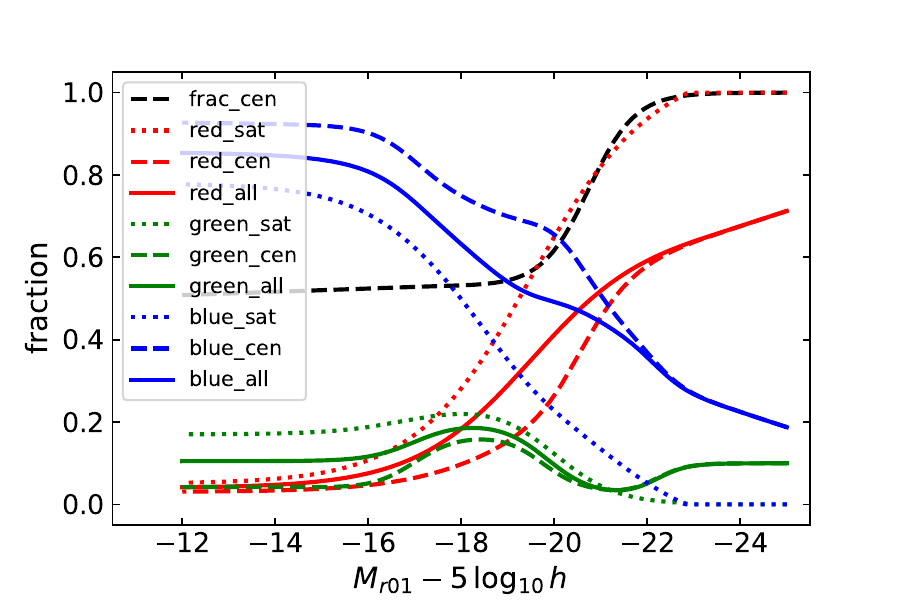}
\caption{\label{fig:fractions_vs_absmagr01} Galaxy fractions as a function of absolute magnitude.  The black dashed line represents the fraction of centrals, while the rest of the lines present combinations of galaxy type and galaxy colour type. The line style distinguishes centrals (dashed lines) from satellites (dotted lines) and all galaxies (solid lines). The galaxy colour types are presented with different colours, red, green, and blue.
}
\end{figure}

\begin{figure}
\center
\includegraphics[width=0.9\columnwidth]{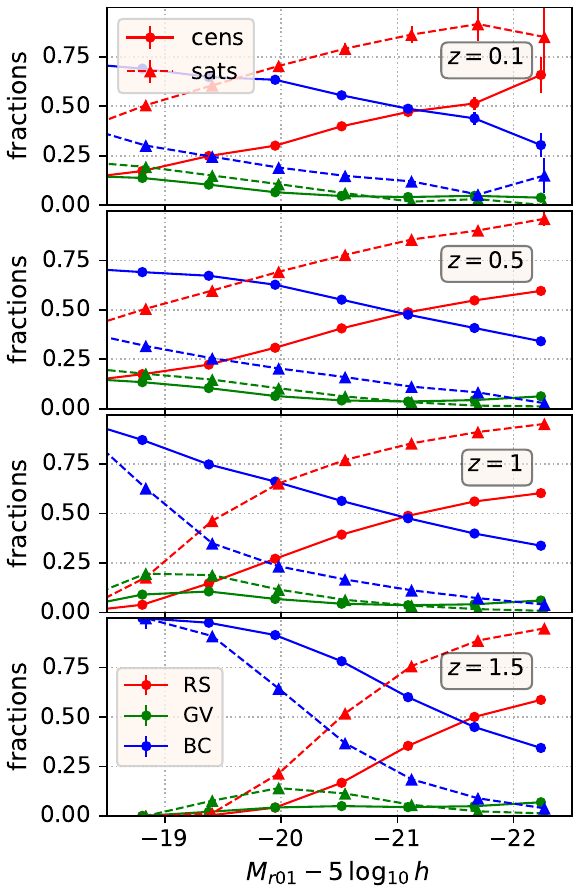}
\caption{\label{fig:color_kind_fractions_vs_absmag_censat} Fraction of galaxies of different types as a function of absolute magnitude for different redshift slices ($\Delta z=\pm0.1$), split between Red Sequence (`RS'), Green Valley (`GV'), and Blue Cloud (`BC'), and between centrals (`cens') and satellites (`sats'). 
The abscissae have been slightly shifted for clarity.
}
\end{figure}

We adopt the procedure of~\cite{Skibba:09} and followed in~C15 to compute the $(g01-r01)_{\rm HOD}$ colour based on the galaxy type (central or satellite). We also differentiate the galaxies into three colour types defined by the Gaussian fits of the CMD. We assign the parameter {\tt color\_kind} to each galaxy in the catalogue to indicate to which colour population it belongs to: red, green, or blue. We define a function to determine the fraction of satellites that belong to the red population as a function of luminosity and another one for the green population. The HOD and the colour-magnitude diagram then determine the fraction of satellites that belong to the blue population and the fraction of centrals that belong to the three populations (see Eqs. 37-43 in C15). 
These functions were optimised in MICE to reproduce the clustering as a function of colour and luminosity of the SDSS sample~\citep{Zehavi:11}. For Flagship we modified these functions slightly and extended them to lower luminosities. However, we did not run a proper minimisation exercise, in part due to the flexibility in the functional form that we have allowed. Figure~\ref{fig:fractions_vs_absmagr01} shows these functions as implemented in the Flagship catalogue. All the brightest galaxies are centrals. Then, there is a transition around the LF characteristic luminosity, $L_{\star}$, where the fraction diminishes to become approximately constant around 50\% for lower luminosities. The overall fraction of red, green, and blue galaxies is constrained by the CMD, with many more red galaxies at bright luminosities and blue galaxies dominating the population at faint luminosities. At all luminosities, there is a higher fraction of red galaxies that are satellites than centrals, and the other way around for blue galaxies. The fraction of green galaxies is always low compared to the red and blue ones. Note that brighter than absolute magnitudes $M_{r01} -5\,\logten h \simeq -21.5$, the galaxies tagged as green are indeed redder than the main red population. The fraction of green satellites is larger than that of green centrals at low luminosities, but the trend is reversed at bright luminosities, where all this extra population of the reddest galaxies (tagged as green) are centrals.

Fig~\ref{fig:color_kind_fractions_vs_absmag_censat} displays the fraction of {\tt color\_kind} values as a function of luminosity at four redshifts, splitting between centrals and satellites.
There are no more local minima or maxima  in the fractions vs. luminosity as were seen in Fig.~\ref{fig:fractions_vs_absmagr01}, because of the split in redshift bins. Flagship predicts that 1)  luminous galaxies are redder, consistent with what is observed at low redshifts \citep{Sandage&Visvanathan78}; 2) satellites are more likely to be red than centrals of the same luminosity, as observed at low redshifts \citep{vandenBosch+08}; 3) centrals and satellites (in particular low-luminosity ones) are more likely to be blue at higher redshifts, in conformity with the Butcher--Oemler effect \citep{Butcher&Oemler78} that galaxies in higher-redshift clusters contain a higher fraction of spiral morphologies (hence bluer colours).

\begin{figure}
\center
\includegraphics[width=0.95\columnwidth]{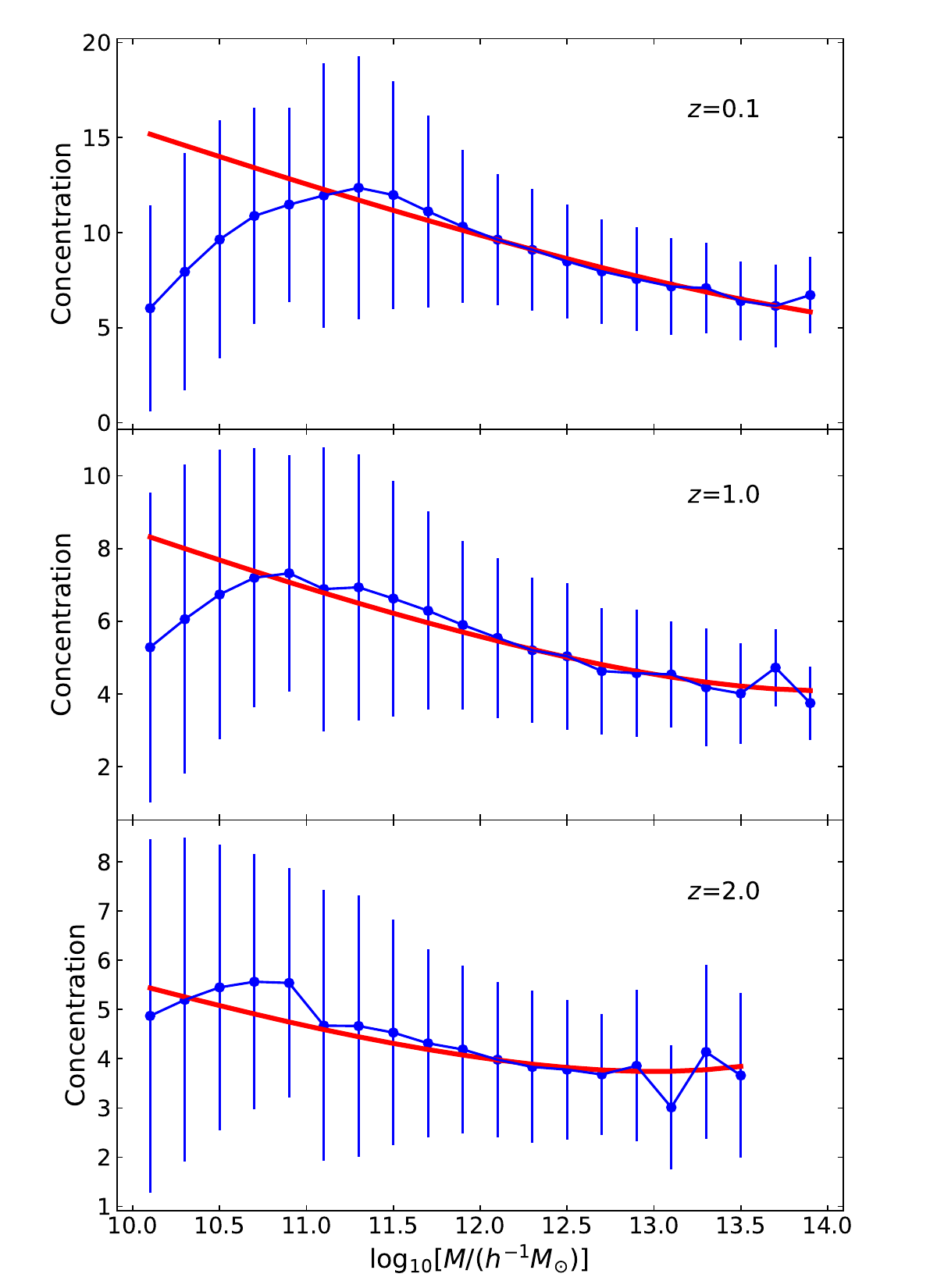}
\caption{\label{fig:concentration_vs_halolm_allz} Concentration index as a function of halo mass at the redshifts marked in each panel. The blue dots correspond to the means of the distributions at each halo mass bin and are linked with a blue line. The vertical blue lines show the standard deviations of the distributions. The red line shows the \citet{Diemer:19} functional form values. 
}
\end{figure}

We place the central galaxies at the centre of their haloes.  
Satellite galaxies are located following a triaxial Navarro, Frenk, and White (NFW;~\citealt{Navarro:97}) profile. We use the virial radius, the concentration parameter, and the vectors of the ellipsoid semi-axes to compute the satellite positions. We obtain the concentration parameter of the haloes using the virial radius and the scale radius from the \texttt{ROCKSTAR} catalogue\footnote{\texttt{ROCKSTAR} computes the scale radius fitting the radial mass profile to a NFW functional form~\citep[for details see][]{Behroozi:13}.} for haloes with masses $\logten [M_{\rm halo} /(h^{-1}\,\si{\solarmass})] > 11.0$, where there are enough particles per halo as to reliably determine the concentration. For this mass range, the concentration parameter agrees 
with a mass dependence that matches that given by~\cite{Diemer:19} as computed with the \texttt{COLOSSUS} code. Below this mass threshold, we use the relation from~\cite{Diemer:19}  to obtain the mean concentration for a given halo mass and redshift. We have computed the ratio of the standard deviation to the mean of the distribution of concentration parameters as a function of halo mass and redshift. This ratio is approximately constant with a value of 1/3. We therefore assign the concentration parameter for the haloes below the mass threshold $M_{\rm halo} = 10^{11}\;h^{-1}\,\si{\solarmass}$, drawing a realisation of a Gaussian distribution with the mean coming from the relation of~\cite{Diemer:19} and a standard deviation that is one-third of the value of the concentration mean. 
Figure~\ref{fig:concentration_vs_halolm_allz} shows with blue dots the mean of the concentration parameter as a function of halo mass at three different redshifts before correcting the values at low halo mass. The blue lines indicate the standard deviation of the concentration parameters. The red line shows the~\cite{Diemer:19} relation. The recomputed mean concentration values at low halo mass coincide with the red line.

The other ingredient for assigning positions to the satellites within their haloes is the triaxial shape of the haloes.
\texttt{ROCKSTAR} outputs the position vectors of two of the ellipsoid semi-axes and the axis ratios. We compute the third position vector as the cross product of the other two. We take the virial radius in the semi-major axis direction as the virial radius provided by the catalogue and in the other directions we multiply the virial radius by the corresponding axis ratios. We assign the position of satellite galaxies within haloes following the procedure of~\cite{Robotham:18},\footnote{\url{https://github.com/CullanHowlett/NFWdist}} which we modify to compute random positions in a triaxial NFW keeping the appropriate density profile, axis ratio values, and ellipsoid orientation. We compute the positions in each halo coordinate system given by the position vectors of their semi-axes and then transform them to comoving coordinates taking into account the orientation of the ellipsoid with respect to the observer's lightcone.  

We implement colour segregation in the satellite galaxy distribution assuming a different concentration parameter for each of the {\tt color\_kind} samples~\citep[e.g.,][]{McDonough:22}. Red galaxies have the same concentrations of the dark matter in the haloes. Blue galaxies have a concentration one fourth of the concentration of the red galaxies (following \citealt{Collister&Lahav05}), while green galaxies have a concentration half of that of the red galaxies. Following C15 who placed galaxies beyond the virial radius to adjust the galaxy clustering in the MICE catalogue, we also place satellites beyond the halo virial radius up to three times its value, resulting in approximately 15\% of the satellite galaxies being outside the virial radius of their haloes.

\subsection{Galaxy velocities \label{subsec:velocities}}

Once the galaxy positions are determined, we compute their velocities within the halo. We also use the parameter {\tt color\_kind} to produce line-of-sight velocity dispersion profiles modulated by the type of galaxy.

Central galaxies are assumed to be at rest at the centre of the halo, so their velocity is the same as the centre of mass of the halo. The satellite galaxy velocities are built by solving the spherical, stationary, Jeans equation of local dynamical equilibrium,
\begin{equation}
\frac{{\rm d} \left (\rho \sigma_r^2\right)}{{\rm d}r} 
+ \frac{2\,\beta(r)\,
\left (\rho \sigma_r^2\right)}{r} 
= - \rho(r) \frac{G\,M(r)}{r^2} \ ,
\label{eq:jeans}
\end{equation}
for given radial profiles of mass $M(r)$, number density $\rho(r)$, and velocity anisotropy $\beta=1-\sigma_\theta^2/\sigma_r^2$, with $\sigma_\theta$ and $\sigma_r$ being the velocity dispersion in the tangential and radial directions, respectively.
The velocity anisotropy is assumed to follow the \cite{Tiret+07} model:
\begin{equation}
    \beta(r) = \beta_\infty\,\frac{r}{r+r_\beta} \ ,
    \label{eq:betaofr}
\end{equation}
which is a reasonably good approximation to what is seen in $\Lambda$CDM haloes.
Using the parameters found for massive clusters at $z=0.05$ by \cite{Mamon+19}, assuming that ellipticals, S0s and spirals respectively trace red sequence, green valley, and blue cloud galaxies as defined by the {\tt color\_kind} parameter, leads to the parameters given in Table~\ref{tab:vparams}.

\begin{table}
\caption{Assumptions for deriving velocity dispersions}
\centering
\tabcolsep=3pt
\begin{tabular}{lccc}
\hline
\hline
\noalign{\vskip 1pt}
{\tt color\_kind} & $c/c_{\rm halo}$ & $\beta_\infty$ &$r_\beta/r_{\rm vir,halo}$ \\
(1) & (2) & (3) & (4) \\
\hline
\noalign{\vskip 1pt}
Red Sequence & 1 & 0.27 & 0.25\\
Green Valley & 0.5 & 0.44 & 0.5 \\
Blue Cloud & 0.25 & 0.82 & 1.0 \\
\hline
\end{tabular}
\tablefoot{The columns are:
(1) the {\tt color\_kind}; 
(2) the ratio of the concentration of the galaxy population of a given {\tt color\_kind} over that of the dark matter;
(3) the asymptotic velocity anisotropy (Eq.~\ref{eq:betaofr}) for the given {\tt color\_kind}; 
and (4) the anisotropy radius (Eq.~\ref{eq:betaofr}) in units of the halo virial radius.}
\label{tab:vparams}
\end{table}
We proceed as follows:
\begin{enumerate}
\item \label{item:sigr} Pre-compute the radial velocity dispersions in virial units, $\sigma_r(r)/v_{\rm vir}$ on a two-dimensional grid of geometrically spaced $r/r_{\rm vir}$ and concentrations, where the $\sigma_r$ are obtained by solving (following Eq.~\ref{eq:rhosigr2}) the Jeans equation in Eq.~\eqref{eq:jeans} using the assumptions of Table~\ref{tab:vparams}; 
\item \label{item:sigrfit} pre-fit three two-dimensional 5th-order polynomial approximations (one per {\tt color\_kind}) for $\logten (\sigma_r/v_{\rm vir})$ in terms of $\logten(r/r_{\rm vir})$ and $\logten c_{\rm dark}$;
\item determine the full set of radial velocity dispersions using these polynomial approximations (extrapolating beyond the fit limits with the constant value at each limit);
\item determine the tangential components of the velocity dispersions, using $\sigma_\theta = \sigma_\phi = \sqrt{1-\beta(r)}\,\sigma_r$, where the first equality comes from the assumption of spherical symmetry ($\sigma_\theta$ and $\sigma_\phi$ are the azimuthal and latitudinal components respectively of the tangential velocity component);
\item derive the three velocity components in spherical coordinates, assuming locally Gaussian velocity distribution functions;
\item \label{item:subtractHub} convert these physical velocities in the halo frame to 
peculiar velocities, by subtracting the Hubble flow, $H(z)\,r$, from the radial velocity component;
\item convert these peculiar halo-frame velocities in spherical coordinates to Cartesian coordinates;
\item convert these peculiar halo-frame Cartesian coordinates into the box frame;
\item derive the line-of-sight velocities $V_{\rm LOS} = ({\vec R}\cdot {\vec V}) / R$ (where $\vec R$ and $\vec V$ are the position and velocity vectors measured relative to the observer and $R$ the modulus of the $\vec R$ vector) in the box frame, assuming that the observer is at rest in the box;
\item derive the redshift from the line-of-sight velocities using the relation 
\begin{equation}
1+z = \left(1+z_{\rm cos}\right)\,\left(1+\frac{V_{\rm LOS}}{c}\right)\ ,
\end{equation}
where $z_{\rm cos}$ is the cosmological (true) redshift, $V_{\rm LOS}$ is the line-of-sight velocity component, and neglecting the peculiar motion of the observer as well as the 2nd-order transverse Doppler term which incorporates the multiplicative Lorentz factor $(1-V^2/c^2)^{-1/2}$ \citep[e.g.,][]{Tatum:85}.
\end{enumerate}
The details of steps Nos.~\ref{item:sigr} and \ref{item:sigrfit} are provided in Appendix~\ref{sec:veldetails}.
In a forthcoming (already coded) version of Flagship, we inserted after item No.~\ref{item:subtractHub} the addition of a radial infall pattern, obtained by fitting a nonlinear function of redshift and log halo mass to the radial motions inside haloes of dissipationless cosmological simulations.

\subsection{Spectral energy distributions \label{subsec:galaxySED}}

As described in Sect.~\ref{subsec:luminosities}, we assign a 
$(g01-r01)_{\rm HOD}$ colour to each galaxy randomly sampling the colour distributions derived from the CMD for our three {\tt color\_kind} populations. This colour corresponds to the rest-frame colour of a galaxy in the $g01$ and $r01$ filters (see Sect.~\ref{subsec:positions}) or equivalently the $g-r$ colour at redshift $z=0.1$. 

Next, we want to assign a spectral energy distribution (SED) to each galaxy. We take as a template basis the COSMOS SED library used in~\cite{Ilbert:09} that originally comes from the SEDs of~\cite{Polletta:07}, complemented with templates from~\cite{Bruzual:03} to expand the range of star-formation histories included. In order to increase the SED coverage, each galaxy SED is computed as a linear combination of two of the templates in the SED basis.  

Following ~\cite{Ilbert:09}, our chosen SED library is composed of 31 templates. We add extinction to some of the templates to expand the colour space coverage and better represent observations. We do not apply any extinction to the first ten reddest templates. The next thirteen templates represent spectral distributions of spiral galaxies and are assumed to have the Pr\'evot Small Magellanic Cloud extinction law~\citep{Prevot:84}. The last eight bluest templates represent starburst galaxies and have the Calzetti extinction law~\citep{Calzetti:00}. We build a sample of 136 templates using the 31 SEDs and the two extinction laws that we sample at six values of the $E(B-V)$ colour excess (or reddening) from 0 to 0.5 in steps of 0.1. Table~\ref{tab:sed_templates} summarises the original templates used with their order number, the extinction law applied for those templates, the reddening values sampled for each of those templates, and the total number of resulting SEDs that constitute our SED template basis. 

\begin{table}
\caption{SED template basis }
\centering
\tabcolsep=3pt
\begin{tabular}{lccc}
\hline
\hline
\noalign{\vskip 1pt}
Original template & & &Number of final\\ 
number & Extinction law & $E(B-V)^{a}$  &SED templates\\
\hline
\noalign{\vskip 1pt}
0--9 & None & 0.0 & 10\\
10--22 & Pr\'evot & 0.0--0.5 & 78\\
23--30 & Calzetti & 0.0--0.5 & 48\\
\hline
\end{tabular}
\tablefoot{$^a$ The reddening values sampled for the Pr\'evot and Calzetti extinction laws are 0.0, 0.1, 0.2, 0.3, 0.4, and 0.5 }
\label{tab:sed_templates}
\end{table}

\begin{figure}
\center
\includegraphics[width=\columnwidth]{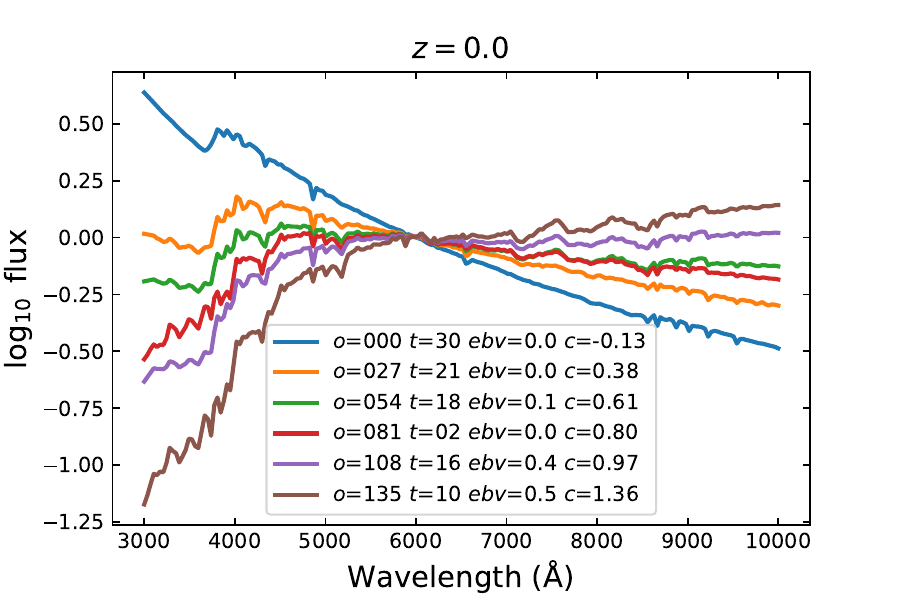}
\caption{\label{fig:templates_ordered_subset_z00} Subset of six SED templates out of the 136 described in the text at redshift $z=0.0$. The templates are normalised to have the same flux at a wavelength of 6000\,\AA. The figure inset indicates the colour-ordered template number (o), the original template number in \cite{Ilbert:09} ($t$), the $E(B-V)$ reddening value applied to the template (ebv) and the value of the $g01-r01$ colour ($c$).}
\end{figure}

For these 136 templates, we compute the $g01-r01$ and the COSMOS $g-r$ colours at seven discrete redshift values covering the redshift range of the lightcone simulation, $0<z<3$, in intervals of $\Delta\, z =0.5$. 
Figure~\ref{fig:templates_ordered_subset_z00} shows a subset of six of these 136 templates at redshift $z=0.0$.

\begin{figure}
\center
\includegraphics[width=\columnwidth]{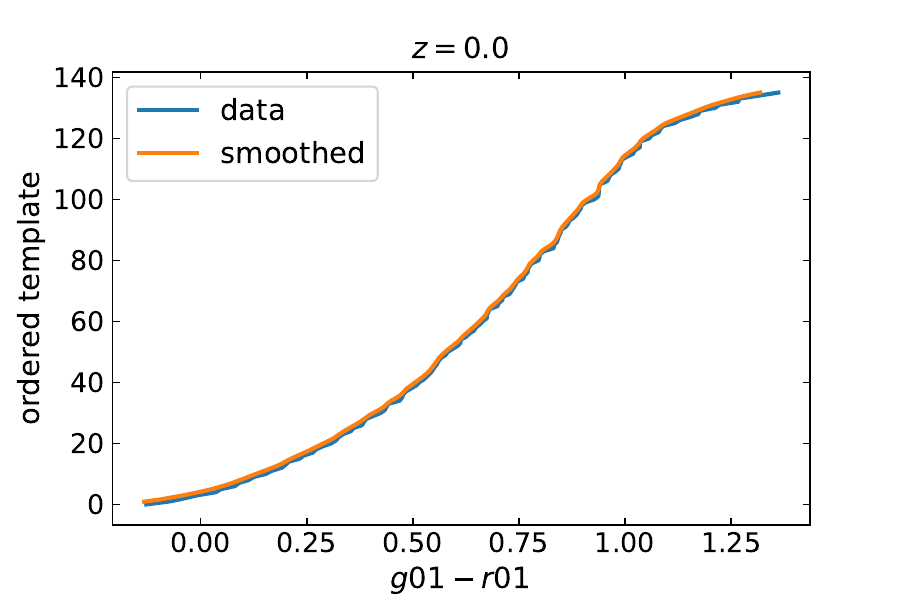}
\caption{\label{fig:g01r01_ordered_template_z00} Relation between the $g01-r01$ colour and the colour-ordered template (which is zero-ordered from 0 to 135) at redshift $z=0$. The blue line connects the values of the colour and the ordered template for the 136 templates. The orange line is a smoothed version of the relation connecting the data points. }
\end{figure}

\begin{figure}
\center
\includegraphics[width=\columnwidth]{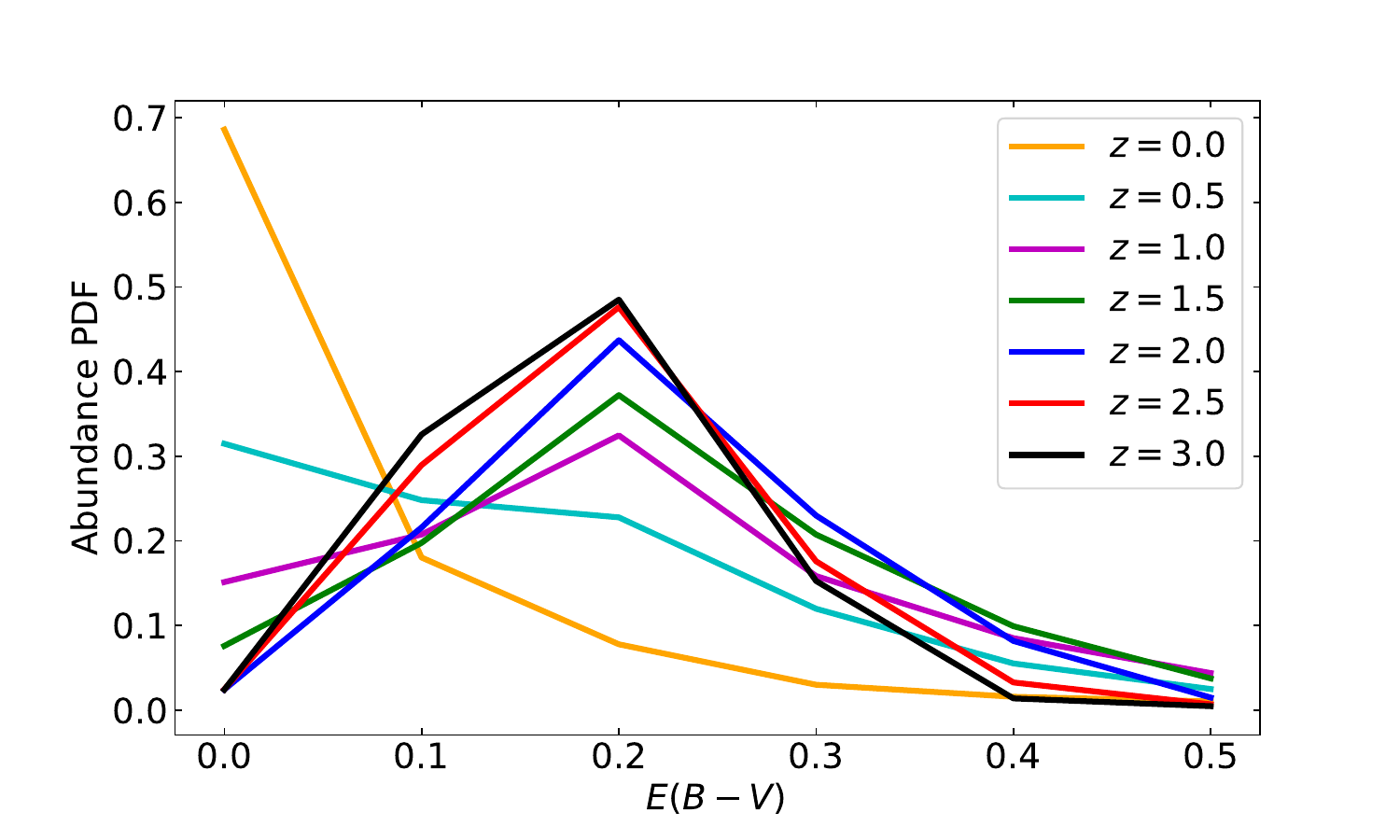}
\caption{\label{fig:ebv_pdf_cosmos_zall} Distribution of $E(B-V)$ values in the COSMOS2020 catalogue for a magnitude range $i < 24.5$. The distributions are normalised so that the total abundance for all $E(B-V)$ values at each redshift adds up to 1. }
\end{figure}

We rank order the templates according to their colour at these redshift values.\footnote{$z\in \{0.0,0.5,1.0,1.5,2.0,2.5,3.0\}$\label{footnote:z_values}} We use the $g01-r01$ colour at $z=0$ and the COSMOS $g-r$ at the rest of the redshift values for the ordering. We compute a relation between the template order number, treated as a real number but sampled at integer values, and the values of the colour at each redshift value. We smooth this relation to make the function monotonic and avoid degenerate values in the relation that will give the same template order number for different values of the colour. Figure~\ref{fig:g01r01_ordered_template_z00} shows this relation for the $g01-r01$ colour at redshift $z=0$. We also establish relations between the $g01-r01$ and $g-r$ colours at all non-zero redshift steps. To obtain these relations we abundance match the $(g01-r01)_{\rm HOD}$ colour distribution in the Flagship catalogue to the $g-r$ colour distribution of the COSMOS2020 catalogue~\citep{Weaver:22} at each redshift value considered (see footnote~\ref{footnote:z_values}). We take all galaxies in a range of $\Delta z =0.2$ centred at the mean redshift value to build the samples for the abundance matching. The only exception is for the last redshift value, $z=3.0$, where we take galaxies in the range $2.8<z<3.0$ as there are no Flagship galaxies beyond that redshift. We also compute the $E(B-V)$ distributions from the COSMOS2020 catalogue at the same redshift values, in this case in a redshift range of width $\Delta z =0.1$. Figure~\ref{fig:ebv_pdf_cosmos_zall} shows the normalised distribution of $E(B-V)$ values in the COSMOS2020 catalogue for a sample with a magnitude limit in this catalogue of $i<24.5$, equivalent to the 10$\,\sigma$ magnitude limit of the Euclid Wide Survey in the $\IE$ filter. The distribution transitions from being dominated by galaxies with no internal extinction at $z=0$ to peaking at values of $E(B-V)\sim0.2$ at redshifts $z\gtrsim 1$. 

Our procedure to assign an SED to a galaxy consists of the following steps. First, we determine the two redshift values immediately lower, $z_1$, and larger, $z_2$, than the redshift of the galaxy from the list in footnote~\ref{footnote:z_values}, so that $z_1<z<z_2$. We determine two SEDs from our 136 template basis, one at each of these two redshifts, ${\rm SED}\,(z_1)$ and ${\rm SED}\,(z_2)$, and compute the final SED of the galaxy as a linear combination of the two templates (Eq.~\ref{eq:SED_combination}) with weights, $w_1$ and $w_2$ proportional to their redshift difference to the reference redshift as in Eq.~\eqref{eq:SED_combination_weights},
\begin{equation}
\label{eq:SED_combination}
\textit{\rm SED}\, (z) = w_1\, \textit{\rm SED}\,(z_1) + w_2\, \textit{\rm SED}\,(z_2) \;,
\end{equation}
\begin{eqnarray}
\label{eq:SED_combination_weights}
w_1 &=& \frac{z_2-z}{z_2 - z_1} \;,\\
w_2 &=& \frac{z-z_1}{z_2 - z_1}\nonumber \;.
\end{eqnarray}

We derive the SED at each of these two redshifts from the $(g01-r01)_{\rm HOD}$ colour of the galaxy. We obtain 
the ordered template number from the colour using the relations found previously (e.g., Fig.~\ref{fig:g01r01_ordered_template_z00}). At $z=0$, we use the relation with the $(g01-r01)_{\rm HOD}$ colour. At higher redshift we first obtain the COSMOS $g-r$ colour from the $(g01-r01)_{\rm HOD}$ colour using the abundance matched relations described before and then use this $g-r$ colour to obtain the ordered template number. We take the six closest ordered templates to this ordered template number from our 136 SED template basis
and compute the colour difference from the original colour, either $(g01-r01)_{\rm HOD}$ or $g-r$, depending on the redshift, and the colour of these six ordered templates. We compute a probability based on this colour difference for each of these six ordered templates, $P_{\rm colour}$. This probability is assumed to be a Gaussian distribution with mean the original colour and standard deviation half the mean value of the colour differences for the six ordered templates. If the standard deviation is lower than 0.015, we set it to this minimum value. We then compute the probability of the template given its $E(B-V)$ value, $P_{E(B-V)}$, using the distributions of $E(B-V)$ values from the COSMOS2020 catalogue obtained before (Fig.~\ref{fig:ebv_pdf_cosmos_zall}). We compute the final probability of each ordered template as the product of these two probabilities in colour and colour excess, $P_{\rm ot} = P_{\rm colour} \, P_{E(B-V)}$. We generate the cumulative distribution of this probability and draw a random number from which we obtain the ordered SED template using the inverse of the cumulative probability distribution.
We perform this procedure at the redshifts $z_1$ and $z_2$ and obtain the final SED following Eqs.~\eqref{eq:SED_combination} and~\eqref{eq:SED_combination_weights}.

\subsection{Galaxy shapes and sizes \label{subsec:shapes_sizes}}

Galaxy shapes and sizes are assigned using  phenomenological prescriptions similar to those of~\cite{Miller:13} with calibration data coming from {\it Hubble} Space Telescope (HST) observations. We use two main data sets as calibrators: the Cosmic Assembly Near-Infrared Deep Extragalactic Legacy Survey\footnote{\url{http://arcoiris.ucolick.org/candels/}}~\citep[CANDELS;][]{Grogin:11,Koekemoer:11} reductions of~\cite{Dimauro:18} and the Advanced Camera for Surveys\footnote{\url{https://science.nasa.gov/mission/hubble/observatory/design/advanced-camera-for-surveys/}} (ACS) reductions of the GOODS-South field of Niraj Welikala (hereafter referred to as HST\_GS, private communication).

\begin{figure}
\center
\includegraphics[width=\columnwidth]{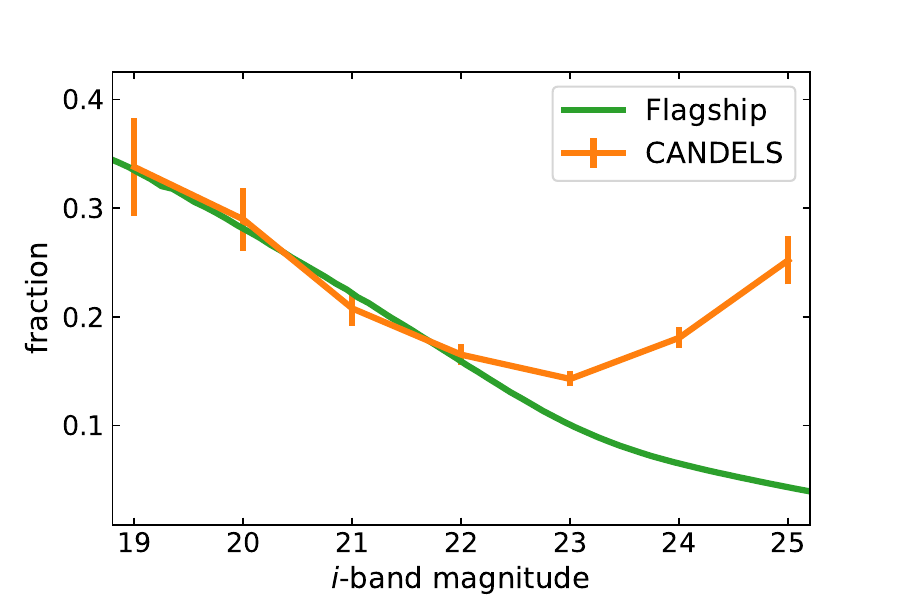}
\caption{\label{fig:fraction_only_bulge} Fraction of galaxies modelled with one component model with respect to the total as a function of $i$-band magnitude. The green line is the resulting fraction in the Flagship catalogue when the relation in Eq.~(\ref{eq:one_morph_component_selection}) is used. The orange points with error bars represent the fraction of galaxies modelled with a one-component model in the CANDELS catalogue of~\cite{Dimauro:18}.}
\end{figure}

Most galaxies are assumed to have two components: a S\'ersic profile~\citep{Sersic:63} component, which we will refer to as bulge, and an exponential disk component. The exception are galaxies fulfilling the following relation:
\begin{align}
\label{eq:one_morph_component_selection}
(g01-r01)_{\rm HOD} &> (0.96 + 0.04\,z) \\
&\notag \quad - 0.015\, [(M_{r01} -5\,\logten h)+ 24.0] \, ,
\end{align}
where $z$ is the galaxy redshift and $M_{r01} - 5\,\logten h$ is the absolute magnitude in the $r01$ filter, that are described by a single S\'ersic profile component, which we will also call bulge. We identify this modelling difference with the parameter {\tt dominant\_shape} in the catalogue which is set to 0 for the one-component (bulge only) profile galaxies and to 1 for the two-component (bulge and disk) galaxies. The selection in Eq.~\eqref{eq:one_morph_component_selection} implies that the reddest galaxies at each magnitude are the ones modelled with a one-component (bulge only) profile. The fraction of these galaxies to the total number of galaxies varies as a function of magnitude as shown in Fig.~\ref{fig:fraction_only_bulge}, with the fraction getting smaller as magnitudes get fainter. We chose to use a simple magnitude-dependent colour cut to separate the way to model the galaxies. This cut reproduces well the fraction in the CANDELS catalogue of~\cite{Dimauro:18} at magnitudes brighter than $i<22$, but it is lower at fainter magnitudes. For these fainter galaxies, their smaller sizes and lower signal-to-noise detections make the distinction of two profile components harder.   
The upward trend seen in CANDELS at $i>23$ can probably be
attributed to this effect.
Taking this into account and as the one- and two-component models are not very different for these small sizes for the \Euclid pixel size, we decided not to change this simple selection cut for faint magnitudes. As the catalogue is dominated by faint galaxies, the overall fraction of bulge-only galaxies is approximately 10\% of the total galaxy population in the catalogue.

\begin{figure}
\center
\includegraphics[width=\columnwidth]{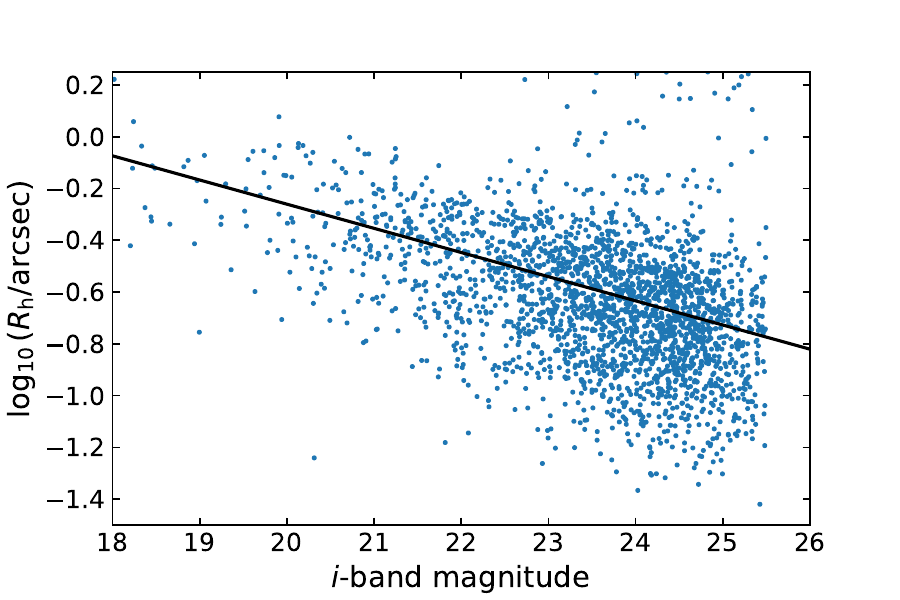}
\caption{\label{fig:scalelength_imag_goods} Scale length values in the HST\_GS sample (blue dots). %
The black line is the fit to the median values of the distribution as a function of the $i$-band magnitude as given in Eq.~(\ref{eq:scalelength_median}) }
\end{figure}

We start by assigning a scalelength, $R_{\rm h}$, to the galaxies based on their $i$-band magnitude.\footnote{This scalelength, $R_{\rm h}$, is defined as the radius at which a galaxy defined by an exponential light profile $I(R) = I_o \exp(-R/R_{\rm h})$ is a factor $e$ less bright than at its centre.} We compute the median scalelength $R_{\rm h}^{\rm median} $ in arcsec using the relation fit from the HST\_GS catalogue, which is similar to one obtained in~\cite{Miller:13},
\begin{equation}
\label{eq:scalelength_median}
\frac{R_{\rm h}^{\rm median} }{ \ang{;;1}} = \exp{[-1.675-0.215\, (i -25.0)]}\,.
\end{equation}

The observed distribution of scalelength values in magnitude bins can be described by a power law and an exponential decay with parameters depending on magnitude. The cumulative function of such a functional form is an incomplete gamma function. Therefore, we assign the value of the scalelength as
\begin{equation}
\label{eq:scalelength_assignment}
R_{\rm h} = \frac{R_{\rm h}^{\rm median}}{1.13} \left[\Gamma^{-1}\left(\frac{1+\beta}{\alpha}, \mu\right)\right]^{\frac{1}{\alpha}}\;,
\end{equation}
where $\Gamma^{-1}$ is the inverse of the incomplete gamma function,\footnote{Implemented with the python \texttt{scipy} function  \\ {\tt scipy.special.gammaincinv}} $\alpha$ is the exponent of the exponential decay argument, which depends on magnitude as $\alpha = 2.0 - 0.08 \,(i - 18)$ with $i$ being the $i$-band magnitude, $\beta$ is the exponent of the power-law, which we fix at  $\beta = 1.5$ and $\mu$ is a uniformly distributed random number in the range [0,1].

We then set the luminosity fraction in the S\'ersic profile, $F^{\sersic}$, which we refer to as bulge in the catalogue, using the following relation:
\begin{equation}
\label{eq:bulge_fraction}
F^{\sersic} = a \left[\left(1 + \frac{1}{a}\right)^{\mu} -1 \right]\;,
\end{equation}
where $a = 0.01 + 0.0015\,(25 - i)$ and $\mu$ is a uniformly distributed random number in the range [0,1]. We set a floor of 0.007 to the minimum value of $a$.
Note that this bulge component can either be the one-component profile for galaxies with ${\tt bulge\_fraction} =1$ and ${\tt dominant\_shape} =0$ or the S\'ersic profile bulge component for the two-component galaxies (${\tt bulge\_fraction} < 1$ and ${\tt dominant\_shape} =1$).
\begin{figure}
\center
\includegraphics[width=\columnwidth]{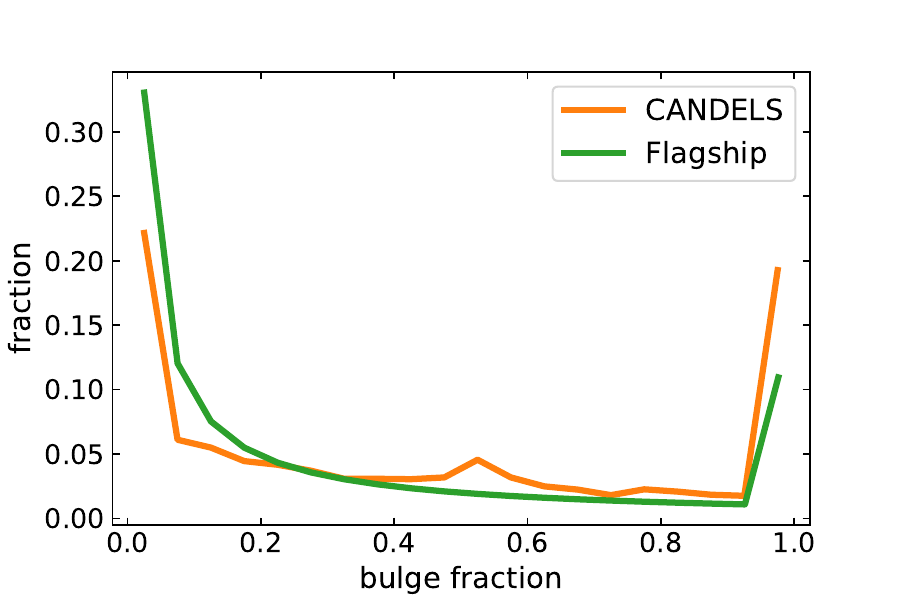}
\caption{\label{fig:bulge_fraction_i245} Bulge fraction distribution  for galaxies brighter than $\IE<24.5$ in the Flagship catalogue compared to the CANDELS catalogue of~\cite{Dimauro:18} for the same magnitude range. }
\end{figure}

We set the disk profile as a S\'ersic profile with index $n^{\sersic}=1$. We set the half-light radius of the disk as $R_{50}^{\rm disk} = 1.678\, R_h$. An exponential profile contains $\sim$26\% of the light within the scalelength radius. The factor 1.678 comes from the ratio of radii containing 50\% and 26\% of the light in a S\'ersic $n^{\sersic}=1$ profile.

\begin{figure}
\center
\includegraphics[width=\columnwidth]{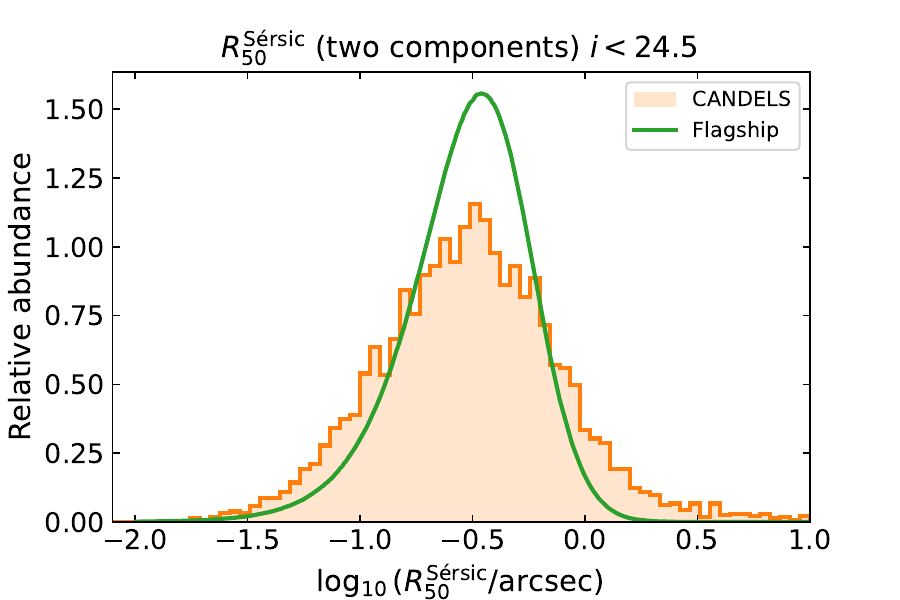}
\caption{\label{fig:bulge_r50_composite_i245} Distribution of half-light radius, $R_{50}^{\sersic}$, in the Flagship catalogue for the S\'ersic 
bulge
component of galaxies simulated as two components compared to the CANDELS  sample.}
\end{figure}

We set the S\'ersic half-light radius, $R_{50}^{\sersic}$, of the S\'ersic profile from the disk half-light radius, $R_{50}^{\rm disk}$, with 
\begin{equation}
\label{eq:bulge_2c_r50}
\begin{split}
\logten \left(\frac{R_{50}^{\sersic}}{\ang{;;1}}\right) = -1.81 & + 
\frac{2.6}{1+\exp \left\{ -2.15 \left[ \logten \left(\frac{R_{50}^{\rm disk}}{\ang{;;1}}  \right)   \right] +0.3 \right\}} \\
  & + \frac{0.7}{1+\exp \left\{ -15.0 \left[ \logten \left(\frac{R_{50}^{\rm disk}}{\ang{;;1}}  \right)   \right] -0.1 \right\}}\;.
\end{split}
\end{equation}
This equation tries to fit the relation between these radii in the CANDELS calibration catalogue. In Fig.~\ref{fig:bulge_r50_composite_i245}, we show the S\'ersic half-light radius, $R_{50}^{\sersic}$, distribution for composite profiles and compare it to the CANDELS. The Flagship distribution of $R_{50}^{\sersic}$ values is similar to the CANDELS distribution, although somewhat narrower. The HST\_GS sample has also measured values of the bulge component half-light radius in composite profiles. However, those values are approximately a factor of five smaller than the ones in CANDELS. The measuring methods were different in both samples. The galaxies in CANDELS were fit individually to two components, while in HST\_GS galaxies were combined and then the bulge half-light radius of the combined galaxy was fit as a multiplicative factor of the scale length, $R_{\rm h}$. Given the individual measurement employed, we deem the CANDELS values more reliable and decided to fit to their distribution. But, one should keep in mind the systematic uncertainty in these measurements depending on the methodology employed when using these values.

\begin{figure}
\center
\includegraphics[width=\columnwidth]{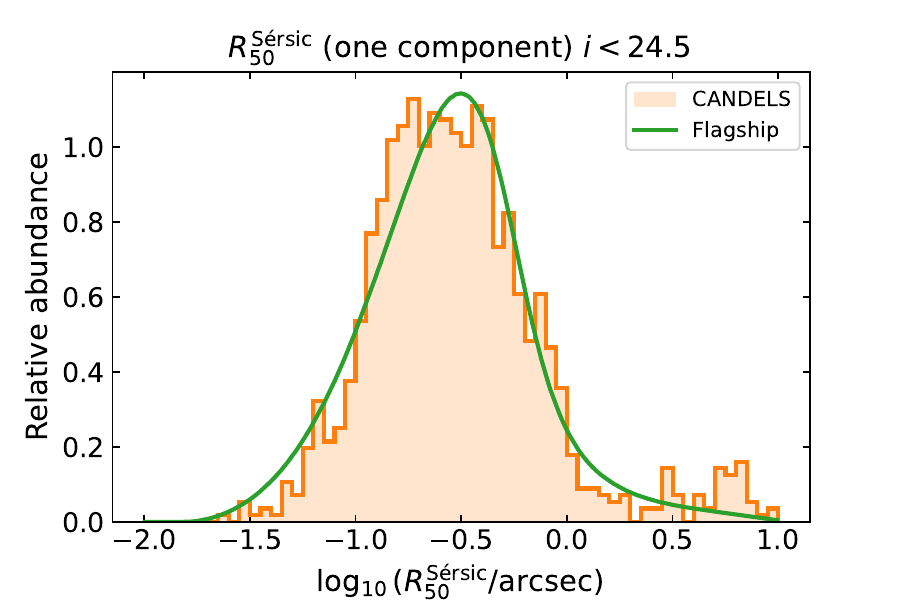}
\caption{\label{fig:sersic_r50_one_component_i245} Distribution of half-light radius, $R_{50}^{\sersic}$, in the Flagship catalogue for one-component galaxies compared to the CANDELS sample.}
\end{figure}

For one-component profiles, we assume the half-light radius of the bulge to be the same as the scalelength radius. Fig.~\ref{fig:sersic_r50_one_component_i245} shows the distribution of these $R_{50}^{\sersic}$ values for galaxies with $i<24.5$ in the catalogue compared to the CANDELS sample.

We obtain the S\'ersic index for the S\'ersic profile components with functions that try to reproduce the distribution of S\'ersic indices in the CANDELS calibrating sample.
For the composite profile, we draw S\'ersic indices with
\begin{equation}
\label{eq:sersic_n_composite}
n^{\sersic} = [(a + n^{\sersic}_{\rm min})\, \nu_{\rm max}] - a\;,
\end{equation}
where
\begin{equation}
\label{eq:nu_max_sersic_n_composite}
\nu_{\rm max} = \left(\frac{a+n^{\sersic}_{\rm max}}{a+n^{\sersic}_{\rm min}}\right)^{\mu}\,,
\end{equation}
$a=0.2$, the minimum and maximum S\'ersic indices, $n^{\sersic}_{\rm min} = 0.5$ and $n^{\sersic}_{\rm max} = 6.0$, and $\mu$ is a uniformly distributed random number in the range [0,1].

\begin{figure}
\center
\includegraphics[width=\columnwidth]{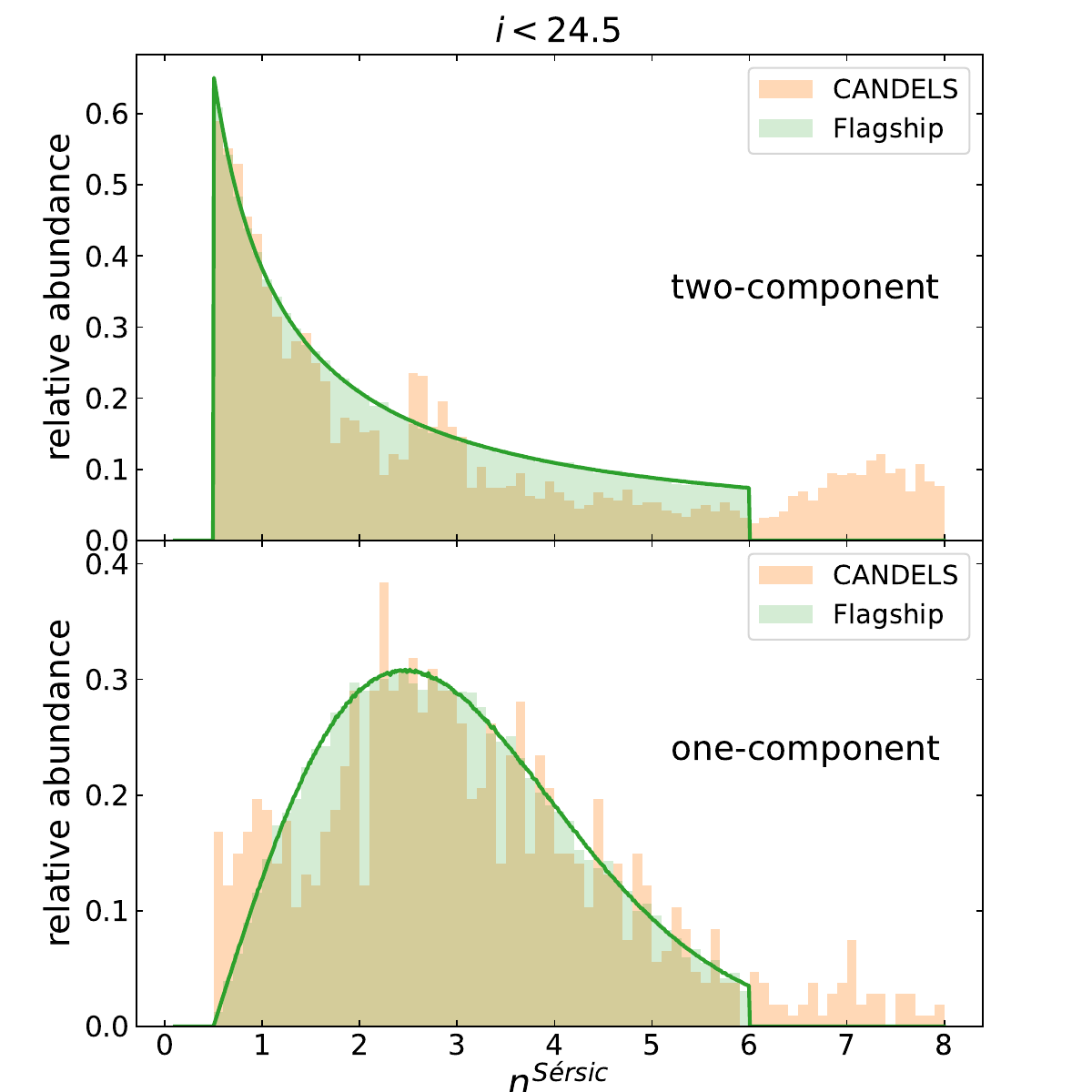}
\caption{\label{fig:nsersic_both_i245} Distribution of S\'ersic indices, $n^\sersic$, in the Flagship catalogue for the two-component galaxies (top) and the one-component galaxies (bottom) compared to the CANDELS sample for galaxies with a magnitude limit of $i<24.5$.
}
\end{figure}

For the single component, we draw the S\'ersic index from
\begin{equation}
\label{eq:sersic_n_onecomponent}
n^{\sersic} = n^{\sersic}_{\rm off} + a\left[\Gamma^{-1}\left(\frac{1+\beta}{\alpha}, \nu\right)\right]^{\frac{1}{\alpha}}\;,
\end{equation}
with $n^{\sersic}_{\rm off} = 0.5$, $a=2.75$, $\alpha=1.9$, $\beta=1.0$ and $\Gamma^{-1}$ the inverse of the incomplete gamma function as in Eq.~(\ref{eq:scalelength_assignment}).
The parameter $\nu$ is given by
\begin{equation}
\label{eq:nu_sersic_n_onecomponent}
\nu = \Gamma_{n^{\sersic}_{\rm min}} - \mu\,(\Gamma_{n^{\sersic}_{\rm min}} - \Gamma_{n^{\sersic}_{\rm max}}) \;,
\end{equation}
with
\begin{subequations}
\begin{flalign}
\label{eq:gammas_sersic_n_onecomponent}
\Gamma_{n^{\sersic}_{\rm min}} &= \Gamma\left[\frac{1+\beta}{\alpha},\left(\frac{n^{\sersic}_{\rm min} - n^{\sersic}_{\rm off}}{a}\right)^{\alpha}\right] \;, \\
\Gamma_{n^{\sersic}_{\rm max}} &= \Gamma\left[\frac{1+\beta}{\alpha},\left(\frac{n^{\sersic}_{\rm max} - n^{\sersic}_{\rm off}}{a}\right)^{\alpha}\right] \;,
\end{flalign}
\end{subequations}
where $\Gamma$ is the incomplete gamma function;\footnote{Implemented with the python scipy function \\ {\tt scipy.special.gammainc}} $n^{\sersic}_{\rm min} = 0.5$ and $n^{\sersic}_{\rm max} = 6.0$, as before, and $\mu$ is a uniformly distributed random number in the range [0,1]. Figure~\ref{fig:nsersic_both_i245} shows the distribution of S\'ersic indices for the bulges of the two-component (top) and the one-component (bottom) galaxies compared to the CANDELS sample. We have restricted the comparison to a magnitude limit of $i<24.5$. We have imposed a minimum and a maximum value to the S\'ersic indices as indicated above because the \Euclid pixel simulation pipeline~\citep[OU-SIM,][]{Serrano:24} and subsequent Science Ground Segment pipelines were very inefficient in dealing with very extended profiles. 
The maximum value is smaller than the maximum value allowed in the CANDELS measurements. As the distributions shown in Fig.~\ref{fig:nsersic_both_i245} are normalised to have an integral equal to 1, the Flagship curves have larger abundances compared to CANDELS as they sample a smaller range.

\begin{figure}
\center
\includegraphics[width=\columnwidth]{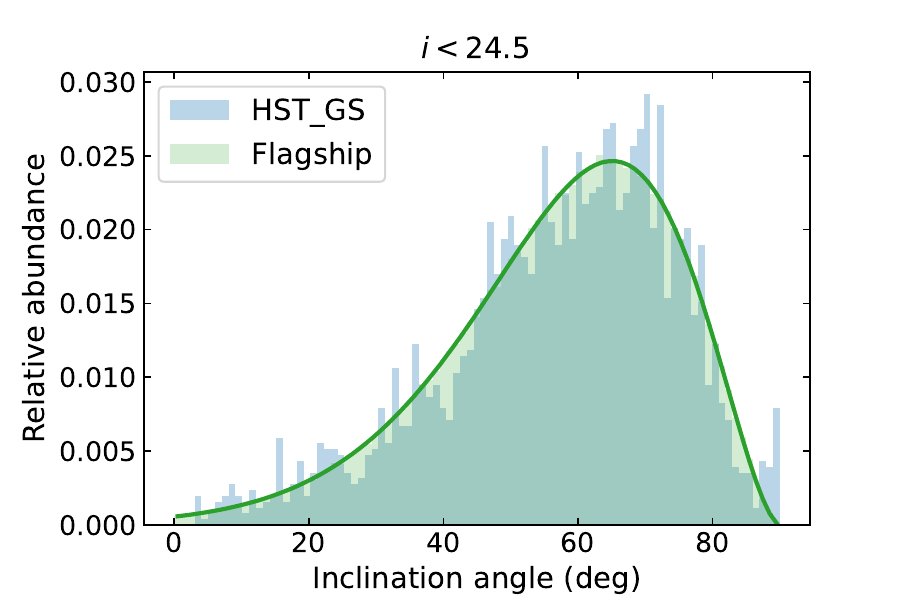}
\caption{\label{fig:inclination_angle_ds1_i245} Distribution of inclination angles, $\theta_{\rm incl}$, in the Flagship catalogue for the disk component of the two-component galaxies compared to the HST\_GS sample for galaxies with a magnitude limit of $i<24.5$.}
\end{figure}

In the \Euclid pixel simulation pipeline, we render the disk component of the two-component galaxies with the inclined disk model of Galsim\footnote{\url{https://github.com/GalSim-developers/GalSim}}~\citep{Rowe:15}. 
We compute this inclination angle, $\theta_{\rm incl}$, with
\begin{equation}
\label{eq:inclination_angle}
\theta_{\rm incl} = 90 - a\left[\Gamma^{-1}\left(\frac{1+\beta}{\alpha}, \nu\right)\right]^{\frac{1}{\alpha}}\;, 
\end{equation}
where $\theta_{\rm incl}$ is in degrees and with
\begin{equation}
\label{eq:inclination_angle_nu}
\nu = \mu\, \Gamma_{\theta_{\rm incl}^{\rm max}}\;,
\end{equation}
and
\begin{equation}
\label{eq:inclination_angle_gamma_max}
\Gamma_{\theta_{\rm incl}^{\rm max}} = \Gamma\left[\frac{1+\beta}{\alpha},\left(\frac{\theta_{\rm incl}^{\rm max}}{a}\right)^{\alpha}\right]\;,
\end{equation}
with $a=22^{\circ}$, $\alpha=1.39$, $\beta=1.65$, $\theta_{\rm incl}^{\rm max} = 90^{\circ}$ and $\mu$ a random number uniformly distributed in the range [0,1].
Figure~\ref{fig:inclination_angle_ds1_i245} shows the distribution of inclination angles for the disk component of the galaxies simulated with two components. The distribution of inclination angles in the HST\_GS sample is shown for comparison.

For the disk component, we calculate an axis ratio, $q=b/a$, where $b$ and $a$ are the semi-minor and semi-major axis respectively, from the inclination angle with
\begin{equation}
\label{eq:axis_ratio_from_inclination_angle}
q = \sqrt{\left(\cos{\theta_{\rm incl}}\right)^a + \left(c\, \sin{\theta_{\rm incl}}\right)^b} \;,
\end{equation}
with parameters $a=2.0153$, $b=4.3161$ and $c=0.35696$ for the disk component of the two component galaxies. 

We compute the ellipticity from the axis ratio as
\begin{equation}
\label{eq:ellipticity_from_axis_ratio}
\epsilon = \frac{1-q}{1+q}\,.
\end{equation}

We compute the axis ratio of the bulge of the two components profiles with the same Eq.~(\ref{eq:axis_ratio_from_inclination_angle}), but with parameter values that depend on the S\'ersic index as
\begin{subequations}
\begin{flalign}
\label{eq:axis_ratio_from_inclination_angle_param_a}
a &= 1.905 + \frac{0.25}{1+\exp \left[0.95\,\left(n^{\sersic} -0.80\right)\right]}\;,\\
\label{eq:axis_ratio_from_inclination_angle_param_b}
b & = 1.18 + \frac{0.50}{1+\exp \left[-2.0\,\left(n^{\sersic} -0.80\right)\right]} \nonumber \\
&\qquad\qquad + \frac{0.95}{1+\exp \left[3.8\,\left(n^{\sersic} -2.45\right)\right]}\;,\\
\label{eq:axis_ratio_from_inclination_angle_param_c}
c &= 0.465 + \frac{0.12}{1+\exp \left[-1.0\,\left(n^{\sersic} -1.80\right)\right]}\nonumber \\
&\qquad\qquad+ \frac{0.155}{1+\exp \left[2.0\,\left(n^{\sersic} -2.66\right)\right]} \nonumber \\
&\qquad\qquad+ \frac{0.06}{1+\exp \left[-4.5\,\left(n^{\sersic} -3.6\right)\right]}\;.
\end{flalign}
\end{subequations}

\begin{figure}
\center
\includegraphics[width=\columnwidth]{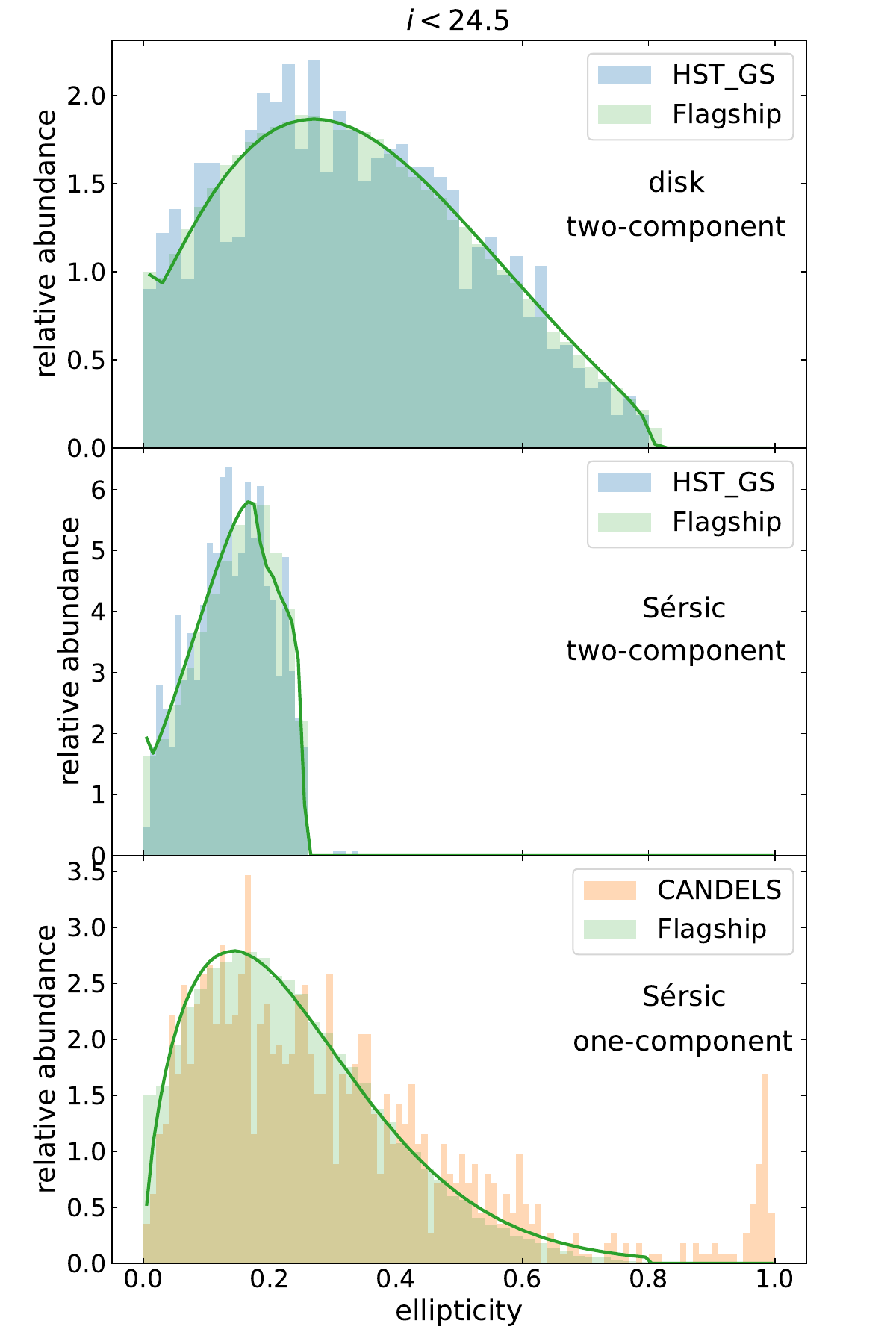}
\caption{\label{fig:ellipticity_3plots_i245} Distribution of ellipticity values, $\epsilon$, in the Flagship catalogue for the disk component (top) and S\'ersic component (middle) of the two-component galaxies compared to the HST\_GS sample; and the S\'ersic component of the one-component galaxies (bottom) compared to the CANDELS sample for galaxies with a magnitude limit of $i<24.5$. 
}
\end{figure}

We compute the ellipticity of the bulge one-component profiles with relations similar to those of Eqs.~(\ref{eq:inclination_angle}), (\ref{eq:inclination_angle_nu}), and (\ref{eq:inclination_angle_gamma_max}),
\begin{equation}
\label{eq:ellipticity_sersic_onecomponent}
\epsilon = a\,\left[\Gamma^{-1}\left(\frac{1+\beta}{\alpha}, \nu\right)\right]^{1/\alpha} \ ,
\end{equation}
where
\begin{equation}
\label{eq:ellipticity_sersic_onecomponent_nu}
\nu = \mu\, \Gamma_{\epsilon_{\rm max}}\;,
\end{equation}
and
\begin{equation}
\label{eq:ellipticity_sersic_onecomponent_gamma_max}
\Gamma_{\epsilon_{\rm max}} = \Gamma\left[\frac{1+\beta}{\alpha},\left(\frac{\epsilon_{\rm max}}{a}\right)^{\alpha}\right]\,,
\end{equation}
with $a=0.26$, $\alpha=1.5$, $\beta=0.6$, $\epsilon_{\rm max} = 0.8$ and $\mu$ a random number uniformly distributed in the range [0,1]. 
Fig~\ref{fig:ellipticity_3plots_i245} shows the resulting ellipticity distributions for the disk and bulge components of the galaxies simulated as two-components and one-component.  Finally, we set the orientation angle of the galaxy randomly.

\subsection{Galaxy intrinsic alignment}
\label{subsec:ia}
In addition to the shape modelling described in the previous subsection, we model
shapes and orientations in a separate process in order to simulate
the intrinsic alignment (IA) signal, as detailed in Hoffmann et al. (in prep.).

Our IA model consists of two steps. Approximating each galaxy as a 3D ellipsoid,
we assign in the first step two 3D axis ratios to each object, taking into account its redshift, absolute magnitude, and rest-frame colour. The parameters of the shape model
component have been calibrated such that the distribution of projected 2D axis ratios
matches the observed distribution from the COSMOS survey.
The 3D orientations of the galaxies' principal axes are modelled in a second step,
using two different methods for central and satellite galaxies. Centrals are
aligned with the principal axes of their host halo, while satellites are pointed
towards their host halo centre. These initial orientations are then randomised
depending on each galaxy's redshift, magnitude, and colour, which allows for
calibrating the dependence of the resulting IA signal on galaxy properties.
Once the 3D axis ratios and orientation have been assigned, the 2D intrinsic
shear components are obtained via projection along the observer's line-of-sight.

A novel feature of the IA model, in comparison to its predecessors, is its calibration
against alignment signals derived from multiple constraining datasets through a thorough
exploration of the simulation parameter space.
At redshifts below $z=0.36$ we calibrate against the IA signal measured from
luminous red galaxies in the LOWZ sample of the Baryon Oscillation Spectroscopic Survey~\citep{Singh:16}
as well as samples of red and blue galaxies from the
Sloan Digital Sky Survey~\citep{Johnston:19}. These low-redshift constraints
are complemented by alignment measurements in three magnitude-limited samples
at $z=1.0$ from the Horizon AGN simulation~\citep{Dubois+14}.

Validations of the alignment model are presented in Hoffmann et al. (in prep.).
A comparison of the resulting signal against theory predictions are studied in
Paviot et al. (in prep.), whereas predictions on the IA parameters for \Euclid-like samples
are investigated in Tutusaus et al. (in prep.).

\subsection{Physical parameters \label{subsec:physical_properties}}

The \Euclid galaxy clustering main probe will select emission line galaxies as the tracers to sample the large-scale structure of the Universe and conduct cosmological inference. In order to compute the emission line fluxes, we first need to estimate other physical properties of the galaxies that the emission lines
depend on.

We start by computing the galaxy ultraviolet (UV) photon flux density from the galaxy SED. We integrate the galaxy SED, without the extinction component, derived in Sect.~\ref{subsec:galaxySED}:
\begin{equation}
\label{eq:photon_flux_UV}
    f_{\rm UV}^{\rm no-ext} = \frac{\int f_{\rm SED}^{\rm no-ext}\, R_{\rm UV}(\lambda)\, \lambda^2\, \diff\lambda}{c \int R_{\rm UV}(\lambda) \,\diff\lambda}  \;,
\end{equation}
where $f_{\rm SED}^{\rm no-ext}$ is the SED flux density, $c$ is the speed of light and $R_{\rm UV}$ is a top-hat filter response in the UV with full transmission in the wavelength range from 1500 to 2300 {\AA} and zero transmission otherwise.

The star-formation rate (SFR) is computed from the UV flux, $f_{\rm UV}^{\rm no-ext}$ in Eq.~(\ref{eq:photon_flux_UV}) following the relation of~\cite{Kennicutt:98} for a Chabrier initial mass function~\citep[IMF,][]{chabrier:03},
\begin{equation}
    \frac{\rm SFR}{M_{\odot}\, {\rm yr}^{-1}} = 0.8\times 10^{-28}\,L_{\rm UV} = 9.57\times 10^{21} f_{\rm UV}^{\rm no-ext}\, d_\mathrm{L}^2 \;,
\end{equation}
where the constants are for $L_{\rm UV}$ given in ${\rm erg}\, {\rm s}^{-1} {\rm Hz}^{-1}$ units; $f_{\rm UV}^{\rm no-ext}$, in ${\rm erg}\, {\rm cm}^{-2} {\rm s}^{-1} {\rm Hz}^{-1}$; and the luminosity distance, ${d}_{\rm L}$, in \hMpc.
Equivalently,
\begin{equation}
    \logten \left(\frac{\rm SFR}{M_{\odot}\, {\rm yr}^{-1}}\right) = -0.4\,\left(M_{\rm UV}^{\rm no-ext}+18.65\right) \;.
\end{equation}

\begin{figure}
\center
\includegraphics[width=\columnwidth]{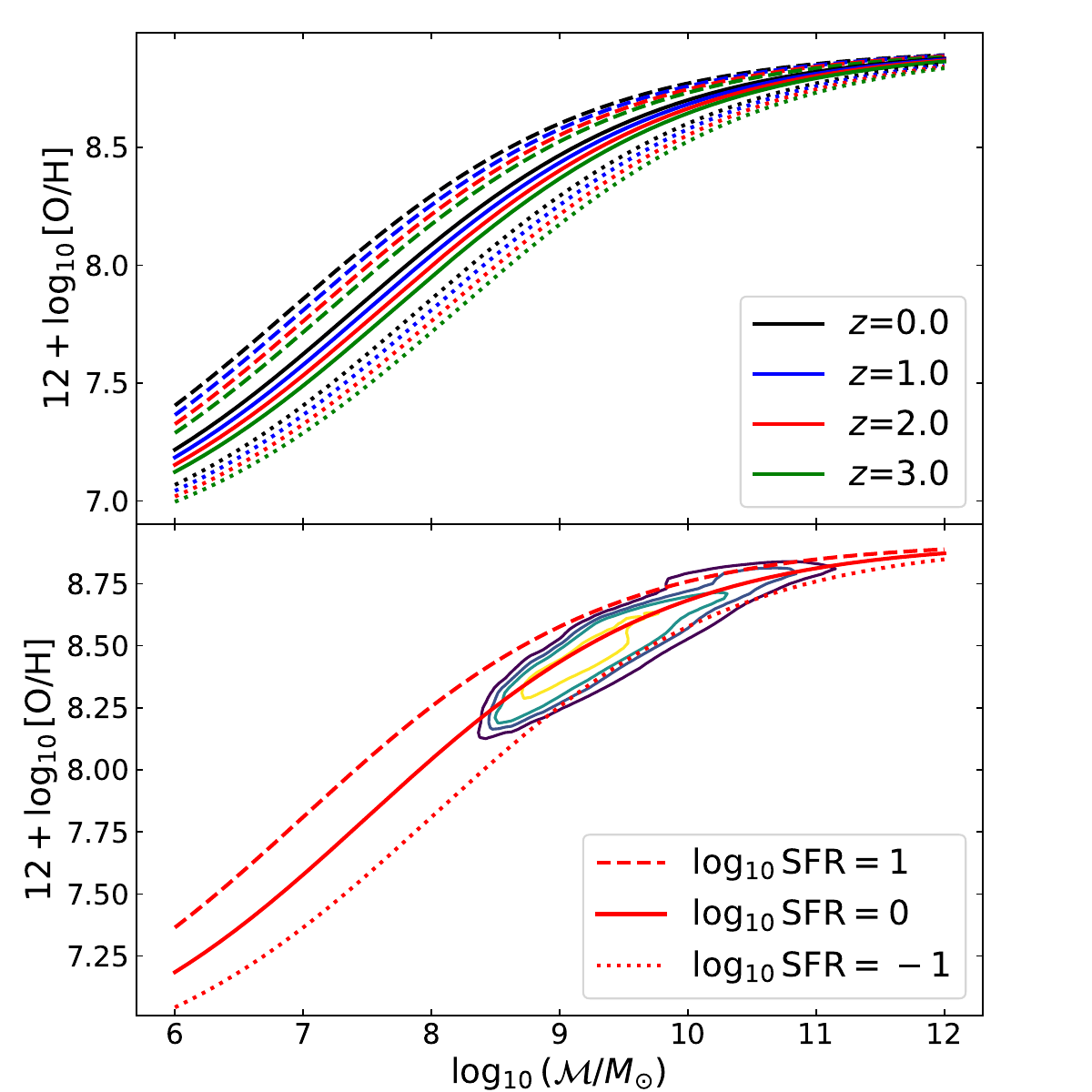}
\caption{\label{fig:MZR_allz_3SFR_and_contour_z10} 
\textit{Top}: Mean values of the metallicity as a function of the stellar mass for different values of redshift (colour code in the figure legend) and SFR (dashed, solid, and dotted lines correspond to $\logten [{\rm SFR}/(M_{\odot}\,
 {\rm yr}^{-1})] = 1, 0 ,-1$, respectively). \textit{Bottom}: The contours show the distribution of metallicity values as a function of stellar mass for a sample at redshift $z=1.0$ with magnitude limit $i<24.5$. The red lines indicate the mean value of the metallicity-stellar mass relation, Eq.~\eqref{eq:mass_metal_relation}, 
for the values of SFR indicated in the legend (that are the same as those in the top panel). }
\end{figure}

We derive the galaxy stellar mass, ${\cal M}$, from the galaxy SED and its luminosity. First, we estimate the stellar mass-to-light ratio from the galaxy colours and SED. As our reference band to compute luminosities is the $r01$ filter (see Sect.~\ref{subsec:luminosities}), we estimate the stellar mass-to-light ratio in this band. The stellar mass is then obtained simply by multiplying its $r01$-band luminosity by the stellar mass-to-light ratio.  
The galaxy metallicity is computed from the galaxy stellar mass following the relation of~\cite{Curti:20}, which we slightly modify and extrapolate to cover the whole stellar mass range covered by our sample. We assume a relation for the mean value of the stellar mass-metallicity given by
\begin{equation}
\label{eq:mass_metal_relation}
    12 + \logten \left(\frac{{\rm O}}{{\rm H}}\right) ({\cal M},{\rm SFR},z) = 6.7 + \frac{2.22}{1\!+\!\exp \left[-0.85\left({\cal M}\!-\!{\cal M}_{\rm ref}\right)\right]}\;,
\end{equation}
where ${\cal M}$ is the stellar mass and ${\cal M}_{\rm ref}$ is a reference value given by
\begin{equation}
    {\cal M}_{\rm ref} ({\rm SFR},z) = 7.5 + 0.5 \logten\left(\frac{\rm SFR}{M_{\odot}\, {\rm yr}^{-1}}\right) + 0.1 (z-1.0)\ ,
\end{equation}
where SFR is the star formation rate and $z$, the galaxy redshift. Figure~\ref{fig:MZR_allz_3SFR_and_contour_z10} shows the stellar mass-metallicity relation (MZR) as a function of redshift and SFR in the top panel. In the bottom panel, we show the contours of a galaxy sample selected at $\IE <24.5$ and redshift $z=1$ in stellar mass-metallicity space together with three MZR lines for different SFR values.

For each galaxy, we draw a realisation assuming a Gaussian distribution of the metallicity values around its mean, with a standard deviation~\citep[e.g.,][]{Tremonti:04,Curti:20}
\begin{equation}
    \sigma = 0.01 + 0.03 \, \{8.92 - [12 + \logten({\rm O/H})]_{\rm mean}\}\;,
\end{equation}
where $[12 + \logten({\rm O/H})]_{\rm mean}$ is the mean value of the metallicity given by Eq.~\eqref{eq:mass_metal_relation}.

\subsection{Galaxy emission lines \label{subsec:emissionlines}}

The determination of the galaxy redshift with the \Euclid NISP instrument~\citep{EuclidSkyNISP} in slitless mode rely mostly on the identification of emission lines. Therefore, 
the computation of emission line fluxes is important to understand the completeness and purity of the emission line-selected galaxies that will be used for the clustering cosmological probe.

First, we compute the H${\alpha}$ line flux, $f_{{\rm H}\alpha}$, from the star-formation rate  using the~\cite{Kennicutt:98} relation adapted to the Chabrier IMF, using the un-extincted UV absolute AB magnitude 
\begin{equation}
\label{eq:halpha_flux}
\logten \left(\frac{f_{\rm H\alpha}}{{\rm erg}\,{\rm cm}^{-2}\,{\rm s}^{-1}}\right) = -0.4\, M_{\rm UV}^{\rm no-ext} -2\, \logten \left(\frac{d_\mathrm{L}}{\si{\hMpc}}\right) - 16.44\;,
\end{equation}
where $f_{{\rm H}\alpha}$ is the flux in the H$\alpha$ line, $M_{\rm UV}^{\rm no-ext}$ is the unextinguished UV absolute AB magnitude, and $d_L$, the luminosity distance. We add a random scatter following a Gaussian distribution of standard deviation of 0.05 in the logarithm of the flux to the computed logarithmic value of the flux.

\begin{figure}
\center
\includegraphics[width=\columnwidth]{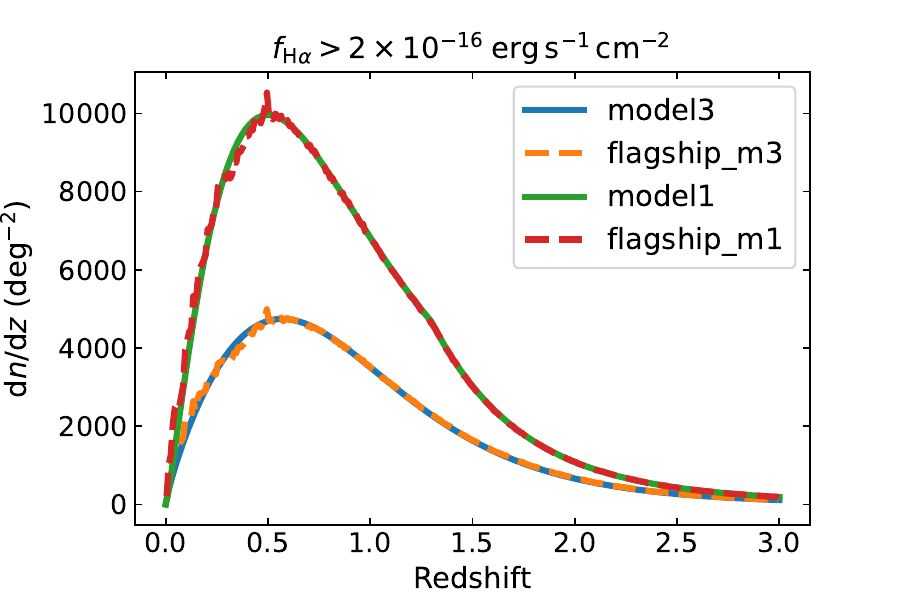}
\caption{\label{fig:dndz_halpha157_m1_m3} 
Redshift distribution (galaxies per deg$^{2}$ per unit redshift) of samples selected with a flux limit cut of  $f_{\rm H \alpha}=2\times10^{-16}$ erg s$^{-1}$ cm$^{-2}$. The sold lines show the redshift distributions from models 1 (light blue) and 3 (light green) of \cite{Pozzetti:16}. The dashed lines are the distributions in the Flagship catalogue for the same flux cut for the model 1 (orange) and 3 (red) simulated fluxes. 
}
\end{figure}

Our H$\alpha$ fluxes are estimated from the UV fluxes that rely on the SED assignment procedure, which is based on observed optical colours (see Sect.~\ref{subsec:galaxySED}). We then compute the dust-extinguished H${\alpha}$ flux. We get the stellar continuum extinction at the H${\alpha}$ wavelength using the extinction law and colour excess value $E(B-V)$ of the galaxy SED (Sect.~\ref{subsec:galaxySED}). We then convert the stellar continuum extinction to a nebular emission line extinction using the redshift-dependent factor $0.44 + 0.2\,z$~\citep{Calzetti:00, Saito:20}, which we clip at 0.44 and 1.0 as minimum and maximum values, respectively, and apply this extinction factor to the H$\alpha$ flux. We provide both the extinguished and un-extinguished fluxes in our catalogue. Finally, we calibrate the resulting $f_{{\rm H}\alpha}$ distribution to the~\cite{Pozzetti:16} models applying a correction to the computed H${\alpha}$ fluxes with abundance matching to the models. Figure~\ref{fig:dndz_halpha157_m1_m3} shows the redshift distributions for the simulated model 1 and model 3 fluxes.  

In the SED assignment procedure, we compute each SED as a linear combination of SEDs at fixed redshift values. In particular, we assign the extinction with distributions computed at particular redshifts given in footnote~\ref{footnote:z_values}. The H$\alpha$ flux is computed from the unextinguished UV flux. In the process of subtracting the extinction from the SEDs, we introduce discontinuities in the H$\alpha$ flux-redshift distribution at those particular redshifts. As a temporary fix, we compute a correction of the H$\alpha$ fluxes to smooth the distributions at those redshifts. In  Fig.~\ref{fig:dndz_halpha157_m1_m3}, we can see the performance of this smoothing procedure that was applied at redshifts $z>0.5$, but not at $z=0.5$ as there is not enough volume in the simulation lightcone to do it properly.

\begin{figure}
\center
\includegraphics[width=\columnwidth]{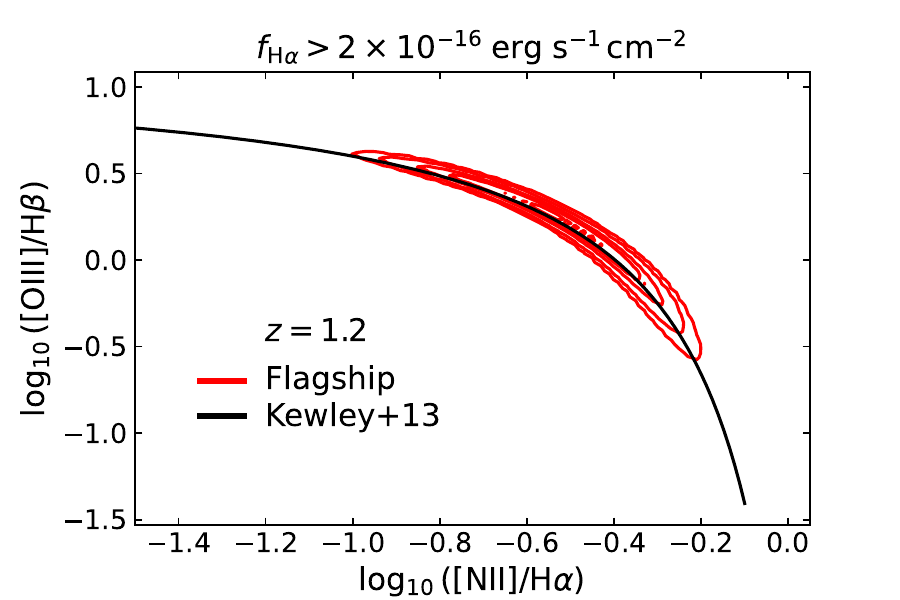}
\caption{\label{fig:BPT_z120} 
BPT diagram at $z=1.20$. The red contours show the density of Flagship galaxies in this diagram. The black solid line is the relation of~\cite{Kewley:13}.  %\gam{Perhaps zoom in.}
 }
\end{figure}

The other hydrogen lines of the Balmer and Paschen series are computed from the H$\alpha$ flux assuming case B recombination from~\cite{Osterbrock:06}. The [\ion{O}{ii}] and [\ion{O}{iii}] fluxes are computed from the H$\beta$ fluxes and the metallicity of the galaxy taking into account the relations of the [\ion{O}{ii}]/H$\beta$ and [\ion{O}{iii}]/H$\beta$ flux ratios as a function of metallicity of~\cite{Curti:20}. The [\ion{N}{ii}] fluxes are calculated from the [\ion{O}{iii}]/H$\beta$ flux ratio and the H$\alpha$ flux following the Baldwin-Phillips-Terlevich (BPT) diagram~\citep{Baldwin:81} using the relations of~\cite{Kewley:13} that include redshift evolution.\footnote{Note that we use the relation of the star formation sequence given in~\cite{Kewley:13}. This relation is not to be confused with the relation of the classification line that separates the star forming sequence and the active galactic nuclei (AGN) region given in~\cite{Kewley:13b}.}
The [\ion{S}{ii}] flux is computed from the H$\alpha$ flux and the metallicity following the relation of~\cite{Curti:20}. We obtain the [\ion{S}{iii}] flux
from the [\ion{S}{ii}] flux and a fit to the relation between the [\ion{S}{iii}]/[\ion{S}{ii}] and [\ion{O}{iii}]/H$\beta$ flux ratios shown by~\cite{Mannucci:21} in their Fig.~2.
We generate a realisation of all fluxes adding scatter to the value obtained with the relations used. As an example, Fig.~\ref{fig:BPT_z120} shows the position of the Flagship galaxies at $z=1.2$ and with H$\alpha$ flux $f_{\rm H \alpha}>2\times10^{-16}$ erg s$^{-1}$ cm$^{-2}$ in this diagram, together with the~\cite{Kewley:13} relation at this redshift.

\begin{figure*}
\includegraphics[width=\textwidth]{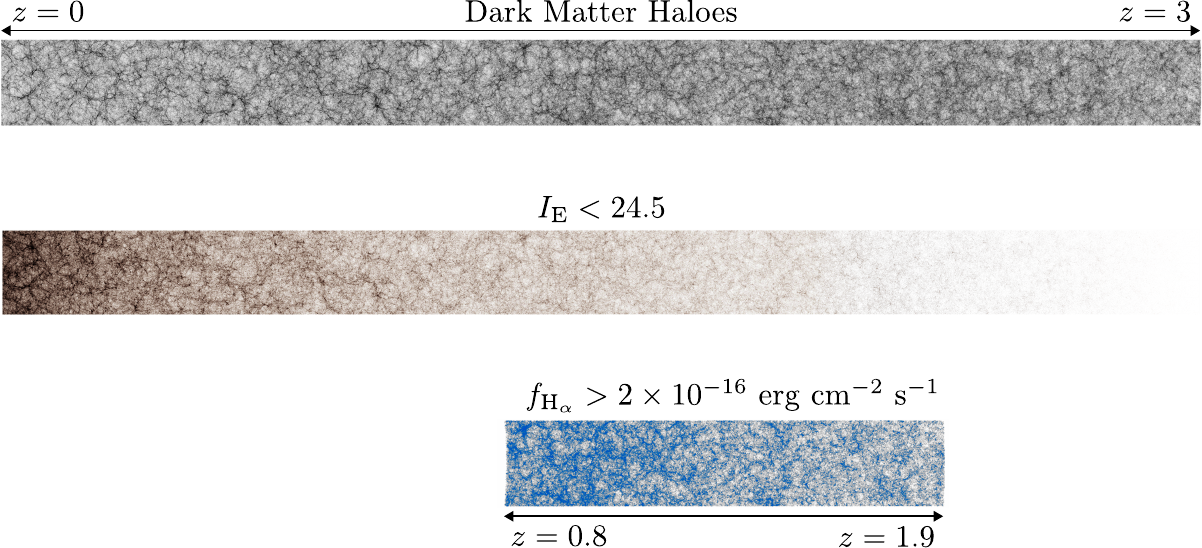}
    \caption{Flagship galaxy catalogue: stripes show sections of the lightcone for $z<3$ for dark matter haloes (top panel), the photometric sample for $\IE <24.5$ (middle), and an H${\alpha}$ spectroscopic galaxy sample for a flux cut $f_{{\rm H}\alpha} > 2 \times 10^{-16}\,{\rm erg}\,{\rm cm}^{-2}\,{\rm s}^{-1}$ (bottom), which selects galaxies only within the range $0.9<z<1.8$ given the NISP red grism wavelength coverage.
    The stripe abscissae share the same scaling in comoving distance. 
    The colour scaling in each panel depends on the density of the sample.
    %The lower stripe (NISP) is shown with lower density cuts relative to the middle one, for clarity. 
    %\gam{Could you use a lighter weight for the three titles and use scientific \LaTeX\ notation for the H$\alpha$ min flux?}
    }
    \label{fig:flagship_mock}
\end{figure*}

\subsection{Galaxy lensing }
\label{sec:gallens}

The implementation of lensing properties on mock
galaxies uses the all-sky dark matter lensing maps discussed in Sect.~\ref{sec:lens}. Given this set of 2D maps covering the entire redshift range of the lightcone output of the simulation, our procedure is the same as the one originally implemented in the MICE simulations \citep{Fosalba:15b}. This is a simple 3-step algorithm that we describe below.

\begin{enumerate}

\item{
for a given galaxy at the 3D
position in the lightcone $(\hat{\vec{n}}, z)$, where $\hat{\vec{n}}$ gives its angular
  position in the sky and $z$ its redshift, find the
  corresponding 3D pixel in the {\it discretised} lightcone, with pixel center coordinates,
  $(\hat{\vec{n}}_i, z_j)$, where the galaxy sits in (i.e., the 3D pixel in the suite of
  `onion slices' or all-sky lensing maps in \texttt{HEALPix} tessellation
  described in  Sect.~\ref{sec:lens})},

\item{
get the lensing values for this 3D pixel using the
dark matter all-sky lensing maps, $\vec{L}_{i,j}
    \equiv \vec{L} (\hat{\vec{n}}_i, z_j)$, where the components of
    the lensing vector are ${\vec L} = (\kappa, \gamma_1,\gamma_2)$, which is
    convergence and shear, and}

\item{
assign these pixelised dark  matter lensing values, $\vec{L}_{i,j}$, to the mock galaxy.}
\end{enumerate}

The resulting implementation of galaxy lensing using the above method is thus limited by the pixel resolution used, \texttt{HEALPix} $N_{\rm side}=8192$, which corresponds to a pixel scale of $\ang{;0.43;}$.  Therefore, we only expect to accurately model lensing observables down to $\sim \ang{;0.5;}$ scales, as we will discuss in detail when we validate the mock properties below (see Sect.~\ref{sec:galwl}). A further limitation intrinsic to this method is that different galaxies falling within a given 3D pixel in the `onion universe' grid pattern of the lightcone will have identical lensing properties. These two limitations can be overcome using the same approach but using a finer pixel scale (i.e., higher $N_{\rm side}$) and/or using interpolation schemes.

\section{Mock validation
\label{sec:validation}
}

Fig~\ref{fig:flagship_mock} shows a small rectangular portion of the Flagship catalogue in the lightcone.\footnote{This figure has used the Splotch package: \url{http://www.mpa-garching.mpg.de/~kdolag/Splotch}} The figure illustrates the breath of the simulation and covers a region of $4368\,\hMpc$ (horizontal direction) $\times$ $300\,\hMpc$ (vertical direction) with a depth of $80$, $160$ and $200\,\hMpc$ for the top, middle and bottom panels respectively, with the observer located in the left-bottom corner. In the top panels we show the dark matter haloes where one can see the growth of structure progressing from right, $z=3$, to left, $z=0$. The middle and lower panels present likely galaxy samples to be used in the \Euclid cosmological analysis. The middle panel shows the galaxies selected with an $\IE <24.5$ magnitude cut, which we expect to be representative of the weak lensing sample. In the lower panel, we show a sample selected with an H$\alpha$ flux cut  $f_{{\rm H}\alpha} > 10^{-16}$ $\rm erg\, cm^{-2}\, s^{-1}$, trying to mimic the galaxy clustering sample. 

\subsection{General galaxy properties}

In order to assess the performance and possible limitations of the Flagship galaxy catalogue, we compare the galaxy properties, distributions, and relations to observations.
In particular, we compare the mock galaxies to observed quantities relevant for the \Euclid cosmological probes, including observed galaxy number densities, colours, emission lines fluxes and ratios, as well as sizes at different redshifts.

\begin{figure}
	\includegraphics[width=\columnwidth]{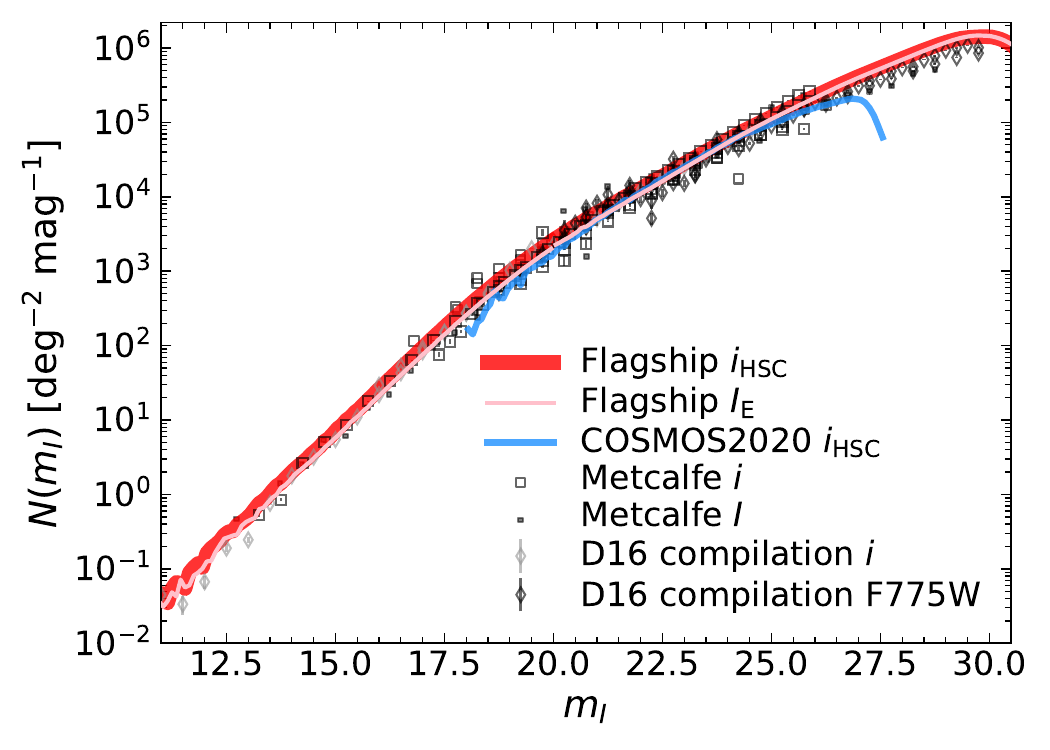}
    \includegraphics[width=\columnwidth]
    {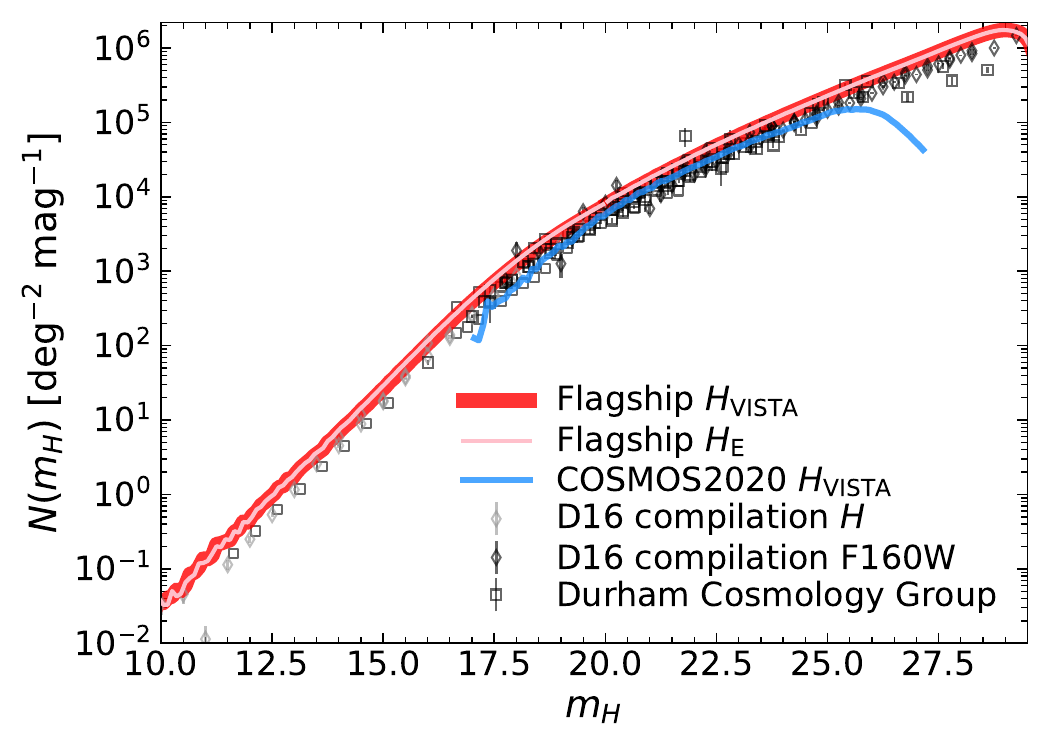}
    \caption{Galaxy number densities in the \IE\ and \HE\ bands compared to literature data: COSMOS2020 \citep{Weaver:22} counts in $H_\mathrm{VISTA}$ and $i_\mathrm{HSC}$ are derived from the `farmer' version of the photometric catalogue, after removing masked regions and objects classified as stars; \cite{Driver:16}; Durham Cosmology Group's counts are from the compilation available at the dedicated web page (\url{http://star-www.dur.ac.uk/\textasciitilde nm/pubhtml/counts/counts.html}).
    }
    \label{fig:counts}
\end{figure}

In Fig.~\ref{fig:counts}, we compare the number counts of the Flagship simulation to literature data and COSMOS2020 \citep{Weaver:22} counts of magnitudes corrected for Galactic extinction. We show the number density per unit area and magnitude in the \IE\ and \HE\ bands, as well as in similar filters observed in COSMOS, without applying further corrections to account for the differences in the filter transmissions. Also,  fluxes and magnitudes from Flagship are intrinsic, while the observed ones are affected by the photometric noise, and their distribution could, therefore, be broader.

In general, we find excellent agreement with observations in the optical band from the brightest to the faintest objects in the mock.  We find, however, some differences in the \HE band, where the galaxy number density in the mock is about $40\%$ higher than in the literature. At the EWS limit of $\HE<24.0$ (detection limit at 5$\,\sigma$ for point sources) the integrated number density in the mock is about $188\,000$ deg$^{-2}$ compared to $\sim 131\,000$ deg$^{-2}$ in the COSMOS field. 

\begin{figure}
 \includegraphics[width=\columnwidth]{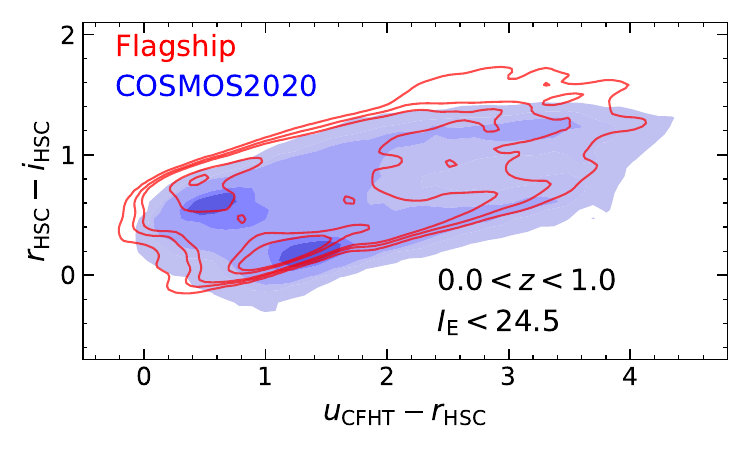}

 \vspace{-2.5mm}
 	\includegraphics[width=\columnwidth]{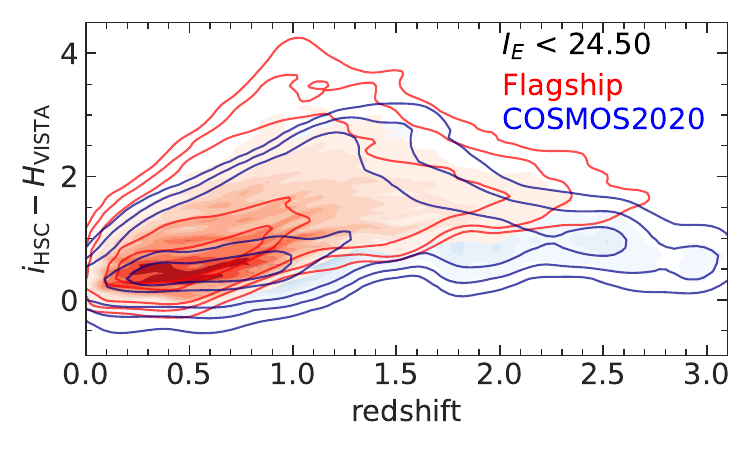}
    \caption{Galaxy colour-colour and colour-redshift diagrams compared to COSMOS2020 data.}
    \label{fig:colors}
\end{figure}

In Fig.~\ref{fig:colors}, we also explore how galaxy colours compare with the COSMOS2020 catalogue: the optical $(r-i)$ vs. $(u-r)$ diagram is quite well reproduced at the EWS magnitude limit, with a small offset. The evolution with redshift of the $i_\mathrm{HSC}-H_\mathrm{VISTA}$ galaxy colour (a proxy for $\IE-\HE$), not used in the input calibration,  also reproduces well the trend observed in COSMOS2020 (for which we are using the photometric redshift obtained with LePhare; \citealp{Ilbert:06}). The lack of galaxies with the bluest colours in Flagship compared to the COSMOS2020 distribution is likely to be the cause of the overestimation in the $H$-band counts. 

\begin{figure}
	\includegraphics[width=\columnwidth]{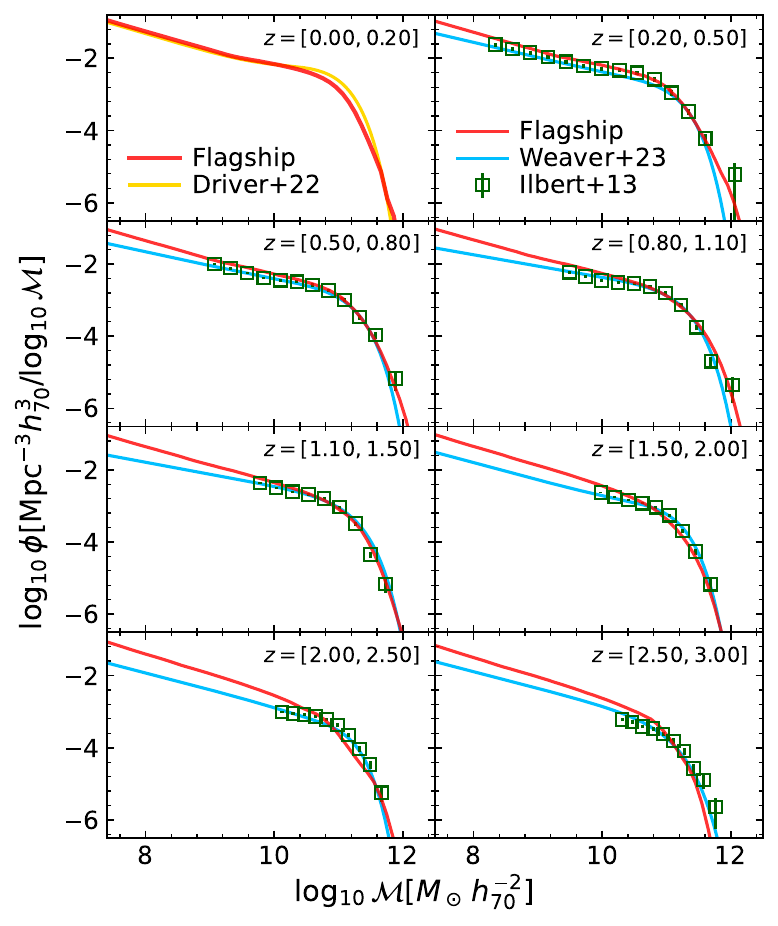}
    \caption{Galaxy stellar mass function at different redshifts. In red the GSMF derived from the Flagship lightcone compared to GSMF derived at $z<0.1$ in GAMA \citep[][yellow line]{Driver:22} and at higher redshifts in COSMOS from the ULTRAVISTA sample \citep[][green squares]{Ilbert:13} and the COSMOS2020 release \citep[][blue lines]{Weaver:23}. Note that we have transformed the Flagship stellar masses to units of the Hubble constant of $H = 70\, h_{70}\, {\rm km}\,{\rm s}^{-1}\,{\rm Mpc}^{-1}$ to bring the units of the Flagship catalogue to the ones used by the other surveys.}
    \label{fig:GSMF}
\end{figure}

In Fig.~\ref{fig:GSMF}, we show the galaxy stellar mass function (GSMF) derived for the full octant of our Flagship lightcone in various redshift bins. The high resolution of the simulation allows us to derive the GSMF down to very low mass limits ($\sim 10^8 M_\odot$) up to $z=3$. Comparison to observations in the COSMOS field~\citep{Ilbert:13} shows a good consistency. In particular, FS2 is always consistent at high mass while at $z>1.5$, it shows an excess below the characteristic galaxy mass of $10^{10} M_\odot$.

\begin{figure}
	\includegraphics[width=\columnwidth]{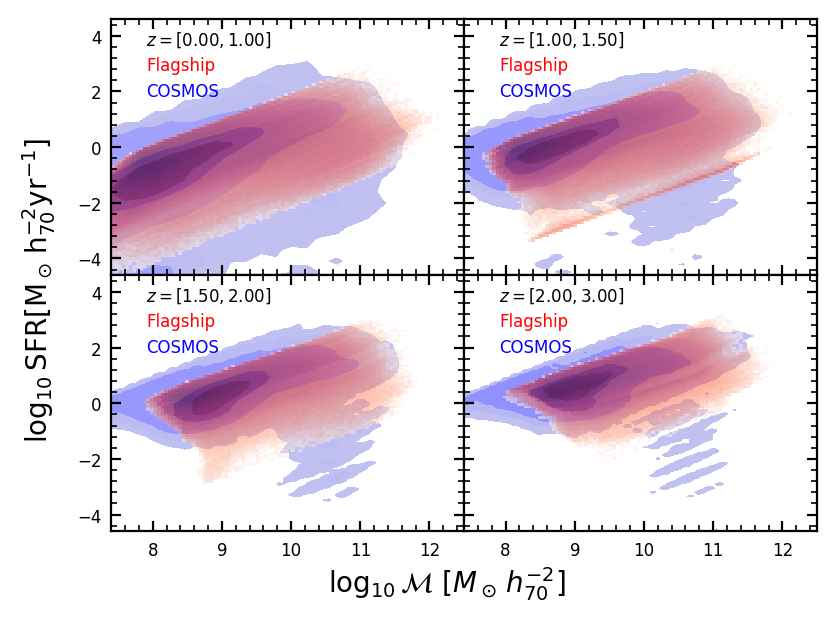}
    \caption{Galaxy star-formation rate -- stellar mass relation at different redshifts. The Flagship lightcone catalogue (red shaded area) is compared to COSMOS2020 \citep[][blue contours]{Weaver:23}.  }
    \label{fig:SFR_M}
\end{figure}

In Fig.~\ref{fig:SFR_M}, we show the SFR mass relation at different redshifts from the simulation. The comparison to observational data in the COSMOS field is fairly satisfactory, with a small remaining offset and broader relation in COSMOS, due to observational errors. The wide \Euclid survey area will allow us to map this relation up to very high mass where previous observations are limited by area and statistics.

\begin{figure}
	\includegraphics[width=\columnwidth]{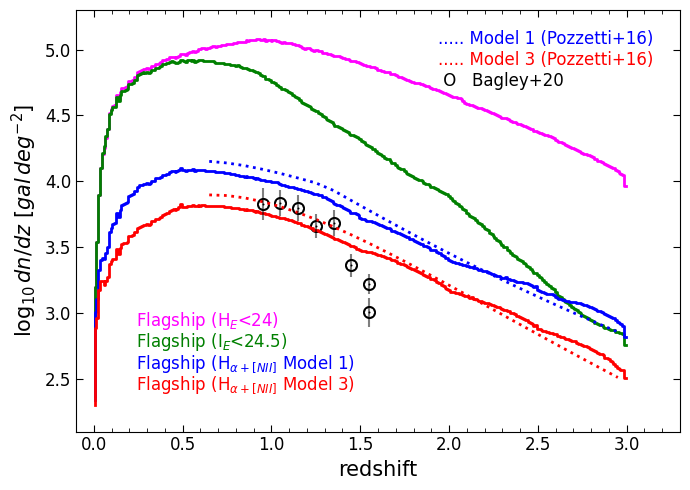}
    \caption{Expected number densities for photometric and spectroscopic emission line flux selected samples as a function of redshift in the \Euclid Wide survey. Different colours show various selection cuts for the EWS (in $\IE$, $\HE$ and in the 2 model calibrations for the line fluxes of $f_\mathrm{H\alpha+[\ion{N}{ii}]}$ $> 2\times 10^{-16}$ erg s$^{-1}$ cm$^{-2}$). The empirical model of~\cite{Pozzetti:16} and the data %from the WISP survey 
    based on slitless HST spectroscopy~\citep{Bagley:20} are also shown. }
    \label{fig:Ha_dNdz}
\end{figure}

Besides photometric properties, the most important quantities to validate for the galaxy clustering \Euclid cosmological probe are the emission line fluxes and number densities.
Fig~\ref{fig:Ha_dNdz} shows the emission line galaxy densities for a flux limit of  $f_\mathrm{H\alpha+[\ion{N}{ii}]}>2\times 10^{-16}$\,erg\,s$^{-1}$\,cm$^{-2}$, together with the global EWS limit of $\IE<24.5$ ($10 \,\sigma$) and $\HE<24$ ($5\,\sigma$). As expected, the H$\alpha$ emitters spectroscopically observed by \Euclid will be only a few percent of the total EWS photometric sample.
Finally, as described in Sect.~\ref{subsec:emissionlines}, we verify that the Flagship mock has been calibrated to the empirical models of~\cite{Pozzetti:16}. 
Some differences are still present, when we consider the combination of H$\alpha$+[\ion{N}{ii}], mainly due to [\ion{N}{ii}] contribution to the total flux, which is assumed constant in the models and not in the Flagship catalogue. 
In particular, the redshift distribution of fluxes calibrated on model 3 is fairly consistent with the one from~\cite{Bagley:20} when we consider a sample limited in H$\alpha$+[\ion{N}{ii}] fluxes to $2\times 10^{-16}$ erg s$^{-1}$ cm$^{-2}$. %(see the lower panel of Fig.~\ref{fig:Ha_dNdz}). 
From the Flagship catalogue, calibrated on the models 3 and 1 of~\cite{Pozzetti:16}, we expect that the EWS will map a density of about $3500$--$6300$  deg$^{-2}$ H$\alpha$+[\ion{N}{ii}] emitters with fluxes above $2\times 10^{-16}$ erg s$^{-1}$ cm$^{-2}$ in the redshift range $0.9<z<1.8$, among which about $2000$--$3800$ deg$^{-2}$ have H$\alpha$ flux brighter than that same flux limit. In the Euclid Deep Survey (EDS), these numbers increase to about $38\,000$--$55\,000$ deg$^{-2}$ in the redshift range $0.4<z<1.8$ with H$\alpha$+[\ion{N}{ii}] fluxes above $5\times 10^{-17}$ erg s$^{-1}$ cm$^{-2}$. 

\begin{figure}
	\includegraphics[width=\columnwidth]{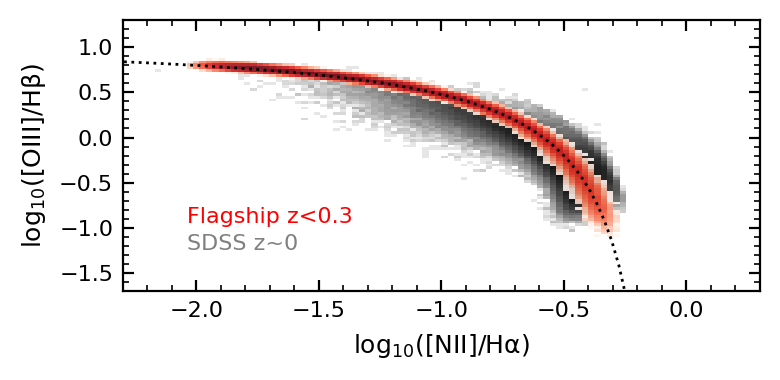}
 \includegraphics[width=\columnwidth]{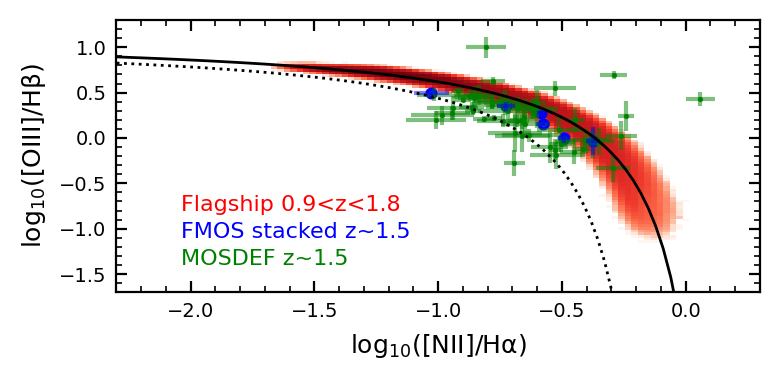}
    \caption{BPT diagram. {\it Top panel:} Emission line ratios in the redshift range $z<0.5$ for Flagship (colour map) compared to SDSS (in grey) main target sample for objects classified as galaxies. {\it Bottom panel:} Emission line ratio in the redshift range $0.9<z<1.8$ compared to FMOS stacked data in COSMOS at $z\sim 1.5$ from ~\cite{Kashino:19} and MOSFDEF data from \cite{Kriek15} and \cite{Reddy15}. The black lines are the relations by~\cite{Kewley:13} at $z\sim0$ (dotted lines) and $z\sim1.5$ (solid lines).}
    \label{fig:BPT}
\end{figure}

\begin{figure}
	\includegraphics[width=\columnwidth]{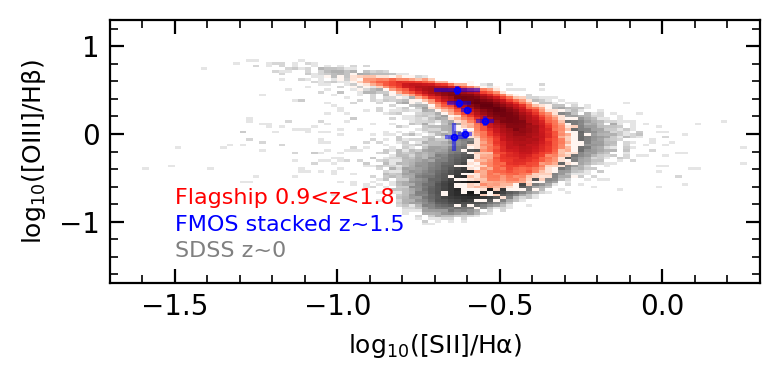}
 
 \includegraphics[width=\columnwidth]{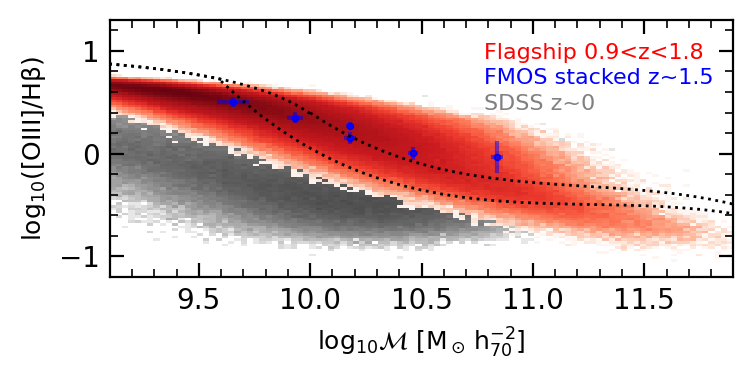}
    \caption{Emission lines ratios in the Flagship simulation. Flagship galaxy fluxes (colour map) compared to SDSS fluxes (grey map) from the main target sample for objects classified as galaxies in the local Universe. In blue, we also show FMOS stacked data in COSMOS from ~\cite{Kashino:19}. {\it Top panel:}  [\ion{O}{iii}]/H$\beta$ versus [\ion{S}{ii}]/H$\alpha$.  {\it Bottom panel:} [\ion{O}{iii}]/H$\beta$ versus stellar mass; the thin dotted curves indicate the divisions between star-forming/composite galaxies and AGN at $z\simeq 0$ \citep{Juneau2014}, with AGN populating the region of larger [\ion{O}{iii}]/H$\beta$ above the dotted lines.}
    \label{fig:lines}
\end{figure}

We underline, however, that the measurements of galaxy redshift rely mainly on the identification of emission lines and on their ratios.
We show in Fig.~\ref{fig:BPT} how our mock catalogue populates the BPT emission line ratio diagram~\citep{Baldwin:81} and how it compares with observations. In particular, we show that we recover quite well the main locus of the BPT diagram from the SDSS galaxy main sample at low redshift ($z<0.3$) and also the trend to higher values shown by high-redshift observation in the Fiber Multi-Object Spectrograph (FMOS)-COSMOS survey~\citep{Kashino:19} and by the MOSFIRE Deep Evolution Field (MOSDEF) survey with Keck by \cite{Kriek15} and \cite{Reddy15} in the redshift range $1.4<z<1.7$.
We also show other diagrams with various emission line ratios (Fig.~\ref{fig:lines}), which confirms that the assignment of emission line fluxes is consistent with those observed in SDSS galaxies at low redshift and the FMOS-COSMOS and MOSDEF surveys at higher redshifts.

\begin{figure}
	\includegraphics[width=\columnwidth]{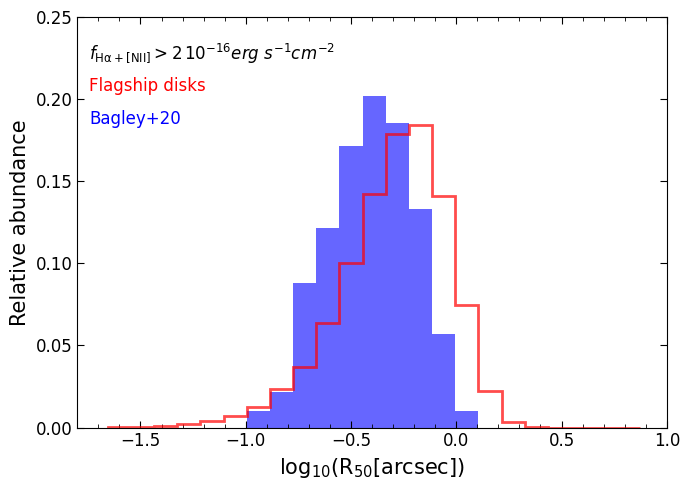}
    \caption{Size distribution. Disk half-light radius distribution of an H$\alpha$ selected sample, at the limit of the EWS in the redshift range $0.9<z<1.6$, compared to data from \cite{Bagley:20}.   %\gam{1) I would use log scaling for the $y$ axis. 2) Increase the sizes of the axis labels and legends.}
    }
    \label{fig:Size}
\end{figure}

\begin{figure}
	\includegraphics[width=\columnwidth]{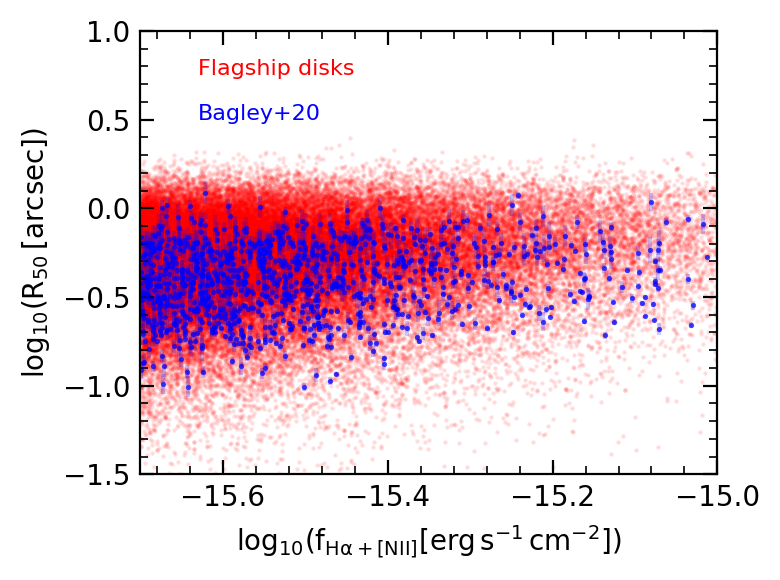}
    \caption{Galaxy half-light radius versus H$\alpha$ flux compared to observed data of ~\cite{Bagley:20}. In red, the data from the Flagship catalogue. In blue, the data from \cite{Bagley:20}}
    \label{fig:Size2}
    % change the cyan colour to something easier to distinguish
\end{figure}

Finally, the weak lensing cosmological probe relies on the measurement of galaxy shapes and sizes. Furthermore, the spectroscopic signal-to-noise ratio of the \Euclid slitless spectra depends on galaxy size and, with it, the probability of detecting and measuring emission lines and galaxy redshifts for the clustering cosmological probe. The procedure to assign galaxy sizes distribution was already described in Sect.~\ref{subsec:shapes_sizes} and its comparison to the calibration data shown in Figs.~\ref{fig:bulge_r50_composite_i245} and \ref{fig:sersic_r50_one_component_i245}. Additionally, in Figs.~\ref{fig:Size} and~\ref{fig:Size2} we show the distribution of disk sizes  
of H$\alpha$ emitters in the redshift range $0.9<z<1.6$ of a sample limited in flux to $f_\mathrm{H\alpha+[\ion{N}{ii}]}>2\times 10^{-16}$ erg s$^{-1}$ cm$^{-2}$ and compare it to the observed distributions from slitless spectroscopic HST Wide Field Camera 3 (WFC3) data sets by~\cite{Bagley:20}. We note that the distribution is realistic, with pretty similar dispersion, even if the median predicted disk size in the FS2 is around \ang{;;0.48}, compared to \ang{;;0.35} in ~\cite{Bagley:20}. This difference can be explained by a selection effect, as the detection probability is dependent on galaxy size, with more compact galaxies having a higher probability of being detected. 
We performed a preliminary comparison of the predicted size distribution under this selection effect, obtaining a consistent shift in the size distribution. The description of the techniques needed to perform this test is beyond the scope of this paper. The details of the selection procedure and its dependence on galaxy properties will be presented in Euclid Collaboration: Monaco et al. (in prep.).

\subsection{Galaxy clustering}

\begin{figure}
    \includegraphics[width=\columnwidth]{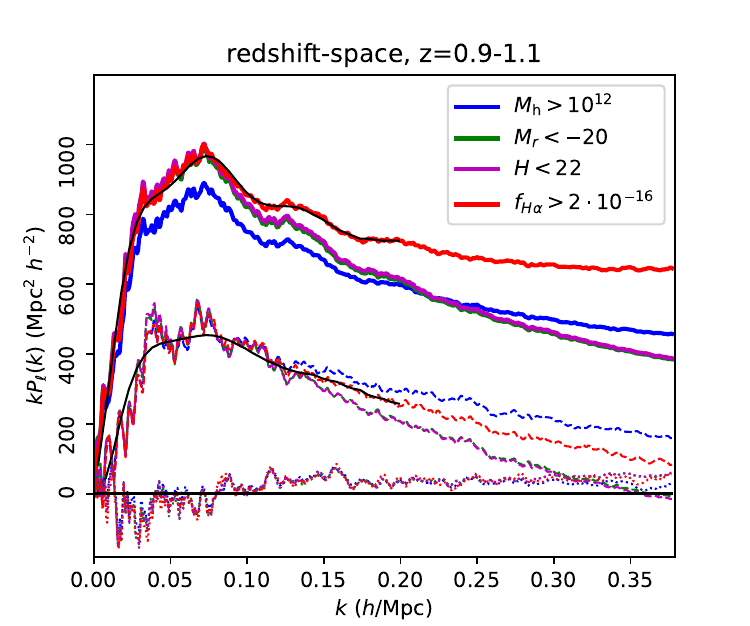}
    \caption{Monopoles (thick continuous lines), quadrupoles (thin continuous lines) and hexadecapoles (dotted lines) of the measured redshift-space \texttt{PK} for the four samples reported in the legend. The monopole includes shot noise. Thin black lines give the best fit model (convolved with the window function) for the monopole and quadrupole of the H$\alpha$ ELG sample.}
    \label{fig:PK}
\end{figure}

\begin{figure}
    \includegraphics[width=\columnwidth]{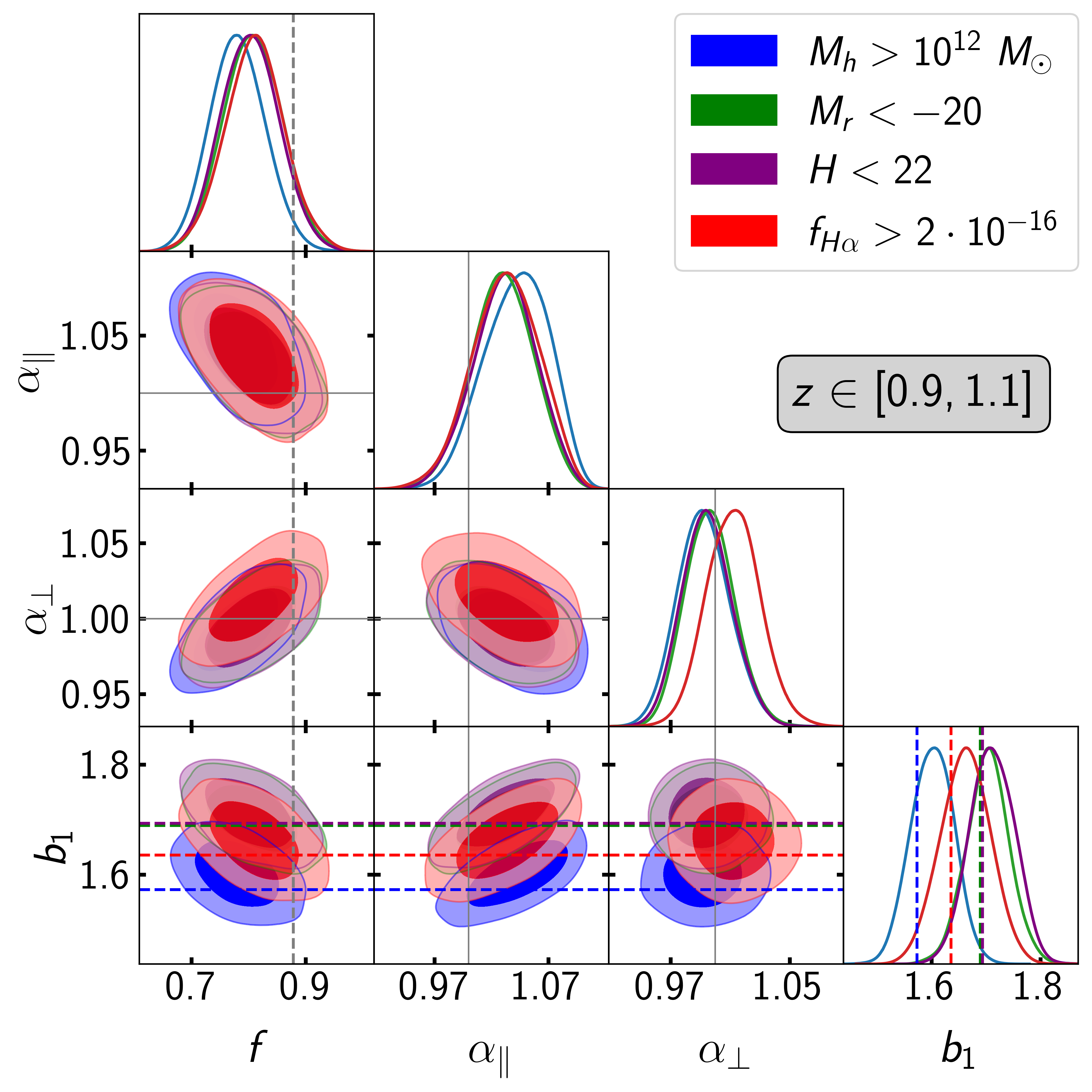}
    \caption{Inferred cosmological parameters (redshift-space $f$ and AP parameters) for the four samples, together with linear bias $b_1$, in the redshift bin $[0.9,1.1]$.
    }
    \label{fig:inferredCosmology}
\end{figure}

\begin{figure}
\includegraphics[width=\columnwidth]{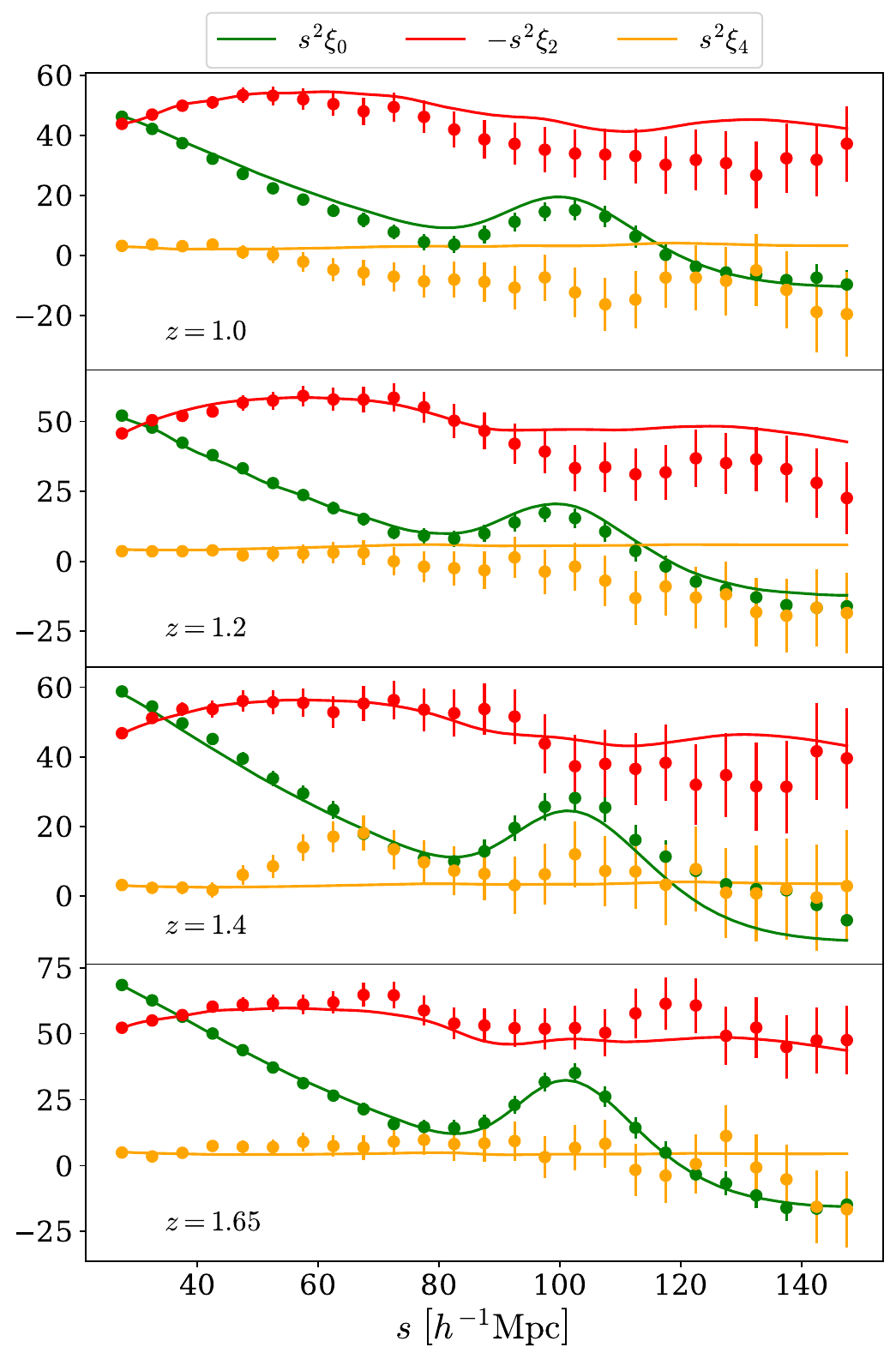}
    \caption{
    Redshift-space two-point correlation function multipoles of H$\alpha$ galaxies, and best-fitting prediction using the EFT model. Green, red, and orange points refer to monopole, quadrupole and hexadecapole respectively.
    Different panels show the results for different redshifts as labelled.}
    \label{fig:fit_multipoles_halpha}
\end{figure}

\begin{figure}
	\includegraphics[width=\columnwidth]{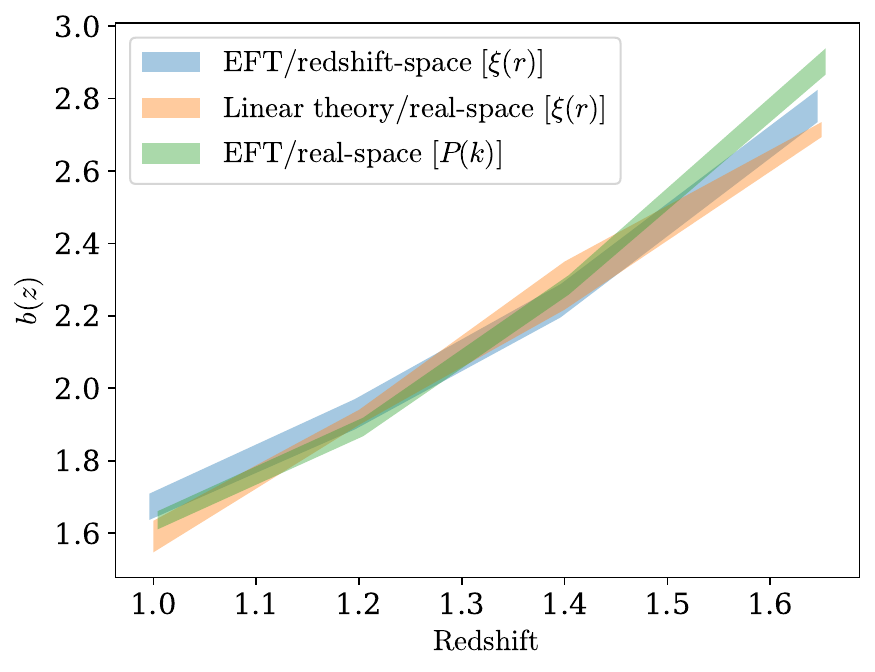}
    \caption{Linear galaxy bias, $b_1$, as function of redshift for H$\alpha$ emitters in the Flagship simulation. In blue we show the estimation from the fit to the two-point correlation function multipoles in redshift space using the EFT model. In orange, we use linear theory in real space to fit the correlation function monopole. Finally, in green we fit the real-space monopole of the power spectrum using EFT at one-loop order.
    }
    \label{fig:linear_biases_halpha}
\end{figure}

The clustering properties of the FS2 galaxies were tested at various levels, to check the stability of clustering for objects produced in various steps along the pipeline. To this aim we selected four classes of objects, identified so as to follow the main step of the catalogue construction (the thresholds are chosen to have a similar level of clustering in the first of the redshift bins defined below):
(i) dark matter haloes on the lightcone more massive than $M_{\rm h}=10^{12}\;h^{-1}\, M_{\odot}$ (the starting point), (ii) galaxies with absolute $r$-band magnitude brighter than $M_r=-20$ (after application of HOD+AM, Sect.~\ref{subsec:luminosities}), (iii) galaxies with apparent 
magnitude 
$\HE <22$ (after SED assignment, Sect.~\ref{subsec:galaxySED}),\footnote{
We select galaxies with $\HE<22$ here to sample the main photometric population that is expected to host H$\alpha$ ELGs; the photometric galaxy sample will correspond to $\HE<24$ galaxies, but few galaxies with $\HE$ between 22 and 24 are expected to be in the spectroscopic sample.
} (iv) emission line galaxies (ELGs) with H$\alpha$ line flux greater than $f_{\rm H \alpha}=2\times10^{-16}$ erg s$^{-1}$ cm$^{-2}$ (after emission lines assignment, Sect.~\ref{subsec:emissionlines}). 
For each selected catalogue we consider the four redshift bins of the spectroscopic sample (namely $[0.9,1.1)$, $[1.1,1.3)$, $[1.3,1.5)$ and $[1.5,1.8)$). 
To isolate the effect of peculiar velocities in redshift-space distortions, we select comoving angles (instead of magnified angles, which are shifted according to the gravitational lensing displacement field), observed redshift containing the contribution from background expansion and first-order peculiar velocities and an unmagnified flux cut. This ensures that our measurements will not be affected by lensing effects, which have been shown to be non-negligible for high-redshift 3D galaxy clustering \citep{Jelic-Cizmek:21, Breton:22, Jelic-Cizmek:23} and whose effect is outside the scope of this paper.
For each catalogue, we generate an associated random catalogue constructed by replicating the galaxy catalogue 50 times, keeping the redshift while assigning random sky positions within the angular footprint of the mock survey (an octant of the sky). This way, the average number density of the random reproduces exactly 50 times that of the galaxy catalogue. \\

First, we report our validation of the galaxy clustering probe in Fourier space using the official code developed to compute the power spectrum (\texttt{PK}) in \Euclid 
({Euclid Collaboration: Sefusatti et al., in prep.}), based on the Yamamoto-Bianchi estimator \citep{Bianchi2015}.
The code accepts a catalogue with (cosmology-independent) angular coordinates and redshift; we computed the monopole of the real-space \texttt{PK}, obtained by providing the true redshift of the galaxy, and the first three odd multipoles of the redshift-space \texttt{PK}, obtained by providing the `observed' redshift that includes peculiar velocities.
The same \texttt{PK} code performs a measurement of the power spectrum of the randoms, which is used to create a model of the window function.

The measured power spectra were fitted with a model based on a standard perturbation theory (SPT) at 1-loop with effective field theory (EFT) counter-terms using the \texttt{PBJ} code presented by \cite{Moretti2023}. 
Cosmological parameters were fixed to those used in the simulation. This model contains the following parameters for galaxy bias, shot noise and EFT that were used as nuisance parameters:  $b_1$ (linear bias), $b_2$ (quadratic bias), $b_{{\cal G}_2}$ (second order Galileon bias), $c_0$ and $c_2$ (the EFT counter-terms), $\alpha_P$ (deviation from the Poisson shot-noise), $\epsilon_{0,k^2}$ ($k^2$ dependent shot-noise) and $c_{\nabla^4 \delta}$ (Finger-of-God counter-term). In redshift space we left the growth rate $f$ free and added Alcock--Paczynski (AP) free parameters $\alpha_\parallel$ and $\alpha_\perp$ to the fiducial power spectrum. Indeed, the aim here is not to perform accurate cosmological inference from the spectroscopic catalogue
but to check that we can infer the input cosmological parameters consistently when using the different catalogues along the mock production pipeline. 
The model was convolved with the window function 
following the matrix multiplication method of~\cite{d'Amico:20}. 
For the covariance matrix we used a simple analytic Gaussian, leading-order covariance; to minimise the impact of this naive choice and the lack of a proper convolution with the window function, we binned the power spectrum in bins of 8 times the fundamental frequency ($k_{\rm f}$) of the box used for the measurement ($L_{\rm box} = 10\;h^{-1}$\,Gpc), so that $8 \, k_{\rm f} \sim \smash{V_{\rm eff}^{1/3}}$, where $V_{\rm eff}$ is the effective survey volume. The model range of validity is thus from $k_{\rm min}=8 k_{\rm f}=0.005\;h\,\mathrm{Mpc}^{-1}$ to $k_{\rm max}=0.2\;h\,\mathrm{Mpc}^{-1}$. For the likelihood, we neglected the hexadecapole, which had a negligible effect on the posteriors.

First, the real-space \texttt{PK} was fitted by the model with fixed cosmological parameters, with the aim of obtaining the $b_1$ linear bias parameter of the four classes of objects. 
In the second stage, we fitted the redshift-space \texttt{PK} of the four catalogues, leaving $f$, $\alpha_\parallel$ and $\alpha_\perp$ free. Figure~\ref{fig:PK} shows the \texttt{PK} measurement of the monopole, quadrupole and hexadecapole of the four samples in the first redshift bin, $[0.9,1.1]$; we also report the best-fit model \texttt{PK} of the H$\alpha$ ELGs. Figure~\ref{fig:inferredCosmology} shows the resulting cosmological parameters obtained from the four samples, together with linear bias $b_1$; posteriors are marginalised over the other nuisance parameters. The cosmological posteriors are very stable when varying the sample and are compatible to within 1$\,\sigma$ with the true values, amounting to 1 for the AP parameters and to $f=0.9$ for the growth rate (computed for the median redshift $z=1$); the small bias obtained for some of the parameters raises no concern as we are using a single realisation and a simplified treatment of the covariance. The variation of parameter error bars for the four catalogues is very limited, and this is due to the fact that at $k=0.2\;h\, \mathrm{Mpc}^{-1}$ the power spectrum is still well above the shot noise level in all cases, so the different number densities of the samples do not influence much the parameter error bars in this regime.
Linear bias, $b_1$, values are compared with those obtained from real space (dashed lines). In this case, we also obtain a very moderate bias in this measurement, which is very consistent for all the samples. In conclusion, the various steps in the definition of the galaxy sample do not bias the inferred cosmological parameters.

We now focus on the multipoles of the two-point correlation function for H$\alpha$ galaxies in redshift space. 
To estimate the correlation function we use the LS estimator \citep{Landy:93}
\begin{equation}
    \xi(s,\mu) = \frac{{\rm DD}(s,\mu) - 2\,{\rm DR}(s,\mu) + {\rm RR}(s,\mu)}{{\rm RR}(s,\mu)}\,,
\end{equation}
where $\rm DD$, $\rm DR$ and $\rm RR$ respectively stand for data-data, data-random and random-random pair counts that we estimate with {\tt Corrfunc}\footnote{\url{https://corrfunc.readthedocs.io/en/master/api/Corrfunc.html}}~\citep{Sinha:20}, in bins of $s$ the comoving pair separation (between 25 and 150\,\hMpc) and $\mu$ the cosine of its angle (using the mid-point definition) with respect to the line of sight. To estimate the multipoles of the correlation function we integrate the 2D correlation function as
\begin{equation}
    \xi_\ell(s) = (2\ell + 1)\int_0^1 \xi(s, \mu)\,{\cal L}_\ell(\mu)\,\diff\mu\,,
\end{equation}
where $\ell$ is the multipole and ${\cal L}_\ell$ the $\ell$-th order Legendre polynomial. Note that we integrate $\mu$ from 0 to 1 due to symmetry along the line of sight when using auto-correlations and the mid-point definition, that is $\mu = \boldsymbol{r}\cdot\boldsymbol{s}/\left(rs\right)$, where $\boldsymbol{r} = (\boldsymbol{r}_1+\boldsymbol{r}_2)/2$ and $\boldsymbol{s} = \boldsymbol{r}_2 - \boldsymbol{r}_1$, with $\boldsymbol{r}_1$ and $\boldsymbol{r}_2$ the vectors of the galaxy positions.

We compare our measurements to a theoretical prediction based on EFTofLSS 
\citep{Ivanov:20, d'Amico:20} as implemented in {\tt COMET} \citep{Eggemeier:23}. To obtain the prediction, we first produced Gaussian covariance matrices for each redshift bin following the recipe from \cite{Grieb:16}. Once we compute a covariance with a model which fits the data, we run Monte Carlo Markov chains (MCMCs) using \texttt{PyMultinest} \citep{Buchner:14}. 
Consistent with the power spectrum analysis presented above,
we performed a fixed-template fit assuming the fiducial cosmology of the simulation and varying $f$ the growth rate, $\alpha_\parallel$ and $\alpha_\perp$ the dilation parameters, the linear and quadratic biases (higher-order biases are fixed assuming the local-lagrangian approximation) and EFT counter-terms. In principle, it is possible to vary $\sigma_{12}$ (the variance of density fluctuations in spheres of 12\,Mpc, see \citealt{sanchez2020arguments}), but we instead fix it to its fiducial value so as to avoid large degeneracies in the estimation of the linear bias.

The simulation measurements and best-fit model shown in Fig.~\ref{fig:fit_multipoles_halpha} display a very good agreement, especially for the highest-redshift bins (with mean redshift $z = 1.4$ and $z = 1.65$). For the low-redshift bins, it seems that the quadrupole is underestimated by the model at scales above 90\,\hMpc. This does not look particularly worrying given that these scales are not independent and that the discrepancy stays between 1 and 2$\,\sigma$. We also note that for $z = 1$, the monopole is close to zero at 80\,\hMpc, which could be attributed to sample fluctuations in the redshift bin for our simulation realisation.

Finally, we show a simulation-based estimation of the linear bias as a function of redshift in Fig.~\ref{fig:linear_biases_halpha}. Different models (linear theory or EFT) in different configurations (real or redshift space, Fourier or configuration space) give very similar results. Because the measurement of $b_1$ from a galaxy sample is not a trivial task, this result can be considered as a validation not only of the simulation itself but also of our analysis pipeline. Nevertheless, our estimate of the bias in either real or redshift space appears to be higher than others reported in the literature, such as the simulation-based forecasts for future missions in \cite{Merson_2019} and \cite{Zhai_2021a, Zhai_2021b}. Those results are derived from semi-analytic models of galaxy formation and evolution, which have been calibrated to match the number density of H$\alpha$ emitters. The methodology employed in these studies differs significantly from the approach used in constructing the Flagship simulation in this work. The extrapolation to clustering measurements may introduce additional uncertainty into the estimated bias, and thus we do not expect great consistency between the analyses. However, the collection of ELG data by Euclid is expected to provide tight constraints on these parameters in the near future.

\subsection{Galaxy weak lensing
\label{sec:galwl}
}

\subsubsection{Main observables:  shear 2-point statistics and galaxy-galaxy lensing}
Galaxy weak lensing is one of the two main cosmological probes of \Euclid. The standard summary statistics used for weak lensing include the shear 2-point correlation functions, $\xi_{\pm}$, which in turn are related to the
tangential and cross-component of the shear,

\begin{equation}
    \xi_{\pm}(\theta) = \langle\gamma_\mathrm{t}(\Vec{\vartheta})\,\gamma_\mathrm{t}(\Vec{\vartheta'})\rangle \pm \langle\gamma_\times(\Vec{\vartheta})\,\gamma_\times(\Vec{\vartheta'})\rangle\;,
\end{equation}
where $\theta$ is the separation between $\Vec{\vartheta}$ and $\Vec{\vartheta'}$ and $\gamma_\mathrm{t}$ and $\gamma_\times$ are the tangential and cross-component of the shear, defined by $\gamma_\mathrm{t}(\Vec{\vartheta}) = -\Re[\gamma(\Vec{\vartheta})\mathrm{e}^{-2\mathrm{i}\phi}]$ and $\gamma_\times(\Vec{\vartheta}) = -\Im[\gamma(\Vec{\vartheta})\mathrm{e}^{-2\mathrm{i}\phi}]$, where $\phi$ is the polar angle of $\Vec{\vartheta}-\vec{\vartheta}'$.

In the weak lensing limit these shear correlations are related to the gradient or E-mode component of
the shear angular power spectrum (i.e, the B-mode component is typically negligible: see \citealp{Bartelmann:01,Hilbert2009AA...499...31H}),
\beq
\xi_{\pm}(\theta) = \frac{1}{2\pi} \int \mathrm{d} \ell \,
\ell \,
J_{0/4}(\ell \theta) \, C_{\ell}^{\gamma \gamma} \;,
\label{eq:shear2pcf}
\eeq 
being $C_{\ell}^{\gamma \gamma} = C_{\ell}^{\kappa \kappa}$, $J_{0/4}$ are the Bessel functions of the first kind of order $0$ and $4$ respectively.
where we have assumed no B-modes, and the validity of the Limber approximation~\citep[e.g.,][]{LoVerde:08}.
Another basic observable is given by the galaxy-galaxy lensing, or
the average tangential shear of a background galaxy sample produced by the foreground matter distribution, $\langle \gamma_\mathrm{t} \rangle$, which is
directly related to the cross power spectrum of the convergence field
of the background galaxies and the foreground galaxy number
counts, $C_{\ell}^{\kappa g}$ (see \citealp{jeong09}),
\beq
\langle\gamma_\mathrm{t}\rangle(\theta) = \frac{1}{2\pi} \int  \mathrm{d} \ell \,
\ell \,
J_{2}(\ell \theta) \, C_{\ell}^{\kappa g} \;,
\label{eq:gammat}
\eeq 
where we have taken the flat-sky limit, which is very accurate for
practically all angular scales (e.g., $\theta <$ few degrees). The exact
expression can be obtained by replacing the  the integral over $\ell$ by a sum over discrete $\ell$ modes, and features Wigner d matrix elements (\citealt{Chon04, deputter10}).

In this section we provide a validation of these basic weak-lensing statistics which are, along with galaxy clustering correlations, the two main cosmology probes of \Euclid.
\begin{figure}
    \centering
    \vspace{-10 pt}
    \includegraphics[width=0.9\hsize]{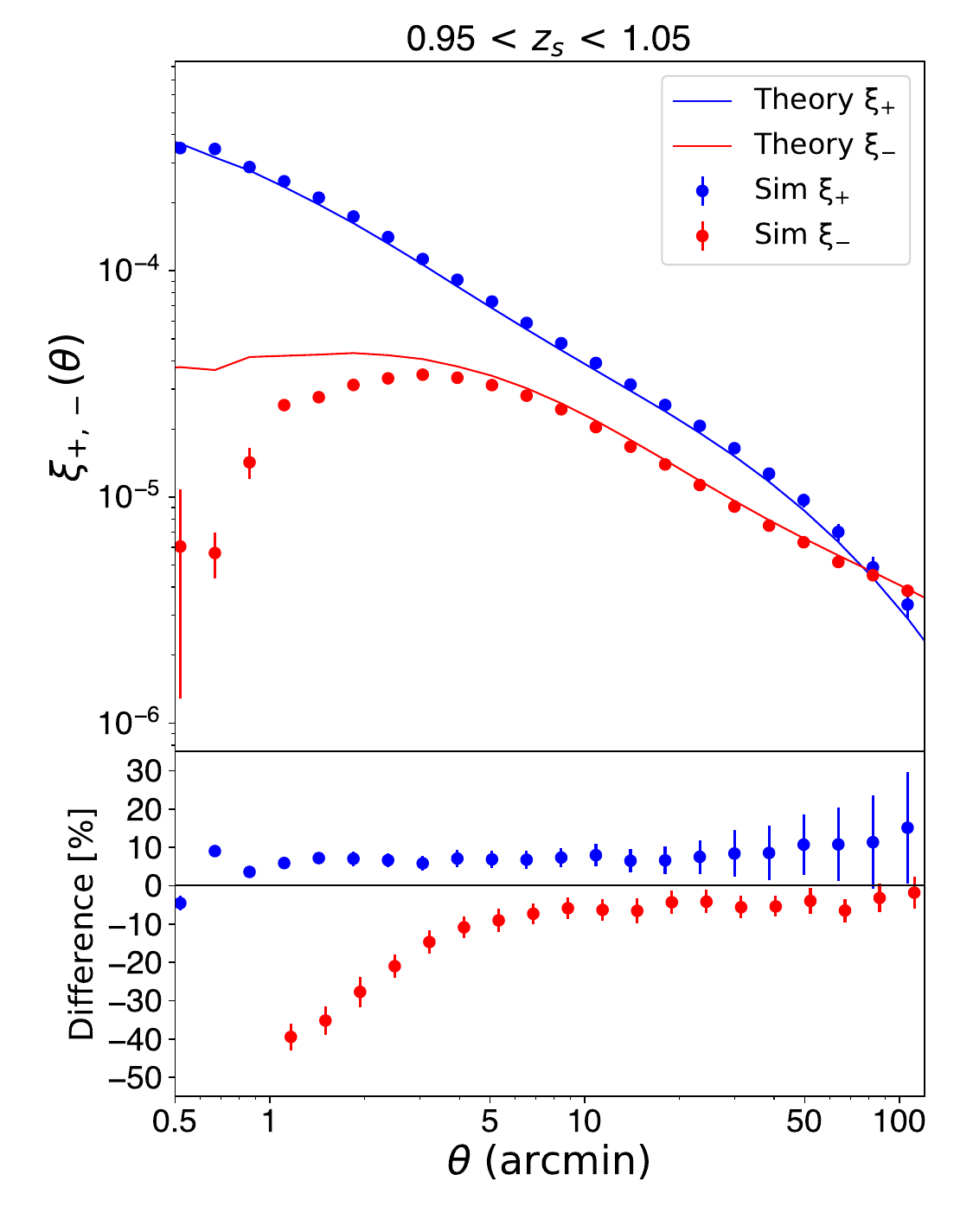}%

      \vspace{-3mm}
      \includegraphics[width=0.9\hsize]{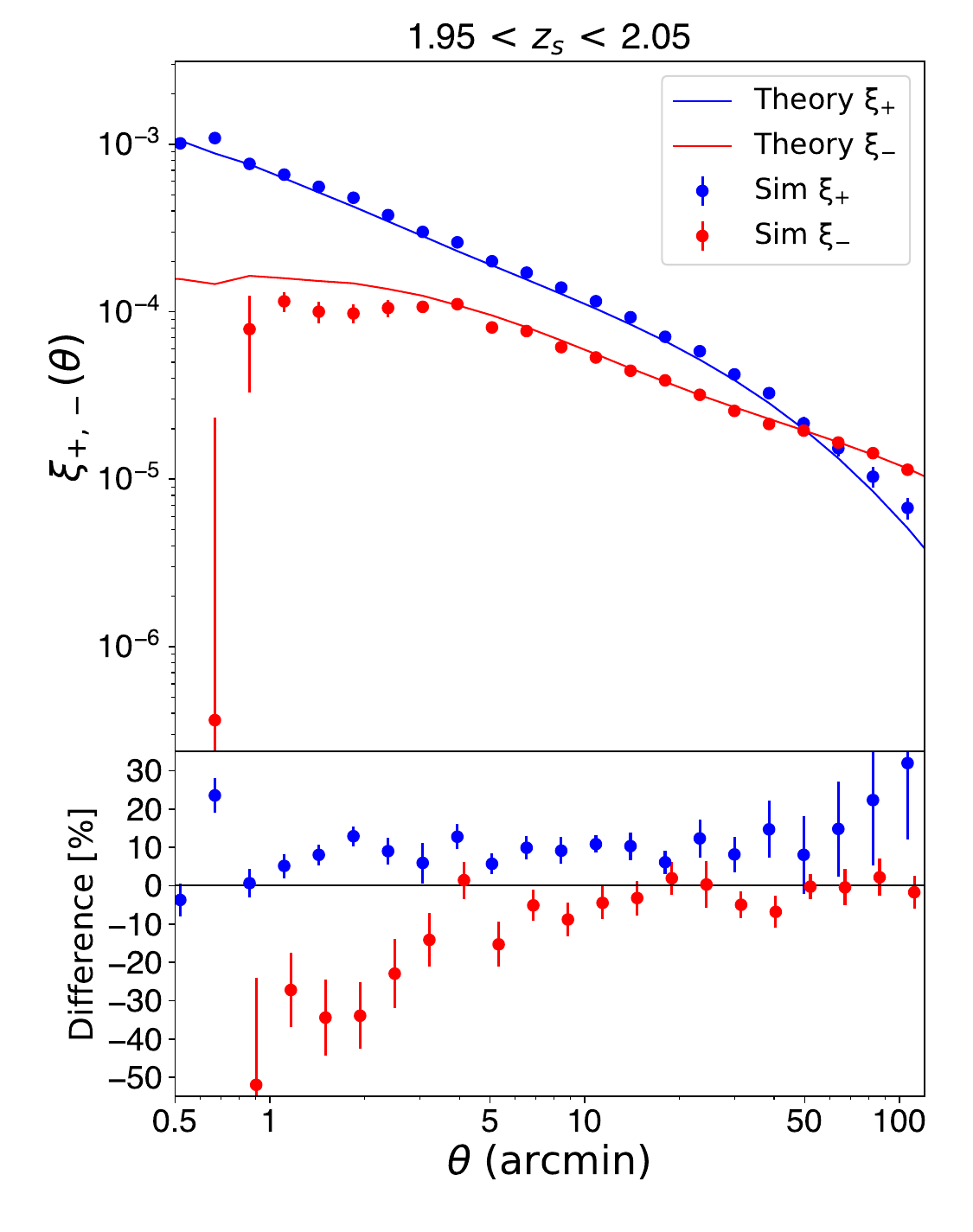}%

    \caption{Shear correlation functions for source samples at $z=1$ (top), $z=2$ (bottom). Simulation measurements (symbols) are compared to linear and nonlinear (\texttt{Halofit}) theory predictions (lines). Errors are obtained from jackknife resampling. The lower sub-panels show fractional differences between simulations and nonlinear theory.}
    \label{fig:xipm}%
\end{figure}
Figure~\ref{fig:xipm} shows the simulation measurements of the shear correlation functions at source redshifts $z=1,2$, using the \texttt{athena} code\footnote{\url{https://www.cosmostat.org/software/athena}} \citep{kilbinger:14} compared to theory predictions from \texttt{Halofit} \citep{Takahashi:20}. The mock galaxy sample has been cut at the nominal magnitude limit of the photometric galaxy sample, $\IE < 24.5$.  The measured correlations are in agreement with theory predictions to within $10\%$ down to highly nonlinear scales, i.e, \ang{;1;} for $\xi_+$ (\ang{;5;} for $\xi_-$). A similar level of agreement between measurements and theory is found for other redshift slices $0.95<z<1.05, 1.95<z<2.05$, and $2.9<z<3$, as displayed in Fig.~\ref{fig:xipmzbins}. We have also compared mock measurements of the galaxy-galaxy lensing, obtained with \texttt{athena}, with theory predictions that use \texttt{Halofit} for the nonlinear matter power spectrum re-scaled with a simple linear galaxy bias model. 
As shown in Fig.~\ref{fig:ggl} we get good agreement within 10--15$\%$ for this simple model even on small (nonlinear) scales. 
Figure~\ref{fig:gglzbins} shows a comparison of the galaxy-galaxy lensing estimator for different lens-source $z$-bin pairs across the simulation redshift range. It shows that overall there is good agreement between the simulation and nonlinear theory predictions, except for the smallest angular scales, where the linear galaxy bias model is expected to break down.

\begin{figure*}
    \centering
    \vspace{-10 pt}
    %\subcaptionbox{}{
    \includegraphics[width=0.8\hsize]{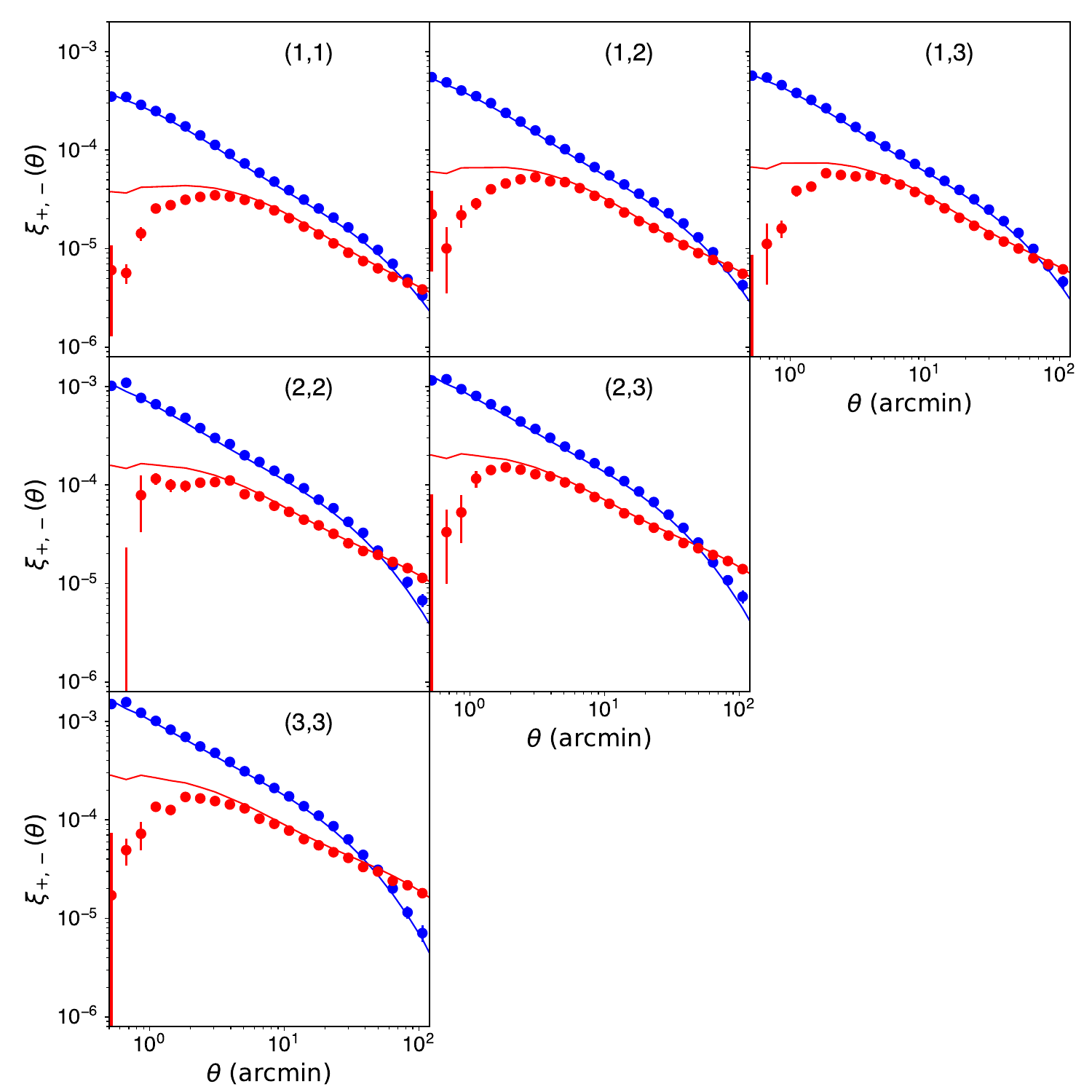}
    %}%
     %\qquad
      %\vspace{-10 pt}
    \caption{Shear cross-correlation functions for samples at three different source bins $0.95<z<1.05, 1.95<z<2.05$, and $2.9<z<3$. Panels show all possible cross-correlation $z$-bin pairs. We denote a $z$-bin pair with e.g., $(1,2)$ when correlating shear amplitudes from the $z\simeq 1$ and $z \simeq 2$ source bins.
    }
    \label{fig:xipmzbins}
\end{figure*}

\begin{figure}
    \centering
    \vspace{-10 pt}
    \includegraphics[width=0.88\hsize]{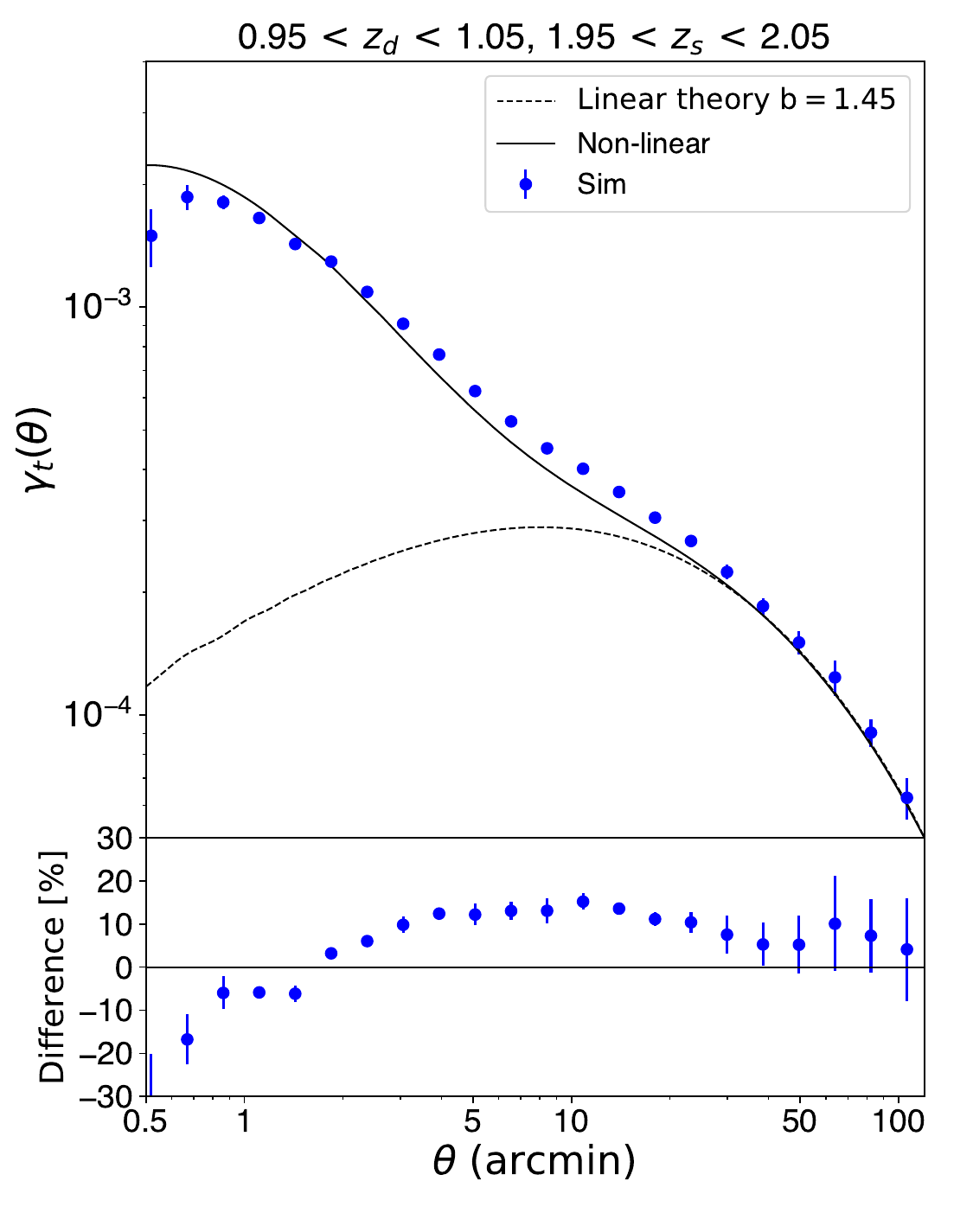}%

     \vspace{-1mm}
      \includegraphics[width=0.88\hsize]{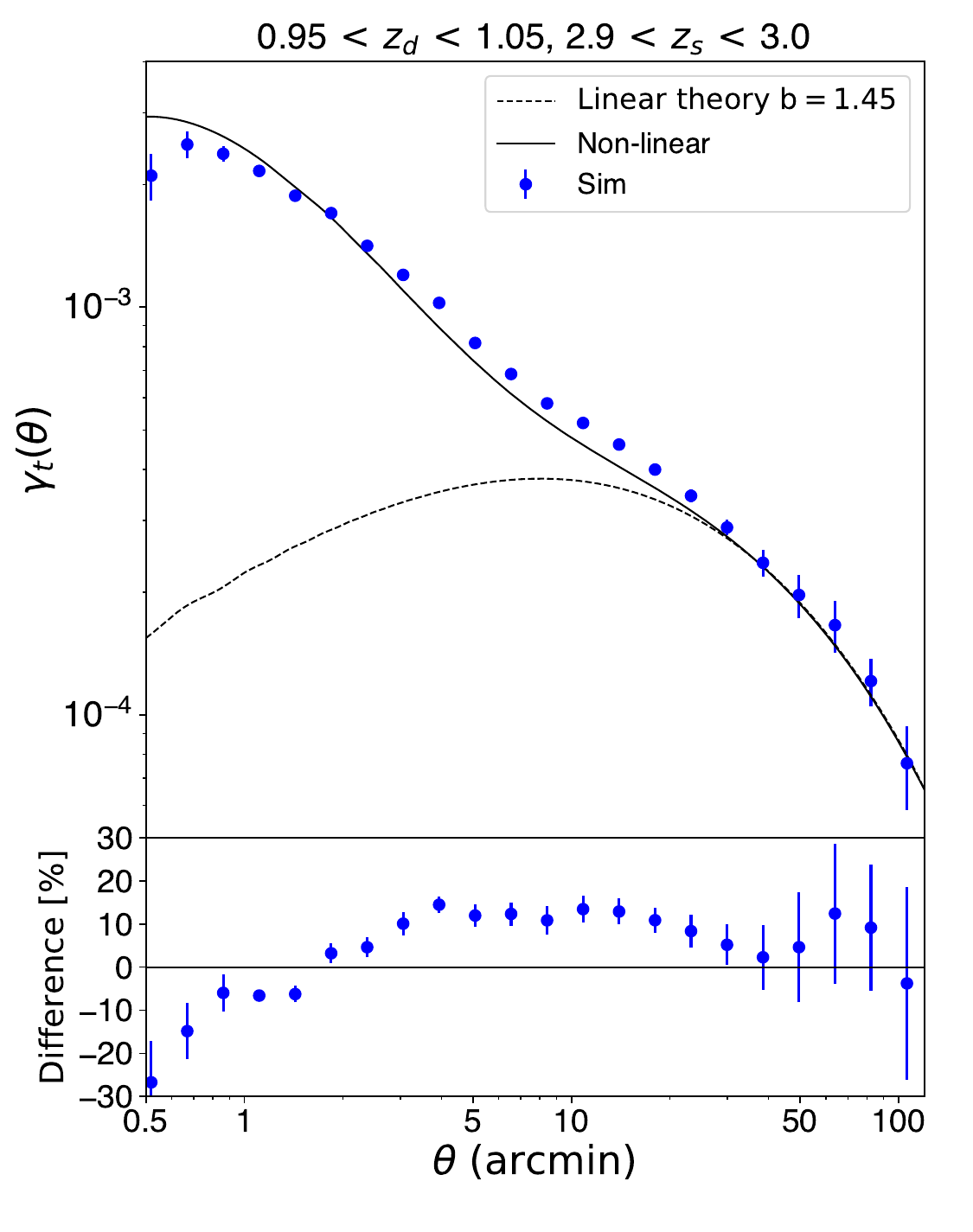}%

      \vspace{-1mm}
    \caption{Galaxy-galaxy lensing:  cross-correlations between lens positions and source shear for different source redshift planes ($z_{\rm s}\simeq2$ in the top panel, and $z_{\rm s}\simeq3$ in the bottom panel, respectively) at fixed lens (deflector) redshift, $z_{\rm d}\simeq1$. Simulation measurements (filled blue circles) are compared to linear and nonlinear theory predictions (lines). Lower panels show fractional differences between simulation and nonlinear theory.}
    \label{fig:ggl}%
\end{figure}

\begin{figure*}
    \centering
    \vspace{-10 pt}
 \includegraphics[width=0.8\textwidth]{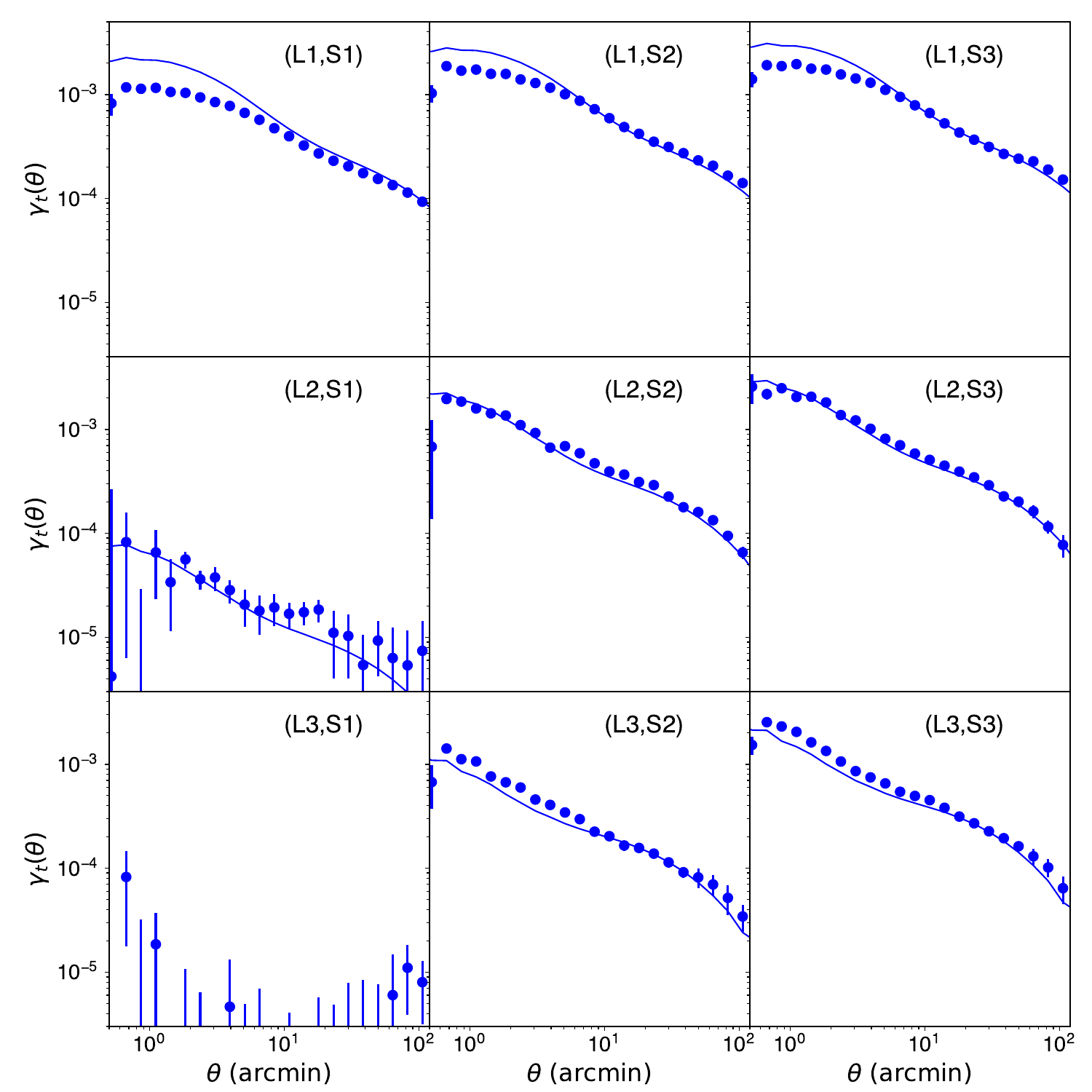}
    \caption{Average tangential shear for source samples in redshift bins $(\mathrm{S1,S2,S3}) = (0.95<z<1.05, 1.95<z<2.05, 2.9<z<3)$, around the lens redshift slices $(\mathrm{L1,L2,L3}) = (0.45<z<0.55, 0.95<z<1.05, 1.45<z<1.55$. Panels show all possible combinations of cross-correlations between lens sample positions and source galaxy shear. Lens sample redshifts increase from top to bottom panels, whereas source redshift increases from left to right. Solid lines show a fiducial theory prediction for the nonlinear dark-matter power spectrum re-scaled with a best-fit linear galaxy bias for the lens samples, $b(\mathrm{L1}) = 1.2, b(\mathrm{L2}) = 1.45, b(\mathrm{L3}) = 2.0$}
    \label{fig:gglzbins}
\end{figure*}

\subsubsection{Magnification bias}

Gravitational lensing by large-scale structures in the Universe changes the number density of background sources
and thus it induces a cross-correlation signal between background and foreground galaxy populations \citep{Moessner:98,Bartelmann:01}.
For a magnitude limited survey,  the cumulative number of galaxies
above a flux limit $f$ scales as $N_{0}(>f) \sim A f^{\alpha}$, where $A$
is the area of the survey, and $\alpha $ is the power-law slope of
the background number counts. Lensing preserves the surface
brightness of galaxies by increasing the observed survey depth
(i.e, decreasing the effective flux limit) and the effective
survey area by the same amount:  $f \rightarrow f/\mu$, $A \rightarrow A/\mu$,
where $\mu$ is the magnification.
These two competing effects induce the so-called {\it magnification
  bias} in the cumulative number of background sources,
\beq
N(>f) \sim  \frac{1}{\mu} A \left(\frac{f}{\mu}\right)^{-\alpha} = \mu^{\alpha-1} N_{0}(>f)\;.
\label{eq:cumnumgal}
\eeq
We define the logarithmic slope of the background
number counts at redshift $z$, for a magnitude limit $m$, as
\beq
s = \frac{2}{5}\,\alpha \equiv \frac{ \diff\, {\logten}N(<m,z)}{\diff m}.
\label{eq:scounts}
\eeq 

In the weak-lensing limit, $\mu = 1 + \delta_{\mu}$ where
$|\delta_{\mu}| \ll 1$, and we can Taylor expand, $\mu^{\alpha-1} \approx 1 + (\alpha
-1) \delta_{\mu}$, and therefore the magnified overdensity of
background sources is given by,
\begin{eqnarray}
\delta_{\rm all} &=& \frac{N-N_{0}}{N_{0}} = \delta_{\rm m} + \delta_{\rm p} \, \\
&=&  (\alpha-1)\,\delta_{\mu} = (2.5s-1)\,\delta_{\mu} = (5s-2)\,\delta_{\kappa} \, ,
\label{eq:deltanumgal}
\end{eqnarray}
where in the last equality, we have identified $\delta_{\mu} =
2\,\delta_{\kappa}$, that is valid in the weak-lensing limit.
Note that in $\delta_{\rm all}$  we have defined the two qualitatively
different contributions:
\begin{enumerate}
\item{{\it magnified magnitudes}, 
$\delta_{\rm m} = \alpha \,\delta_{\mu}$} \; ,
\item{{\it magnified or lensed positions}, 
$\delta_{\rm p} =-\delta_{\mu}$} \; .
\end{enumerate} 
These two contributions cannot be separated observationally, but we
define two different galaxy samples accordingly in our simulation in order to validate
the two magnification contributions separately.

The net magnification from these two competing effects depends on
how the loss of sources due to the area dilution, $\delta_{\rm p}$, is compensated by the
gain of sources from the flux magnification, $\delta_{\rm m}$. Number counts
for source populations with flat luminosity functions, such as faint
galaxies, decrease due to magnification, 
whereas sources with steep luminosity functions, such as
quasars, increase.
Note that, in the particular case when $s=0.4$, then $\alpha=1$, and there is no net
magnification effect.

For the implementation of the magnification in the flux (or magnitudes) and the galaxy positions, we follow \citet{Fosalba:15b} and we refer to that paper for further details. Below we just provide the main definitions.

\begin{enumerate}
\item{{\it Magnified magnitudes:}
flux magnification makes the
mock galaxy magnitudes, $m$, brighter by an amount
\begin{equation}
\Delta m = \frac{5}{2} \logten\mu = 2.5 \logten (1-\delta\mu) \simeq 2.174\, \kappa \;,
\label{eq:deltam}
\end {equation}
where in the last equality we have Taylor 
expanded $\logten (1-\delta\mu)$ and used the fact that $\delta\mu \simeq 2\,\kappa$ in the weak-lensing limit.
Therefore, knowing the value of the convergence, $\kappa$, at a given
point in the source plane, it is straightforward to compute the flux
magnification induced, which in turn produces the change in the
background number counts, $\delta_{\rm m}$.} 

\item{{\it Magnified or lensed positions:} the `observed' or lensed
position, $\vec\beta$, of a light ray is shifted from the `true' or unlensed
position, $\vec\theta$, by an angle given by the scaled deflection vector,
$\vec\alpha$, according to the {\it lens equation} on the source
plane~\citep[see e.g.,][]{Bartelmann:01}.
In the single-plane (or Born) approximation, 
the lens equation reads
\begin{equation}
\vec\theta = \vec\beta + \vec\alpha\;, 
\label{eq:lenseq}
\end{equation}
where the deflection vector, $\vec\alpha$ is a tangent vector at the
{\it unlensed position} of the light ray, and the lensed position is
found by moving along a geodesic on the sphere in the
direction of this tangent vector and for an arc length given by the scaled deflection angle,
$\vec\alpha$. If we denote the unlensed position on the sphere by
$(\theta,\phi)$, then the lensed position, $(\theta^{\prime},\phi +
\Delta \phi)$, can be simply derived by using identities of spherical
triangles \citep{Lewis:05}
}.
\end{enumerate}

\begin{figure}
	\includegraphics[width=\linewidth]{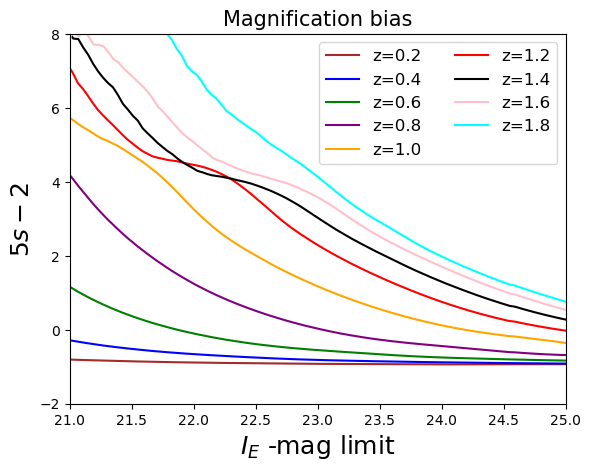}
    \caption{Magnification bias for magnitude limited source
galaxy samples with redshift bin-width $\diff z = 0.1$.} %\gam{$y$-axis label `5s-2' is unclear.}}
    \label{fig:magbias}
\end{figure}

Figure~\ref{fig:magbias} shows the measured magnification bias $b = 5*s -2$, where $s$ is the slope of the background galaxy number counts, see Eq.~\eqref{eq:scounts}, as a function of magnitude limit in the visible ($\IE$) and of redshift. We find that the smooth evolution of $s$ with the magnitude limit applied can be well fitted with a third-order polynomial, $s = s_{0} + s_{1}\,\IE + s_{2}\,\IE^2 + s_{3}\,\IE^3$ for a given source redshift bin. 
Table~\ref{tab:magbiasfit} gives the coefficients of the polynomial fit for a set of source redshift bins with width $\Delta z = 0.1$ up to $z=1.8$.

\begin{table}
\caption{Fitting functions for the slope of the galaxy number counts vs. magnitude cut, $s = s_{0} + s_{1}\,\IE + s_{2}\,\IE^2 + s_{3}\,\IE^3$, for different source redshift bins, $z_{\rm s}$.}
\centering
\tabcolsep=4pt
\begin{tabular}{lrrrr}
\hline
\hline
\noalign{\vskip 1pt}
\multicolumn{1}{c}{$z_{\rm s}$} & \multicolumn{1}{c}{$s_{0}$} & \multicolumn{1}{c}{$s_{1}$} & \multicolumn{1}{c}{$s_{2}$} & \multicolumn{1}{c}{$s_{3}$}\\ 
\hline
\noalign{\vskip 1pt}
0.2 & $5.879086$ & $-0.667647$ & $0.026146$ & $-0.000340$\\
0.4 & $41.633433$ & $-5.118432$ & $0.211170$ & $-0.002908$\\ 
0.6 & $157.448038$ & $-19.570285$ & $0.812851$ & $-0.011264$\\
0.8 & $327.832860$ & $-40.459089$ & $1.667851$ & $-0.022944$\\
1.0 & $38.859528$ & $-2.483457$ & $0.013260$ & $0.000978$\\
1.2 & $-77.665962$ & $12.006749$ & $-0.582611$ & $0.009089$\\
1.4 & $254.313858$ & $-31.150105$ & $1.285380$ & $-0.017824$\\
1.6 & $222.837780$ & $-26.927314$ & $1.098698$ & $-0.015094$\\
1.8 & $237.449373$ & $-27.163814$ & $1.044348$ & $-0.013473$\\
\hline
\end{tabular}
%\tablefoot{}
\label{tab:magbiasfit}
\end{table}

\begin{figure}
    \centering
    \vspace{-10 pt}
    \includegraphics[width=0.45\textwidth]{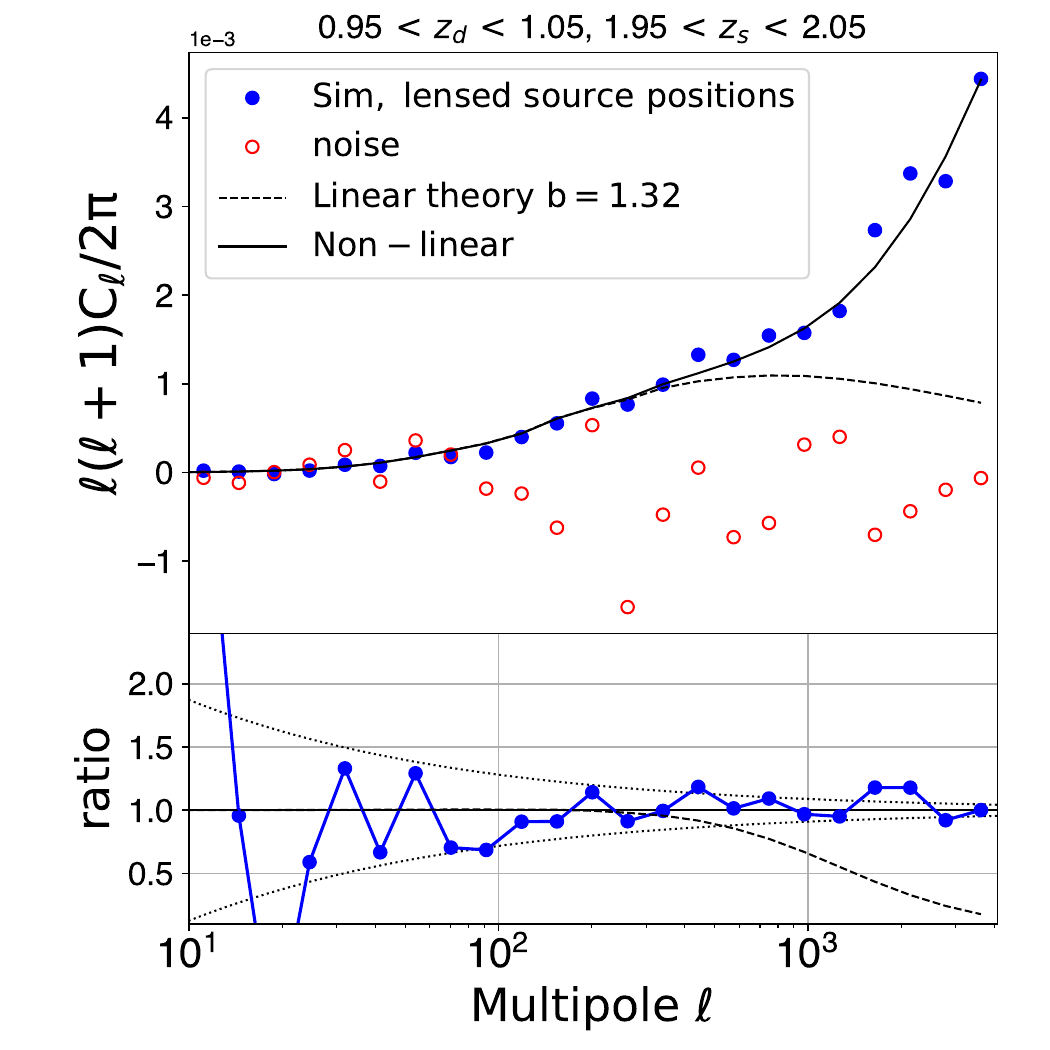}
     \qquad
      \includegraphics[width=0.45\textwidth]{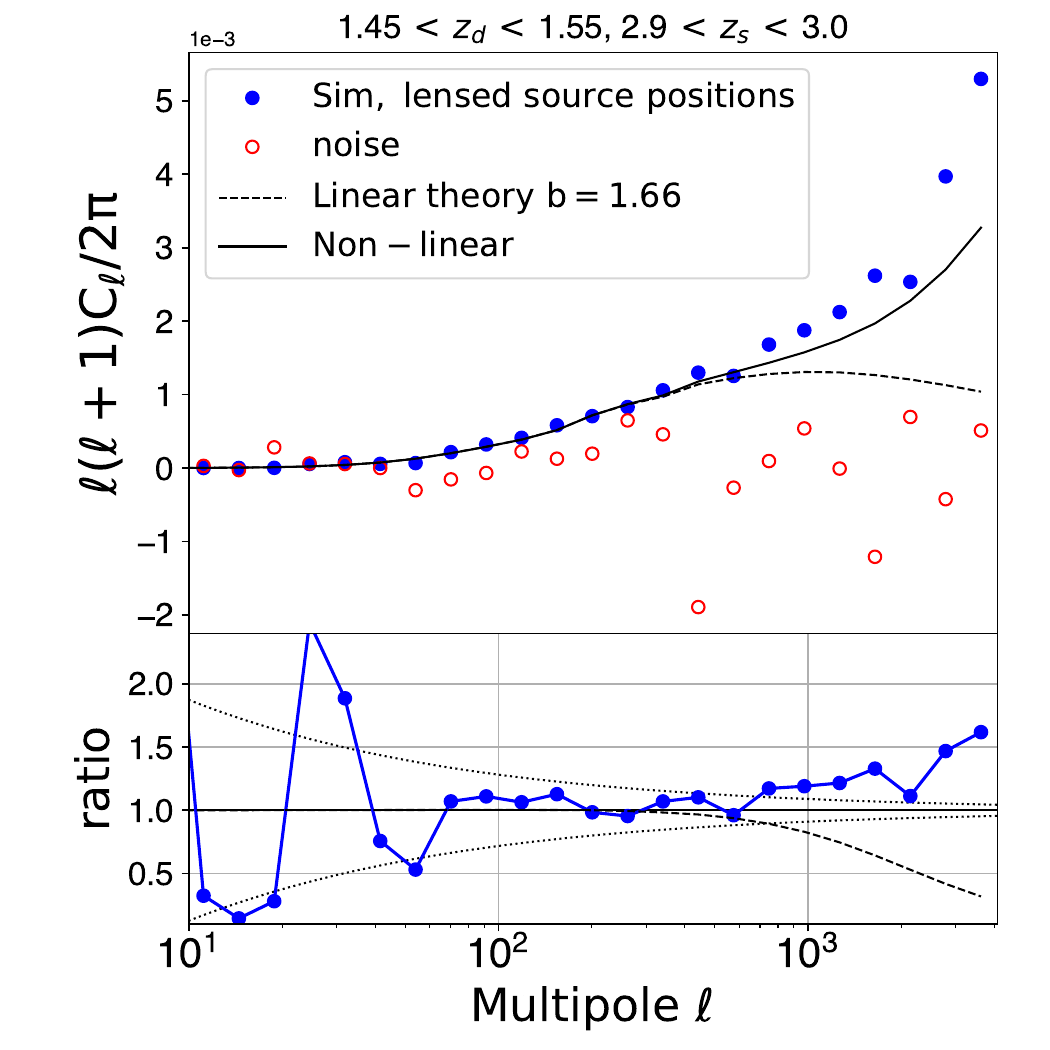}
     \qquad

    \caption{Lens-source cross-correlations as a test of source galaxy position deflections. We show two different lens-source $z$-bin pairs for galaxy samples cut at $\IE < 26$: a case with $z_{\rm d}\simeq1$ and $z_{\rm s}\simeq2$ in the \emph{top} panel, $z_{\rm d}\simeq1.5$ and $z_{\rm s}\simeq3$ in the \emph{bottom} panel, respectively. Flagship measurements (filled blue circles) are compared to linear and nonlinear theory predictions (lines). Lower panels show the ratio of FS2 measurements over nonlinear theory prediction. Sample variance errors around nonlinear theory are displayed as dotted lines. Cross-correlation in absence of magnification (i.e, noise estimate) is shown as open red circles. Note that FS2 measurements use an estimator that roughly cancels sample-variance (see text for details).}
    \label{fig:sourcemag}%
\end{figure}

In order to validate the magnified or deflected positions of mock galaxies, we use the estimator for `sample variance free' cross-correlations between background and foreground mock galaxy samples (see section~5.2 of \citealt{Fosalba:15b}). As shown in Fig.~\ref{fig:sourcemag}, the measured cross-correlations are in good agreement with theory expectations and above the estimated noise level for most of the angular scales probed.

\subsubsection{Higher-order lensing statistics}

We also validate the agreement of the third-order aperture statistics $ \langle{M_\mathrm{ap}^3}\rangle$ in Flagship with theoretical predictions. The $\langle{M_\mathrm{ap}^3} \rangle$ values are the third moments of the aperture mass and therefore depend on the matter bispectrum. They can be inferred in two ways \citep{Schneider:05}: either from maps of the lensing convergence or by first measuring the third-order shear correlation functions and then convolving them with a suitable kernel function. By comparing the results of these two approaches with theory, we can simultaneously test the skewness of the convergence maps and the third-order statistics of the shear catalogues. 

For our model, we use the approach in \citet{Heydenreich:23}, based on the \verb|bihalofit| bispectrum \citep{Takahashi:20} and the aperture filter function by \citet{Crittenden:02}. This model has an expected accuracy of $10\%$ for the aperture radii we consider here \citep{Heydenreich:23}.

We measure $\langle{M_\mathrm{ap}^3} \rangle$ on the convergence map at source redshift $z_{\rm s}=0.996$. For this, the convergence is convolved with the aperture filter for (\ang{;4;}, \ang{;8;}, \ang{;16;}, \ang{;32;}) using Fast Fourier Transforms to obtain aperture mass maps. Then, a border of 
width $4\times \ang{;32;} = \ang{;128;}$
is cut from each aperture mass map to avoid border effects. Finally, the mean of the product of three maps gives $\langle{M_\mathrm{ap}^3} \rangle(\theta_1, \theta_2, \theta_3)$. 

We also measure $\langle{M_\mathrm{ap}^3} \rangle$ from the shear of galaxies at redshifts between $0.984$ and $1.01$ with a \Euclid-like magnitude cut of $\IE\leq 24.5$. For this we measure the shear three-point correlation function with \verb|TreeCorr| \citep{Jarvis:04} for galaxy separations between \ang{;0.1;} and \ang{;400;}, and convert to $\langle{M_\mathrm{ap}^3} \rangle$ according to \citet{Schneider:05}. The correlation function is measured individually for 40 patches, allowing for a jackknife estimate of the variance.

\begin{figure}
    \centering
    \includegraphics[width=\linewidth]{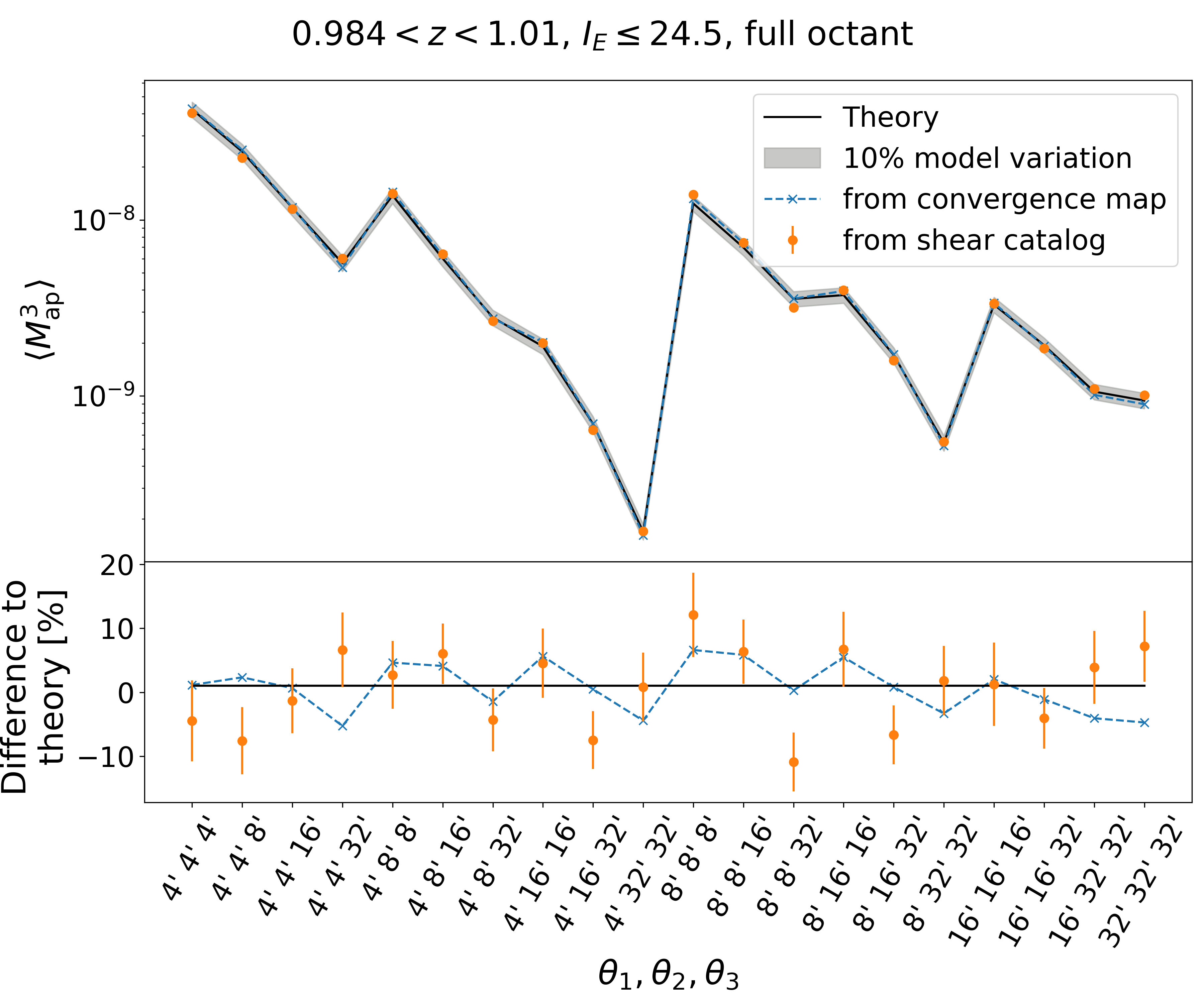}
    \caption{\textit{Top}: Third-order aperture mass statistic, measured from the convergence map at $z_{\rm s}=0.996$ (blue, dashed), measured from the shear of galaxies at $0.984<z_{\rm s}<1.01$ (orange points), and modelled (black, solid). \textit{Bottom}: Relative difference of measurements to theory.
}
    \label{fig: Map3}
\end{figure}

Figure~\ref{fig: Map3} shows the measurement results and the model. The measurements for both methods agree  with the model within the 10\% model accuracy. They also agree with each other within the jackknife uncertainty. This confirms that the third moment of the matter distribution corresponds to theoretical expectations and that the galaxy shear catalogues retain the correct third-order moment. 

\subsection{Galaxy clusters}

In this subsection, we concentrate on the properties of galaxies in clusters. For this purpose, we consider as `clusters' all haloes more massive than $10^{14}\;h^{-1}\,M_\odot$. For computational ease, we concentrate on statistics extracted from a $49\,\rm deg^2$ square region of the Flagship catalogue, which is statistically representative of the whole population for the comparisons we present below.

\begin{figure}
    \centering
    \includegraphics[width=0.9\hsize]{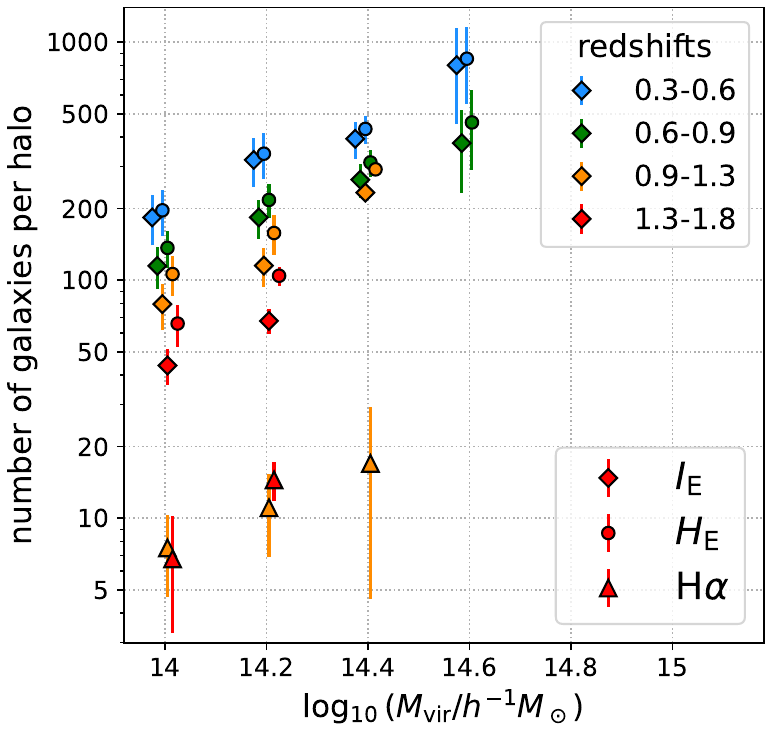}
    \caption{Arithmetic means of the number of galaxies per halo as a function of halo mass. The colours indicate the redshifts, while the symbols indicate the selection: $\IE < 24.5$ (diamonds), $\HE<24$ (circles), and H$\alpha$ flux $>2\times 10^{-16}\,\rm erg\,s^{-1}\,cm^{-2}$ (triangles, limited to the redshift range, $0.9<z<1.8$, where the red grism can see the H$\alpha$ line). We requested at least 2 clusters per cell of redshift and log halo mass. The error bars show the standard deviations of the numbers of galaxies per halo. The abscissae are slightly offset for clarity.  }
    \label{fig:Ngalsperhalo}
\end{figure}

Fig~\ref{fig:Ngalsperhalo} displays the number of galaxies as a function of cluster halo mass, computed in three ways: for the EWS limits of $\IE<24.5$ (diamonds) and $\HE<24$ (circles) and for the H$\alpha$ flux limit of $f({\rm H}\alpha)>2\times 10^{-16}\,\rm erg\,s^{-1}\,cm^{-2}$ (triangles).
At a given redshift, the number of galaxies per halo increases with halo mass, as expected from the HOD approach implemented in the mock. Also, at given halo mass, the number of galaxies per halo decreases with increasing redshift, because of the broadband or emission line flux limits. Also, the number of galaxies with potentially detectable H$\alpha$ emission lines, in the red grism redshift range $0.9 < z < 1.8$ is of the order of 5 to 20 per halo, amounting to roughly 6\% ($0.9<z<1.3$) and 12\% ($1.3<z<1.8$) of the number of galaxies with $\HE<24$ at those redshifts. Finally, there are no H$\alpha$ flux-limited selected galaxies in haloes of mass  $\logten [M_{\rm h}/(h^{-1}\,M_{\odot})] > 14.5$, as there are no star-forming blue galaxies in those massive clusters (see Sect.~\ref{subsec:positions}).

\begin{figure}
    \centering
    % DO NOT ERASE LINE BELOW!
    % profilesbytype in Flagship2_paper.ipynb
    \includegraphics[width=0.9\columnwidth]{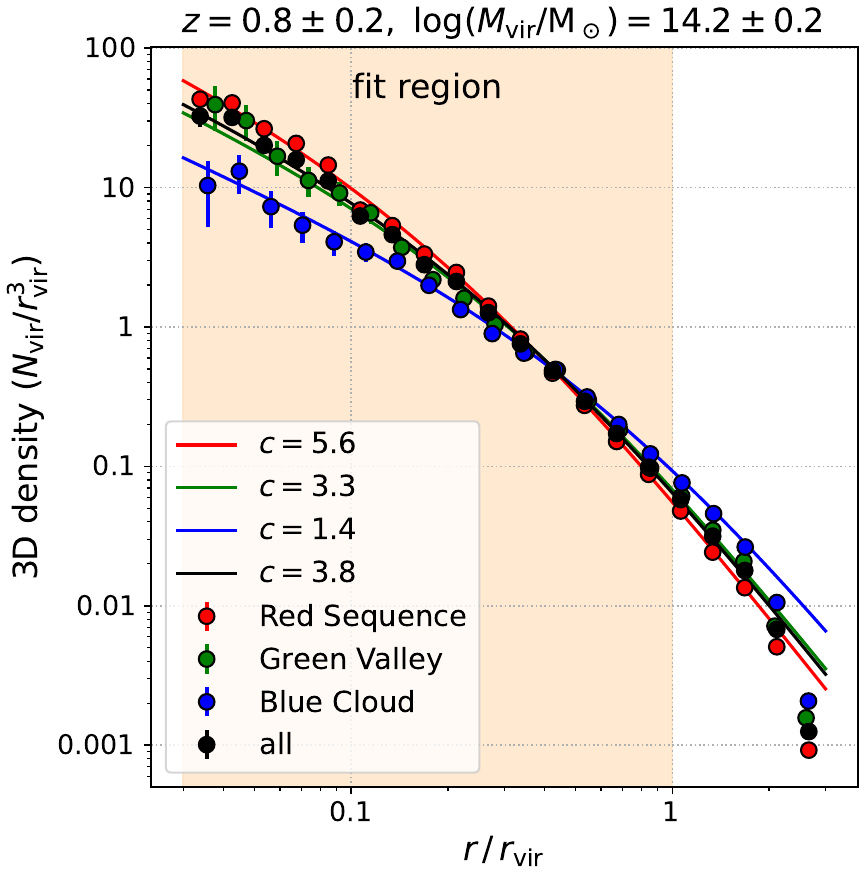}
    % {figs/profile.png}
    \caption{3D galaxy number density profiles of stacks of Flagship haloes, after removing all haloes lying closer than $3\,r_{\rm vir}$ from the edges of the region.
    The maximum likelihood concentration parameters  (virial radius over NFW scale radius), obtained by fitting an NFW model to the distribution of radial distances $r$ (restricted to the shaded region),  are shown in the legend.
    }
    \label{fig:densprofile}
\end{figure}

Figure~\ref{fig:densprofile} shows the 3D number density profile of halo galaxies in a slice of redshift and halo mass. We restricted the analysis to  199 haloes further than three virial radii from the edges of the studied 49\,deg${}^2$ region of Flagship,
containing a total of over $10^4$ member galaxies (with $\IE<24.5$ and $\HE < 24$) within the virial radius: 6545, 1332, and 2926, Red Sequence, Green Valley, and Blue Cloud, respectively.
One clearly sees the different radial distributions for the Red-Sequence, Green-Valley, and Blue-Cloud galaxies, with the latter having a concentration 4 times lower than for the Red-Sequence galaxies, as designed and shown in Table~\ref{tab:vparams}. The stacked radial distribution for all galaxies is also well fit by an NFW model even if it is the composite of three NFW models with very different scale radii. The best-fit concentration for all galaxies is intermediate between the extreme red and blue colour classes. One also sees that the NFW model is very well followed out to $\simeq 1.8\,r_{\rm vir}$, beyond which only a fraction of galaxies are halo members because of the triaxiality of the haloes (Sect.~\ref{subsec:positions}).

\begin{figure}
    \centering
    \includegraphics[width=0.8\hsize]{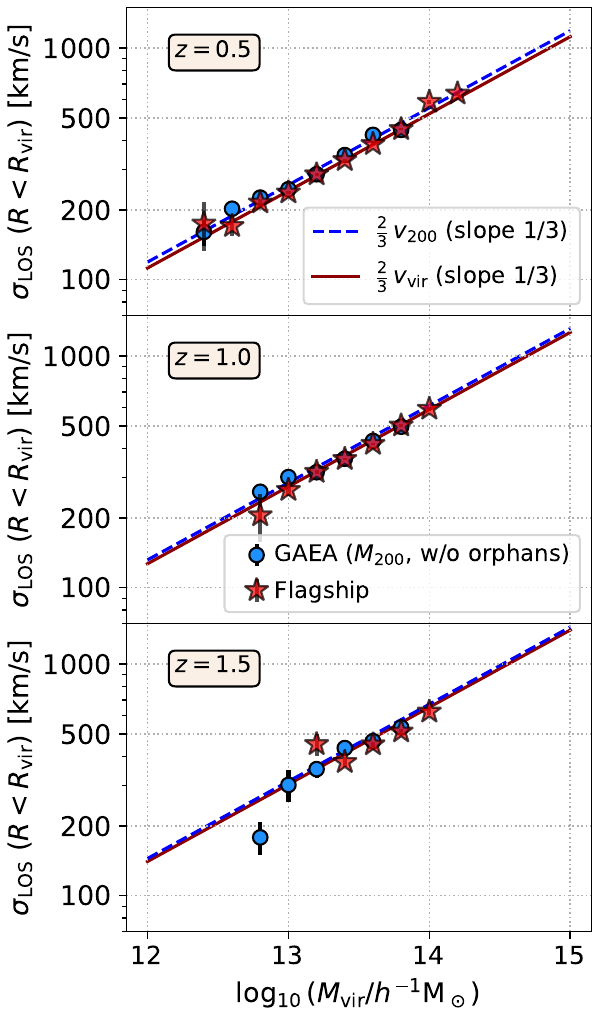}
    \caption{Halo line-of-sight velocity dispersion vs. halo mass, for lightcones from the \texttt{GAEA} semi-analytical model (without orphan galaxies, \emph{blue circles}, $7\,\rm deg^2$) and for Flagship (\emph{red stars}, here for the $49\,\rm deg^2$ region), at three representative redshifts.
    The \emph{lines} are simple non-theoretical representations. 
    Note that while Flagship uses virial quantities, \texttt{GAEA} uses $R_{200\rm c}$ for the virial radius and therefore the $x$~axis is $\logten[M_{200\rm c}/(h^{-1}\,M_{\odot})]$ for \texttt{GAEA}. Only haloes with at least 10 member galaxies with $\IE < 24.5$ and $\HE < 24$ within the virial cylinder are considered, with at least 10 such haloes per mass bin. The velocity dispersions are measured with the gapper method \citep{Wainer&Thissen76}, and the error bars assume Gaussian velocity distributions, i.e., $\epsilon(\sigma) = \sigma/\sqrt{2\,(N-1)}$, where $N$ is the number of galaxies per mass bin.}
    \label{fig:siglosvsM}
\end{figure}

Figure~\ref{fig:siglosvsM} compares the Flagship line-of-sight velocity dispersion as a function of halo mass with that measured in a $7\, {\rm deg}^2$ lightcone extracted from the \texttt{GAEA} semi-analytical model \citep{Hirschmann2016}. The agreement between the Flagship and the semi-analytical model is excellent. This indicates that the velocities of Flagship, which follow analytical prescriptions (Sect.~\ref{subsec:velocities}) are fairly realistic, because the galaxy velocities in \texttt{GAEA} are those of the subhaloes as extracted from the dark matter-only simulation on which the \texttt{GAEA} semi-analytical model was extracted.
This agreement is robust to the different virial definitions (the \texttt{ROCKSTAR} bound halo virial mass following the \citealt{Bryan:98} definition for Flagship vs. $M_{\rm 200c}$ for \texttt{GAEA}).  Note that the $\frac{2}{3}\,v_{\rm vir}$ and $\frac{2}{3}\,v_{\rm 200c}$ lines (which by definition both have slopes of 2/3)  only serve as a reference and are not theoretical expectations.

\begin{figure}
    \centering
    \includegraphics[width=0.8\hsize]{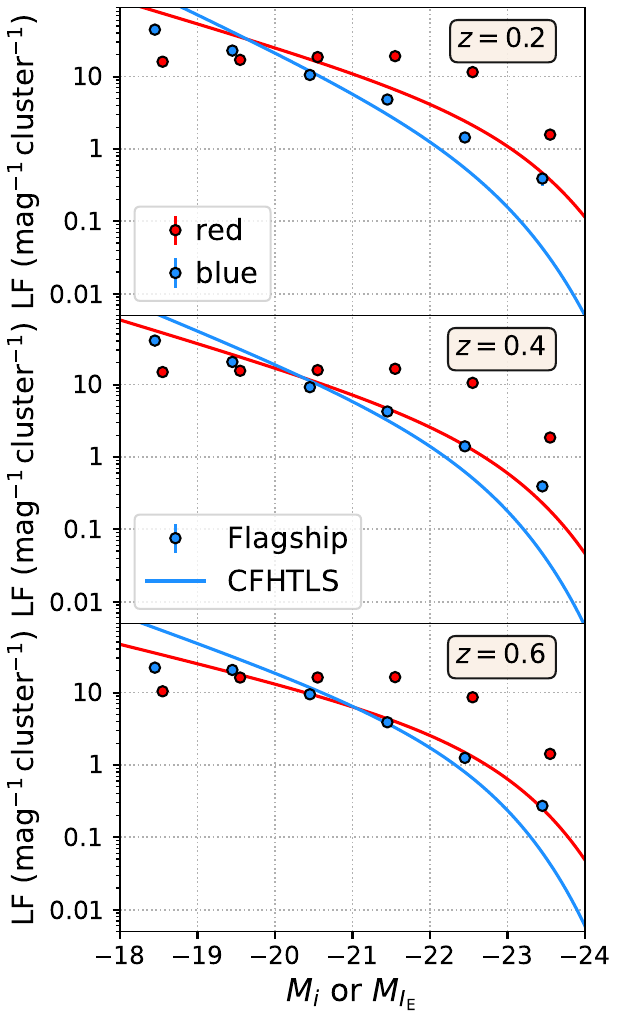}
    \caption{Cluster luminosity functions of red and blue galaxies for Flagship (points) and \cite{Schechter76} function fits by \cite{Sarron+18} to the CFHTLS absolute magnitudes (lines). For Flagship (\IE\ band), the red and blue galaxies are defined by {\tt color\_kind}, while for CFHTLS ($i$ band) they were defined according to $g-i$ vs. $i$ colour-magnitude diagrams, which indicate a colour bimodality at these three redshifts.
    The normalisation in Flagship is per cluster, while for CFHTLS it is per cylinder of  $1\;h^{-1}\,{\rm Mpc}$ 
    % FJC assumed it needs h scaling -> added it
    radius.
    The abscissas of the points are slightly offset for clarity.
    The discrepancies between the CFHTLS analysis and Flagship are discussed in the text.
    }
    \label{fig:ClusterLF}
\end{figure}

Figure~\ref{fig:ClusterLF} compares the $\IE$-band cluster luminosity functions (LFs) of Flagship in three redshift bins with the \cite{Schechter76} function fits 
%using a \cite{Schechter76} function parameterisation 
performed by \cite{Sarron+18} on clusters using $i$-band galaxy luminosities from the Canada France Hawaii Legacy Survey (CFHTLS).
The Flagship LF reproduces reasonably well the observed one, given the intrinsic differences in the LF modelling technique.
In particular, the numbers of faint galaxies (absolute magnitude between --19 and --21) in Flagship are similar to those predicted by the LF fits to the CFHTLS data.
However, one sees several differences.
First, Flagship predicts more luminous galaxies in clusters than observed in CFHTLS, regardless of colour. 
Second, the faint-end slopes for the red (respectively blue) galaxies is much (respectively somewhat) shallower than observed in CFHTLS.
Third, the break
of the red galaxy LF occurs at somewhat lower luminosities for Flagship than for CFHTLS, while it appears similar for the blue LFs of both Flagship and CFHTLS. 

\begin{figure}
\centering
\includegraphics[width=0.9\hsize]{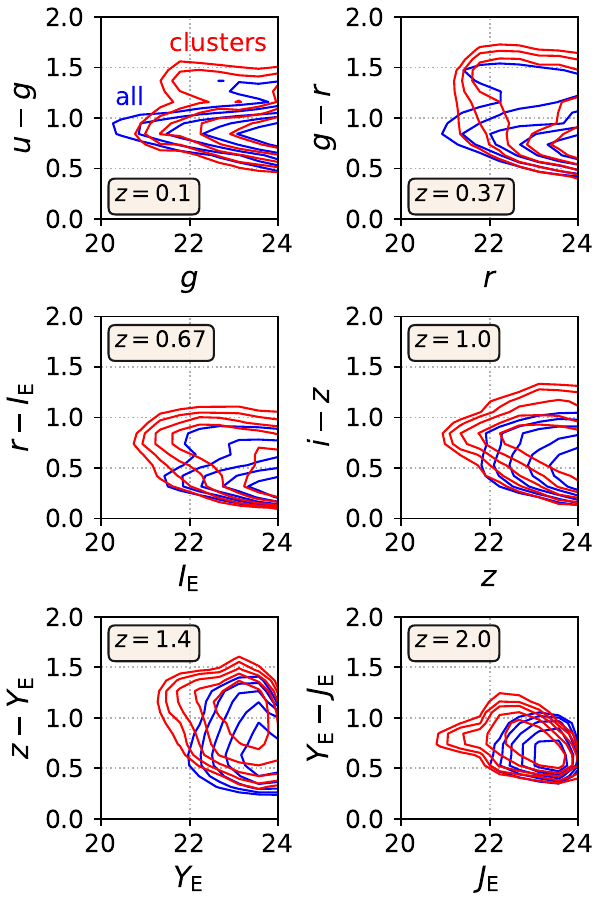}
    \caption{Contours of the colour-magnitude diagrams of Flagship  field galaxies (blue) and cluster ($M_{\rm vir}>10^{14}\;h^{-1}\,M_\odot$) galaxies (red), at selected redshifts. The figure uses proxies for the rest-frame $U-B$ and rest-frame $B$ magnitude. The galaxies are limited to $\IE<24.5$ and $\HE<24$. The redshift bins have widths of $\pm 0.05,0.06,0.07,0.1,0.2$ and 0.2. The contours are spaced  by 0.2 dex and are KDE-smoothed. The figure highlights how the Red Sequence is more prominent in clusters.}
    \label{fig:UmBref_CMDs}
\end{figure}

Figure~\ref{fig:UmBref_CMDs} shows the Flagship colour-magnitude relations for field and cluster galaxies, using proxies for rest-frame $U-B$ colours and rest-frame $B$ magnitudes at selected redshifts where the 4000\,\AA\ break lies in between the two considered wavebands for the colour, so as to emphasize the separation between Red-Sequence and Blue-Cloud galaxies.
One sees that Flagship reproduces nicely the bimodality of colours at $z=0.1$, although the Red Sequence is barely visible for the field galaxies, because the Blue-Cloud galaxies are so dominant. This trend is also visible at higher redshifts, showing that cluster galaxies in Flagship are more likely to be luminous and red at all redshifts. Again, the Red Sequence is present by construction in all galaxies (blue contours), but is overshadowed by the dominant Blue Cloud.

%-----------------------------------------------------------------

\section{Summary and conclusions \label{sec:summary}}

We have presented the Flagship galaxy mock, a large simulated catalogue especially designed for the \Euclid mission. \Euclid will use weak lensing and galaxy clustering as its main observables to infer cosmological parameters. The Flagship catalogue simulates these observables in a self-consistent manner and covers the volume and depth required for \Euclid. We also compute many other properties for each galaxy to increase the science cases that can be addressed with the catalogue.

We ran a 4 trillion particle dark matter $N$-body simulations in a box of $3600\,\si{\hMpc}$ length side at the Swiss National Supercomputing Centre. We generated a lightcone on the fly out to redshift $z=3$ in one octant of the sky containing 31 trillion particle positions and velocities. We have checked that the dark matter clustering properties behave as expected from theoretical predictions. 

We also produce an all-sky lensing map in \texttt{HEALPix} format with $N_{\rm side} = 8192$ resolution, corresponding to $0\farcm{43}$ per pixel. From these maps, we can compute the convergence, shear and displacement at any position in the simulation lightcone. The lensing statistics are in general agreement with expectations from theoretical predictions.

We produce a halo catalogue with the \texttt{ROCKSTAR} halo finder of 15.8 billion main haloes in one octant of the lightcone. In order to push the completeness of the catalogue to faint magnitudes as to be complete for the \Euclid magnitude selection limit at low redshift, we select haloes down to a threshold of 10 particles, before discarding unbound particles. After the unbinding process, some of our haloes end up fewer than 10 particles for the $M_{\rm bound}$ mass definition that we use. While many of these few-particle structures may not be individual virialised haloes, their statistical clustering strength will still be as expected, owing to the uniform clustering strength as a function of mass at this low mass range. We compute the halo mass function for the different mass estimates computed with \texttt{ROCKSTAR}. At low redshift, the HMF roughly agrees with the T08, D16 and C17 HMFs for the same halo mass definition. However, there is a small difference in the HMF slope at higher redshift $z\simeq1.5$. We adopt the virial bound mass definition as our fiducial mass for the catalogue. We nevertheless reassign the halo mass values to correct for completeness and discreteness at low halo masses. We compute the clustering of the haloes in the lightcone and compare them to theoretical expectations, finding reasonable agreement.

The next step in our mock production pipeline is to generate a galaxy mock catalogue from the halo catalogue using a combination of halo occupation distribution and abundance matching  
techniques. We complement these with observed correlations to generate recipes to assign other properties to each galaxy.

We start by generating the galaxy luminosities. First, we compute how many galaxies there are as a function of halo mass threshold. We use the halo mass function and the halo occupation distribution for the calculation of what we call the cumulative galaxy function (Eq.~\ref{eq:CGF}). For the HOD, we assume a simple parameterisation given in Eqs.~\eqref{eq:HOD} and~\eqref{eq:M1}. We compare the CGF to the cumulative luminosity function
to establish the relation between halo mass and luminosity.
In our luminosity assignment procedure, we apply scatter to the luminosities resulting from the AM relation. Therefore,
we compute the unscattered luminosity function, which is the one representing the luminosities that, after being scattered, produce the observed luminosity function (Eqs.~\ref{eq:LF_scatter} and~\ref{eq:LF_scatter_approx}). We generate the relation between halo mass and luminosity for values of these quantities that have the same abundance (Eq.~\ref{eq:AM_ML}). We assume that this redshift-dependent relation (Fig.~\ref{fig:AM_GFLF_allz}) is applicable to central galaxies and generate their luminosities with it. To assign the satellite galaxies' luminosities, we compute the global satellite luminosity function by subtracting the central LF from the LF for all galaxies (Fig.~\ref{fig:dLF_cen_sat_all_z05}). We assume that all haloes share the same functional form of the satellite LF within each halo. We fit the parameters of this universal LF relation to make the sum of the individual LF of all haloes coincide with the global satellite LF. We randomly draw the satellite luminosities from these individual halo LFs that depend on the luminosity of the central galaxy.   

We next assign a galaxy colour. We divide the galaxy population into three colour types: red, green and blue. 
We fit the local SDSS colour-magnitude diagram with three Gaussian distributions, one for each colour type. We define functions determining the fraction of red and green satellites as a function of luminosity. The fraction of blue satellites and red, green and blue centrals is then determined by the CMD and the HOD. We assign a $(g01-r01)_{\rm HOD}$ colour by randomly sampling the distributions.

We place central galaxies at the central position of each halo. Satellites are distributed with the same triaxial NFW profile as the dark matter. We implement colour segregation, changing the concentration index of each population colour type. Red galaxies are assigned NFW profiles with the same concentration index as the DM. Green and blue galaxies are distributed using concentration indices that are 1/2 and 1/4 of the concentration of the DM.

Central galaxies are assumed to be at rest at the centre of the halo. The velocities of the satellite galaxies are assigned by solving the Jeans equation of local dynamical equilibrium with anisotropy parameters coming from observations of local clusters and differentiated according to the colour type.

We choose the spectral library used by~\cite{Ilbert:09} as our basis for spectral energy distribution assignment.
We construct a sample of 136 templates using different extinction laws and values from the original 31 templates in the library. We rank-order them according to their $(g01-r01)_{\rm HOD}$ colour. We compute the $g-r$ colour distributions of the SDSS at low redshift and of the COSMOS catalogue at $z \geq 0.5$. For each galaxy, we choose the template that has the same percentile in the colour distribution in SDSS or COSMOS catalogues as the $(g01-r01)_{\rm HOD}$ colour already assigned. We assign a probability to the six closest templates in the rank-ordered list to this one based on their colour difference. We also assign a probability depending on the extinction value of these six templates compared to the distribution of extinction values in the COSMOS galaxies. We finally choose a template randomly sampling from the probability resulting from the product of these two colour and extinction probabilities. We do this template assignment at two redshifts close to the galaxy redshift. The final SED is an interpolation between these two templates with weights based on their redshift distance.

We assign the values of the convergence and shear at the galaxy positions from the values of the \texttt{HEALPix} maps of these quantities. We also compute a displaced position of the galaxy based on the lensing displacement field.

We assign the shapes and sizes of the galaxies using measured distributions in observed HST fields. We use the GOODS South field and the CANDELS observations as calibrators. We model galaxies either as one component (bulge) or two components (bulge and disk). The bulge component is simulated as a S\'ersic profile. We use a simple cut in the CMD to divide these two options. For all the galaxies, we compute the scale height and the fraction of light coming from the bulge component. For each of the two galaxy components (or just for the bulge), we compute the half-light radius, the S\'ersic index, the inclination angle, the ellipticity and the axis ratio, mimicking the distributions of our calibration samples. 

We compute the SFR of the galaxies from the rest-frame UV luminosity of the SEDs. We compute the stellar mass from the galaxy luminosity and the stellar mass-to-luminosity ratio. We assign a metallicity from the stellar mass, the SFR and the redshift of the galaxy.

We assign the flux of the H$\alpha$ line based on the value of the SFR. We recalibrate the fluxes of the H$\alpha$ as to agree with the \cite{Pozzetti:16} models. The other Balmer lines are computed assuming the hydrogen lines case B recombination ratios. The other usually most prominent lines are assigned based on observed correlations. 

Overall, we compute 199 quantities for each galaxy. These properties include the galaxy positions and velocities, the galaxy fluxes in several bands with and without extinction, 
the lensing properties at the galaxy positions, the galaxy shapes and sizes, the SED, the SFR, the stellar mass and the metallicity, the emission lines fluxes calibrated to both model 1 and model 3 of~\cite{Pozzetti:16} with and without extinction, a photometric redshift estimate, the intrinsic alignment properties and some of the halo properties they belong to.

We validate the catalogue by comparing it to observations. We compute the catalogue number counts. In the optical bands, the Flagship number counts coincide within the envelope spanned by observations. In the near-infrared, however, our number counts are slightly larger than observations. We check the colour-colour and colour redshift distributions against the COSMOS2020 catalogue. Overall, there is good agreement between the two, with the exception of a small deficit of blue galaxies in colours including a near-infrared band. We also compute the stellar mass function as a function of redshift. We find relatively good agreement with the determination from the GAMA survey and the COSMOS field. We compare the emission line fluxes in various diagnostic plots and find them to be consistent with observational data. 

We also validate the two main \Euclid cosmology observational probes. We perform a basic validation of the clustering properties of a few samples using common selection criteria, cutting our sample in halo mass, absolute and apparent magnitude, and emission line flux. We compare our results to theoretical expectations, considering both the power spectrum and multipoles of the two-point correlation function in configuration space. We compute the expected linear bias for an emission line flux-selected sample similar to the one expected in Euclid. We also check that we can recover the input cosmological parameter from this clustering analysis. Overall, all the computed clustering statistics behave as we would expect from theoretical expectations, validating the use of the catalogue for \Euclid analysis. 

We also validate the lensing properties of the catalogue. We compute the two-point shear correlation function and the average tangential shear and compare them to theoretical predictions. Both show good agreement. We also check the lensing magnification properties of the sample against models and found a good agreement. Finally, we also compare the higher-order lensing statistics to models corroborating that the catalogue shows the correct higher-order behaviour.

We check how some galaxy cluster properties are reproduced in the Flagship galaxy catalogue. We show the number of galaxies in clusters as a function of halo mass and redshift for magnitude and emission line flux-limited samples. We check the radial profile of galaxies within clusters as a function of their colour type, showing an NFW distribution for a stack of clusters. We present the relation between halo mass and the line of sight velocity dispersion at different redshifts and compare them to the \texttt{GAEA} semi-analytical model, finding good agreement. We show the LF of galaxies inside clusters for red and blue galaxies at different redshifts and compare them to data from the CFHTLS. We also show the colour-magnitude diagram for cluster galaxies and compare it to that of the overall population at several redshifts. The cluster red sequence is apparent with respect to the rest of the galaxy population.

\subsection{Future developments\label{subsec:future_developments}}

Modelling the properties of a massive and complex galaxy survey like \Euclid necessarily entails making certain assumptions and approximations with respect to the true galaxy samples. Below we discuss some of the possible shortcomings in the approach we have taken in the modelling, how they can impact the different observables and, in some cases, how these could be improved in future releases of the Flagship galaxy mock.

In this paper we have implemented a simple HOD approach to assign galaxies to dark-matter haloes, in which the number of satellite galaxies in each halo depend only on the halo mass. Several studies have shown that other halo properties are needed to accurately simulate the galaxy population within this type of framework. This is commonly referred to as halo assembly bias~\citep[e.g.,][]{Contreras:23}. While our HOD prescription is simple, the way we assign galaxy properties is more complex than the approach of typical HOD and AM models, and in a certain sense our method resembles models that introduce assembly bias as our galaxy properties assignment depends on other parameters. For example, the way we distribute galaxies within haloes depends on galaxy colour which can be related to the assembly history~\citep[e.g.,][]{Hearin:13b}. Moreover, normally observations of a particular galaxy sample, selected with a flux limit or colour criteria as for example luminous red galaxies, are interpreted in the HOD framework fitting parameters of the occupancy distribution. In the Flagship catalogue, the HOD is assigned up front and corresponds to the whole population. Any galaxy subsample selected with colour and/or flux cuts will have a different HOD than the original assignment. 
Similarly, we have not included the effects of gas physics, which are expected to significantly alter the distribution of dark matter and thus of galaxies within galaxy cluster scales~\citep[e.g.,][]{Schneider15, Gebhardt24}. Again, our galaxy properties assignment depends on luminosity and colour in order to fit observations, and therefore it must incorporate the effect that baryons have on the galaxy distributions when compared to dark-matter only simulations. Nevertheless, we are exploring HOD prescriptions that depend on other parameters for implementation in future versions of the catalogue.

Another improvement in the catalogue that we are considering is to increase the resolution of the \texttt{HEALPix} lensing maps. Currently, we use a pixel scale of \ang{;0.43;}, i.e., \texttt{HEALPix} $N_{\rm side}=8192$, without interpolation between pixels. This resolution is enough to measure adequately the lensing correlation functions down to $1$ arcmin (e.g., for $\xi_+$ and the tangential shear), which coincide with the smallest scales probed by \Euclid. Therefore, the resolution of the lensing properties is good enough for the main purpose of the simulation. Moreover, in the MICE-Grand Challenge simulation~\citep{Fosalba:15b}, which used the same assignment scheme, we tested that interpolating linearly between the \texttt{HEALPix} pixels only improved marginally the effective resolution on sub-arcmin scales, and at a high computational cost.

We are also exploring how to improve our SED assignment. Currently, we use linear interpolation between SEDs selected from the COSMOS SED template set (see Sect.~\ref{subsec:galaxySED}). We would like to use a wider set of templates that can cover better the range of observed SEDs as a function of redshift. 
Another shortcoming of our simulation is the lack of AGN. We are working on methods to include them in future versions~\citep[e.g.,][]{Allevato:21}. 

The calibration of our recipes to assign galaxy properties are normally restricted to low redshift observed samples, $z\lesssim1$, which we extrapolate to higher redshifts. We are working on extending the calibration to higher redshifts using more observational data.

The assignment of galaxy positions and velocities is simplistic, relying on analytic profiles without substructure. This is mainly driven by our push to simulate faint galaxies while covering a large volume. While for the relevant scales to be explored with the \Euclid main cosmological probes this level of detail is enough, we are exploring  keeping more information from the original $N$-body simulation to be able to produce more realistic galaxy profiles within haloes and therefore support more extensive cluster science analyses.

We have assigned shapes and sizes to galaxies based on distributions measured with HST. The distributions of each parameter are well reproduced by our catalogue. However, we have not enforced the correlations between different morphological parameters in part due to the usage of two different samples for our calibration. We are working on a multi-parameter calibration for future implementations, that will be based on the \Euclid data themselves which are going to provide a large and homogeneously measured morphological sample.  

\subsection{Data availability\label{subsec:data_availability}}

Given the comprehensive set of validation tests that we have performed on the Flagship galaxy mock catalogue, we believe it is a valuable resource to perform a wide suite of astrophysical analyses beyond its original design goals of supporting the \Euclid weak lensing and galaxy clustering analyses. We make the catalogue publicly available so that everybody can potentially benefit from its usage. We distribute this catalogue from the CosmoHub platform.

%------------------------------

\begin{acknowledgements}
FJC and PF acknowledge support form the Spanish Ministerio de Ciencia, Innovaci\'on y Universidades, projects PID2019-11317GB, PID2022-141079NB, PID2022-138896NB; the European Research Executive Agency HORIZON-MSCA-2021-SE-01 Research and Innovation programme under the Marie Skłodowska-Curie grant agreement number 101086388 (LACEGAL) and the programme Unidad de Excelencia Mar\'{\i}a de Maeztu, project CEX2020-001058-M. We acknowledge the support of the PRACE project (Call 17) "Simulating the Euclid Universe" for the computer time on the Piz Daint supercomputer at CSCS, Lugano, Switzerland, which made possible the world-leading Flagship 2 N-body simulation. The Port d'Informaci\'{o} Cient\'{i}fica (PIC) is maintained through a collaboration between CIEMAT and IFAE. This work has been supported by MCIN/AEI grant EQC2021-007479-P and  European Union NextGenerationEU grant PRTR-C17.I1 and by Generalitat de Catalunya. This work has been partially funded by Premiale MITIC 2015.
%\AckERO
  \AckEC
\end{acknowledgements}

%
% Here comes the reference list, generated via bibtex from
% the bibfile AandA.bib
%

\bibliography{flagship_bibliography,EROplus}

%
% Now you can add appendices.
% In this example, the appendices are in one column mode.
% If that is not requires, comment out \onecolumn
%

\begin{appendix}

\section{Halo mass function with different mass estimates \label{app:hmf_defs}}

The Flagship halo catalogue was produced with the \texttt{ROCKSTAR} code adapted to the lightcone (see Sect.~\ref{sec:halocat}). \texttt{ROCKSTAR} computes different estimates of the halo mass. The mass values that we calculated are
the mass of the particles linked together with a friends-of-friends algorithm of linking length $b=0.2$, $M_{\rm fof}$; the mass contained within the virial radius, $M_{\rm vir}$; the sum of the mass of the bound particles within the virial radius, $M_{\rm bound}$; the mass of the particles within an overdensity of 200 relative to the background density,  $M_{\rm 200b}$; and the mass of the particles within an overdensity of 200 relative to the critical density,  $M_{\rm 200c}$.

In this Appendix, we provide a comparison of the cumulative halo mass function resulting from the different mass estimates and also to the T08 $M_{\rm vir}$ halo mass function as a function of redshift.

\begin{figure}
\center
\includegraphics[width=\columnwidth]{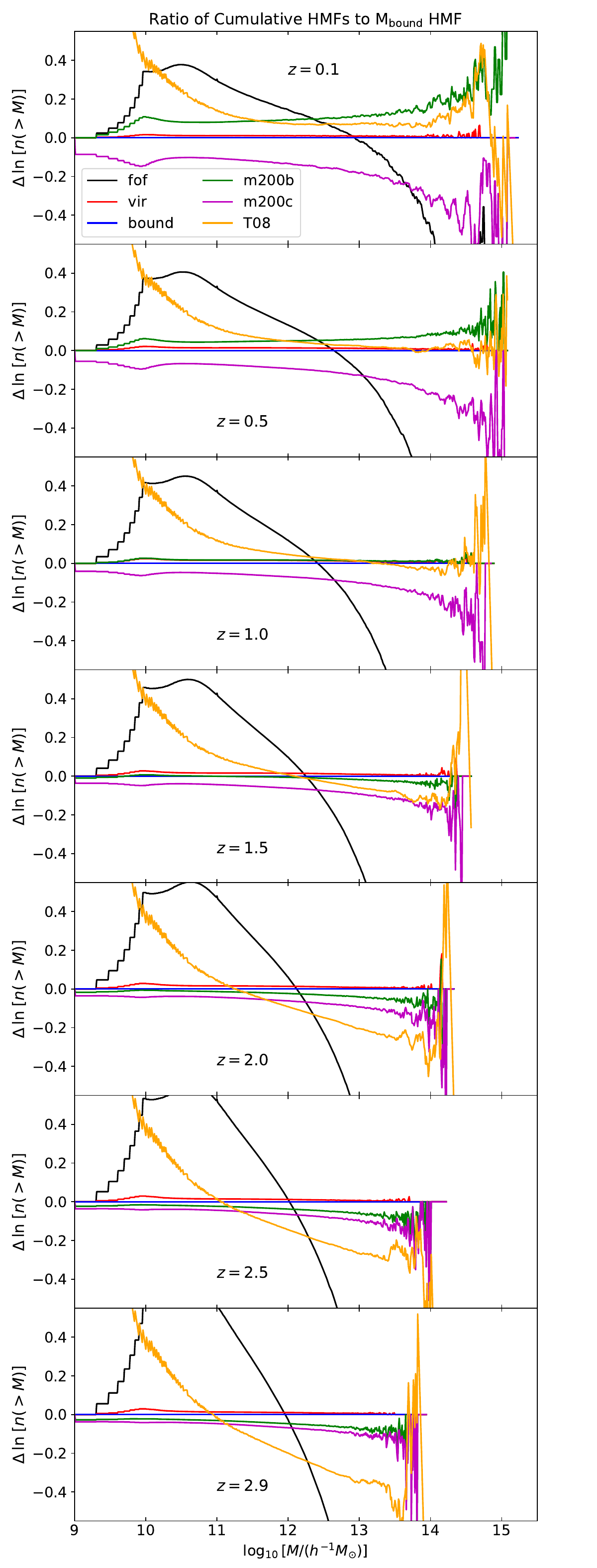}
\caption{\label{fig:ratio_cumHMF_to_bound_allz_ln} Natural logarithm of the ratio of the cumulative halo mass function for all mass estimates computed with \texttt{ROCKSTAR} and also the T08 $M_{\rm vir}$ HMF to the  $M_{\rm bound}$ cumulative HMF as a function of redshift. 
%\gam{Use larger legends and somewhat larger axis labels.}
}
\end{figure}

Figure~\ref{fig:ratio_cumHMF_to_bound_allz_ln} shows the ratio of the cumulative halo mass function for all the mass estimates to the cumulative $M_{\rm bound}$ HMF. It also contains the ratio of the T08 $M_{\rm bound}$ HMF to the $M_{\rm bound}$ HMF for comparison. The $M_{\rm vir}$ and $M_{\rm bound}$ HMFs coincide except for small differences at the low and high mass end due to the unbound particle rejection process.  The $M_{\rm 200b}$ HMF is higher than the $M_{\rm bound}$ HMF at low redshift $z<1$, similar at $1.0<z<1.5$, and lower at $z>1.5$. The $M_{\rm 200c}$ HMF is always lower than the  $M_{\rm bound}$ HMF. The $M_{\rm 200b}$ and $M_{\rm 200c}$ HMFs vary as expected due to the different redshift dependence of the density threshold used to define the mass. The ratio of the background density to the critical density gets lower at low redshift when the effect of $\Omega_{\Lambda}$ starts to be important, therefore increasing the ratio of $M_{\rm 200b}$ to the $M_{\rm 200c}$ values. The $M_{\rm fof}$ HMF is quite different from the $M_{\rm bound}$ HMF. This is expected as the friends-of-friends halo-finding technique and its mass definition is different from the rest. The T08 $M_{\rm vir}$ HMF shows the same trends as described in Sect.~\ref{sec:halo_mass_function}.

\section{The \texttt{SciPIC} algorithm \label{app:SciPIC}} 

\texttt{SciPIC}, first described in~\cite{Carretero:17} and named after Scientific Pipeline at PIC, is a suite of algorithms integrated into a powerful pipeline dedicated to the generation of massive synthetic galaxy catalogues based on halo catalogues coming from $N$-body dark matter cosmological simulations. \texttt{SciPIC} is the algorithm used to generate the Euclid Flagship galaxies. Originally, most of the algorithms were developed to the production of the MICE galaxy catalogues. In the case of MICE, the code was written in C and executed on different desktops. In particular, a significant amount of time was dedicated to input-output tasks, given that several steps were executed sequentially. However, with the Flagship parent halo catalogue containing a much larger volume of data, the methodology employed for running the MICE catalogue is unfeasible. The code has been refactored, optimised and ported to Python. Additionally, it runs on top of Apache Spark, an engine for scaleable computing. \texttt{SciPIC} is executed in the PIC Big Data platform, based on Hadoop, which comprises 20 nodes for a total processing power of 960 CPUs. The code runs efficiently with a fast interface with the CosmoHub portal~\citep{Tallada:20}, where the input halo catalogue is ingested, and the output galaxy catalogues are stored and distributed. The current implementation is able to generate a 15TB catalogue of 5B galaxies in 3 hours. The fact that it takes so little time to generate a mock allows for multiple iterations of progressive refinement. 

\section{Details of the measurement of radial velocity dispersions of galaxies in their haloes}
\label{sec:veldetails}

\begin{figure}
    \centering
    \includegraphics[width=0.9\hsize]{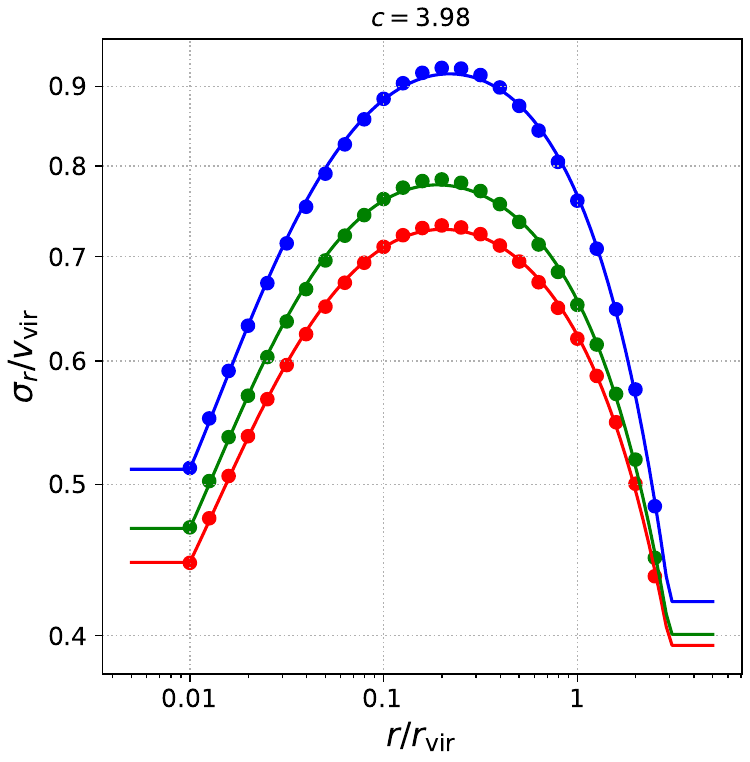}
    \caption{Adopted model for the radial velocity dispersion profiles in virial units, for the case of dark matter concentration of 4, for the three galaxy classes. The symbols indicate the exact values, while the curves indicate the 2D-fifth-order polynomial approximations used in Flagship. One clearly sees the adopted constant values beyond the fit limits of $r/r_{\rm vir}$.}
    \label{fig:sigr}
\end{figure}

We provide here details on the mock velocities.
The Jeans equation~(\ref{eq:jeans}) is solved for the dynamical pressure, $\rho \,\sigma_r^2$, which for the \cite{Tiret+07} velocity anisotropy profile of Eq.~(\ref{eq:betaofr}) is \citep{Mamon+13}
\begin{equation}
\rho(r)\,\sigma_r^2(r) = \frac{G}{ (r+r_\beta)^{2\beta_\infty}} \int_r^{r_{\rm max}}
\frac{{\left(s+r_\beta\right)^{2\beta_\infty}}}{ s^2}\,\rho(s)\,M(s)\,{\rm d}s \ ,
\label{eq:rhosigr2}
\end{equation}
trivially yielding the radial velocity dispersion $\sigma_r$ for our chosen NFW model for $\rho(r)$. For each of the three galaxy colour classes, we pre-computed $\sigma_r$ using Eq.~\eqref{eq:rhosigr2} on a 25$\times$16 grid of $[\logten(r/r_{\rm vir}),\logten(c_{\rm dark})]$, with $-2\leq \logten(r/r_{\rm vir}) \leq 0.4$ and $-1 \leq \logten(c_{\rm dark}) \leq 1.5$, in steps of 0.1 dex. We then fit a two-dimensional 5th-order polynomial to allow Flagship to rapidly determine the radial velocity dispersion at given radial distances.
The rms errors on $\logten \sigma_r$ are 0.0019, 0.0022, and 0.0026 respectively for red, green, and blue galaxies, i.e., better than 0.6\% rms precision on $\sigma_r$ for all colour classes.

\end{appendix}

\label{LastPage}
\end{document}